\documentclass[12pt,preprint]{aastex}

\slugcomment{Revised: 31 May 2006}

\shorttitle{Search for compact extragalactic radio sources}
\shortauthors{Xu et al.}

\begin{document}

\title{Search for compact extragalactic radio sources near massive star forming regions}

\author{Y. Xu\altaffilmark{1,2}, M. J. Reid\altaffilmark{3}, K. M. Menten\altaffilmark{2},
\and X. W. Zheng\altaffilmark{4}} \altaffiltext{1}{Shanghai
Astronomical Observatory Chinese Academy of Sciences, Shanghai
20030, China} \altaffiltext{2}{Max-Planck-Institut f$\rm\ddot{u}$r
Radioastronomie, Auf dem H$\rm\ddot{u}$gel 69, 53121 Bonn,
Germany} \altaffiltext{3}{Center for Astrophysics, 60 Garden
Street, Cambridge, MA 02138, USA} \altaffiltext{4}{Department of
Astronomy, Nanjing University Nanjing 210093, China}

\begin{abstract}
We have used the Very Large Array to search for compact
milliarcsecond-size radio sources near methanol masers in
high-mass star-forming regions. Such sources are required for Very
Long Baseline Interferometry phase-referencing observations.  We
conducted pointed observations of 234 compact sources found in the
NVSS survey and find 92 sources with unresolved components and
synchrotron spectral indexes. These sources are likely the cores
of AGNs and, thus, good candidates for astrometric calibrators.
\end{abstract}


\section{Introduction}

Phase-referenced Very Long Baseline Interferometry (VLBI)
observations can measure accurately the position of a target
source relative to a reference source.  This, for example, permits
one to measure the trigonometric parallax and proper motions of
sources in our galaxy relative to extragalactic sources. We are
now carrying out a large program to do this for methanol masers in
regions of high mass star formation throughout the Milky Way.
Table 1 contains 38 sources showing maser emission in 12.2 GHz
$2_0-3_{-1}E$ line of methanol that we ultimately
intend to observe.

VLBI phase-referenced observations involve a phase calibrator and
a target source.  However, the distribution of known calibrators
is not sufficiently dense enough to find one within
$\approx2^{\circ}$ of any given maser source, which is necessary
for high precision astrometry: Systematic errors are usually the
limiting factor in phase-referenced position measurements, and
these errors scale with the angular separation between the
sources. Therefore, finding calibrators as close as possible to
the maser target sources is mandatory.

For this purpose, we have conducted a survey of NRAO VLA Sky
Survey (NVSS; Condon et al. 1998) sources near our methanol maser
targets. We chose sources whose 1.4 GHz flux densities are greater
than 20 mJy and that are unresolved ($<20''$) at the NVSS
resolution. The NVSS survey typically yields about 15 such compact
sources within $1^{\circ}$ of any target position. Since our maser
sources are in the Galactic plane many of these NVSS sources are
compact HII regions, planetary nebulae (PNe), and, perhaps, compact
supernova remnants (SNRs), and, thus, unsuitable for VLBI
observations. However, some of the unresolved sources are likely
extragalactic synchrotron sources.

In addition to the NVSS sources, we augmented our candidate list
with sources from the literature (Gregory \& Condon 1991; Ma et
al. 1998) and included known compact sources as a check on our
procedures. In total, we observed 234 compact sources within
1$^{\circ}$ of the maser sources.

\section{Observations and Data reductions}

On 2004 April 2 we made snap-shot observations of our 234 sources
using the NRAO\footnote{The National Radio Astronomy Observatory
(NRAO) is operated by Associated Universities Inc., under a
collaborative agreement with the U.S. National Science Foundation.
Very Large Array (VLA)} VLA in its C-configuration at C-band
(4.885 GHz) and U-band (14.939 GHz). Bandwidths were 50 MHz,
observations were in dual circular polarization, and on-source
times were between 30 and 70 seconds. The absolute flux density
scale was determined from observations of 3C286 and 3C48.
Occasionally we missed observations owing to electronic problems
and, thus, for some sources no data could be obtained (see Table
4). The data were reduced with the NRAO Astronomical Image
Processing System (AIPS) using standard procedures. We achieved
rms noise levels of about 0.3 mJy at 4.885 GHz and 0.8 mJy at
14.939 GHz, respectively.

In order to obtain better positional accuracy and check on
compactness, we re-observed 53 sources that had flux densities
higher than 5 mJy in both, the C- and U- bands and were unresolved
in C-band.  These observations were done as low-priority
observations in small gaps in the VLA schedule in the A- or
B-array at U-band on 2004 September 26, 27, 28, October 29, and
2005 May 1, 10, and 26, respectively. We chose U-band as it is the
VLA frequency that is closest to that of the 12.2 GHz masers. The
observations of each object involved 6 short (1 min duration) scans, spread
over a range in time for better $uv$-coverage. To minimize the phase
errors arising from temporal changes in the atmospheric phase
paths, we observed a standard VLA calibrator before and after each
target source. Typical map rms noise levels were about 0.2 mJy.
This allowed us determine positions accurate to better than 0.1
arcsec. It is important to reach that precision for VLBI
astrometry in order to minimize second-order positional errors and
improve image quality.

\section{Results}

We detected about 92 of the 234 candidate sources with flux
densities of 5 mJy at C-band with flattish or falling spectral
indexes. These are likely compact extragalactic sources associated
with AGN activity.

In Table 2 we list weak sources $<5$ mJy at C-band observed in the
C-configuration. These are too weak for current astrometric
observations.

In Table 3 we list the sources that were clearly resolved. These
sources are probably mostly HII regions, PNe, and SNRs. While some
ultra- and hyper-compact HII regions may not be resolved by the
VLA, they should have flat or rising spectra at centimeter
wavelengths (Wood \& Churchwell 1989; Kurtz et al. 1994), that can
be easily recognized in our dual-frequency observations. Although
SNRs have falling spectra, they should be resolved due to their
larger size (Green 2004). The sources detected a C-band, but not
at U-band, might be used as calibrators at lower frequency, e.g.,
for VLBI observations of 6.7 GHz CH$_{3}$OH masers, and/or 1.6 and 6.0 GHz
OH masers.

Table 4 lists the parameters of the 92 sources that are unresolved
at C-band, have flat or falling spectral indexes, and are
potentially strong enough to be calibration sources for VLBI
phase-referenced observations. Their images are shown in Figure 1
which is available in the electronic version of this paper. Except
for a few sources, consisting of two components, most of the
sources appear dominated by a single compact component. Most of
them contain over 85\% of their flux density in an unresolved
component at C- and/or U-band and are likely compact
extragalactic sources. Even though we selected sources with our
C-configuration observations that were nearly unresolved and had
$>5$~mJy at C-band, when observed with the more extended A- or
B-configurations some were not detected. Since these might contain
a compact component that could be used as a phase-reference
calibrator at lower frequencies, we also retain them in Table 4.

Already, source 023016+620937 in Table 4 has proven to be a good
calibrator for Very Long Baseline Array observations (VLBA; Xu et
al. 2006). This source probably has a position uncertainty of
about a few milli arcseconds (mas). Using this and two other
sources we have measured a parallax (to the W3OH masers) with an
accuracy of 0.007 mas. Three of our sources are incluced in the
third extension to the Very Long Baseline Array (VLBI) Calibrator
Survey (VCS3; Petrov et al. 2005) listed there with (sub)mas
accuracies. By comparing the results from different observations
listed in Table 5 we find that our C-array positions deviate from
the VCS3 positions by 0.25, 0.91, and 0.14 arcsec and our A-array
positions by 51 and 11 mas. We, thus,  estimate that the position
uncertainty is about $0.5''$ in snap-shot C-configuration
observations and $0.05''$ -- $0.15''$ in VLA A- and
B-configuration observations, respectively, for high declination
sources. Positional accuracy is likely to degrade considerably at
low declinations and for weak sources (less than 10 mJy).

We calculated the two-point spectral index from 5 to 15 GHz from
C-configuration observations. Most of the sources in Table 4 have
a steep-spectrum with an average spectral index of $\alpha = 0.96
\ (S_{\nu} \sim \nu^{-\alpha}$). One must keep in mind that since
these were often measured with observations within one VLA
configuration, the 5 GHz beam size was about a factor of 10 larger
in solid angle than the 15 GHz beam size.  Thus, if a source is
unresolved at 5 GHz, but partially resolved at 15 GHz, the
spectral index may appear overly steep. Ulvestad et al. (1981)
found a median spectral index of $\alpha = 0.09$ between 1.4 and 5
GHz for unresolved sources in a flux-limited sample. Zhang \& Fan
(2003) classified this sample by properties of galaxies. They
found an average spectral index of 0.22 for blazers, 0.28 for
QSOs, and 1.01 for galaxies, respectively.  As a comparison, we
obtained an average spectral index of $\alpha = 0.83$ between 1.4
and 5 GHz. Of course, since both observations spanned many years,
time variability could, in principle, affect spectral index
results. However general speaking, most of sources are likely to
be radio galaxies.

\section{Conclusions}
A VLA search has resulted in the identification of 92
compact sources most likely of extragalactic nature. These sources
are likely to be suitable calibrators for phase-referencing VLBI
observations of 12.2 GHz CH$_3$OH or/and lower frequency masers.

\begin{deluxetable}{lllll}
\tablecaption{12.2 GHz Methanol Maser Sources Considered} \startdata \hline
\object{G8.68$-$0.37}  & \object{G9.62+0.20}   & \object{G10.47+0.03} & \object{G11.90$-$0.14} &\object{G12.03$-$0.03}  \\
\object{G12.68$-$0.18} & \object{G12.89+0.49}  & \object{G12.91$-$0.26} & \object{G15.03$-$0.68} & \object{G23.01$-$0.41} \\
\object{G23.44$-$0.18} & \object{G24.33+0.14}  & \object{G27.36$-$0.16} &\object{G28.15+0.00}  & \object{G29.86$-$0.05} \\
\object{G29.95$-$0.02} & \object{G30.20$-$0.17}  & \object{G30.78+0.23} & \object{G30.79+0.20} &\object{G31.28+0.06}  \\
\object{G32.74$-$0.08} & \object{G35.20$-$0.74}  & \object{G35.20$-$1.74} & \object{G37.43+1.50} & \object{G49.49$-$0.37} \\
\object{G59.78+0.06} & \object{G109.87+2.11} &\object{G111.54+0.78} & \object{G133.95+1.07}&\object{G188.95+0.89} \\
\object{G232.62+0.99}& \object{G345.01+1.79}\tablenotemark{a}&\object{G348.55$-$0.98} &\object{G348.70$-$1.04} &\object{G351.42+0.64} \\
\object{G353.41$-$0.36}& \object{G354.61+0.47} & \object{G359.61$-$0.24}\tablenotemark{a}\\
\enddata
\tablenotetext{a}{No detected compact extragalactic radio
  sources within $1^{\circ}$ of this source.}\\
\end{deluxetable}

\begin{deluxetable}{lllllll}
\tablecaption{Sources  with C-band Flux Densities $<$ 5 mJy  Observed in
C-Configuration} \startdata \hline
   172149$-$352620 &
   180349$-$203653 &
   181213$-$171318 &
   181345$-$190435 &
   181540$-$175738 & \\
   183408$-$084922 &
   184114$-$045503 &
   184601$-$024601 &
   184612$-$015715 &
   184640$-$021544 & \\
   184656$-$020132 &
   185211+035715 &
   185316+010040 &
   185845+005600 &
   192232+141303 & \\

\enddata
\end{deluxetable}

\begin{deluxetable}{llllll}
\tablecaption{Sources Resolved in C- or U-Band Observations} \startdata \hline

   022649+621220 &
   022703+615232$^{a}$ &
   060915+223503 &
   061032+221514 &
   073139$-$161756 & \\
   073223$-$173012 &
   073257$-$164058 &
   165451$-$393420 &
   165640$-$401334 &
   165646$-$401437$^{a}$ & \\
   171742$-$354155 &
   172051$-$354603$^{a}$ &
   172052$-$345448 &
   172150$-$364115 &
   172201$-$394441 & \\
   172215$-$382857 &
   172804$-$325351 &
   172818$-$350410 &
   172827$-$350730 &
   172919$-$343351 & \\
   172928$-$343630 &
   172937$-$351344$^{a}$ &
   173329$-$324758$^{a}$ &
   174451$-$294039 &
   174828$-$293912 & \\
   180416$-$202042 &
   180506$-$193602 &
   180513$-$195034$^{b}$ &
   180618$-$213731$^{a}$ &
   180816$-$171451$^{a}$ & \\
   180826$-$221653$^{b}$ &
   180827$-$173332 &
   180846$-$192243 &
   180849$-$172219 &
   180859$-$200352 & \\
   180939$-$192116$^{a}$ &
   181018$-$204544 &
   181028$-$195548$^{a}$ &
   181031$-$195629$^{a}$ &
   181050$-$204110 & \\
   181109$-$164738 &
   181132$-$193042$^{a}$ &
   181210$-$184558 &
   181345$-$190435$^{b}$ &
   181401$-$185324$^{a}$ & \\
   181449$-$173243$^{a}$ &
   181644$-$162701 &
   181645$-$182524 &
   181646$-$162035 &
   181651$-$184129 & \\
   183241$-$090446 &
   183321$-$073121 &
   183344$-$071344$^{a}$ &
   183432$-$073917 &
   183628$-$070511 & \\
   183657$-$092859 &
   183729$-$082246$^{b}$ &
   183735$-$085836 &
   183920$-$054509 &
   183921$-$041952 & \\
   183929$-$053553 &
   183932$-$054420$^{a}$ &
   183955$-$053853$^{a}$ &
   184030$-$045722 &
   184037$-$054317 & \\
   184119$-$044418 &
   184206$-$042219 &
   184230$-$021839 &
   184415$-$041757$^{a}$ &
   184600$-$024157 & \\
   184601$-$010146 &
   184621$-$022143 &
   184659$-$020727$^{a}$ &
   184727$-$015504 &
   184735$-$020143 & \\
   184737$-$015856 &
   184811$-$012630 &
   184844$-$013314$^{a}$ &
   184916+001622$^{a}$ &
   184933$-$012907 & \\
   184950$-$022526 &
   185031$-$000155 &
   185102$-$004126 &
   185124+000412 &
   185158+045723 & \\
   185203+000926 &
   185234+044321 &
   185250+005527$^{a}$ &
   185336+042625 &
   185343+004144$^{a}$ & \\
   185351$-$002508 &
   185403+043304 &
   185406+041156 &
   185418+034959 &
   185423+013529 & \\
   185534+021910 &
   185623+022328 &
   185709+013854 &
   185741+044212 &
   185810+013657 & \\
   185908+004150 &
   190014+003556 &
   190018+003719 &
   190042$-$005151 &
   190059+002757 & \\
   190133+020915 &
   190215+012223 &
   190230$-$002831 &
   190236$-$002657 &
   190335+003217 & \\
   192352+135603 &
   192424+142738 &
   192527+151310 &
   194050+230605 &
   194054+230544 & \\
   194251+240520 &
   194318+233032 &
   194411+232659 &
   194429+241937 &
   194552+240957 & \\
   225526+621739 &
   225552+613904 &
   225616+622227 &
   225809+623240 &
   231319+615644 & \\
   231531+610709 & \\
\enddata
\tablenotetext{a}{Known HII region according to SIMBAD Astronomical Database and/or source with rising spectrum.}
\tablenotetext{b}{Planetary nebula according to SIMBAD Astronomical Database.}
\end{deluxetable}

\begin{deluxetable}{lllrrrrlr}
\tablecolumns{1} \tabletypesize{\scriptsize} \tablewidth{0pc}
\tablecaption{Parameters of Possible Calibrators}
\tablehead{ \colhead{Source Name} &
\multicolumn{2}{c}{Source Position}& \multicolumn{1}{c}{1.4 GHz}&
\multicolumn{1}{c}{5 GHz} & \multicolumn{1}{c}{15 GHz}
&Spectral & $\theta$&
$\Delta$\\
& \colhead{$\alpha$ (2000)}   & \colhead{$\delta$
(2000)} & \colhead{(mJy)} &\colhead{(mJy)} &
   \colhead{(mJy)}    &
 Index  & ($''$) & $\circ$}
\startdata
\begin{bf}133.95+1.07\end{bf}   &02\ 27\ 04.1   &$+$61\ 52\ 22           & \multicolumn{3}{c}{\textbf{W3OH}}\\
022515+611912                   &02\ 25\ 15.202 &$+$61\ 19\ 15.86        & 93.5 &24.5    &       &      &2.8                  & 0.72\\
023016+620936                   &02\ 30\ 16.1617&$+$62\ 09\ 37.675       &199.6 &76.0    &15.6   &1.44  &0.6\tablenotemark{a} & 0.85\\
\begin{bf}188.95+0.89\end{bf}   &06\ 08\ 53.7   &$+$21\ 38\ 30          &      &        &       &      &                     &     \\
060759+212945                   &06\ 07\ 59.5656&$+$21\ 29\ 43.733       &45.0  &17.3    &6.6    &1.01  &0.0\tablenotemark{b} & 0.27\\
060831+215535                   &06\ 08\ 31.338 &$+$21\ 55\ 34.59        &39.2  &7.3     &       &      &1.6                  & 0.30\\
061026+213412                   &06\ 10\ 26.593 &$+$21\ 34\ 12.69        &33.3  &7.9     &       &      &1.5                  & 0.39\\
060806+211327                   &06\ 08\ 06.328 &$+$21\ 13\ 26.47        &32.8  &6.0     &       &      &1.5                  & 0.46\\
060718+220452                   &06\ 07\ 18.656 &$+$22\ 04\ 51.28        &34.9  &10.1    &       &      &1.6                  & 0.59\\
J0606+2159\tablenotemark{d}     &06\ 06\ 38.8201&$+$21\ 59\ 28.270       &      &17.9    &9.9    &0.59  &0.1\tablenotemark{b} & 0.66\\
060622+211751                   &06\ 06\ 22.533 &$+$21\ 17\ 50.66        &42.4  &6.1     &       &      &2.2                  & 0.72\\
060717+221819                   &06\ 07\ 17.436 &$+$22\ 18\ 18.32        &67.1  &31.4    &7.8    &0.90  &1.5                  & 0.78\\
J0608+2229\tablenotemark{d}     &06\ 08\ 34.327 &$+$22\ 29\ 42.94        &      &64.3    &60.1   &0.01  &0.5\tablenotemark{a} & 0.85\\
\begin{bf}232.62+0.99\end{bf}   &07\ 32\ 09.6   &$-$16\ 58\ 16          &      &        &       &      &                     &     \\
073111$-$163223                   &07\ 31\ 11.451 &$-$16\ 32\ 23.53       &254.1 &117.5   &55.7   &0.63  &0.5\tablenotemark{a} & 0.50\\
073002$-$162904                   &07\ 30\ 02.496 &$-$16\ 29\ 04.33       & 30.6 &12.1    &       &      &1.7                  & 0.72\\
072924$-$163656                   &07\ 29\ 23.928 &$-$16\ 36\ 56.07       &364.1 &63.5    &6.2    &1.96  &1.0                  & 0.78\\
072912$-$171729                   &07\ 29\ 12.081 &$-$17\ 17\ 29.46       & 20.1 &5.3     &       &      &1.7                  & 0.81\\
072833$-$164825                   &07\ 28\ 33.798 &$-$16\ 48\ 26.70       & 39.5 &12.1    &       &      &1.1                  & 0.91\\
072858$-$173000                   &07\ 28\ 58.175 &$-$17\ 30\ 00.94       & 41.6 &15.7    &       &      &1.5                  & 0.96\\
\begin{bf}348.55$-$0.98\end{bf}   &17\ 19\ 20.4   &$-$39\ 03\ 51          &      &        &       &      &                     &     \\
172216$-$391447                   &17\ 22\ 16.609 &$-$39\ 14\ 49.39       &139.0 &31.6    &6.2    &1.56  &2.1                  & 0.76\\
172101$-$352009                   &17\ 21\ 01.5961&$-$35\ 20\ 09.121      &122.2 &43.2    &7.7    &1.55  &0.4\tablenotemark{c} & 0.96\\
\begin{bf}348.70-1.04\end{bf}   &17\ 20\ 04.0   &$-$38\ 58\ 32          &      &        &       &      &                     &     \\
172216$-$391447                   &17\ 22\ 16.609 &$-$39\ 14\ 49.39       &139.0 &31.6    &6.2    &1.56  &2.1                  & 0.62\\
172101$-$352009                   &17\ 21\ 01.5961&$-$35\ 20\ 09.121      &122.2 &43.2    &7.7    &1.55  &0.4\tablenotemark{c} & 0.91\\
\begin{bf}351.42+0.64\end{bf}   &17\ 20\ 53.5   &$-$35\ 47\ 02          &  \multicolumn{3}{c}{\textbf{NGC 6334-F}}   \\
172101$-$352009                   &17\ 21\ 01.5961&$-$35\ 20\ 09.121      &122.2 &43.2    &7.7    &1.55  &0.4\tablenotemark{c} & 0.45\\
\begin{bf}353.41$-$0.36\end{bf}   &17\ 30\ 26.1   &$-$34\ 41\ 47          &      &        &       &      &                     &     \\
173107$-$343225                   &17\ 31\ 07.6656&$-$34\ 32\ 22.846      & 73.2 &101.6   &75.4   &0.32  &0.1\tablenotemark{c} & 0.23\\
173155$-$344959                   &17\ 31\ 55.4007&$-$34\ 49\ 59.864      &265.1 &89.8    &22.0   &1.26  &0.2\tablenotemark{c} & 0.40\\
\begin{bf}354.61+0.47\end{bf}   &17\ 30\ 17.1   &$-$33\ 13\ 55          &      &        &       &      &                     &     \\
173157$-$331835                   &17\ 31\ 56.9970&$-$33\ 18\ 34.466      &191.5 &50.3    &7.9    &1.61  &0.1\tablenotemark{c} & 0.42\\
\begin{bf}8.68$-$0.37\end{bf}     &18\ 06\ 22.9   &$-$21\ 37\ 03          &      &        &       &      &                     &     \\
180437$-$214709                   &18\ 04\ 38.0246&$-$21\ 47\ 07.564      &285.6 &37.5    &5.2    &1.92  &0.2\tablenotemark{c} & 0.47\\
180806$-$212445                   &18\ 08\ 06.8482&$-$21\ 24\ 45.067      &259.1 &144.0   &63.7   &0.67  &0.2\tablenotemark{c} & 0.48\\
J1811$-$2055                      &18\ 11\ 06.7945&$-$20\ 55\ 03.240      &      &        &93.5   &      &0.1\tablenotemark{c} & 1.31\\
\begin{bf}9.62+0.20\end{bf}     &18\ 06\ 14.8   &$-$20\ 31\ 32          &      &        &       &      &                     &     \\
180650$-$202742                   &18\ 06\ 50.2510&$-$20\ 27\ 40.542      &113.8 &59.6    &5.9    &1.59  &0.2\tablenotemark{b} & 0.16\\
180323$-$203017                   &18\ 03\ 23.7239&$-$20\ 30\ 17.719      & 64.3 &30.8    &16.0   &0.65  &0.2\tablenotemark{c} & 0.71\\
J1811$-$2055                      &18\ 11\ 06.7945&$-$20\ 55\ 03.240      &      &        &93.5   &      &0.1\tablenotemark{c} & 1.21\\
\begin{bf}10.47+0.03\end{bf}    &18\ 08\ 38.2   &$-$19\ 51\ 50          &      &        &       &      &                     &     \\
180650$-$202742                   &18\ 06\ 50.2507&$-$20\ 27\ 40.537      &113.8 &59.6    &5.9    &1.59  &0.2\tablenotemark{b} & 0.75\\
J1811$-$2055                      &18\ 11\ 06.7945&$-$20\ 55\ 03.240      &      &        &93.5   &      &0.1\tablenotemark{c} & 1.16\\
\begin{bf}11.90$-$0.14\end{bf}    &18\ 12\ 11.6   &$-$18\ 41\ 38          &      &        &       &      &                     &     \\
181401$-$181142                   &18\ 14\ 01.9952&$-$18\ 11\ 36.524      & 46.5 &10.7    &4.4    &0.81  &0.1\tablenotemark{c} & 0.68\\
181007$-$181331                   &18\ 10\ 07.9885&$-$18\ 13\ 30.396      & 76.5 &13.8    &2.9    &1.43  &0.3\tablenotemark{c} & 0.70\\
181530$-$183614                   &18\ 15\ 30.3696&$-$18\ 36\ 13.250      & 71.0 &64.6    &56.6   &0.10  &0.1\tablenotemark{c} & 0.83\\
181445$-$191400                   &18\ 14\ 45.066 &$-$19\ 13\ 59.50       &109.3 &27.3    &5.2    &1.12  &2.0                  & 0.84\\
180855$-$182253                   &18\ 08\ 55.488 &$-$18\ 22\ 53.70       &204.1 &137.4   &53.6   &0.78  &0.6\tablenotemark{a} & 0.88\\
180927$-$180451                   &18\ 09\ 27.878 &$-$18\ 04\ 47.84       & 26.3 &20.6    &11.9   &0.44  &0.8\tablenotemark{a} & 0.92\\
\begin{bf}12.03$-$0.03\end{bf}    &18\ 12\ 02.5   &$-$18\ 31\ 57          &      &        &       &      &                     &     \\
181007$-$181331                   &18\ 10\ 07.9885&$-$18\ 13\ 30.396      & 76.5 &13.8    &2.9    &1.43  &0.3\tablenotemark{c} & 0.57\\
181401$-$181142                   &18\ 14\ 01.9952&$-$18\ 11\ 36.524      & 46.5 &10.7    &4.4    &0.81  &0.1\tablenotemark{c} & 0.60\\
180927$-$180451                   &18\ 09\ 27.878 &$-$18\ 04\ 47.84       & 26.3 &20.6    &11.9   &0.44  &0.8\tablenotemark{a} & 0.79\\
180855$-$182253                   &18\ 08\ 55.488 &$-$18\ 22\ 53.70       &204.1 &137.4   &53.6   &0.78  &0.6\tablenotemark{a} & 0.79\\
181530$-$183614                   &18\ 15\ 30.3696&$-$18\ 36\ 13.250      & 71.0 &64.6    &56.6   &0.10  &0.1\tablenotemark{c} & 0.87\\
181445$-$191400                   &18\ 14\ 45.066 &$-$19\ 13\ 59.50       &109.3 &27.3    &5.2    &1.12  &2.0                  & 0.98\\
\begin{bf}12.68$-$0.18\end{bf}    &18\ 13\ 55.1   &$-$18\ 01\ 35          &      &        &       &      &                     &     \\
181401$-$181142                   &18\ 14\ 01.9952&$-$18\ 11\ 36.524      & 46.5 &10.7    &4.4    &0.81  &0.1\tablenotemark{c} & 0.17\\
181530$-$183614                   &18\ 15\ 30.3696&$-$18\ 36\ 13.250      & 71.0 &64.6    &56.6   &0.10  &0.1\tablenotemark{c} & 0.70\\
181007$-$181331                   &18\ 10\ 07.9877&$-$18\ 13\ 30.520      & 76.5 &13.8    &0.2    &1.43  &0.6\tablenotemark{c} & 0.97\\
\begin{bf}12.89+0.49\end{bf}    &18\ 11\ 51.3   &$-$17\ 31\ 29          &      &        &       &      &                     &     \\
181134$-$170325                   &18\ 11\ 34.488 &$-$17\ 03\ 20.54       & 44.9 &6.2     &       &      &1.7                  & 0.47\\
180929$-$170352                   &18\ 09\ 28.906 &$-$17\ 03\ 51.03       & 53.9 &8.6     &       &      &2.3                  & 0.75\\
180927$-$180451                   &18\ 09\ 27.878 &$-$18\ 04\ 47.84       & 26.3 &20.6    &11.9   &0.44  &0.8\tablenotemark{a} & 0.77\\
181007$-$181331                   &18\ 10\ 07.9885&$-$18\ 13\ 30.396      & 76.5 &13.8    &2.9    &1.43  &0.3\tablenotemark{c} & 0.82\\
181401$-$181142                   &18\ 14\ 01.9952&$-$18\ 11\ 36.524      & 46.5 &10.7    &4.4    &0.81  &0.1\tablenotemark{c} & 0.86\\
\begin{bf}12.91$-$0.26\end{bf}    &18\ 14\ 39.1   &$-$17\ 52\ 06          &  \multicolumn{3}{c}{\textbf{W33A}} \\
181401$-$181142                   &18\ 14\ 01.9952&$-$18\ 11\ 36.524      & 46.5 &10.7    &4.4    &0.81  &0.1\tablenotemark{c} & 0.36\\
181530$-$183614                   &18\ 15\ 30.3696&$-$18\ 36\ 13.250      & 71.0 &64.6    &56.6   &0.10  &0.1\tablenotemark{c} & 0.77\\
\begin{bf}15.03$-$0.68\end{bf}    &18\ 20\ 24.7   &$-$16\ 11\ 39          &  \multicolumn{3}{c}{\textbf{M17}}        \\
181723$-$161442                   &18\ 17\ 22.964 &$-$16\ 14\ 39.96       & 71.0 &35.3    &5.2    &1.47  &2.0                  & 0.76\\
\begin{bf}23.01$-$0.41\end{bf}    &18\ 34\ 39.9   &$-$09\ 00\ 44          &      &        &       &      &                     &     \\
183319$-$085527                   &18\ 33\ 19.5819&$-$08\ 55\ 26.211      &273.9 &63.5    &70.8   &0.17  &0.0\tablenotemark{b} & 0.35\\
183558$-$093724                   &18\ 35\ 58.7041&$-$09\ 37\ 29.849      &144.9 &18.0    &3.8    &1.42  &0.1\tablenotemark{c} & 0.69\\
183424$-$095319                   &18\ 34\ 24.198 &$-$09\ 53\ 21.49       &181.2 &51.1    &17.2   &0.79  &1.3\tablenotemark{a} & 0.88\\
183809$-$084048                   &18\ 38\ 09.672 &$-$08\ 40\ 41.45       & 66.5 &7.3     &       &      &2.6                  & 0.93\\
\begin{bf}23.44$-$0.18\end{bf}    &18\ 34\ 39.0   &$-$08\ 31\ 36          &      &        &       &      &                     &     \\
183319$-$085527                   &18\ 33\ 19.5819&$-$08\ 55\ 26.211      &273.9 &63.5    &70.8   &0.17  &0.0\tablenotemark{b} & 0.52\\
183809$-$084048                   &18\ 38\ 09.672 &$-$08\ 40\ 41.45       & 66.5 &7.3     &       &      &2.6                  & 0.89\\
 \begin{bf}24.33+0.14\end{bf}   &18\ 35\ 09.4   &$-$07\ 34\ 58          &      &        &       &      &                     &     \\
183354$-$071106                   &18\ 33\ 54.0032&$-$07\ 11\ 09.436      &200.0 &91.5    &32.2   &0.95  &0.1\tablenotemark{c} & 0.51\\
183214$-$070643                   &18\ 32\ 14.958 &$-$07\ 06\ 45.44       & 99.6 &18.1    &       &      &3.2                  & 0.87\\
183124$-$072551                   &18\ 31\ 24.746 &$-$07\ 25\ 48.83       & 28.0 &13.4    &3.9    &1.11  &1.8                  & 0.95\\
183758$-$065330                   &18\ 37\ 58.039 &$-$06\ 53\ 31.00       &575.7 &235.5   &122.6  &0.48  &0.6\tablenotemark{a} & 0.99\\
\begin{bf}27.36$-$0.16\end{bf}    &18\ 41\ 50.6   &$-$05\ 01\ 47          &      &        &       &      &                     &     \\
184445$-$045425                   &18\ 44\ 45.795 &$-$04\ 54\ 23.75       &107.0 &24.3    &       &      &1.4                  & 0.74\\
\begin{bf}28.15+0.00\end{bf}    &18\ 42\ 40.9   &$-$04\ 15\ 26          &      &        &       &      &                     &     \\
184445$-$045425                   &18\ 44\ 45.795 &$-$04\ 54\ 23.75       &107.0 &24.3    &       &      &1.4                  & 0.83\\
184626$-$040853                   &18\ 46\ 26.874 &$-$04\ 08\ 52.39       & 47.7 &16.7    &no data&      &1.7                  & 0.95\\
\begin{bf}29.86$-$0.05\end{bf}    &18\ 45\ 59.9   &$-$02\ 45\ 12          &      &        &       &      &                     &     \\
184700$-$022751                   &18\ 47\ 00.394 &$-$02\ 27\ 52.27       &800.0 &345.1   &132.6  &0.76  &0.9\tablenotemark{a} & 0.38\\
184236$-$031108                   &18\ 42\ 35.941 &$-$03\ 11\ 08.05       &380.6 &87.8    &16.2   &1.29  &1.3                  & 0.95\\
\begin{bf}29.95$-$0.02\end{bf}    &18\ 46\ 03.6   &$-$02\ 39\ 24          &      &        &       &      &                     &     \\
184700$-$022751                   &18\ 47\ 00.394 &$-$02\ 27\ 52.27       &800.0 &345.1   &132.6  &0.76  &0.9\tablenotemark{a} & 0.31\\
184936$-$021629                   &18\ 49\ 35.1676&$-$02\ 17\ 11.965      &      &18.7    &8.4    &0.61  &0.1\tablenotemark{c} & 0.97\\
\begin{bf}30.20$-$0.17\end{bf}    &18\ 47\ 02.5   &$-$02\ 30\ 38          &      &        &       &      &                     &     \\
184700$-$022751                   &18\ 47\ 00.394 &$-$02\ 27\ 52.27       &800.0 &345.1   &132.6  &0.76  &0.9\tablenotemark{a} & 0.05\\
184936$-$021629                   &18\ 49\ 35.1676&$-$02\ 17\ 11.965      &      &18.7    &8.4    &0.61  &0.1\tablenotemark{c} & 0.68\\
\begin{bf}30.78+0.23\end{bf}    &18\ 46\ 41.6   &$-$01\ 48\ 31          &      &        &       &      &                     &     \\
184614$-$011942                   &18\ 46\ 14.6811&$-$01\ 19\ 40.600      &138.1 &52.6    &6.6    &1.51  &0.1\tablenotemark{b} & 0.49\\
184700$-$022751                   &18\ 47\ 00.394 &$-$02\ 27\ 52.27       &800.0 &345.1   &132.6  &0.76  &0.9\tablenotemark{a} & 0.66\\
184353$-$013415                   &18\ 43\ 53.049 &$-$01\ 34\ 15.37       & 71.7 &30.9    &       &      &1.6                  & 0.74\\
184351$-$012914                   &18\ 43\ 51.280 &$-$01\ 29\ 17.52       &371.3 &52.4    &       &      &1.5                  & 0.78\\
184959$-$013256                   &18\ 49\ 59.3231&$-$01\ 32\ 57.020      &1459.8&no data &47.5   &      &0.1\tablenotemark{b} & 0.86\\
184936$-$021629                   &18\ 49\ 35.1676&$-$02\ 17\ 11.965      &      &18.7    &8.4    &0.61  &0.1\tablenotemark{c} & 0.86\\
\begin{bf}30.79+0.20\end{bf}    &18\ 46\ 48.2   &$-$01\ 48\ 46          &      &        &       &      &                     &     \\
184700$-$022751                   &18\ 47\ 00.394 &$-$02\ 27\ 52.27       &800.0 &345.1   &132.6  &0.76  &0.9\tablenotemark{a} & 0.65\\
184353$-$013415                   &18\ 43\ 53.049 &$-$01\ 34\ 15.37       & 71.7 &30.9    &       &      &1.6                  & 0.77\\
184351$-$012914                   &18\ 43\ 51.280 &$-$01\ 29\ 17.52       &371.3 &52.4    &       &      &1.5                  & 0.81\\
184936$-$021629                   &18\ 49\ 35.1676&$-$02\ 17\ 11.965      &      &18.7    &8.4    &0.61  &0.1\tablenotemark{c} & 0.84\\
184959$-$013256                   &18\ 49\ 59.3231&$-$01\ 32\ 57.020      &1459.8&no data &47.5   &      &0.1\tablenotemark{b} & 0.84\\
\begin{bf}31.28+0.06\end{bf}    &18\ 48\ 12.7   &$-$01\ 26\ 36          &      &        &       &      &                     &     \\
184959$-$013256                   &18\ 49\ 59.3231&$-$01\ 32\ 57.020      &1459.8&no data &47.5   &      &0.1\tablenotemark{b} & 0.46\\
184614$-$011942                   &18\ 46\ 14.6811&$-$01\ 19\ 40.600      &138.1 &52.6    &6.6    &1.51  &0.1\tablenotemark{b} & 0.51\\
184936$-$021629                   &18\ 49\ 35.1676&$-02$\ 17\ 11.965      &      &18.7    &8.4    &0.61  &0.1\tablenotemark{c} & 0.90\\
\begin{bf}32.74$-$0.08\end{bf}    &18\ 51\ 21.8   &$-$00\ 12\ 13          &      &        &       &      &                     &     \\
185146+003532$^{e}$             &18\ 51\ 46.7218&$+$00\ 35\ 32.413      & 839.9&771.2   &625.5  &0.15  &0.0\tablenotemark{b} & 0.40\\
185343+001957                   &18\ 53\ 43.8523&$+00$\ 19\ 57.481       &  39.8&20.5    &9.4    &0.67  &0.1\tablenotemark{b} & 0.61\\
184821+001108                   &18\ 48\ 21.237 &$+$00\ 11\ 02.22        &135.6 &no data &8.1    &      &0.6\tablenotemark{a} & 0.75\\
185152$-$004000                   &18\ 51\ 52.940 &$-$00\ 40\ 01.85       & 388.8&no data &11.1   &      &0.4\tablenotemark{a} & 0.88\\
\begin{bf}35.20$-$0.74\end{bf}    &18\ 58\ 13.0   &$+$01\ 40\ 33          &      &        &       &      &                     &     \\
190035+012119                   &19\ 00\ 35.444 &$+$01\ 21\ 19.23        & 45.6 &23.9    &       &      &2.1                  & 0.67\\
185749+005119                   &18\ 57\ 49.540 &$+$00\ 51\ 19.29        & 73.6 &55.3    &23.8   &0.82  &0.5\tablenotemark{a} & 0.83\\
185515+021054                   &18\ 55\ 15.458 &$+$02\ 10\ 50.55        & 84.6 &19.5    &       &      &1.7                  & 0.90\\
185500+021544                   &18\ 55\ 00.116 &$+$02\ 15\ 40.77        & 58.4 &47.6    &28.7   &0.51  &0.7\tablenotemark{a} & 1.00\\
\begin{bf}35.20$-$1.74\end{bf}    &19\ 01\ 45.5   &$+$01\ 13\ 28          &      &        &       &      &                     &     \\
190035+012119                   &19\ 00\ 35.444 &$+$01\ 21\ 19.23        & 45.6 &23.9    &       &      &2.1                  & 0.32\\
190231+004047                   &19\ 02\ 31.888 &$+$00\ 40\ 47.98        & 31.1 &10.5    &       &      &1.7                  & 0.58\\
190426+011036                   &19\ 04\ 26.3965&$+$01\ 10\ 38.746       & 61.7 &63.5    &48.6   &-0.04 &0.0\tablenotemark{b} & 0.67\\
190353+014526                   &19\ 03\ 53.0610&$+$01\ 45\ 26.351       &196.4 &137.0   &72.9   &0.57  &0.0\tablenotemark{b} & 0.75\\
190227+002155                   &19\ 02\ 27.711 &$+$00\ 21\ 56.69        &216.1 &25.9    &       &      &1.4                  & 0.88\\
190515+010609                   &19\ 05\ 15.789 &$+$01\ 06\ 09.80        & 54.6 &21.2    &       &      &1.3                  & 0.89\\
\begin{bf}37.43+1.50\end{bf}    &18\ 54\ 17.1   &$+$04\ 41\ 09          &      &        &       &      &                     &     \\
185521+050927                   &18\ 55\ 21.9147&$+$05\ 09\ 28.468       &113.1 &29.8    &9.5    &1.08  &0.1\tablenotemark{b} & 0.54\\
185120+050228                   &18\ 51\ 21.111 &$+$05\ 02\ 35.05        &      &7.6     &       &      &1.7                  & 0.66\\
185407+040059                   &18\ 54\ 07.9034&$+$04\ 00\ 59.403       &126.6 &40.9    &14.1   &0.89  &0.1\tablenotemark{c} & 0.67\\
185618+041420                   &18\ 56\ 18.7659&$+$04\ 14\ 22.601       & 74.6 &22.4    &4.1    &1.28  &0.1\tablenotemark{b} & 0.68\\
185513+052158                   &18\ 55\ 13.502 &$+$05\ 21\ 57.09        &111.4 &53.6    &23.9   &0.73  &0.5\tablenotemark{a} & 0.72\\
185650+041520                   &18\ 56\ 50.269 &$+$04\ 15\ 19.32        & 55.5 &16.1    &       &      &0.9                  & 0.77\\
185124+050118                   &18\ 51\ 24.812 &$+$05\ 01\ 18.84        & 19.9 &6.5     &       &      &1.2                  & 0.79\\
185319+035218                   &18\ 53\ 19.625 &$+$03\ 52\ 19.19        & 40.1 &10.5    &       &      &2.0                  & 0.85\\
185244+053534                   &18\ 52\ 44.699 &$+$05\ 35\ 35.43        & 53.2 &6.8     &       &      &2.1                  & 0.99\\
\begin{bf}49.49$-$0.37\end{bf}    &19\ 23\ 39.4 &$+$14\ 31\ 01           & \multicolumn{3}{c}{\textbf{W51 IRS2 (=W51d)}} \\
192233+150448                   &19\ 22\ 33.2725&$+$15\ 04\ 47.541       &115.3 &53.0    &20.7   &0.62  &0.0\tablenotemark{b} & 0.63\\
\begin{bf}59.78+0.06\end{bf}    &19\ 43\ 10.9   &$+$23\ 44\ 03           &      &        &       &      &                     &     \\
194358+233010                   &19\ 43\ 58.1411&$+$23\ 30\ 10.014       &118.8 &34.1    &15.0   &0.88  &0.0\tablenotemark{b} & 0.30\\
194313+241314                   &19\ 43\ 13.699 &$+$24\ 13\ 16.50        & 43.4 &13.3    &       &      &2.0                  & 0.49\\
194111+232805                   &19\ 41\ 11.645 &$+$23\ 28\ 05.22        & 30.8 &7.3     &       &      &2.0                  & 0.56\\
194256+230547                   &19\ 42\ 56.707 &$+$23\ 05\ 49.45        & 20.4 &8.9     &       &      &1.6                  & 0.64\\
194155+230755                   &19\ 41\ 55.1113&$+$23\ 07\ 57.521       & 70.9 &32.6    &8.6    &1.08  &0.0\tablenotemark{b} & 0.68\\
194619+241858                   &19\ 46\ 19.737 &$+$24\ 18\ 57.60        & 52.5 &52.9    &       &      &2.2                  & 0.98\\
J1946+2300\tablenotemark{e}     &19\ 46\ 06.273 &$+$23\ 00\ 05.27        &      &55.0    &27.2   &0.52  &0.7\tablenotemark{a} & 1.03\\
\begin{bf}109.87+2.11\end{bf}   &22\ 56\ 18.1   &$+$62\ 01\ 49          &  \multicolumn{3}{c}{\textbf{Cepheus A}} \\
225713+621509                   &22\ 57\ 13.4508&$+$62\ 15\ 09.881       & 84.7 &26.3    &3.7    &1.70  &0.1\tablenotemark{b} & 0.32\\
J2254+6209\tablenotemark{e}     &22\ 54\ 25.2944&$+$62\ 09\ 38.719       &      &58.6    &24.1   &0.83  &0.1\tablenotemark{b} & 0.45\\
225452+613313                   &22\ 54\ 52.761 &$+$61\ 33\ 13.43        & 70.6 &28.8    &       &      &1.6                  & 0.59\\
225420+613919                   &22\ 54\ 20.925 &$+$61\ 39\ 20.49        & 43.2 &12.3    &       &      &1.7                  & 0.62\\
225838+611402                   &22\ 58\ 38.186 &$+$61\ 14\ 03.49        & 42.5 &9.2     &       &      &1.7                  & 0.99\\
\begin{bf}111.54+0.78\end{bf}   &23\ 13\ 45.4   &$+$61\ 28\ 11          &   \multicolumn{3}{c}{\textbf{NGC 7538 IRS 1}}  \\
231310+611009                   &23\ 13\ 10.297 &$+$61\ 10\ 09.37        & 297.3&43.5    &       &      &2.1                  & 0.33\\
231218+610804                   &23\ 12\ 18.1477&$+$61\ 08\ 02.072       & 32.0 &17.9    &5.3    &0.92  &0.1\tablenotemark{b} & 0.49\\
231544+615109                   &23\ 15\ 43.695 &$+$61\ 51\ 18.21        & 32.0 &11.7    &       &      &2.3                  & 0.63\\
231658+620347                   &23\ 16\ 57.747 &$+$62\ 03\ 48.54        & 85.5 &34.3    &       &      &1.6                  & 1.00\\
\enddata
\tablecomments{The first column
presents source name for the masers (bold face) and the extragalactic radio
sources. The next two columns list their J2000 equatorial
coordinates. For the maser sources the next column gives a common name (if available).
For the calibrator sources columns 4 to 6 give the integrated flux density at L-
(from NVSS), peak flux density at C- and U-bands, respectively.
Column 7 is the spectral index between C- and U- bands ($S_{\nu}
\sim \nu^{-\alpha}$). Column 8 is the upper limits of source sizes
(FWHM of the maximum major axes of the sources as determined from
the Gaussian fits using AIPS task JMFIT). The data come from
C-band observed with C-configuration, except those marked by
signs. The last column gives the separation between masers and
calibrators. Some sources have no data 
due to
failure of the telescope.}
\tablenotetext{a}{U-band data observed with C configuration.}
\tablenotetext{b}{U-band data observed with A configuration.}
\tablenotetext{c}{U-band data observed with B configuration.}
\tablenotetext{d}{Source listed in Gregory \& Condon (1991).}
\tablenotetext{e}{Source listed in VLBA calibrator source (Petrov et al. 2005).}
\end{deluxetable}

\begin{deluxetable}{rccccccccc}
\tablecolumns{1} \tablewidth{0pc} \tabletypesize{\tiny}
\tablecaption{Comparison of Calibrator Position Errors Derived from VLA and VLBI Data}
\tablehead{ \colhead{Source Name} &
\multicolumn{2}{c}{VLA-C Position}& $\Delta$(Y/C$-$VLBI)& \multicolumn{2}{c}{VLA-A Position}&
$\Delta$(Y/A$-$VLBI)&
\multicolumn{2}{c}{VLBI} &Error\\
& \colhead{$\alpha$ (2000)}   & \colhead{$\delta$
(2000)} &(arcsec) &\colhead{$\alpha$ (2000)}   &
\colhead{$\delta$ (2000)} &(arcsec)&\colhead{$\alpha$ (2000)} &
\colhead{$\delta$ (2000)} &(mas)} \startdata

023016+620936&02 30 16.068 &$+$62 09 38.47 &($-$0.58, $-$0.75)&16.1617&37.675&(+0.076, $-$0.045)&16.15090 &$+$37.7200\tablenotemark{a}&\\
185146+003532&18 51 46.729 &$+$00 35 32.59 &(+0.09, +0.23)&46.7218&32.413&($-$0.019, +0.050)&46.72308 &$+$32.3628\tablenotemark{b}&(1.43, 1.66) \\
J1946+2300   &19 46 06.273 &$+$23 00 05.27 &(+0.30, +0.86)&       &         &              &06.25140 &$+$04.4146\tablenotemark{b}&(0.32, 0.57) \\
J2254+6209   &22 54 25.275 &$+$62 09 38.66 &($-$0.13, $-$0.06)&25.2944&38.719&(+0.009, $-$0.006)&25.29305 &$+$38.7247\tablenotemark{b}&(6.13, 4.93) \\

\enddata

\tablecomments {-Listed are, left to right, source name, position
derived from VLA C-array observations, difference between the
latter and the VLBI position (east offset and north offset),
position derived from VLA A-array observations, difference between
the latter and the VLBI position, and the VLBI position and its
uncertainties. Of the coordinates for the latter two positions
only seconds of time and arcseconds are given.}
\tablenotetext{a}{From our VLBA observations (Xu et al. 2006).}
\tablenotetext{b}{From VCS3 (Petrov et al. 2005).}

\end{deluxetable}

\begin{figure*}
\begin{tabular}{ccc}
\includegraphics[width=5cm]{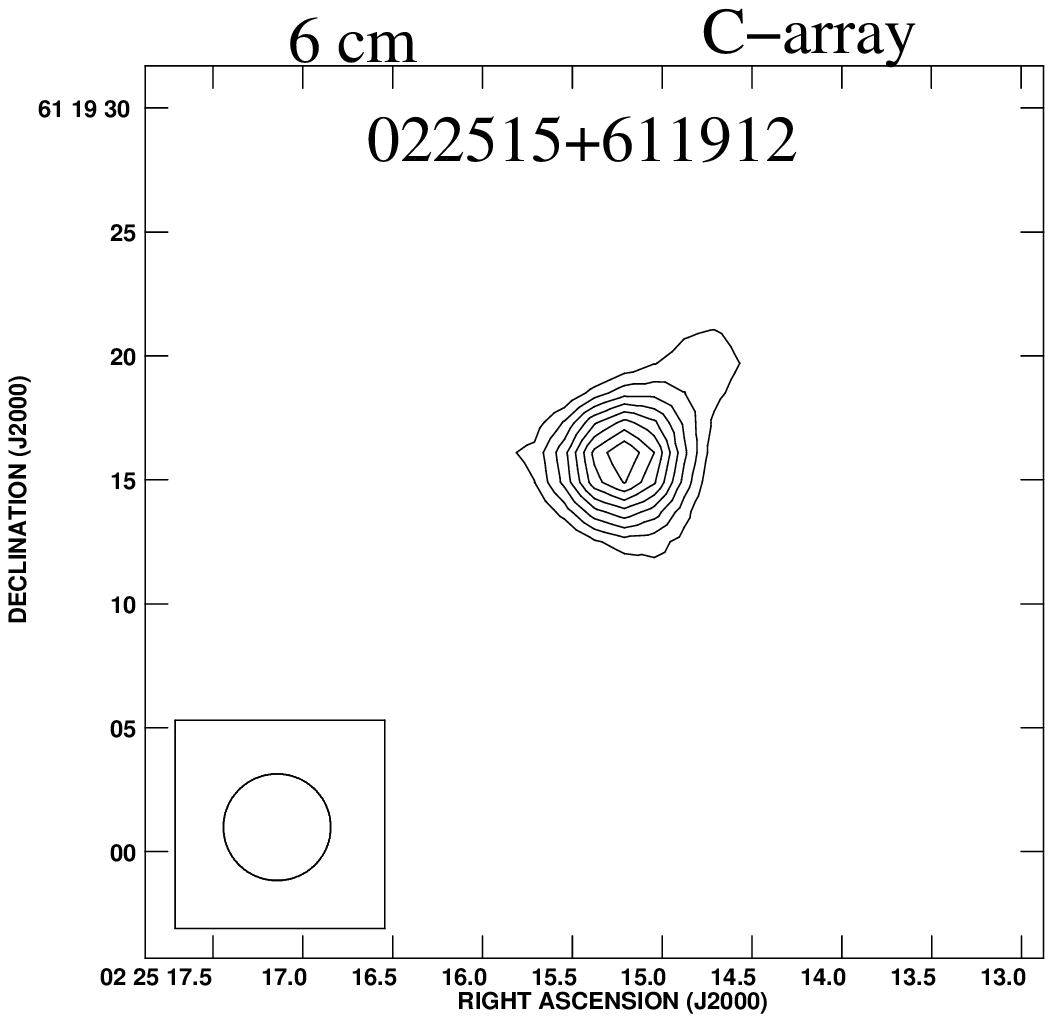} &
\includegraphics[width=5cm]{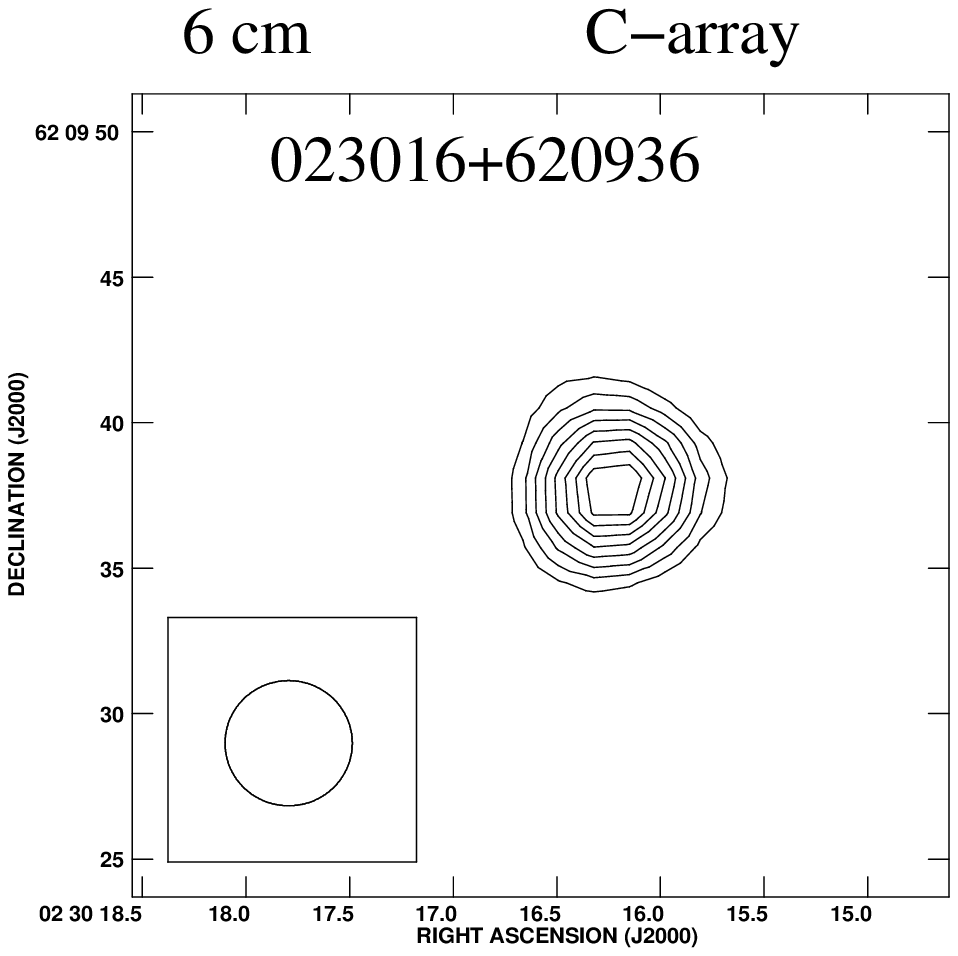} &
\includegraphics[width=5cm]{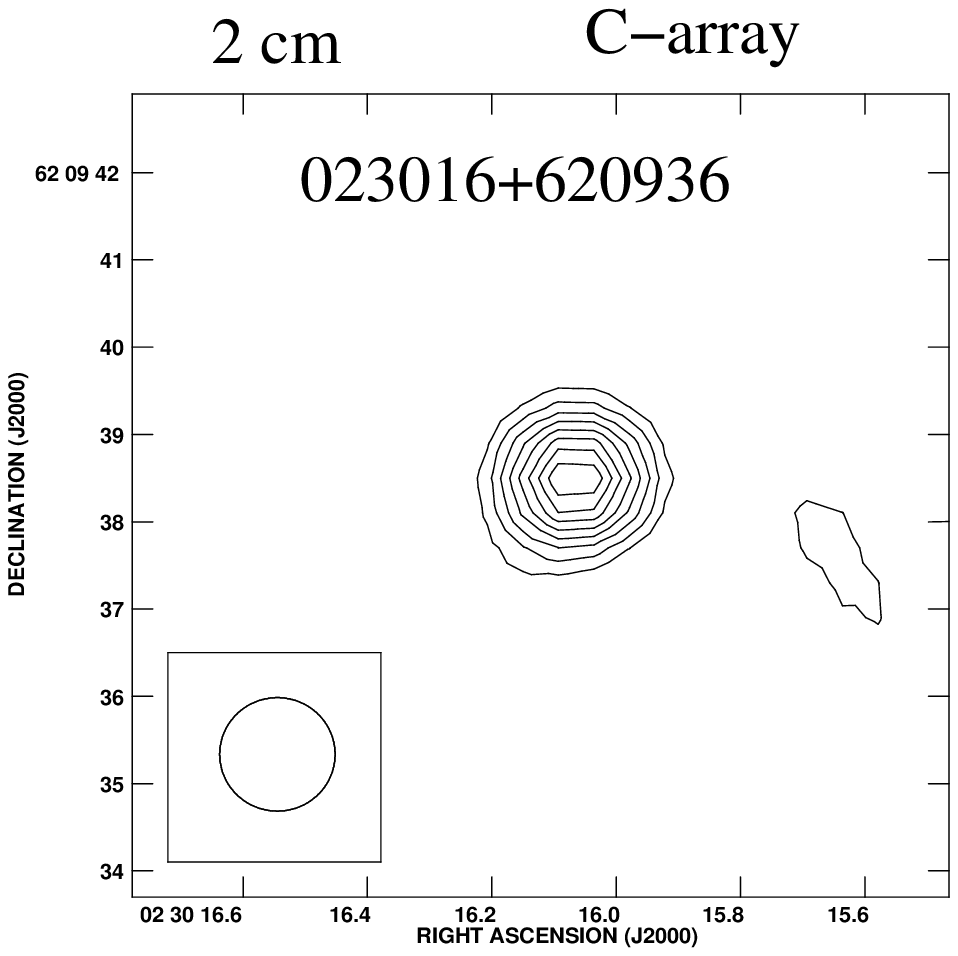}\\
\includegraphics[width=5cm]{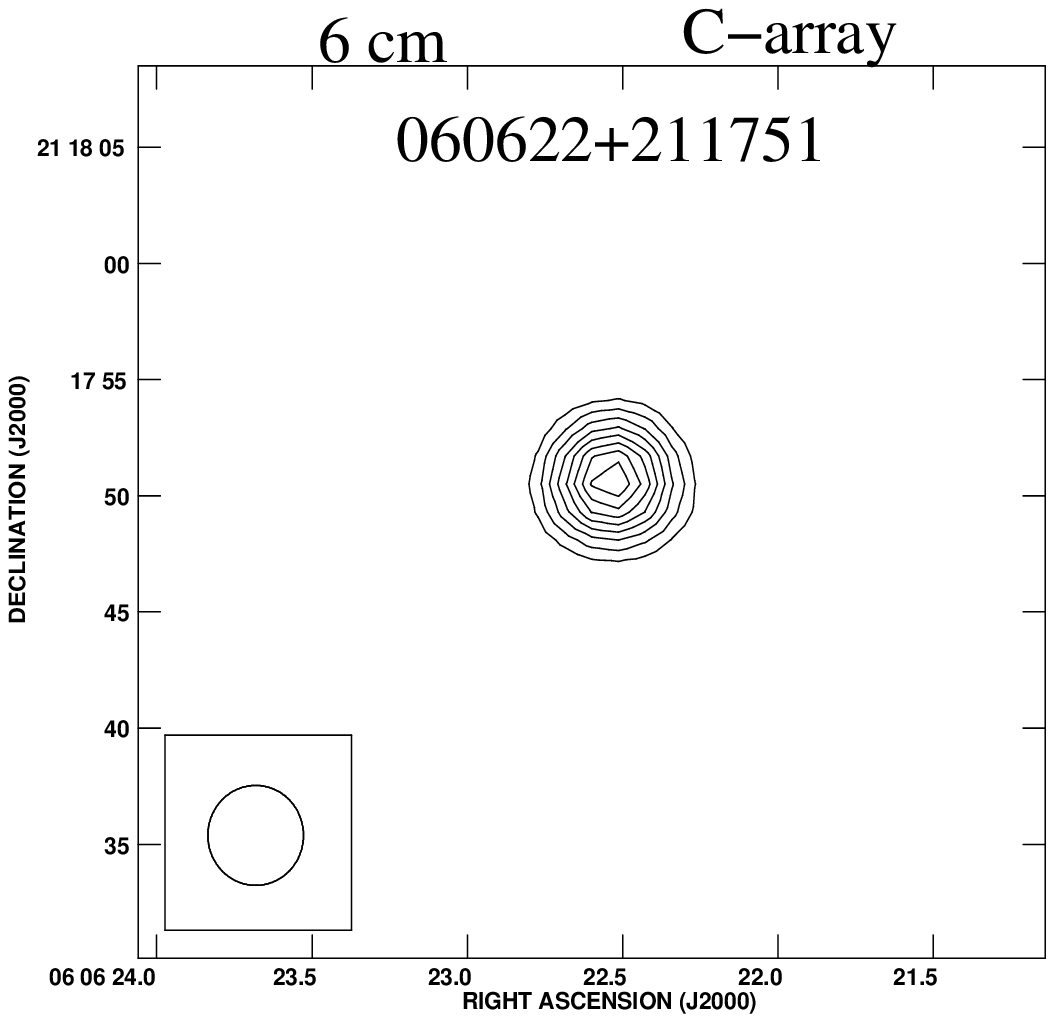} &
\includegraphics[width=5cm]{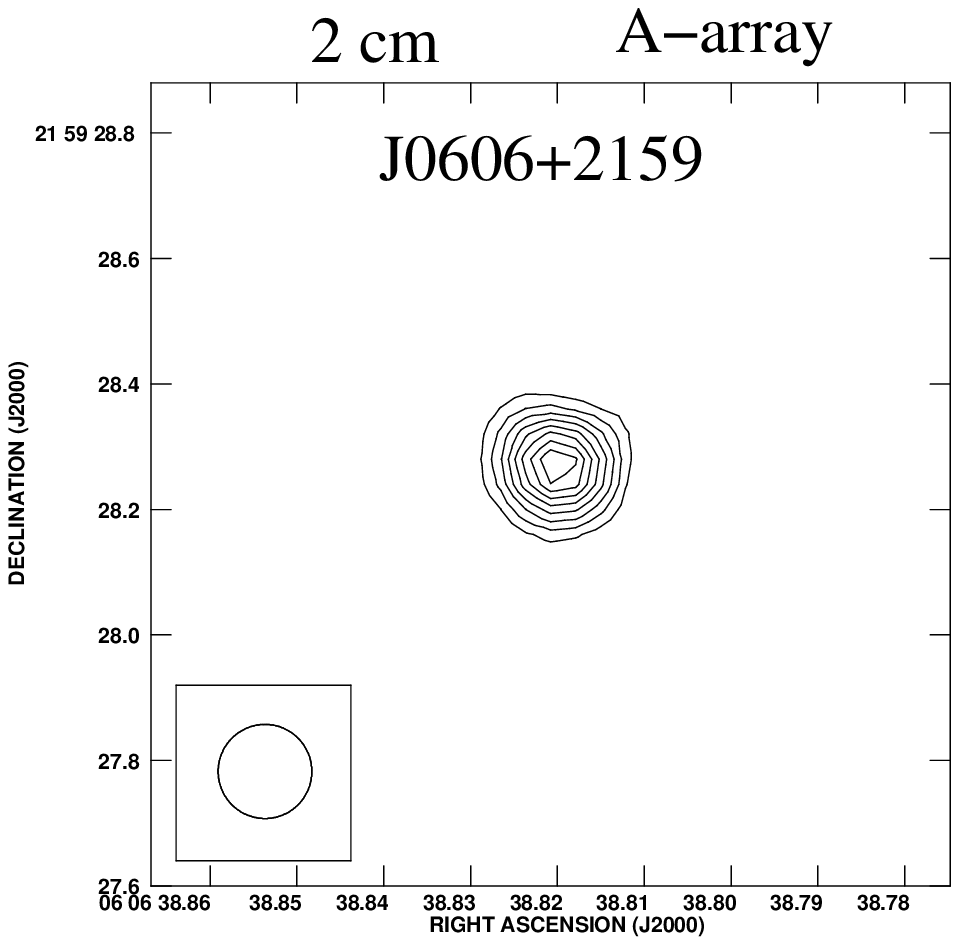} &
\includegraphics[width=5cm]{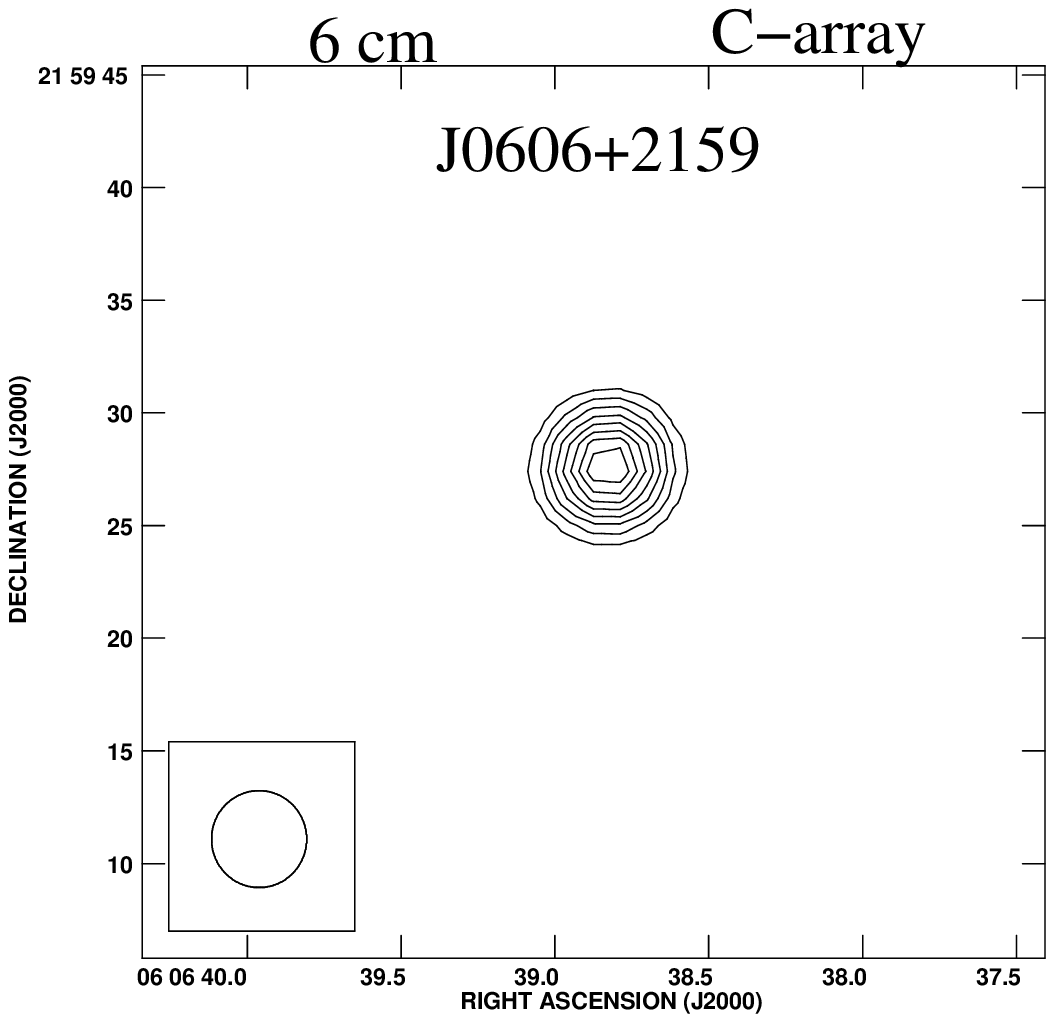}\\
\includegraphics[width=5cm]{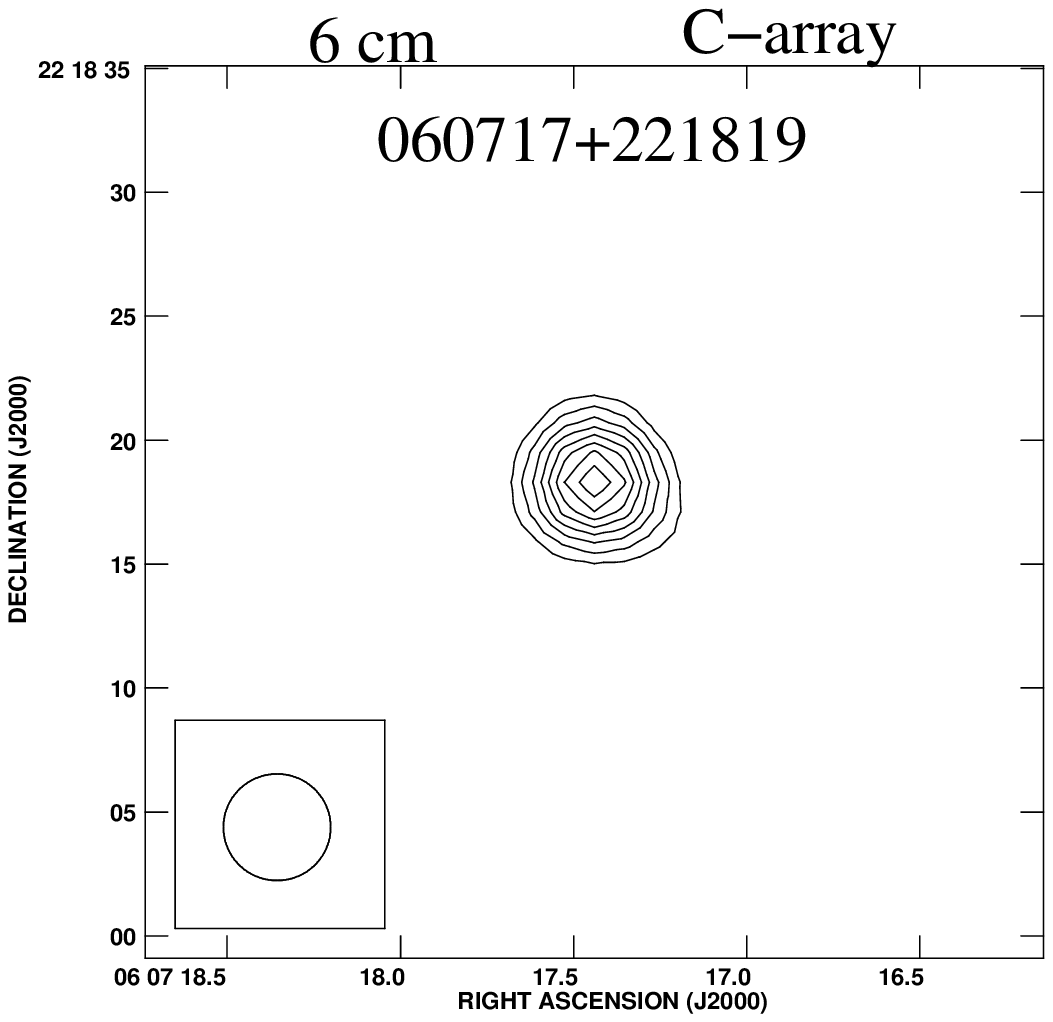} &
\includegraphics[width=5cm]{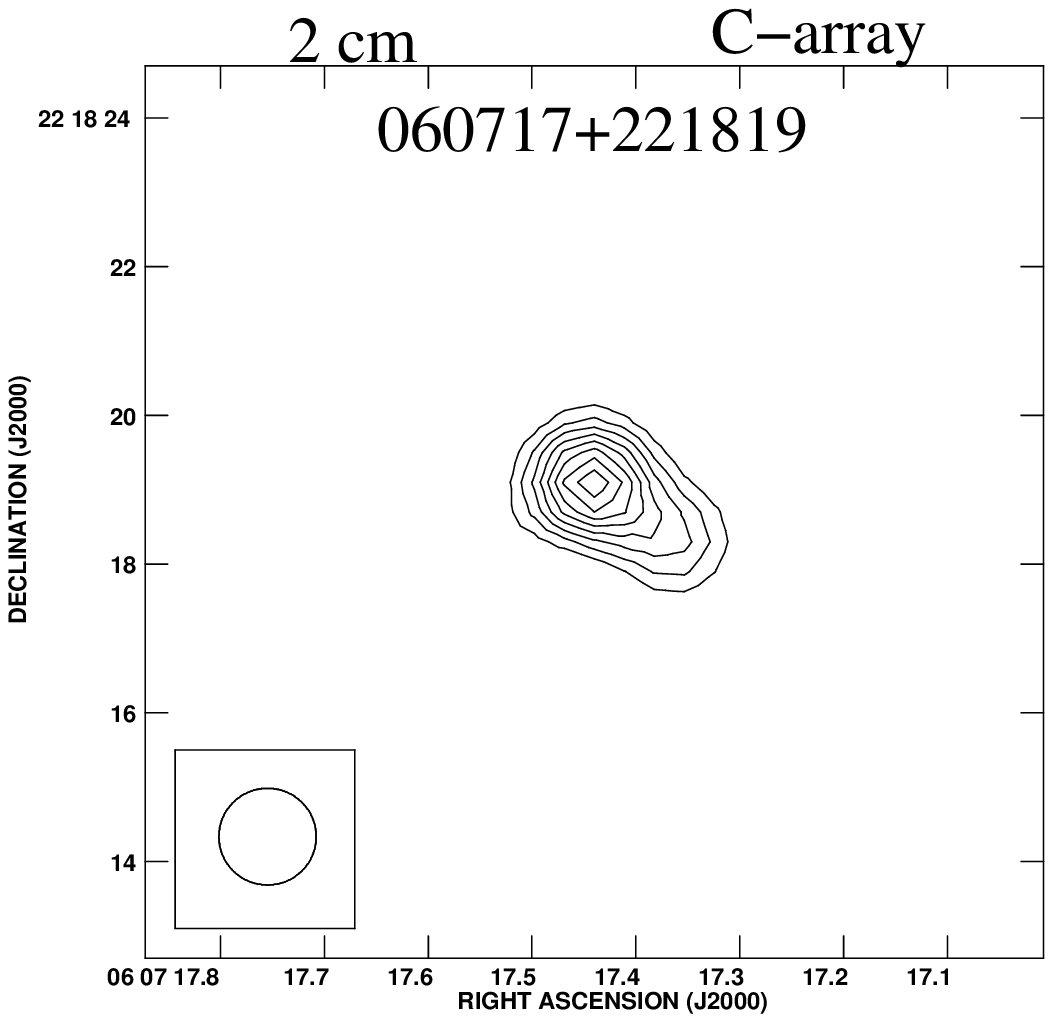} &
\includegraphics[width=5cm]{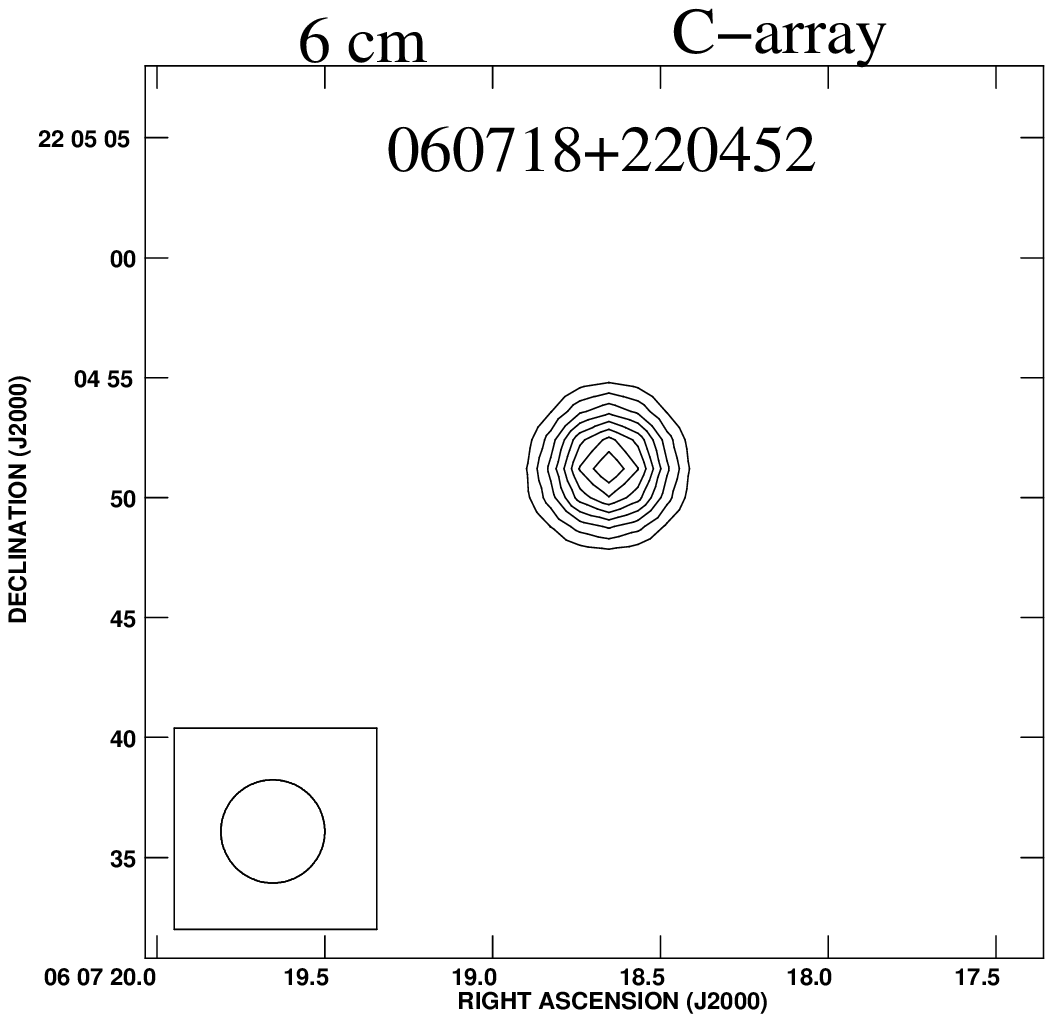}\\
\includegraphics[width=5cm]{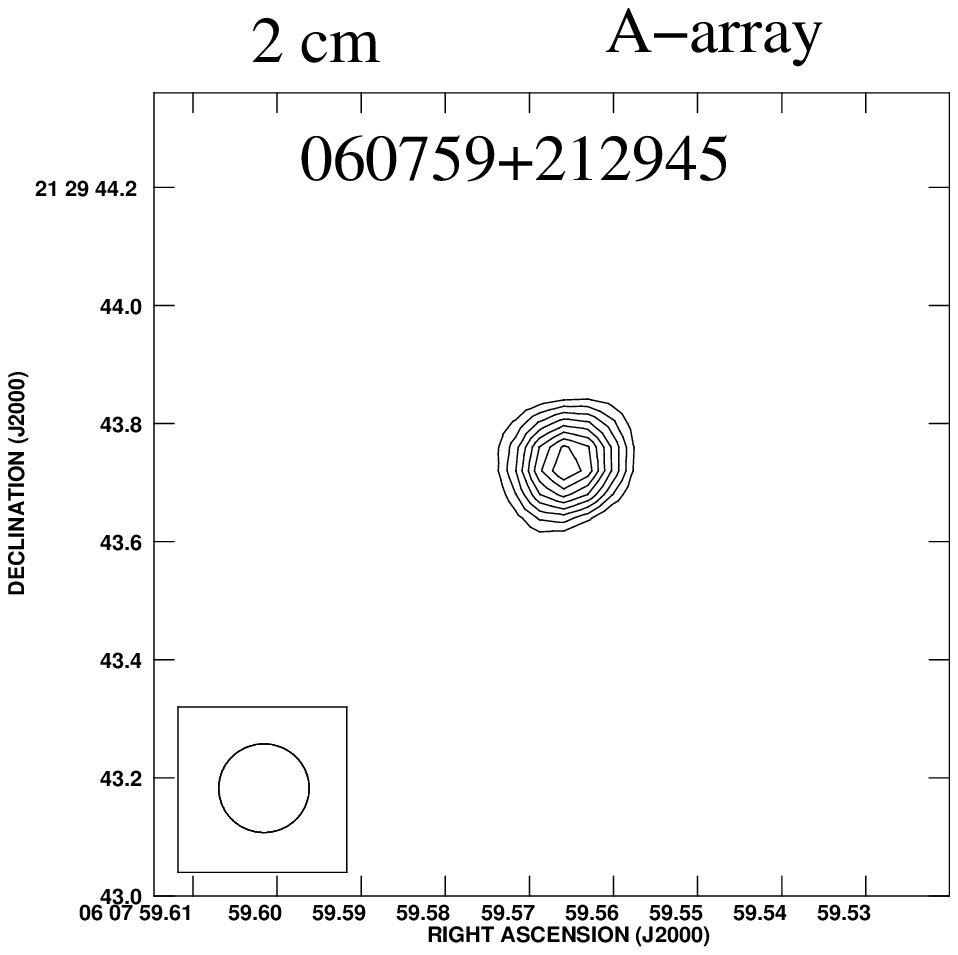} &
\includegraphics[width=5cm]{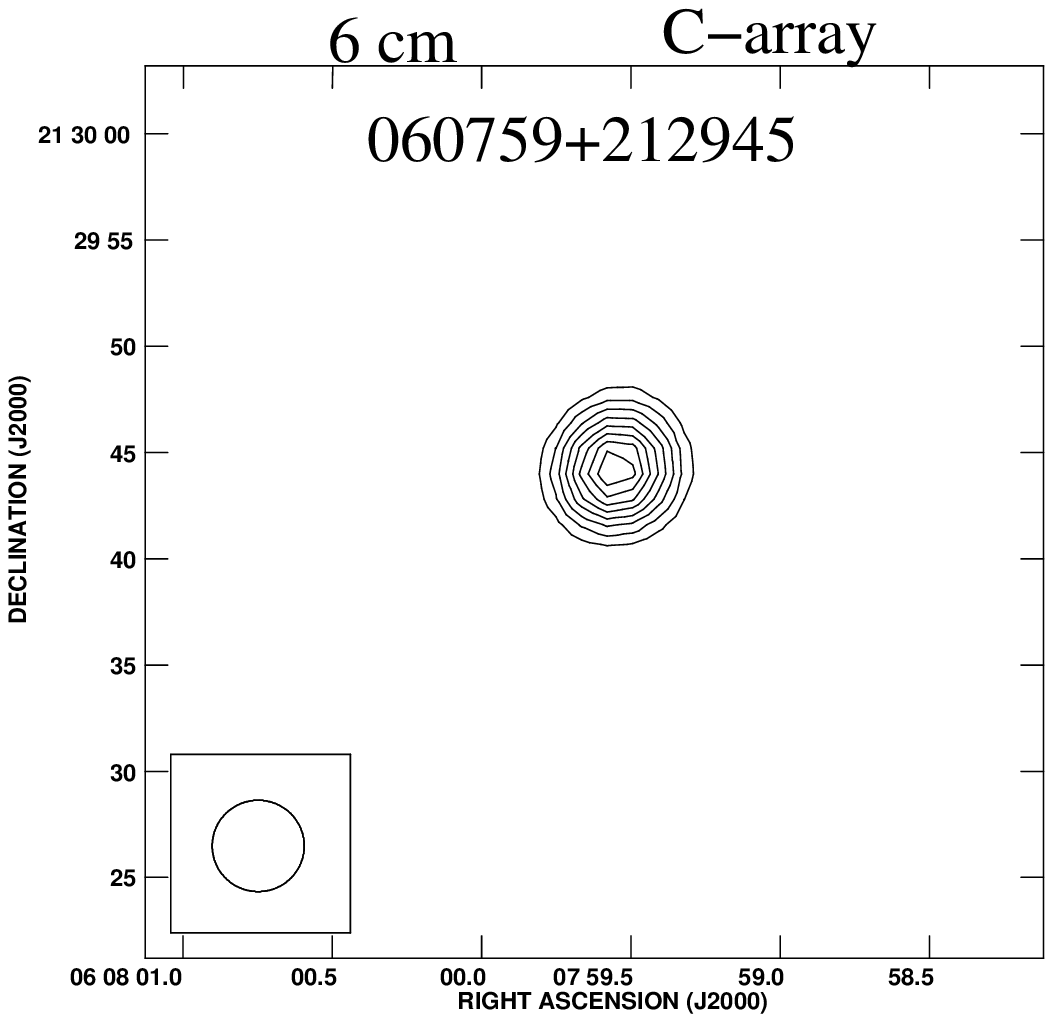} &
\includegraphics[width=5cm]{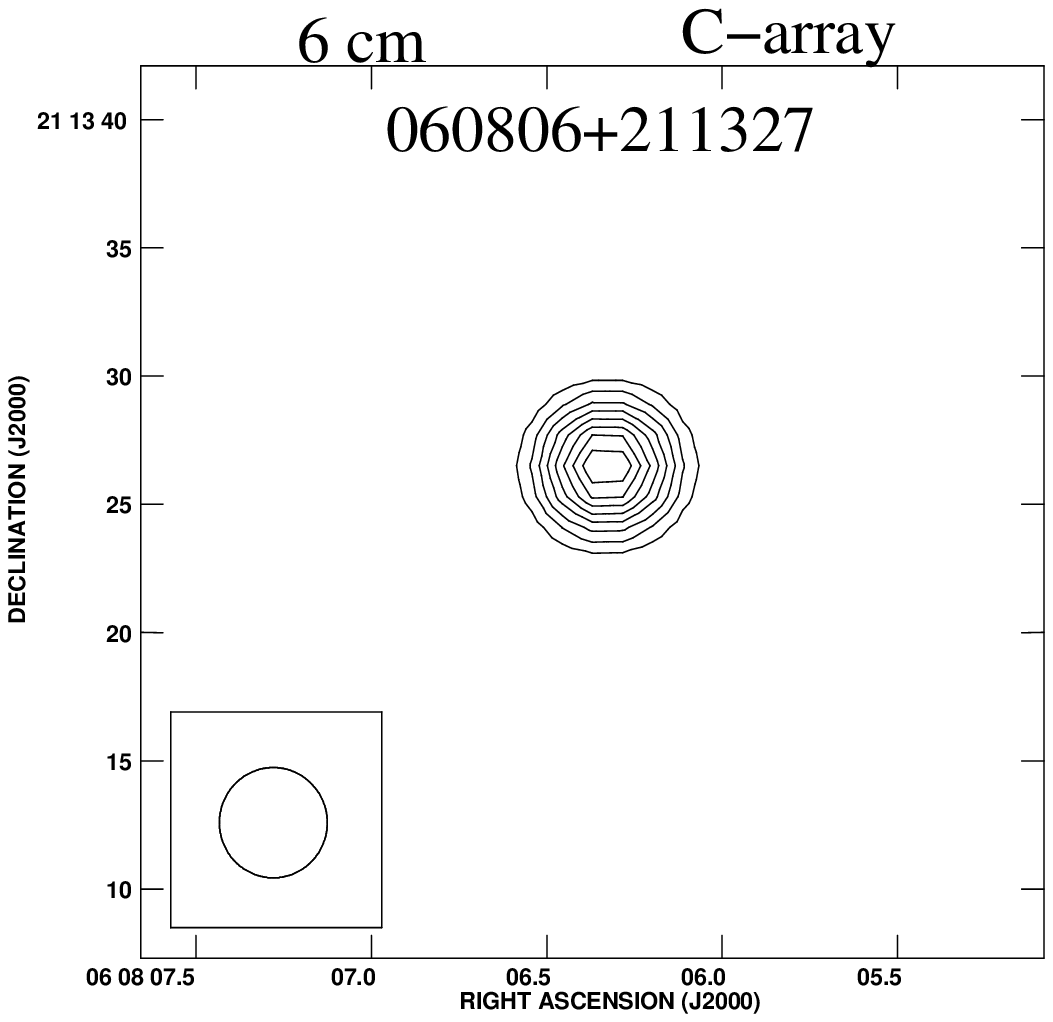}
\\

\end{tabular}
\end{figure*}
\begin{figure*}
\begin{tabular}{ccc}
\includegraphics[width=5cm]{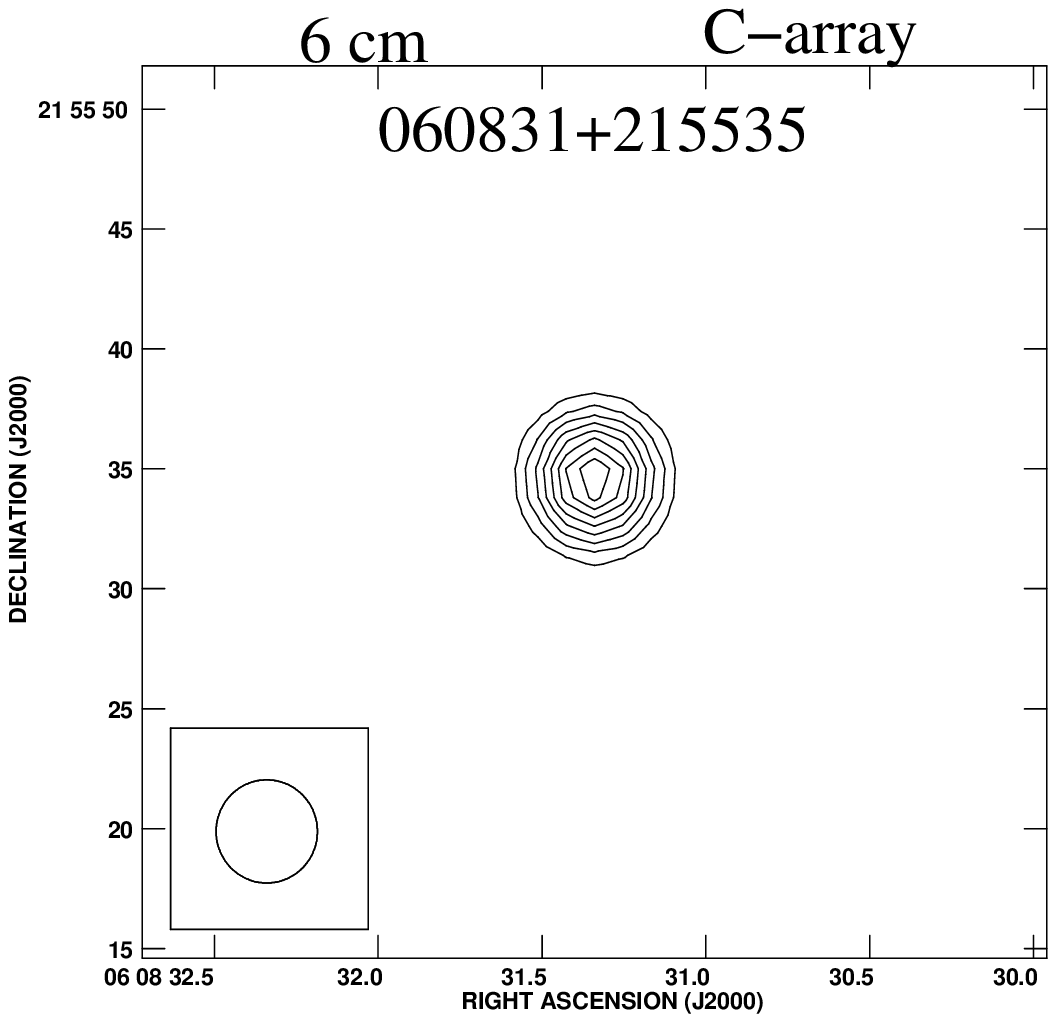} &
\includegraphics[width=5cm]{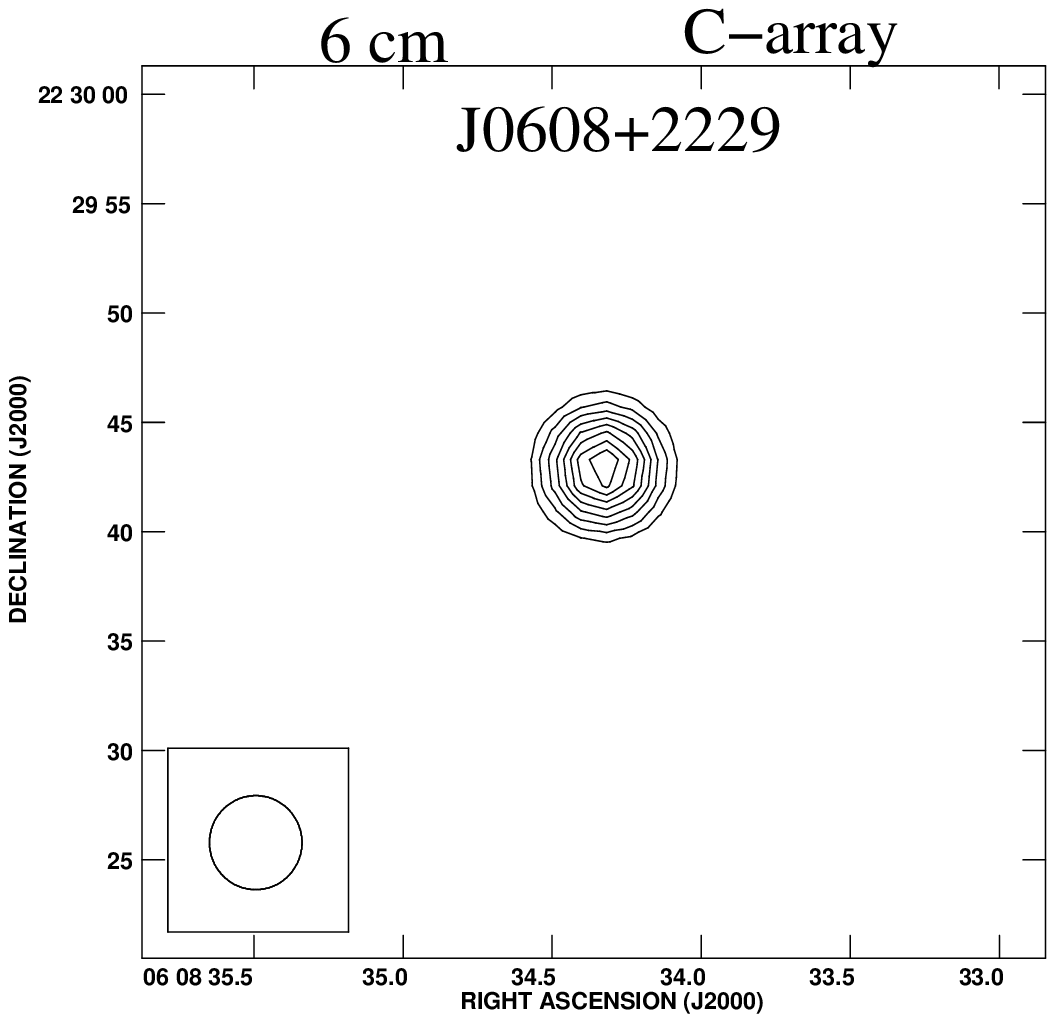} &
\includegraphics[width=5cm]{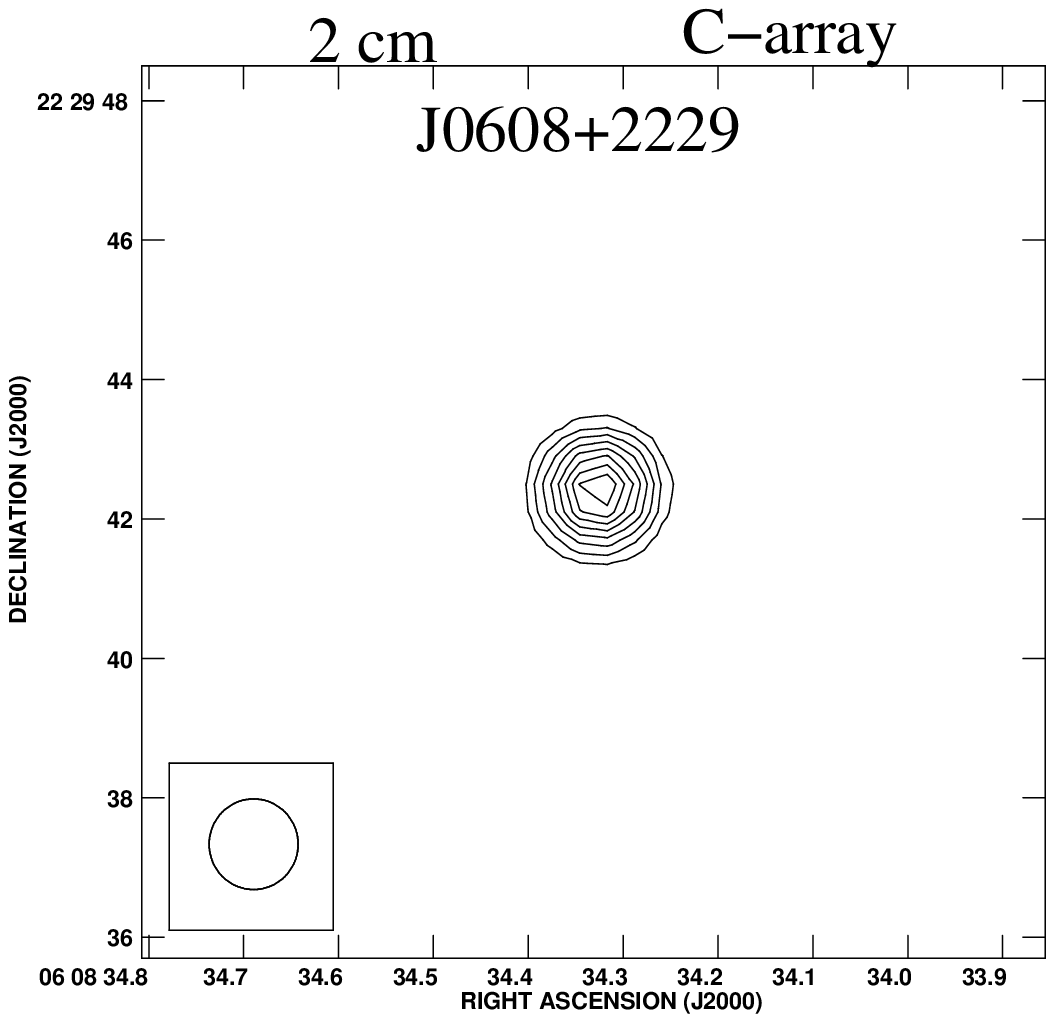}\\
\includegraphics[width=5cm]{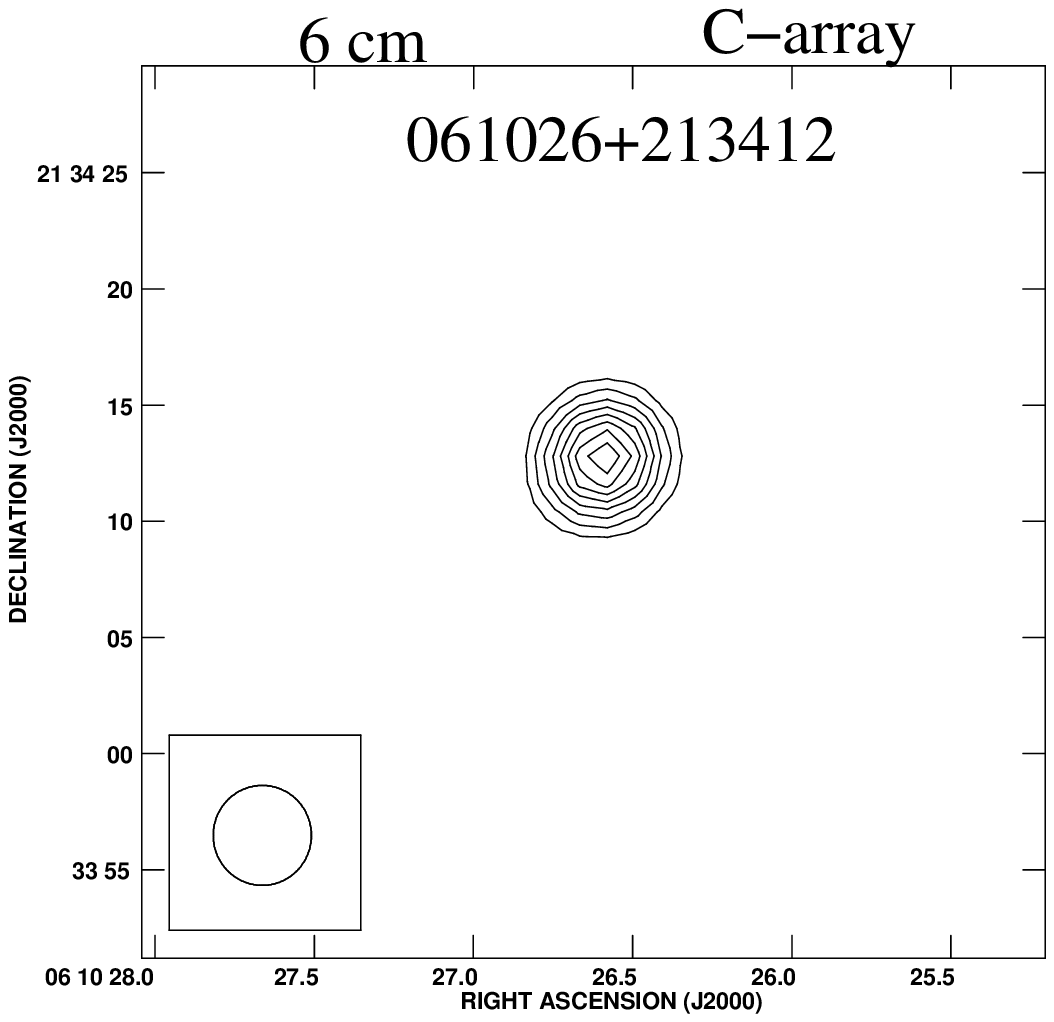} &
\includegraphics[width=5cm]{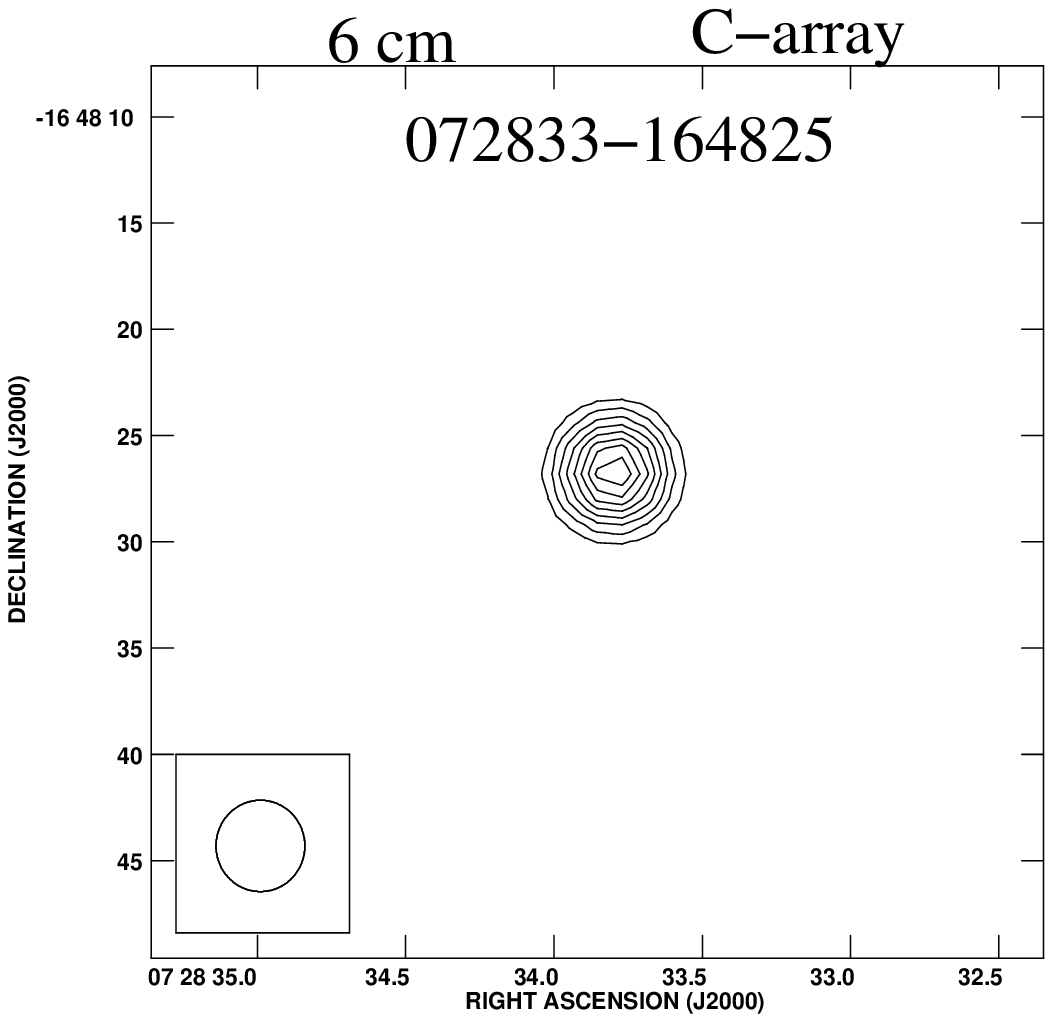} &
\includegraphics[width=5cm]{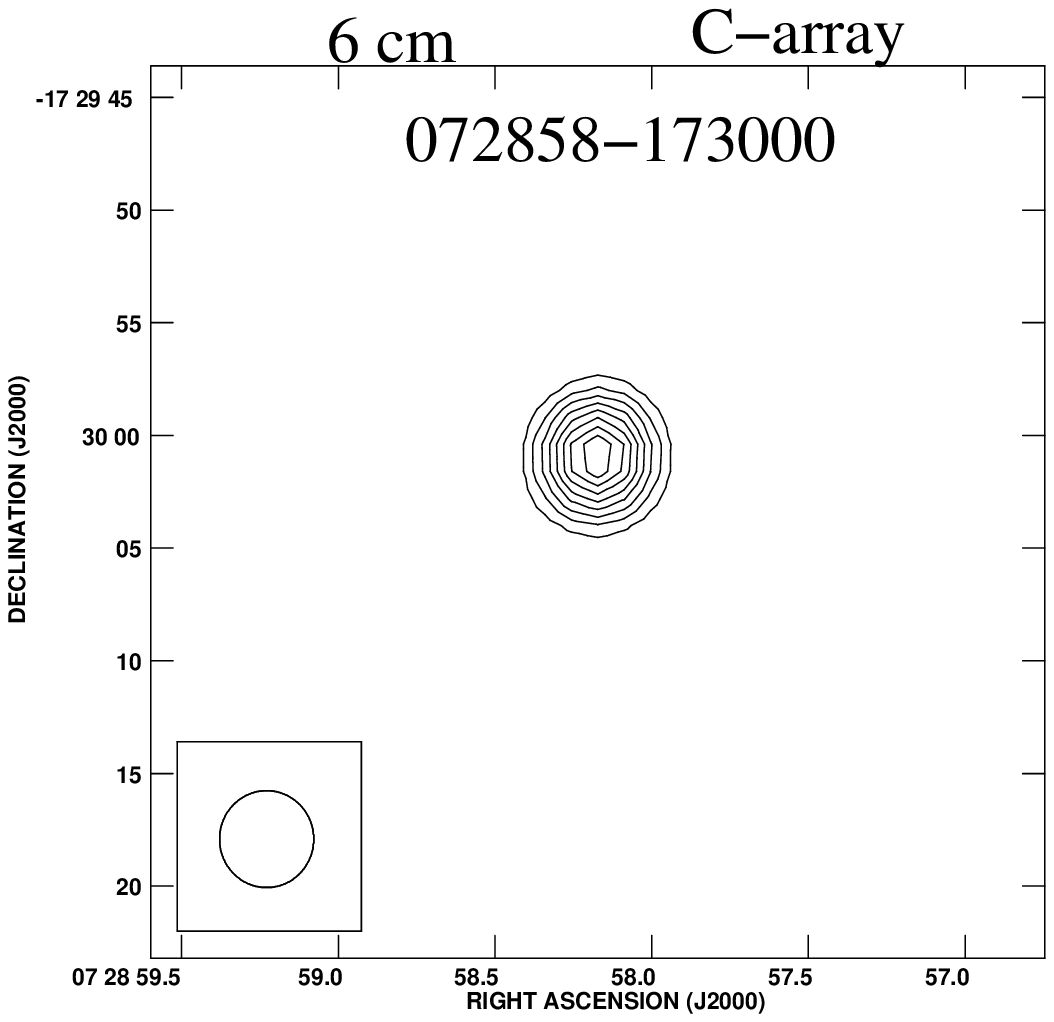}\\
\includegraphics[width=5cm]{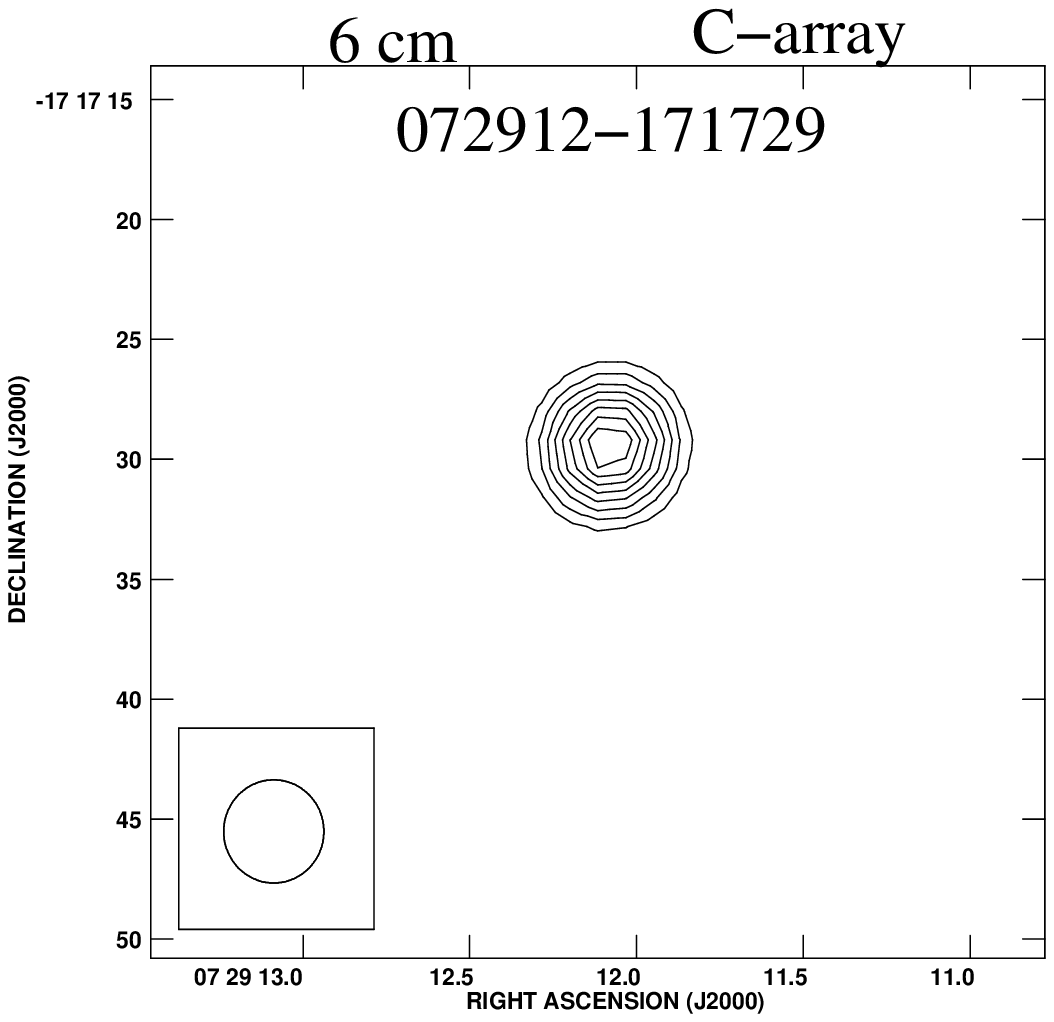} &
\includegraphics[width=5cm]{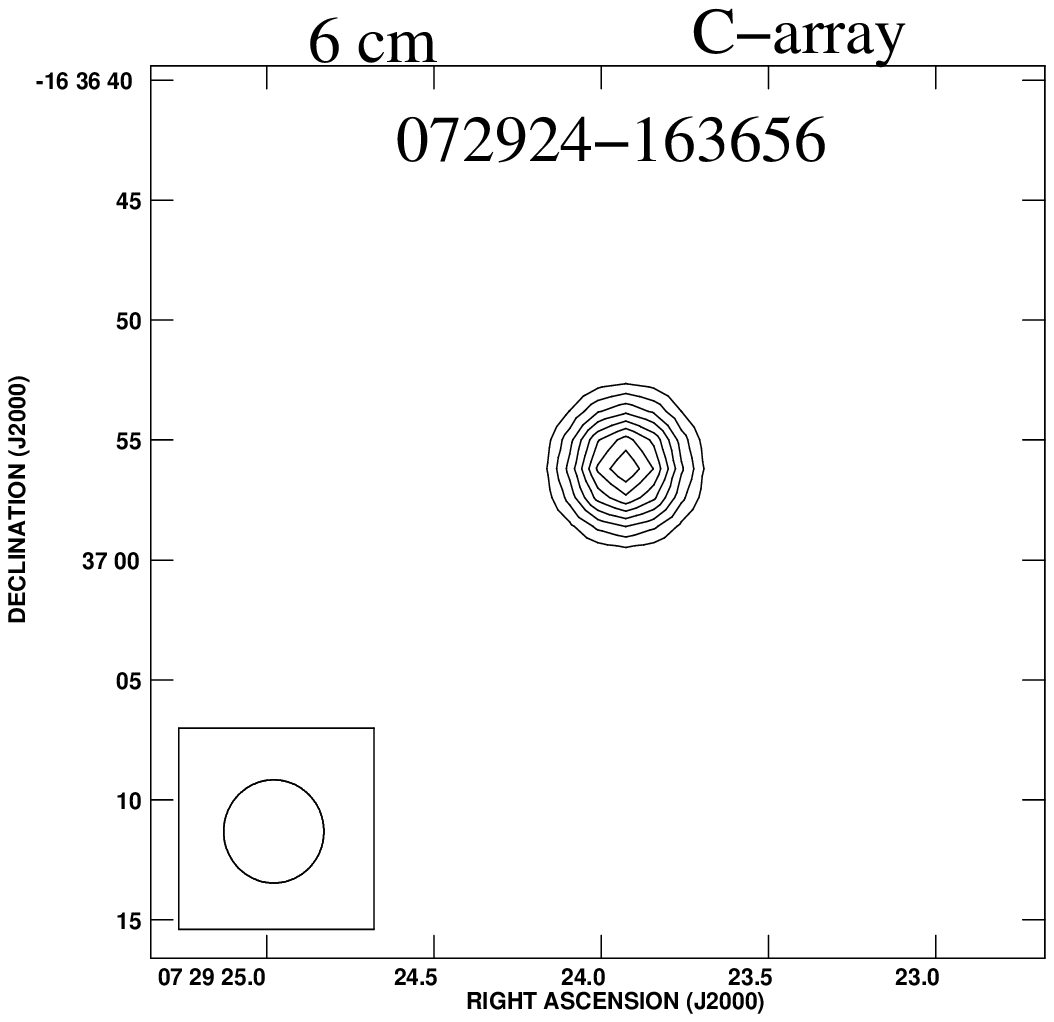} &
\includegraphics[width=5cm]{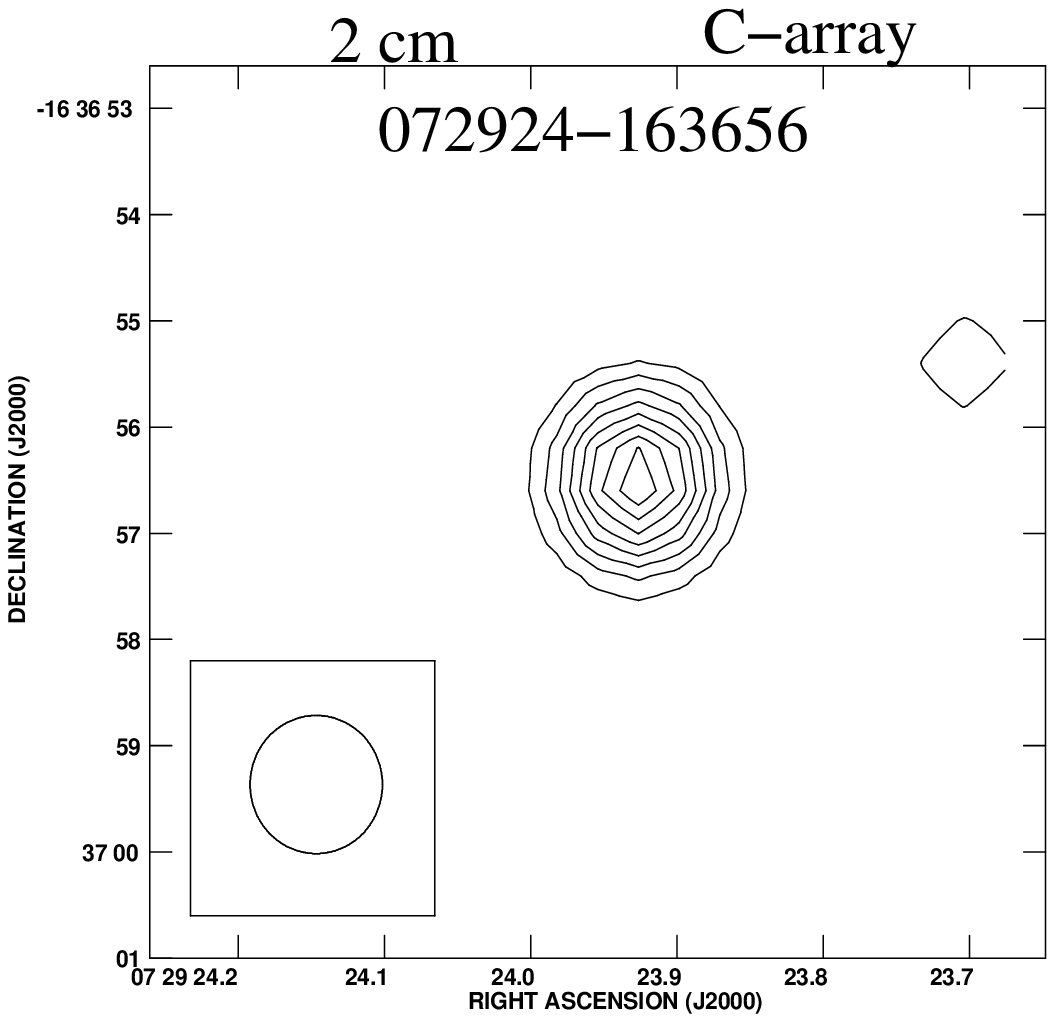}\\
\includegraphics[width=5cm]{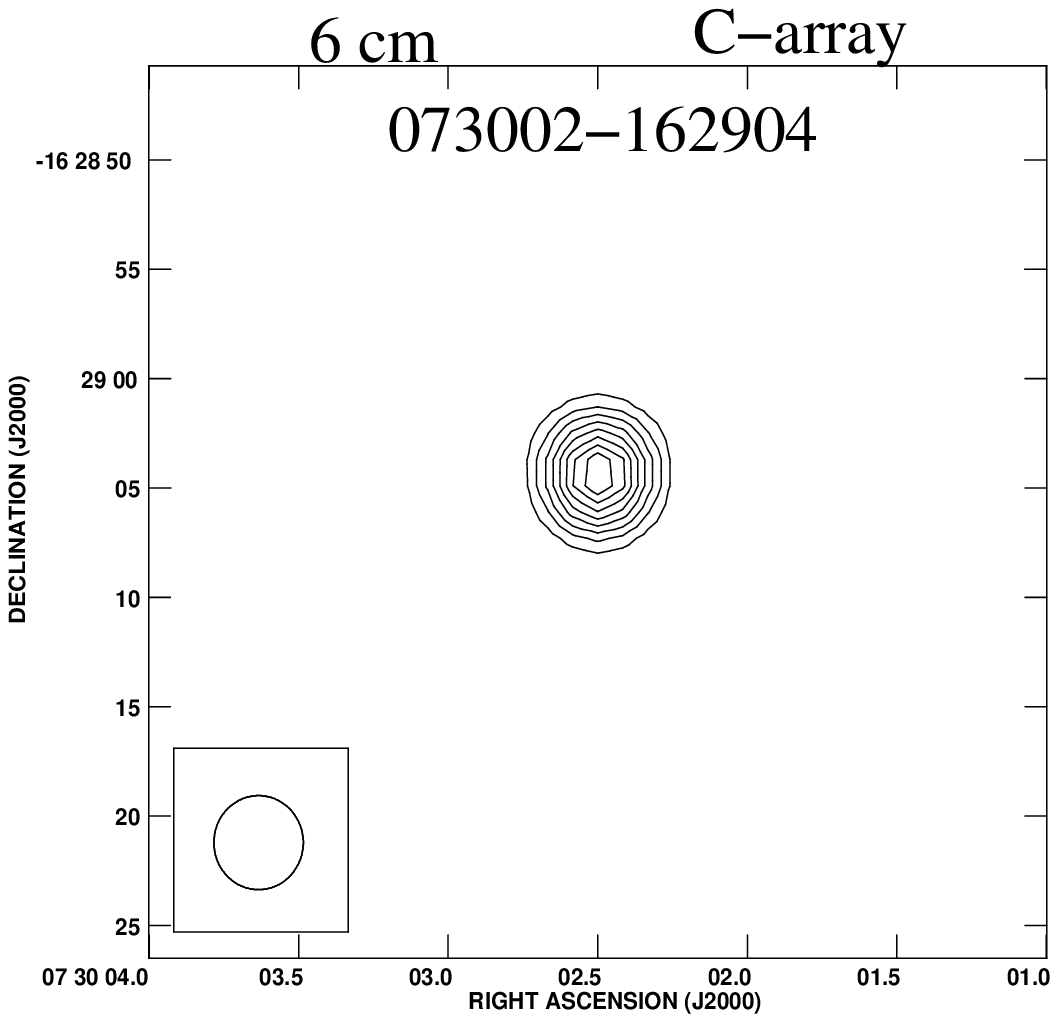} &
\includegraphics[width=5cm]{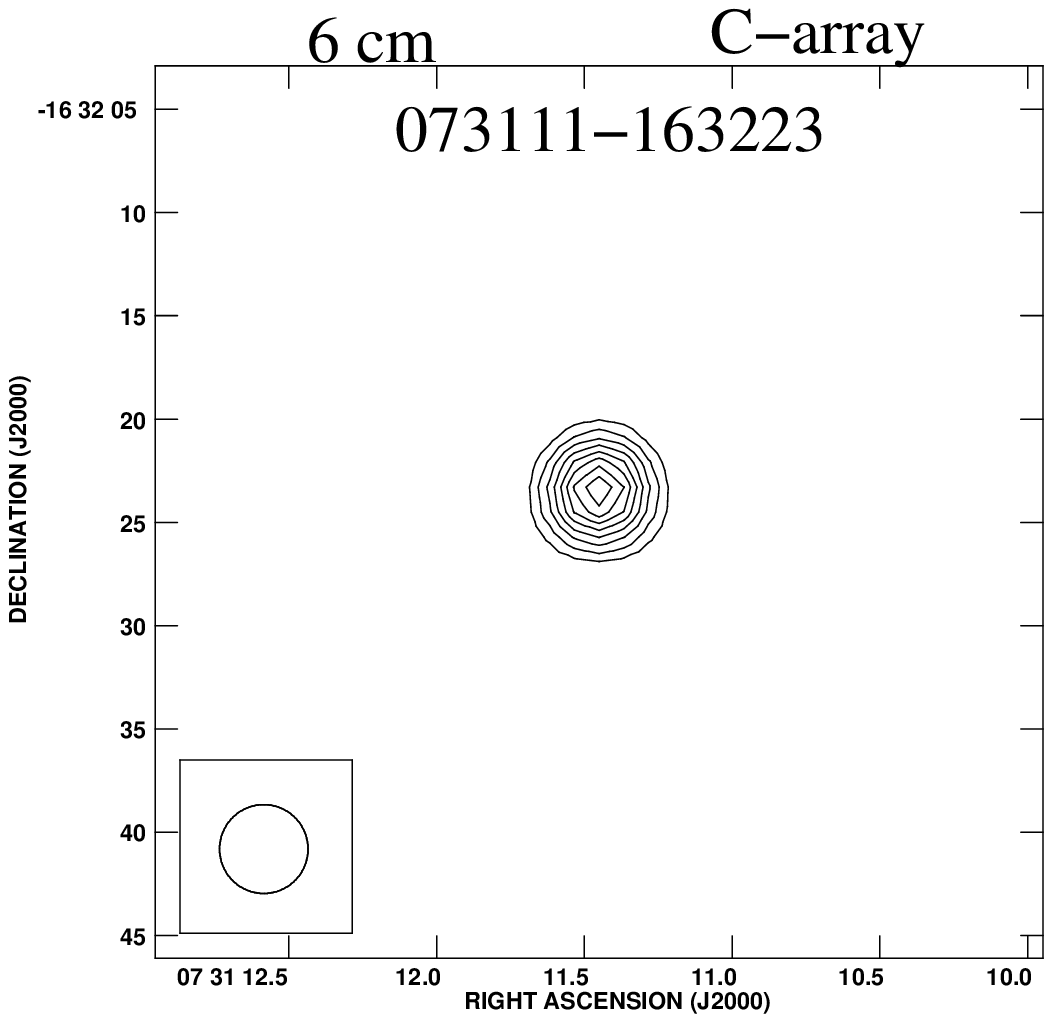} &
\includegraphics[width=5cm]{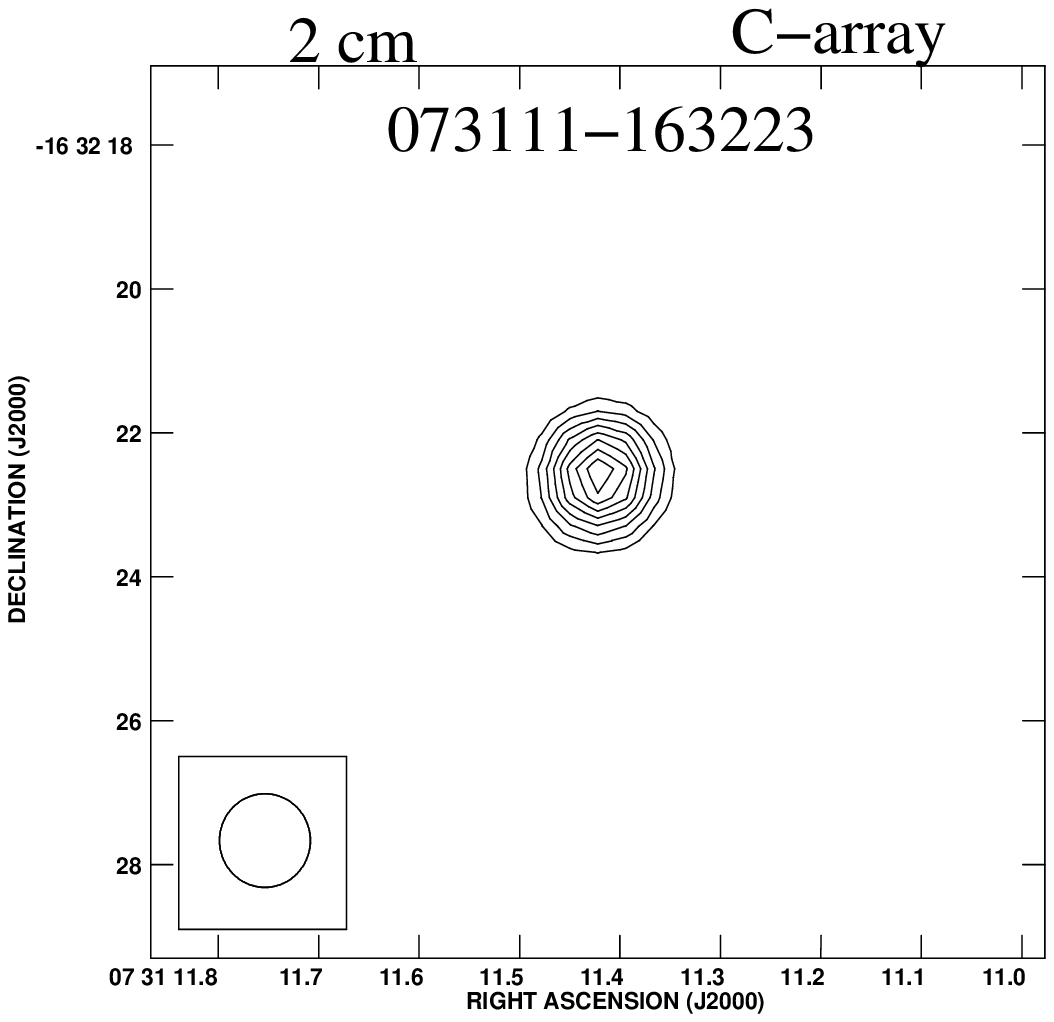}
\\

\end{tabular}
\end{figure*}

\begin{figure*}
\begin{tabular}{ccc}
\includegraphics[width=5cm]{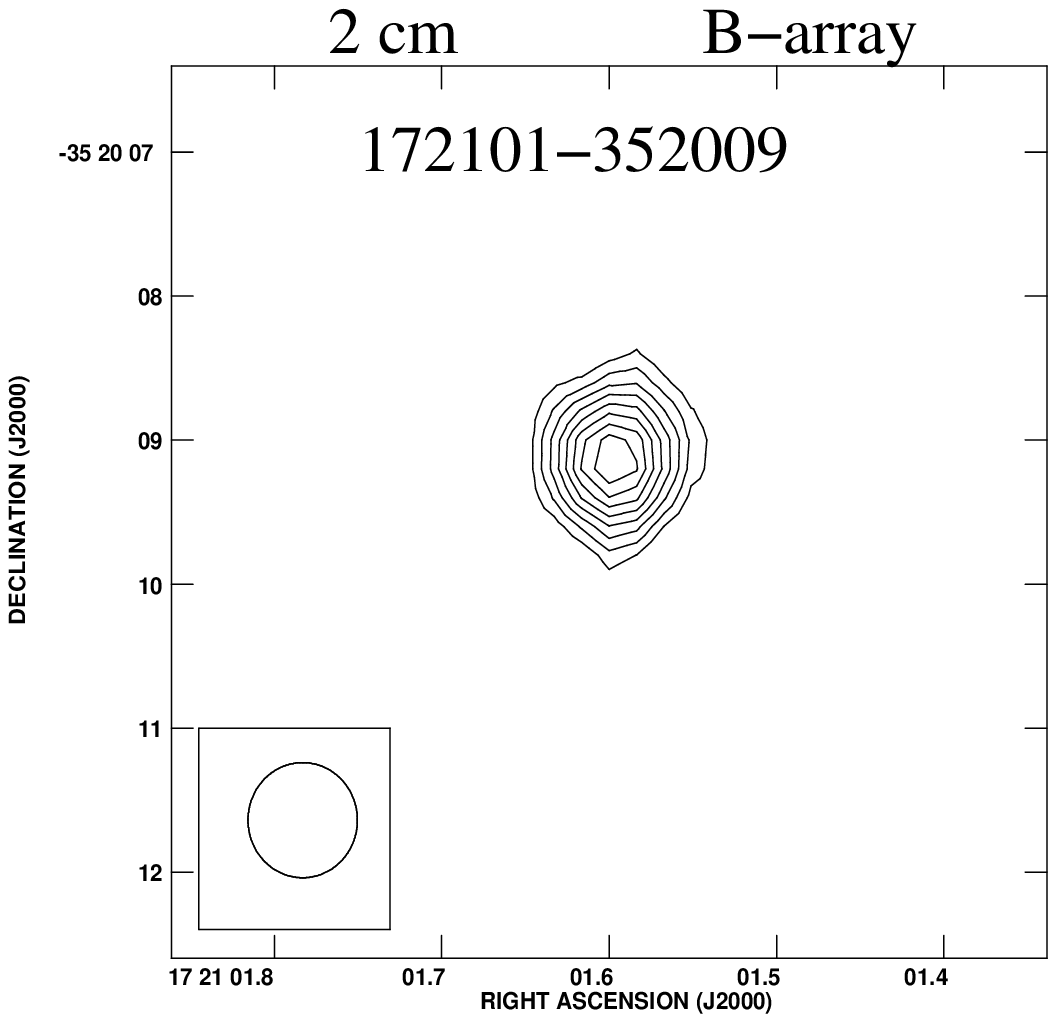}&
\includegraphics[width=5cm]{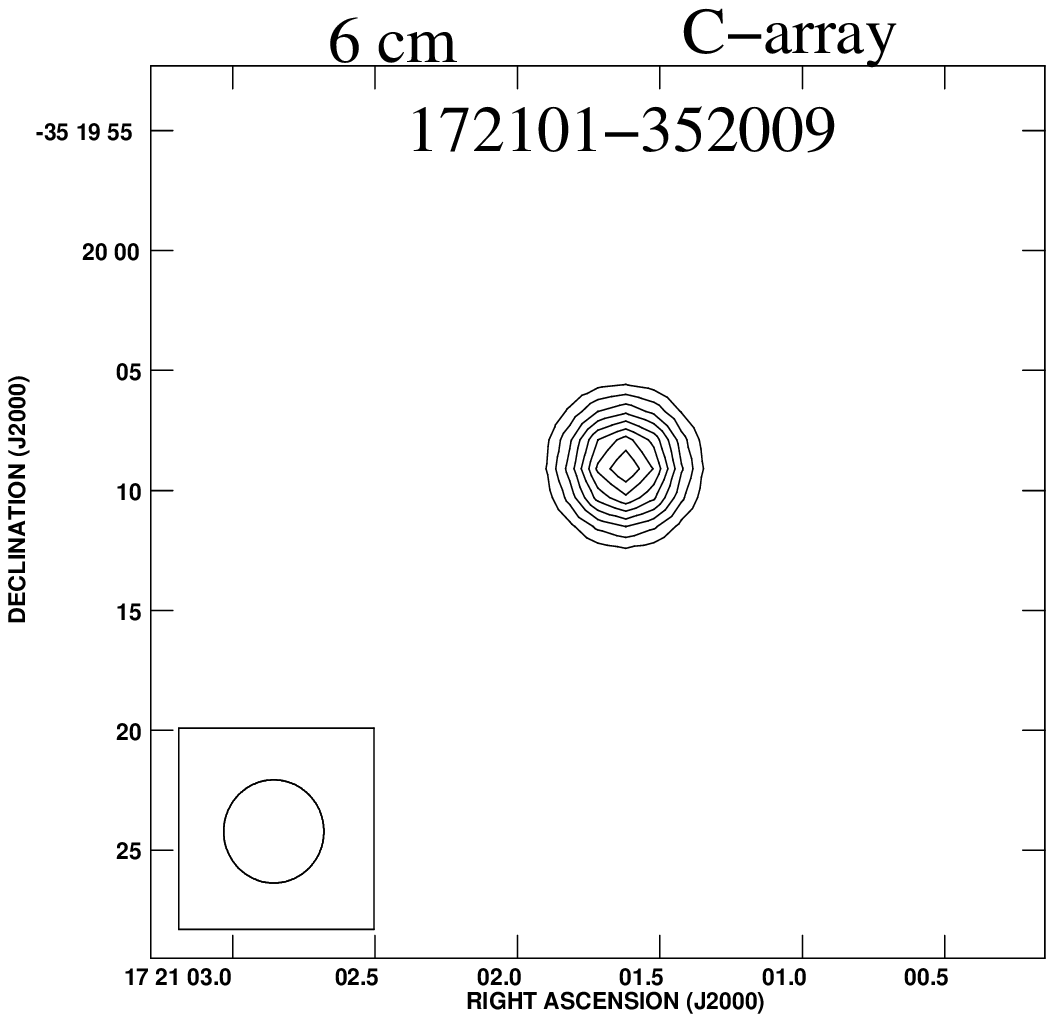} &
\includegraphics[width=5cm]{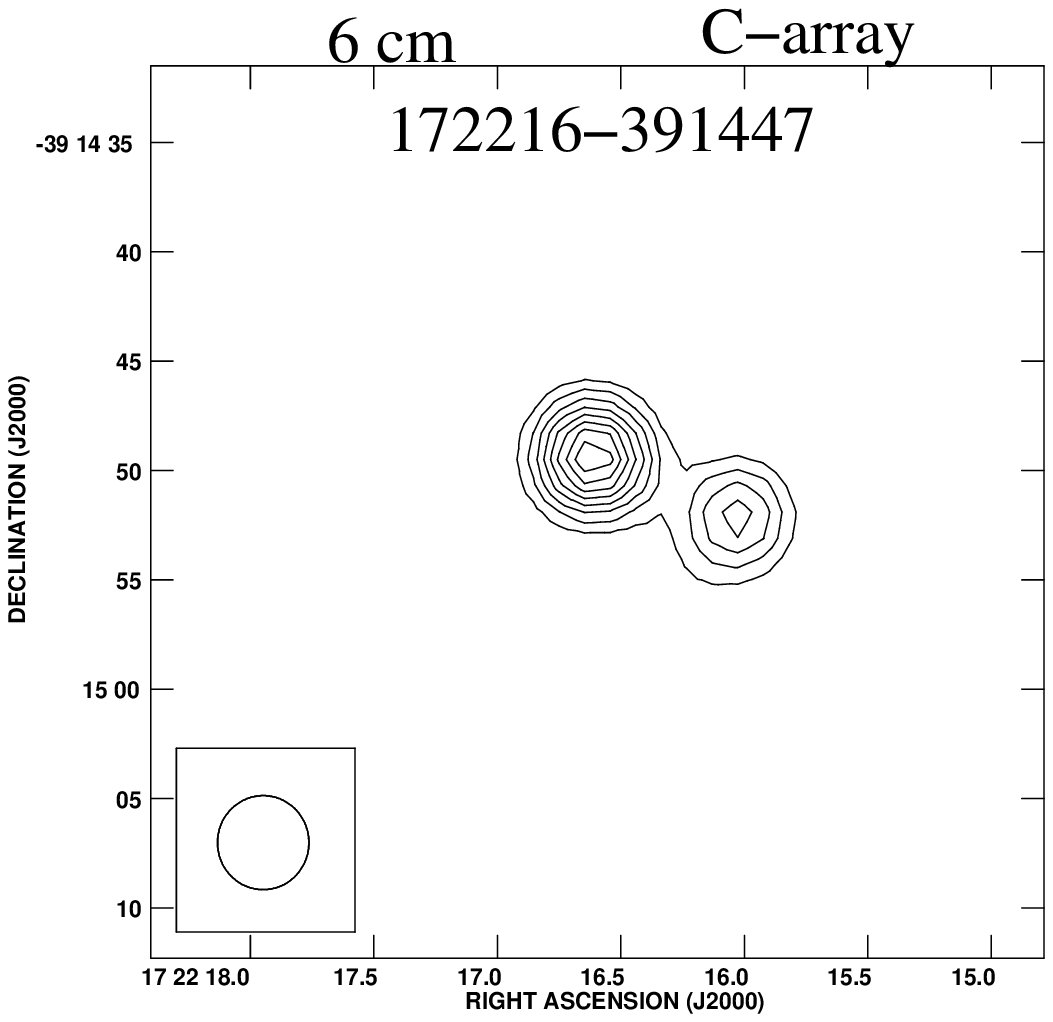}\\
\includegraphics[width=5cm]{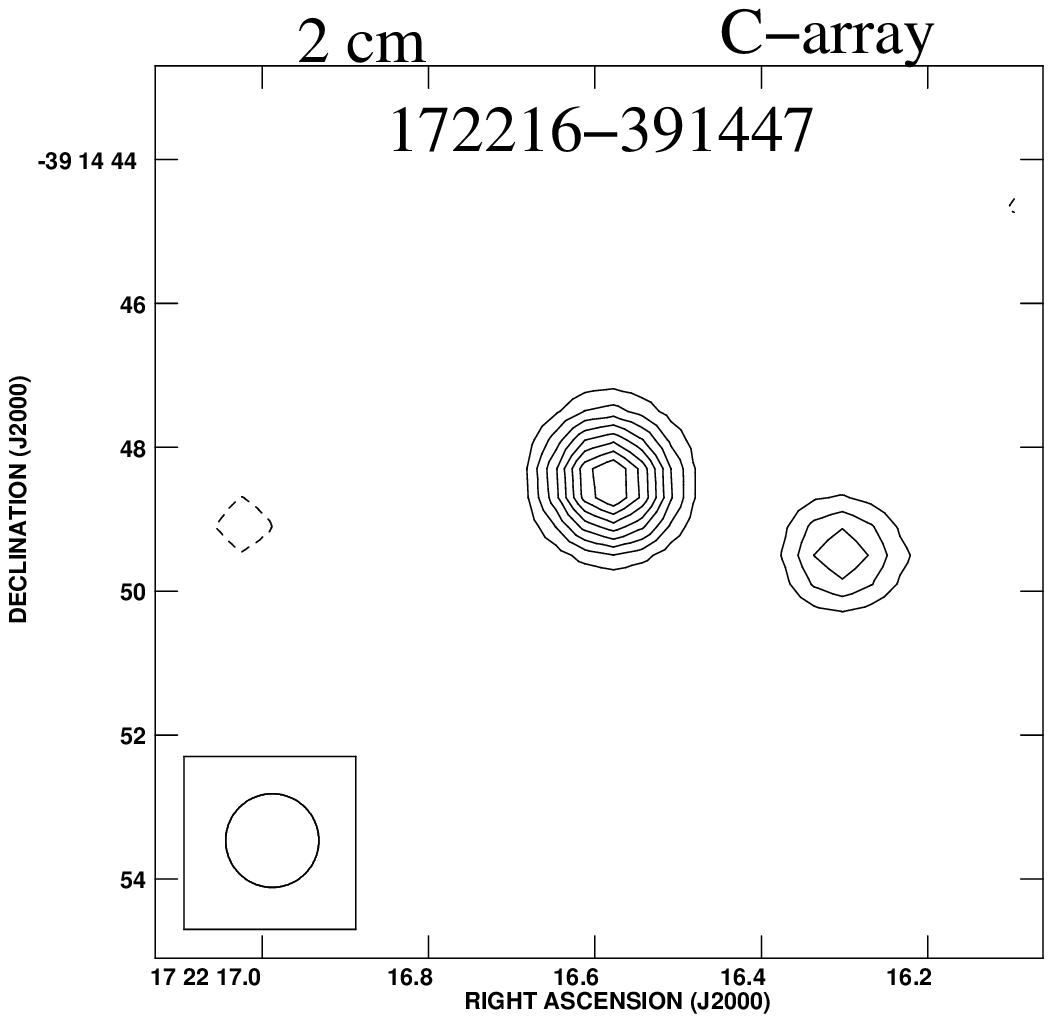}&
\includegraphics[width=5cm]{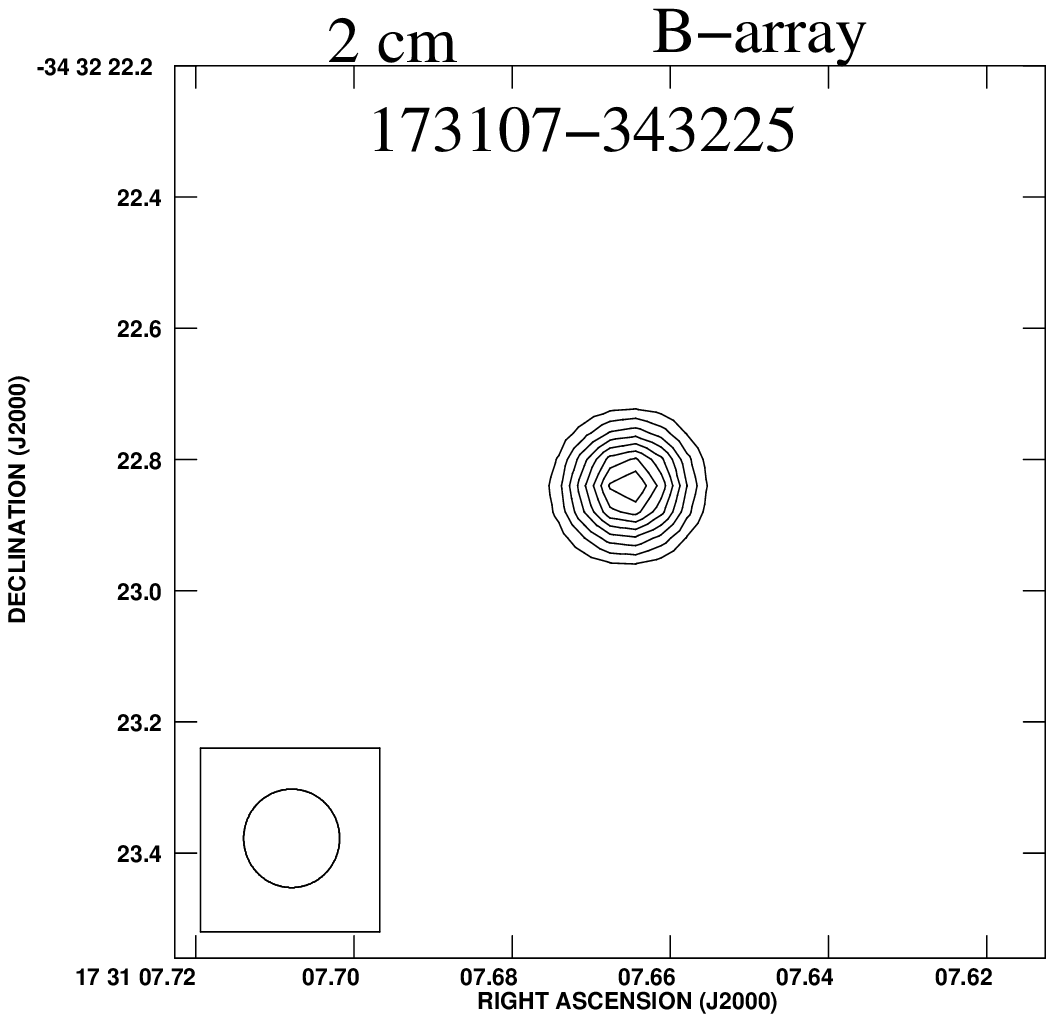} &
\includegraphics[width=5cm]{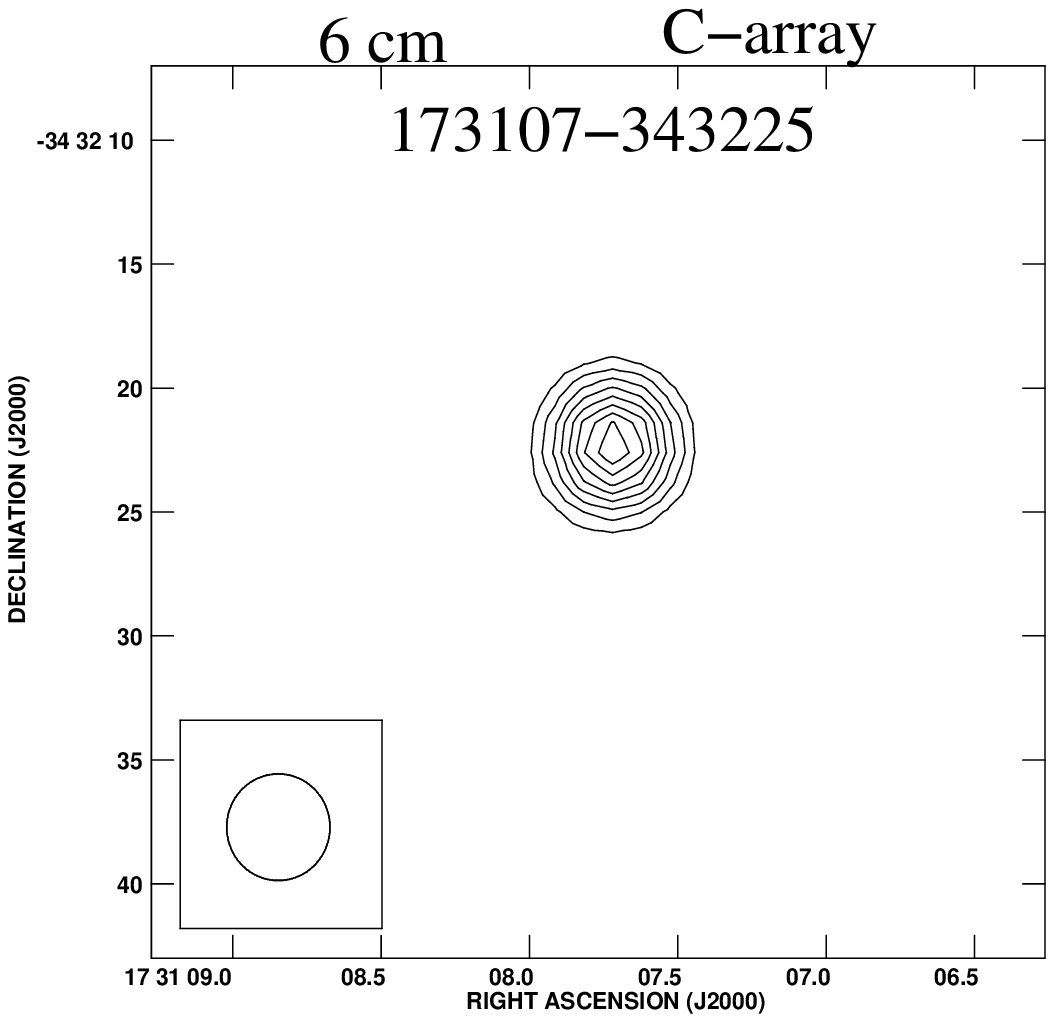}\\
\includegraphics[width=5cm]{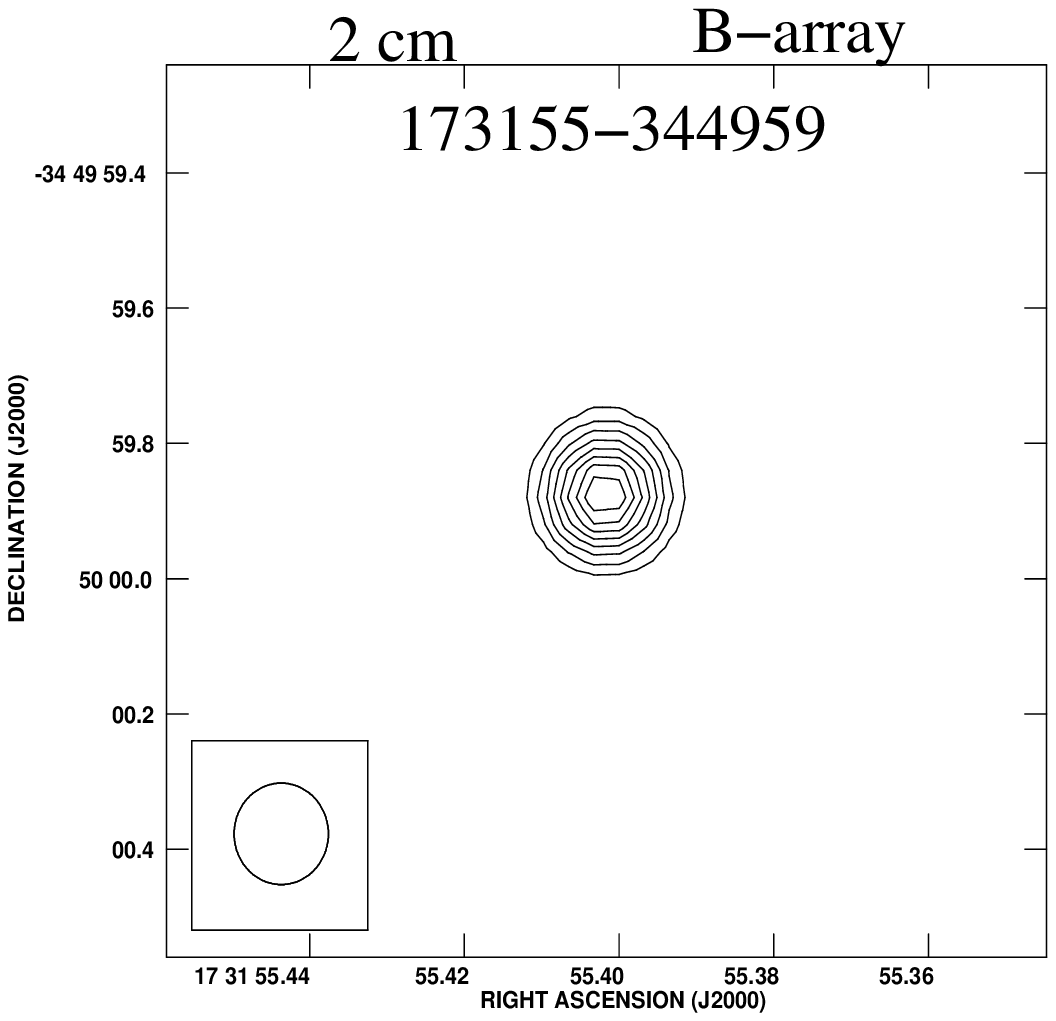}&
\includegraphics[width=5cm]{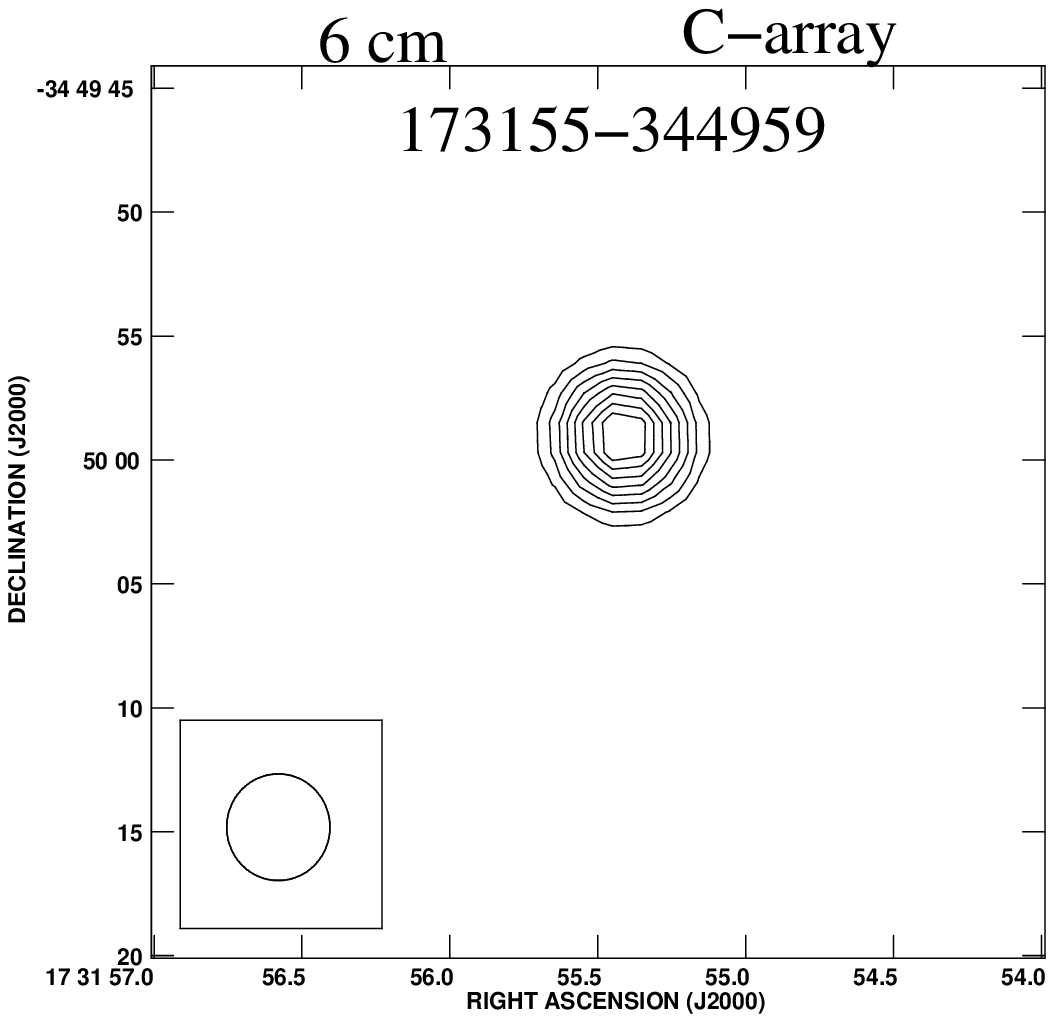} &
\includegraphics[width=5cm]{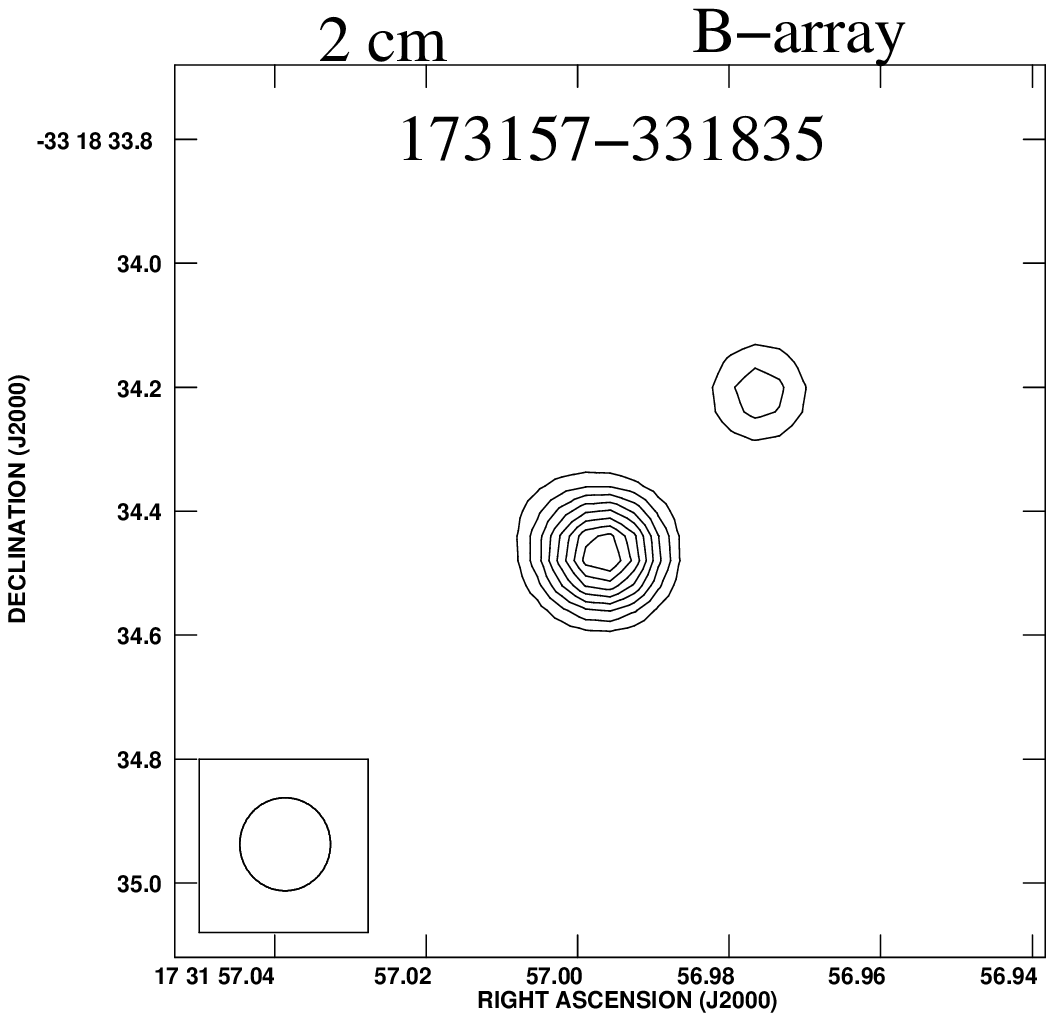}\\
\includegraphics[width=5cm]{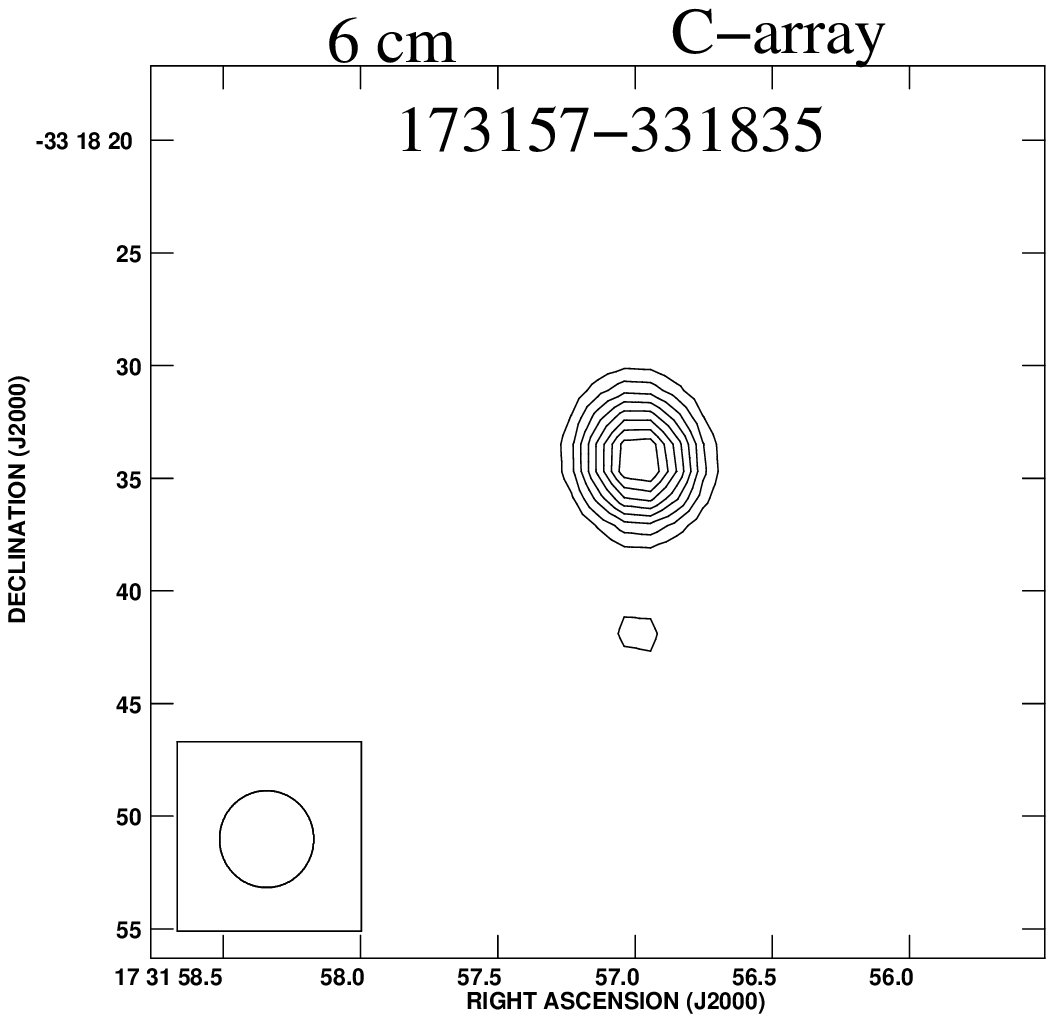}&
\includegraphics[width=5cm]{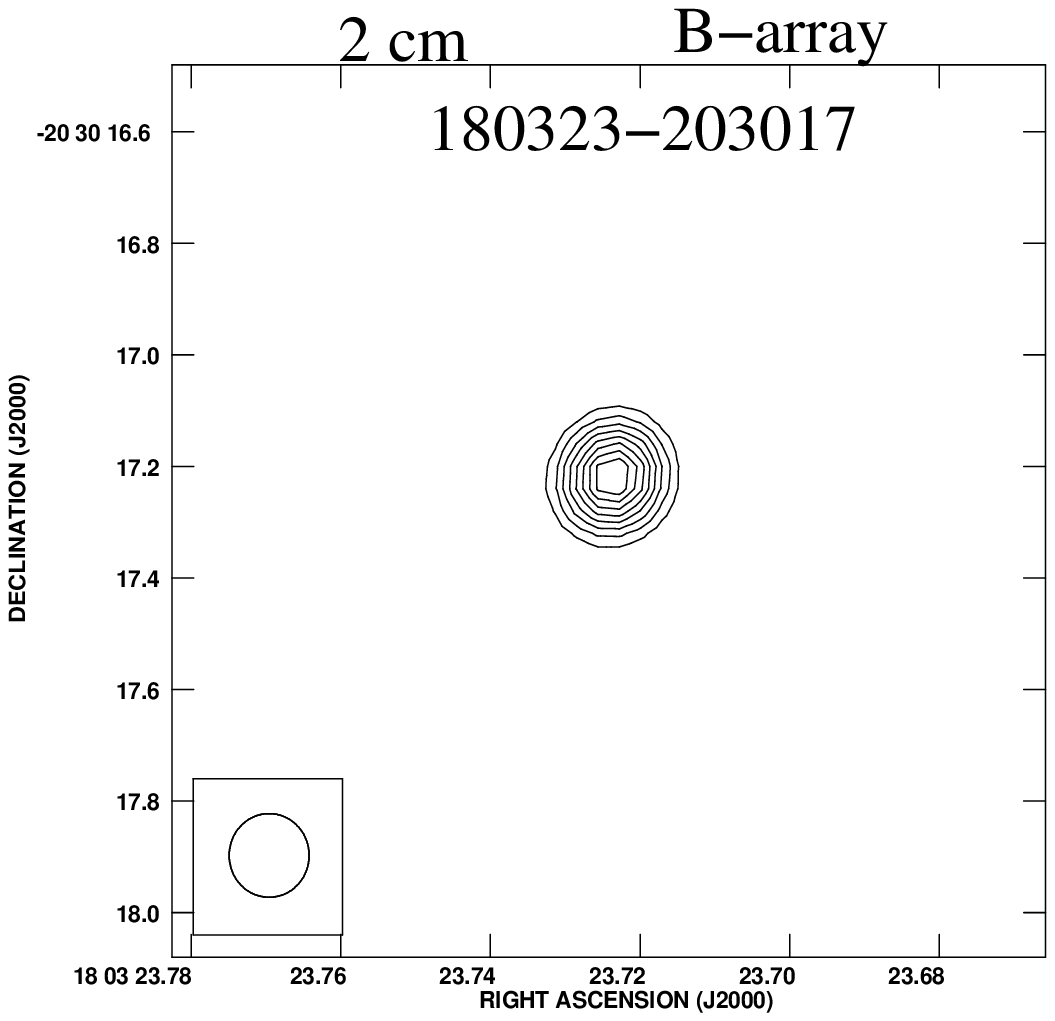} &
\includegraphics[width=5cm]{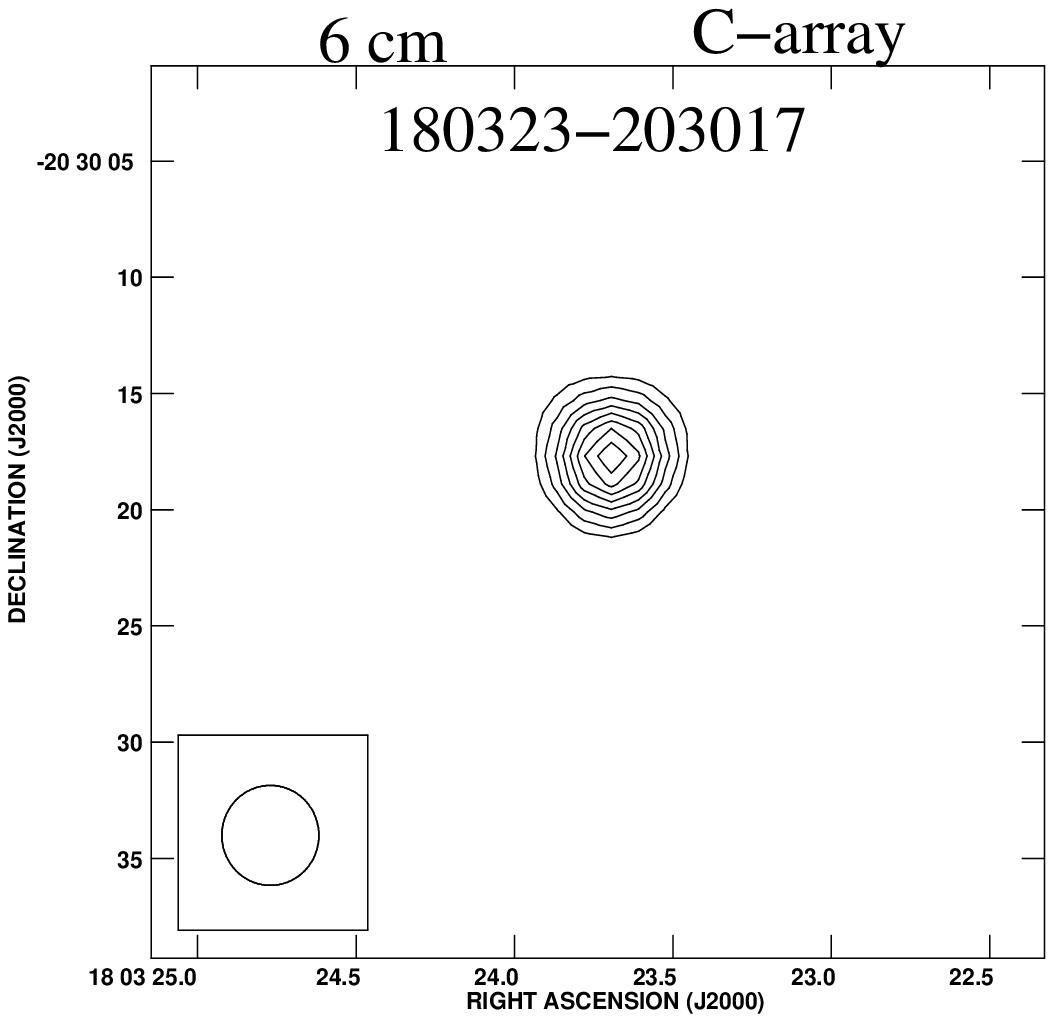}
\\

\end{tabular}
\end{figure*}
\begin{figure*}
\begin{tabular}{ccc}
\includegraphics[width=5cm]{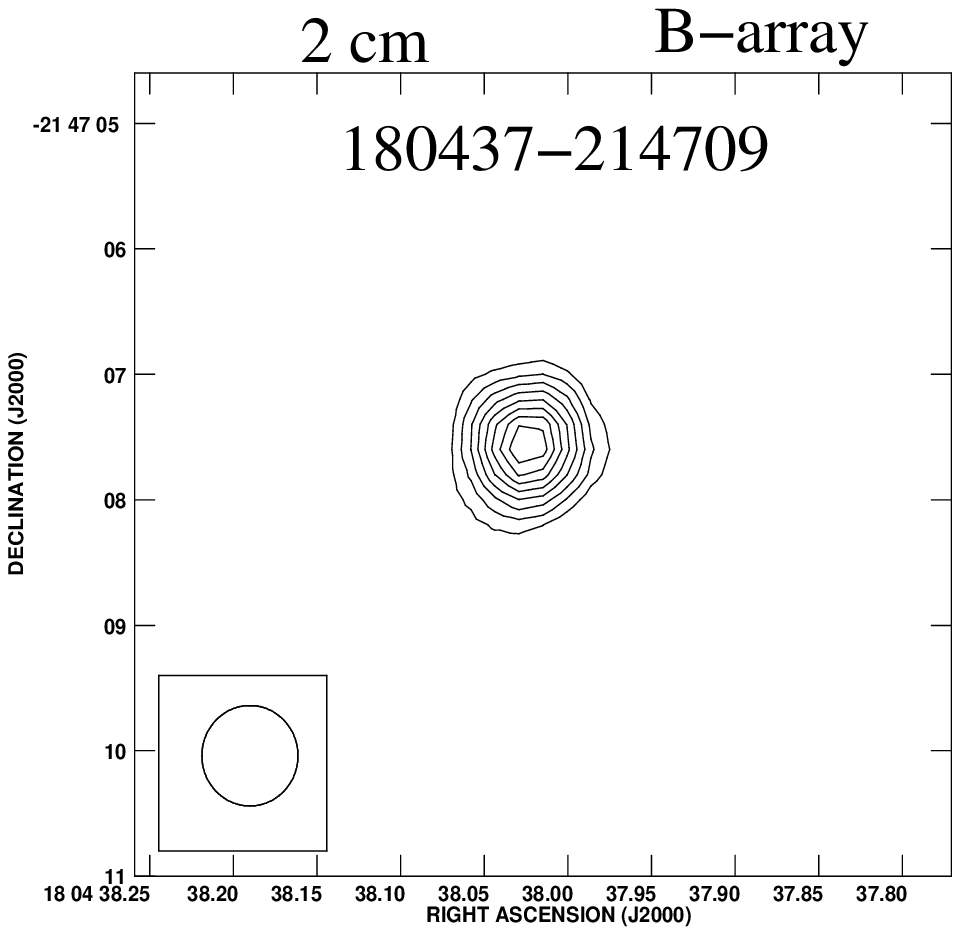}&
\includegraphics[width=5cm]{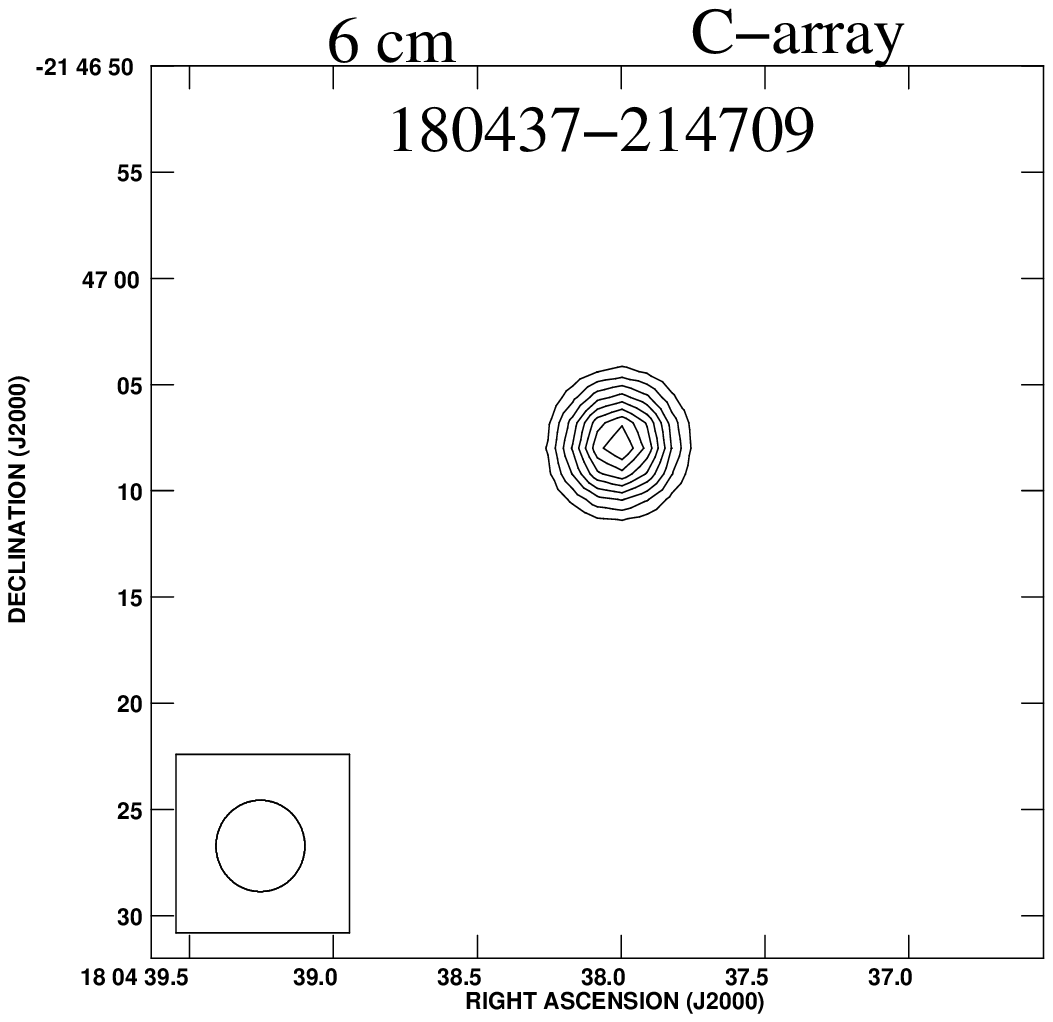} &
\includegraphics[width=5cm]{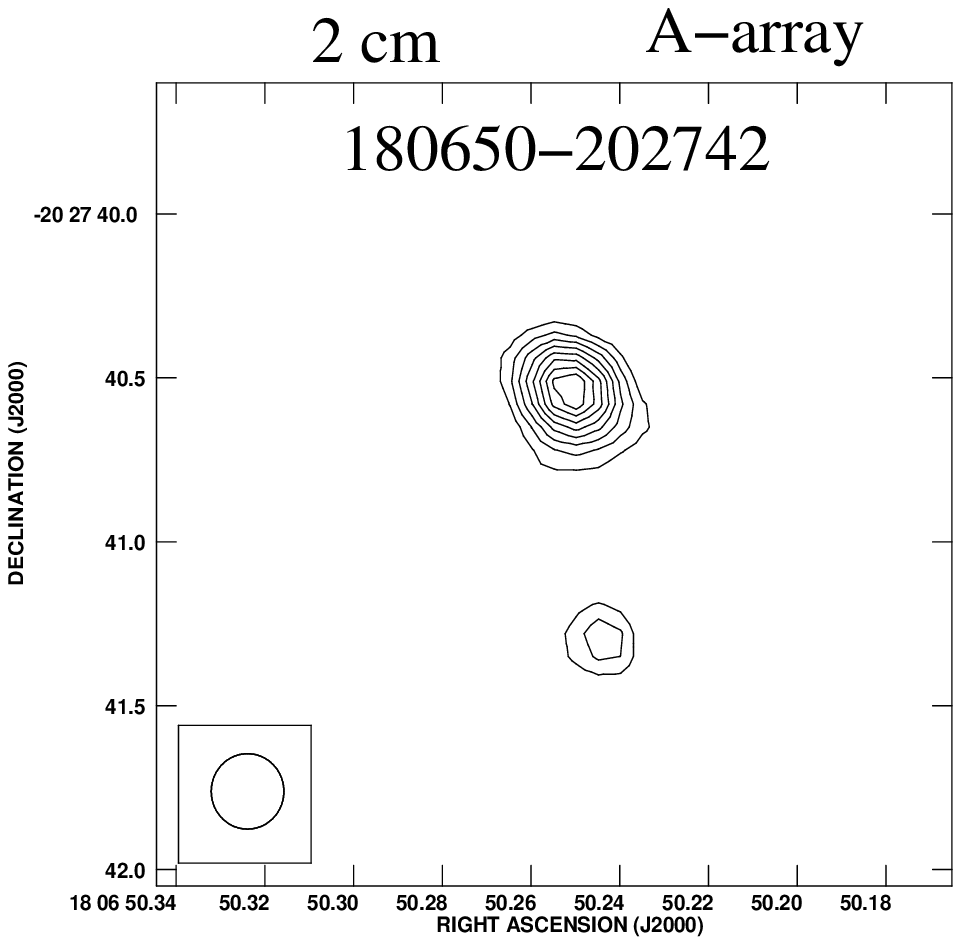}\\
\includegraphics[width=5cm]{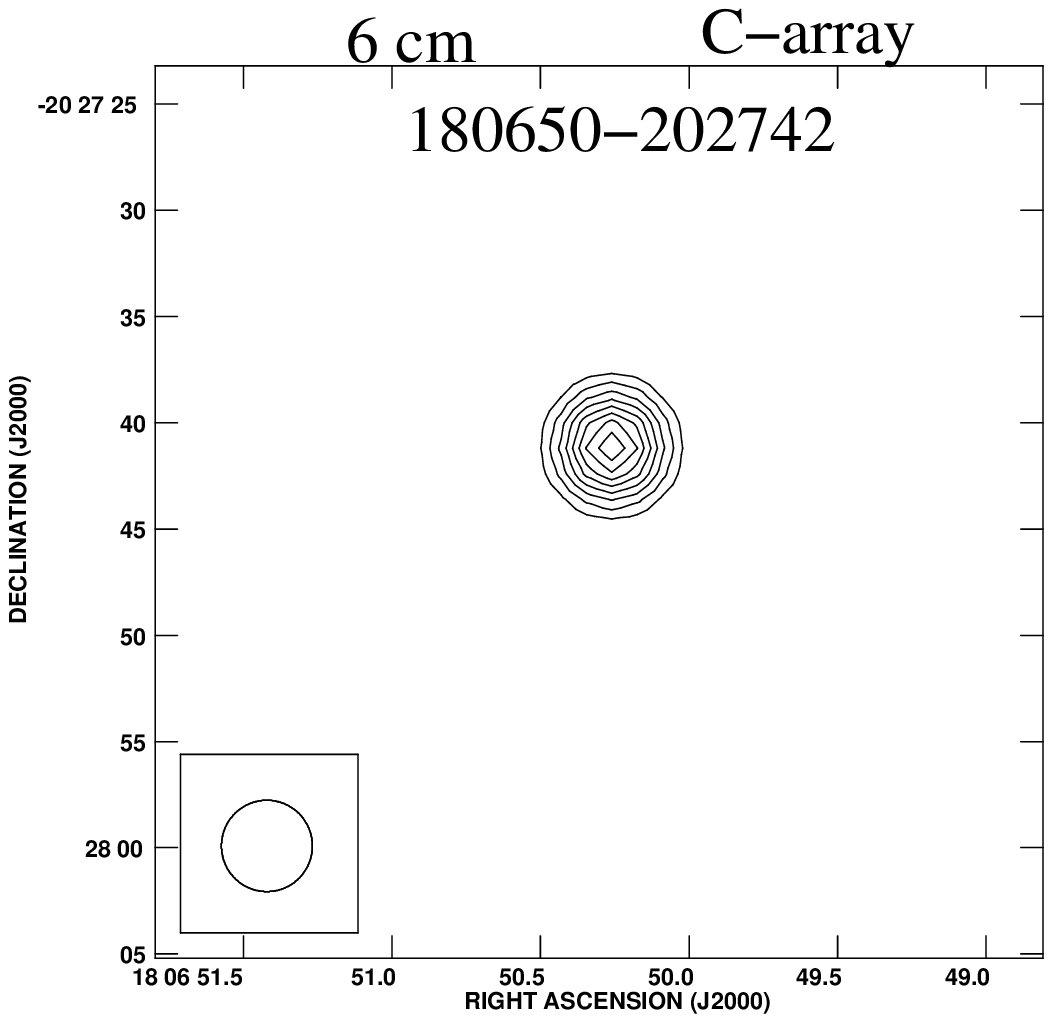}&
\includegraphics[width=5cm]{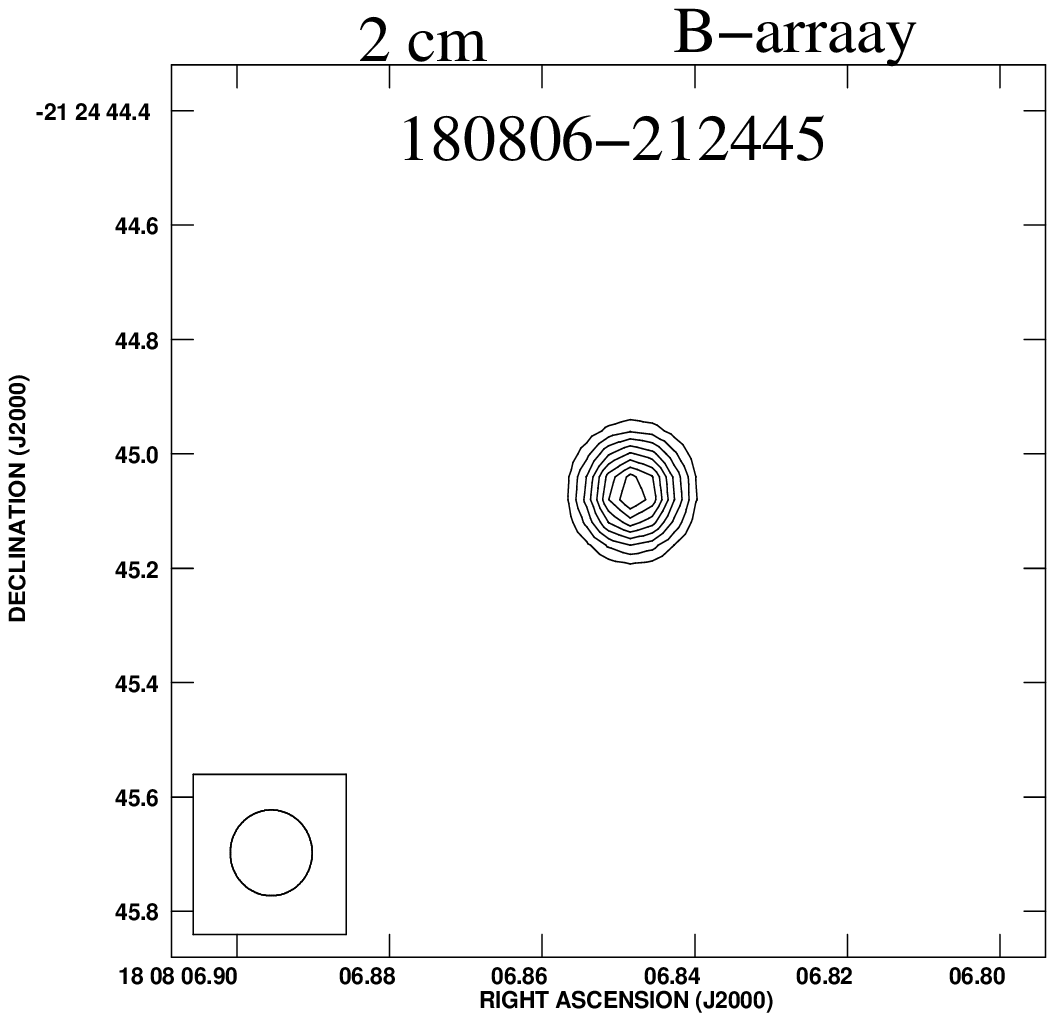} &
\includegraphics[width=5cm]{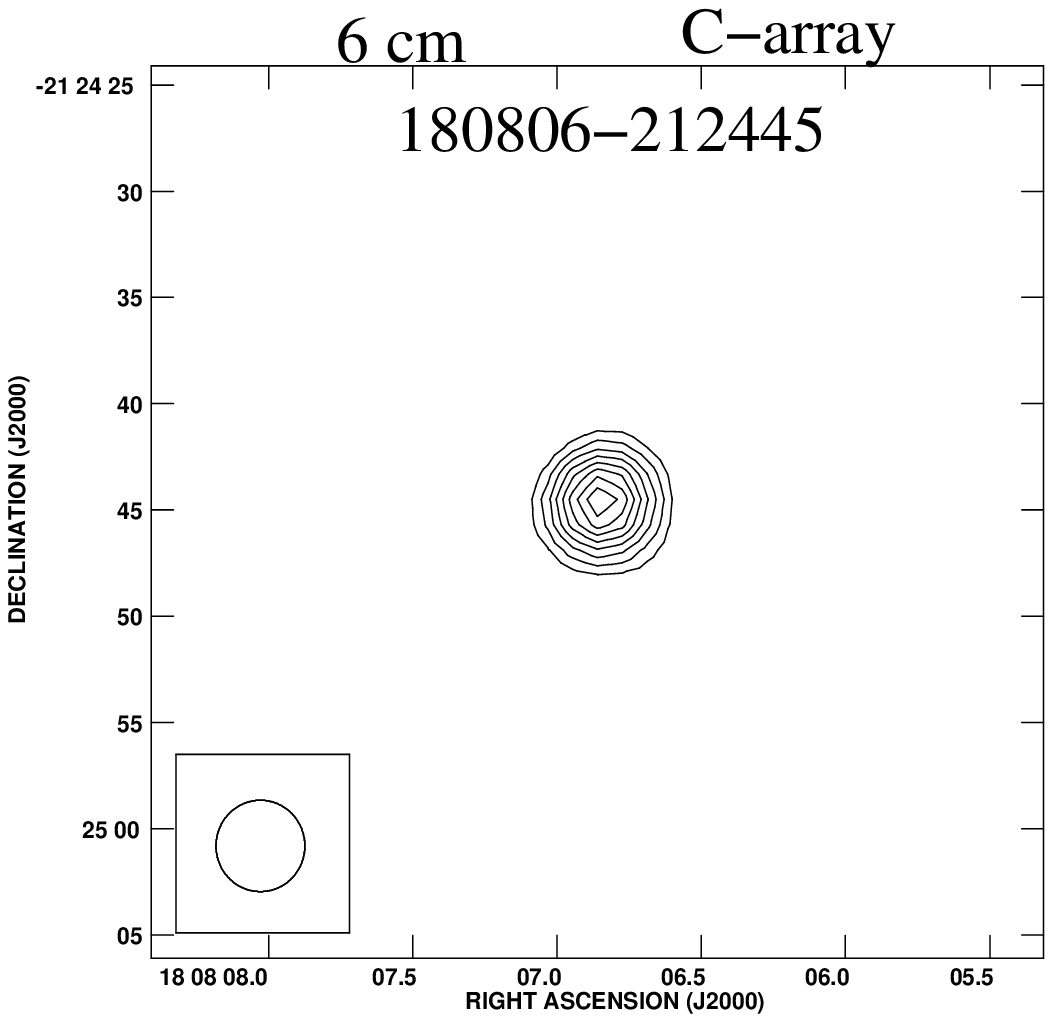}\\
\includegraphics[width=5cm]{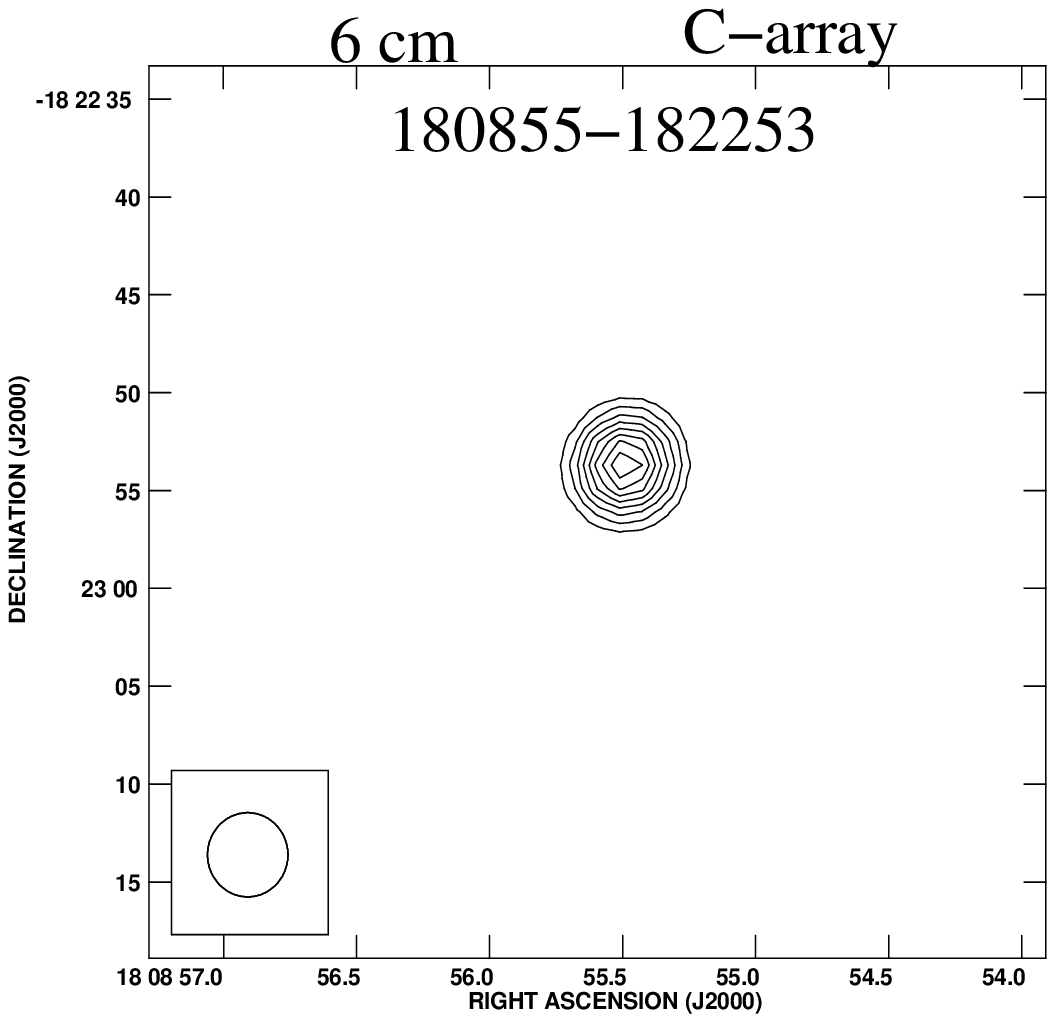}&
\includegraphics[width=5cm]{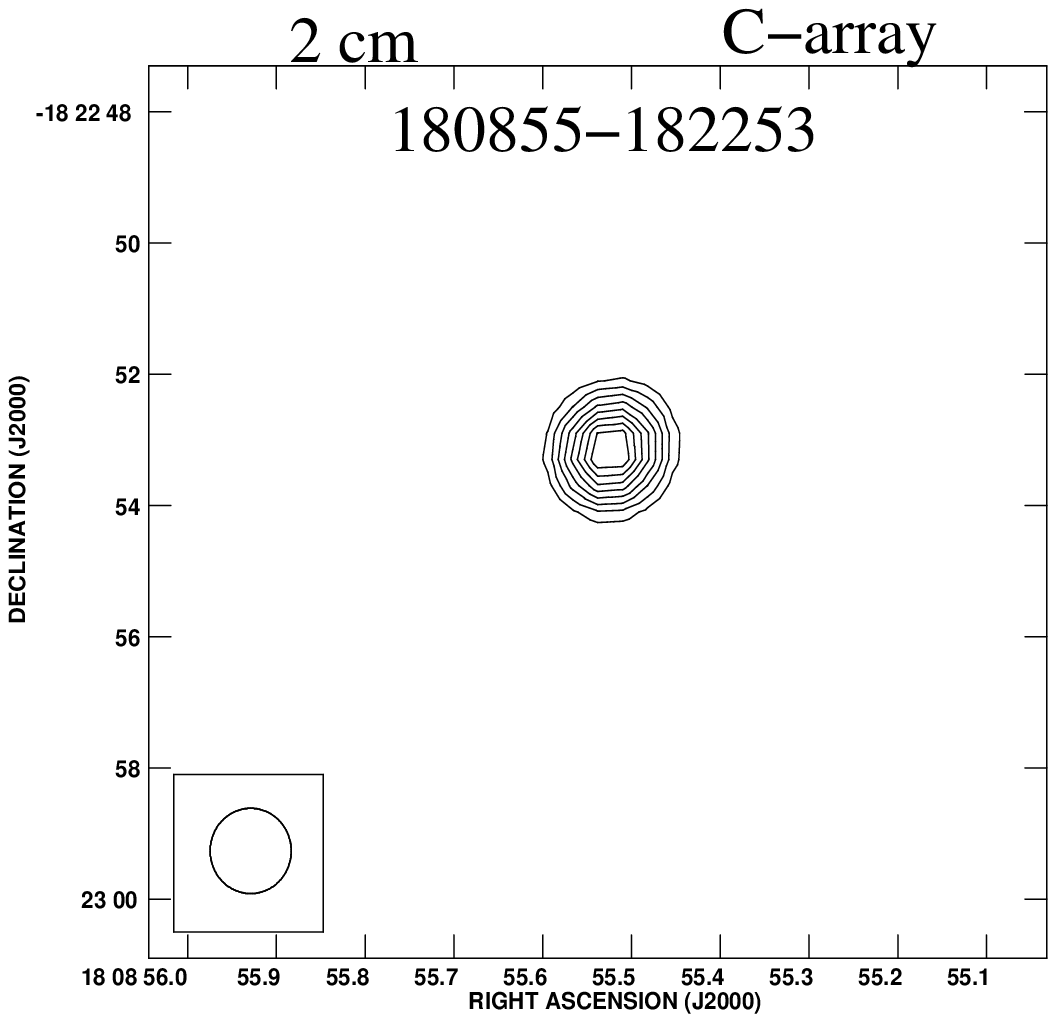} &
\includegraphics[width=5cm]{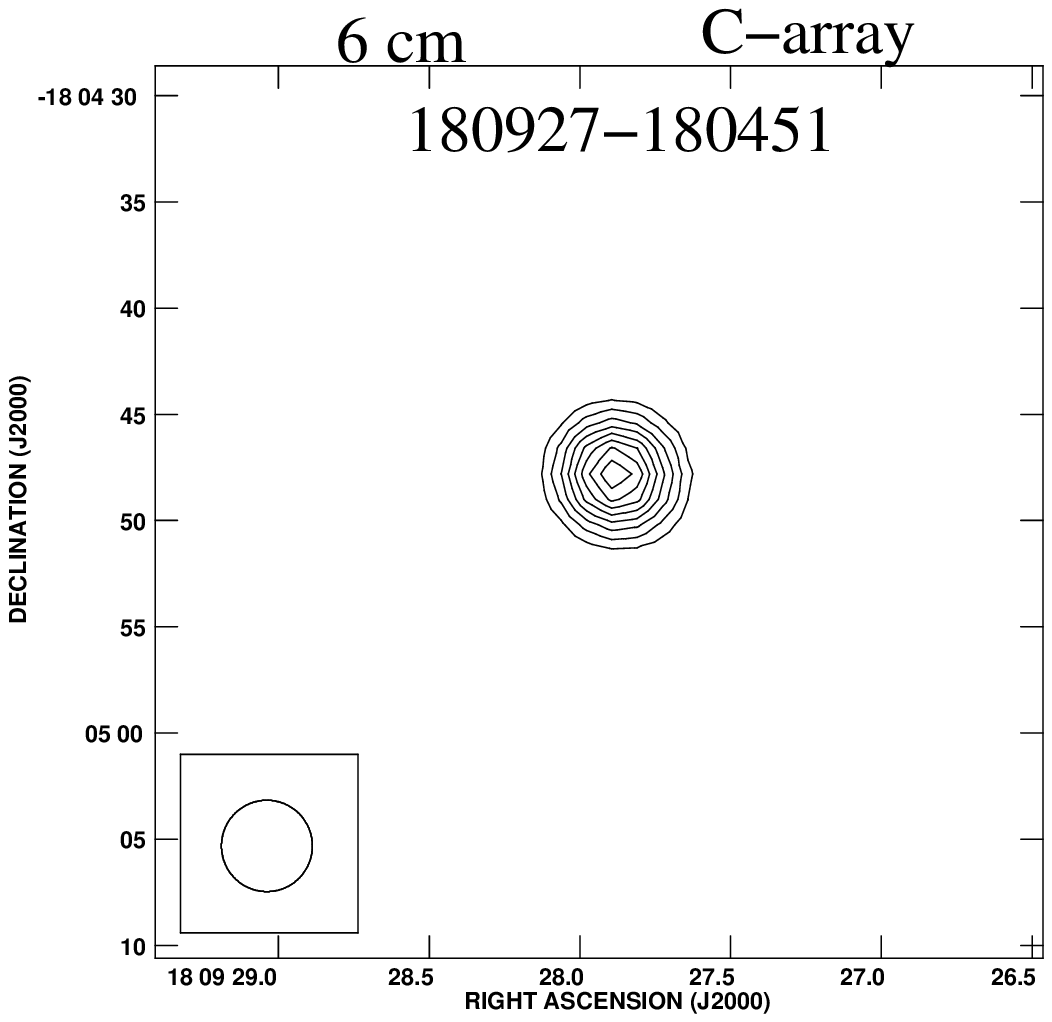}\\
\includegraphics[width=5cm]{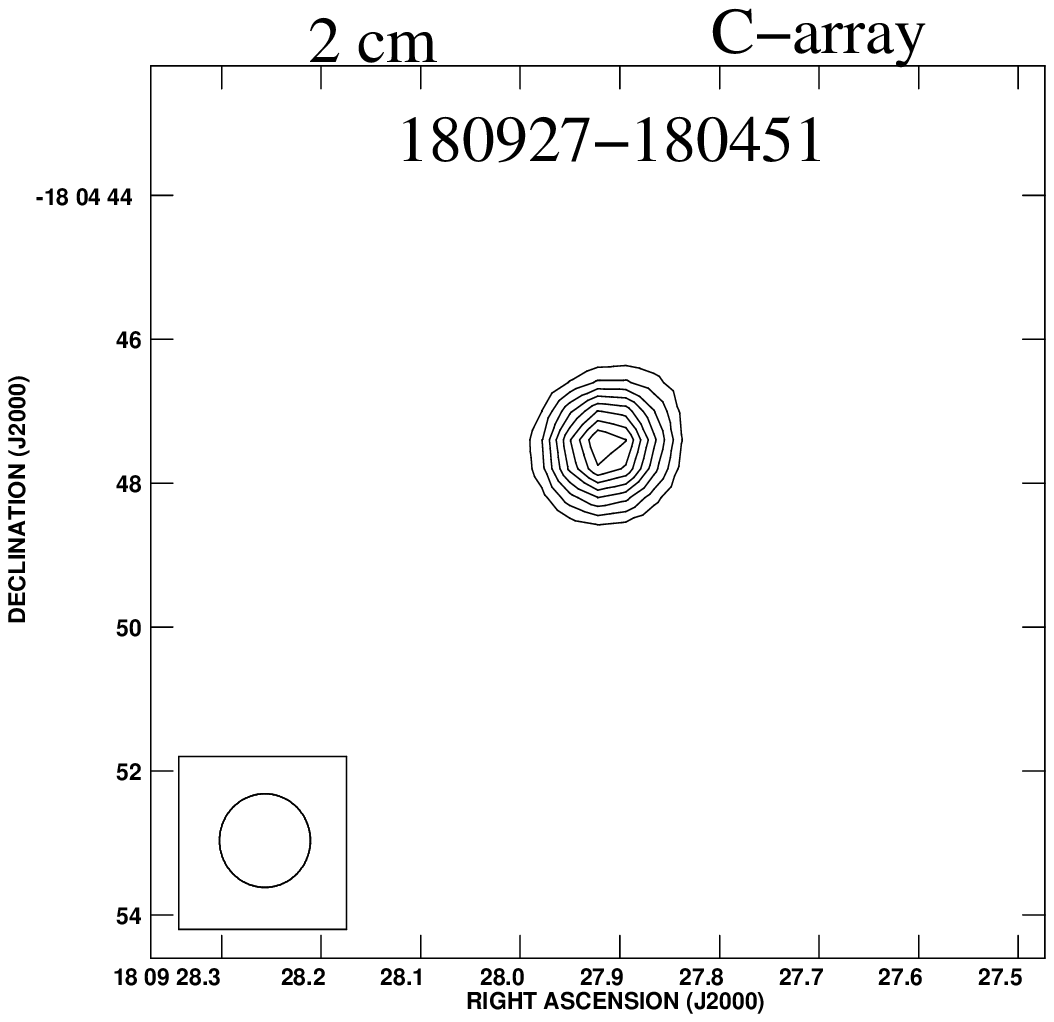}&
\includegraphics[width=5cm]{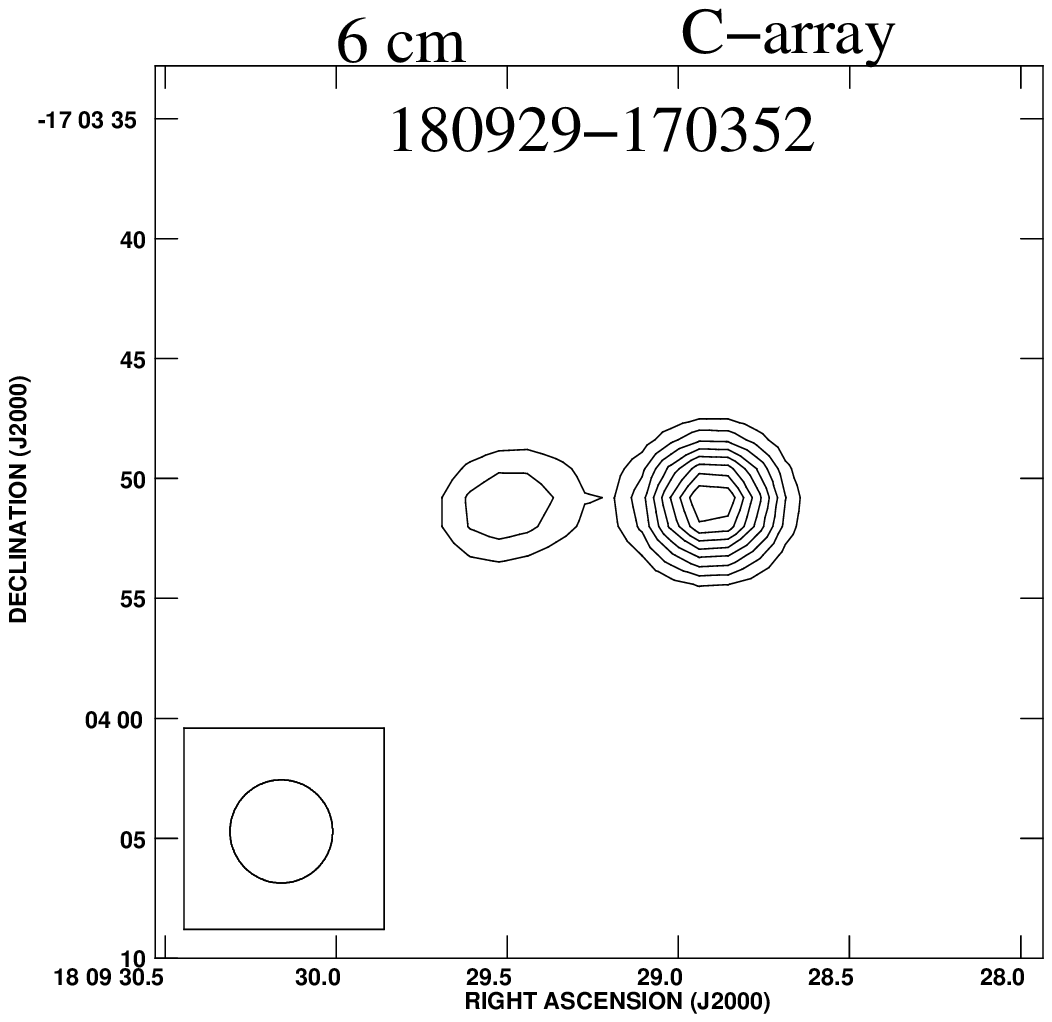} &
\includegraphics[width=5cm]{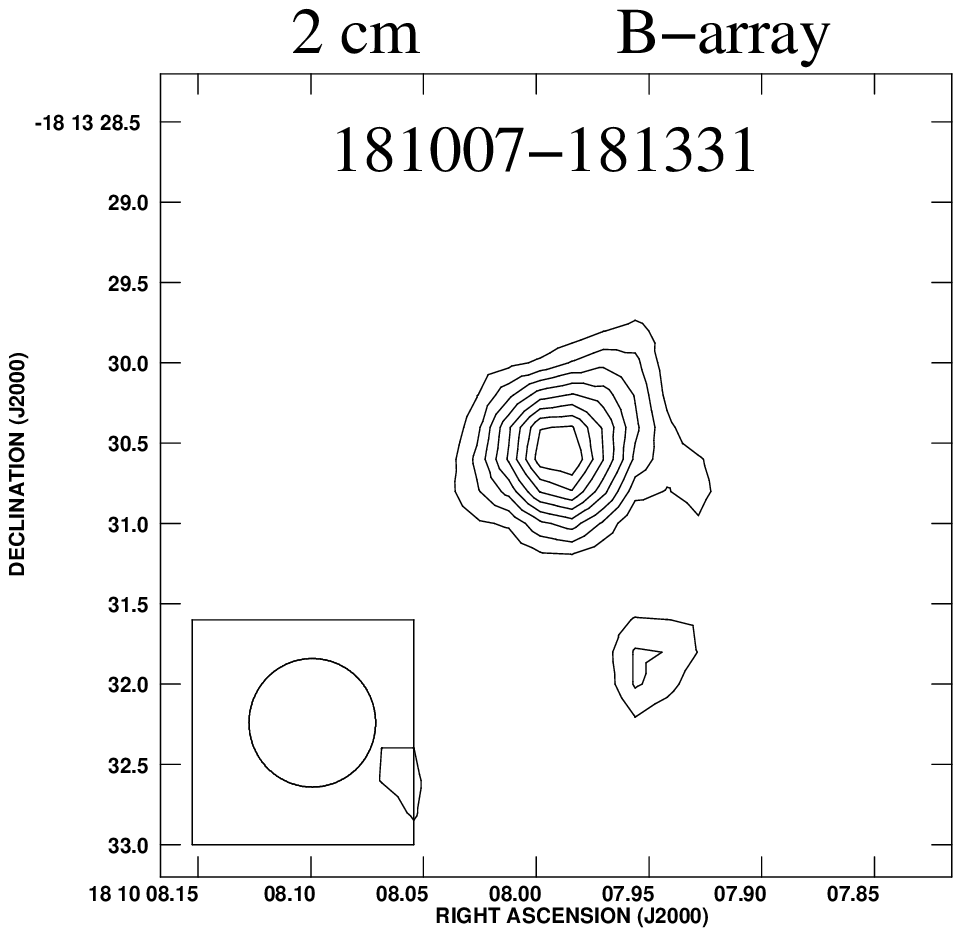}
\\

\end{tabular}
\end{figure*}

\begin{figure*}
\begin{tabular}{ccc}
\includegraphics[width=5cm]{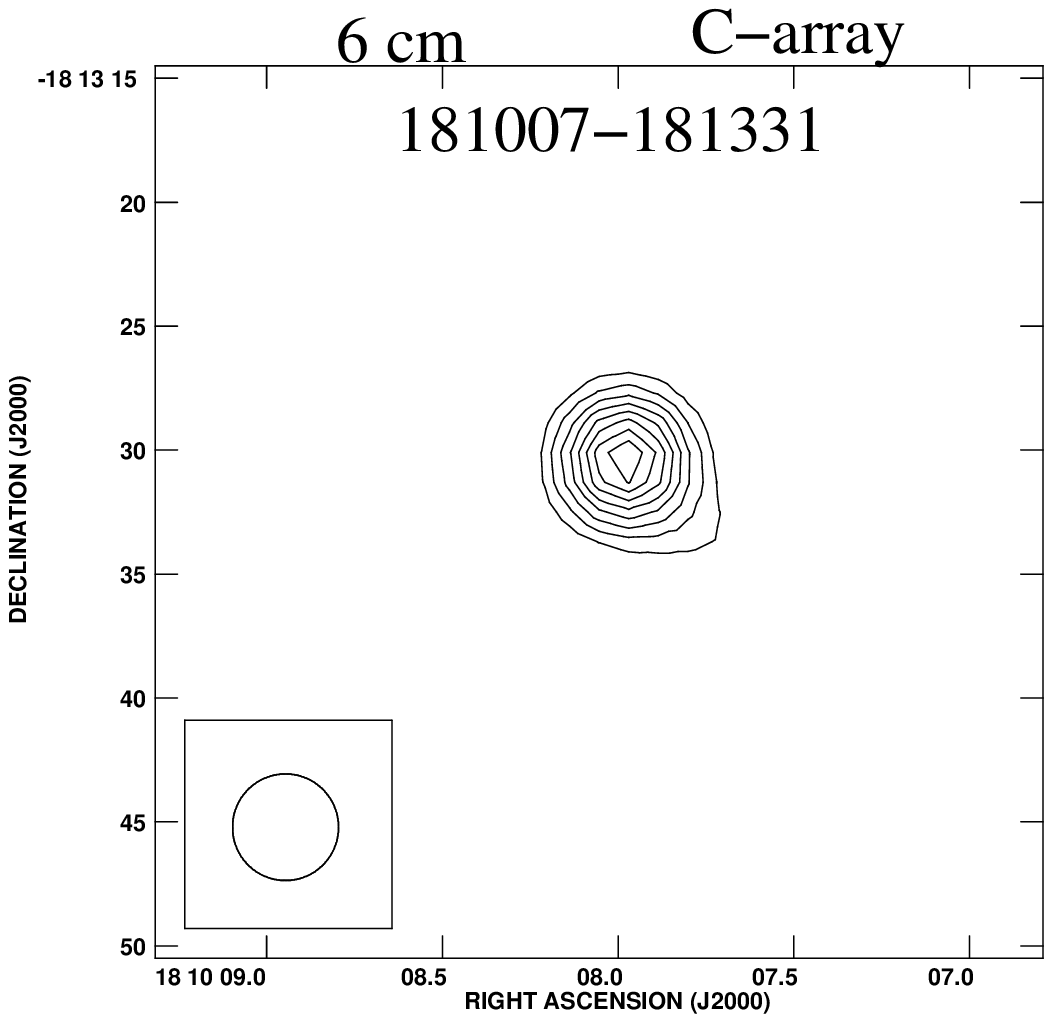}&
\includegraphics[width=5cm]{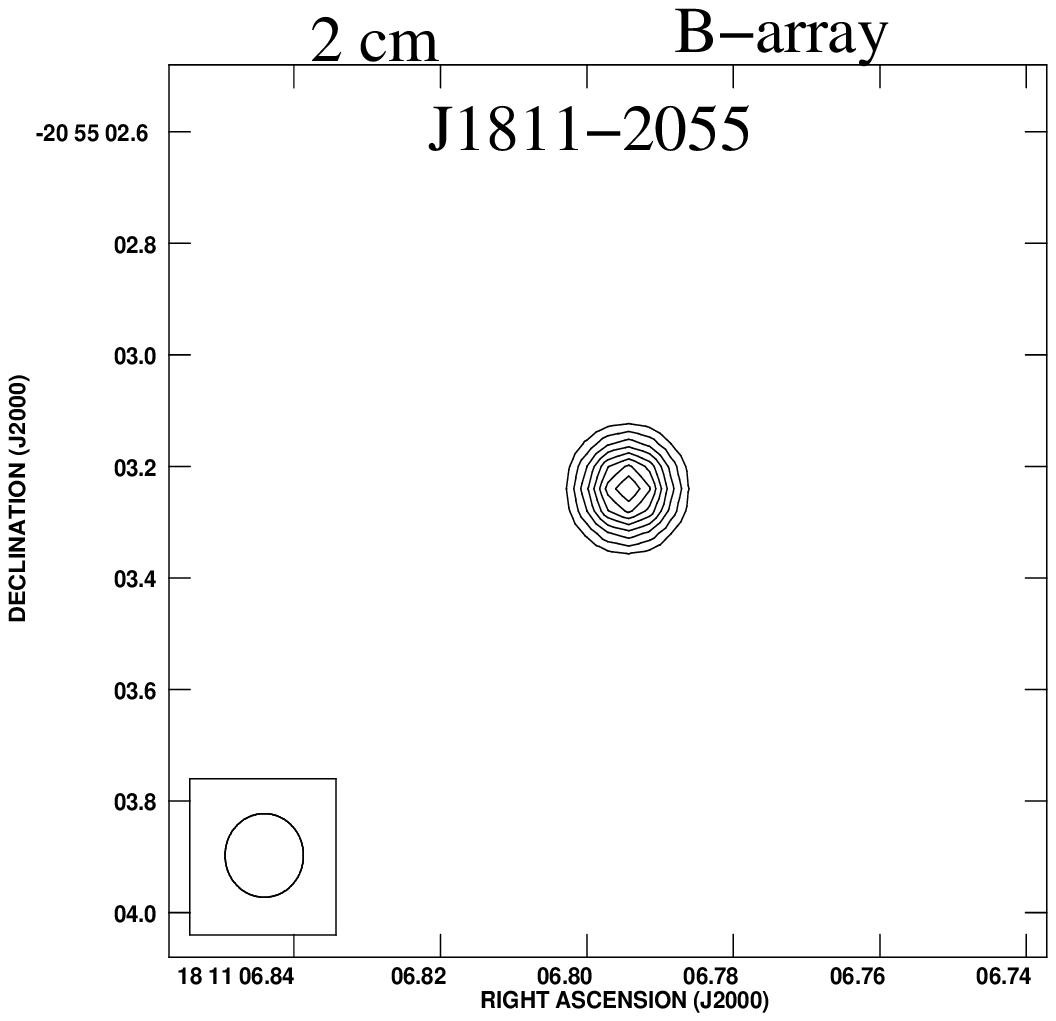} &
\includegraphics[width=5cm]{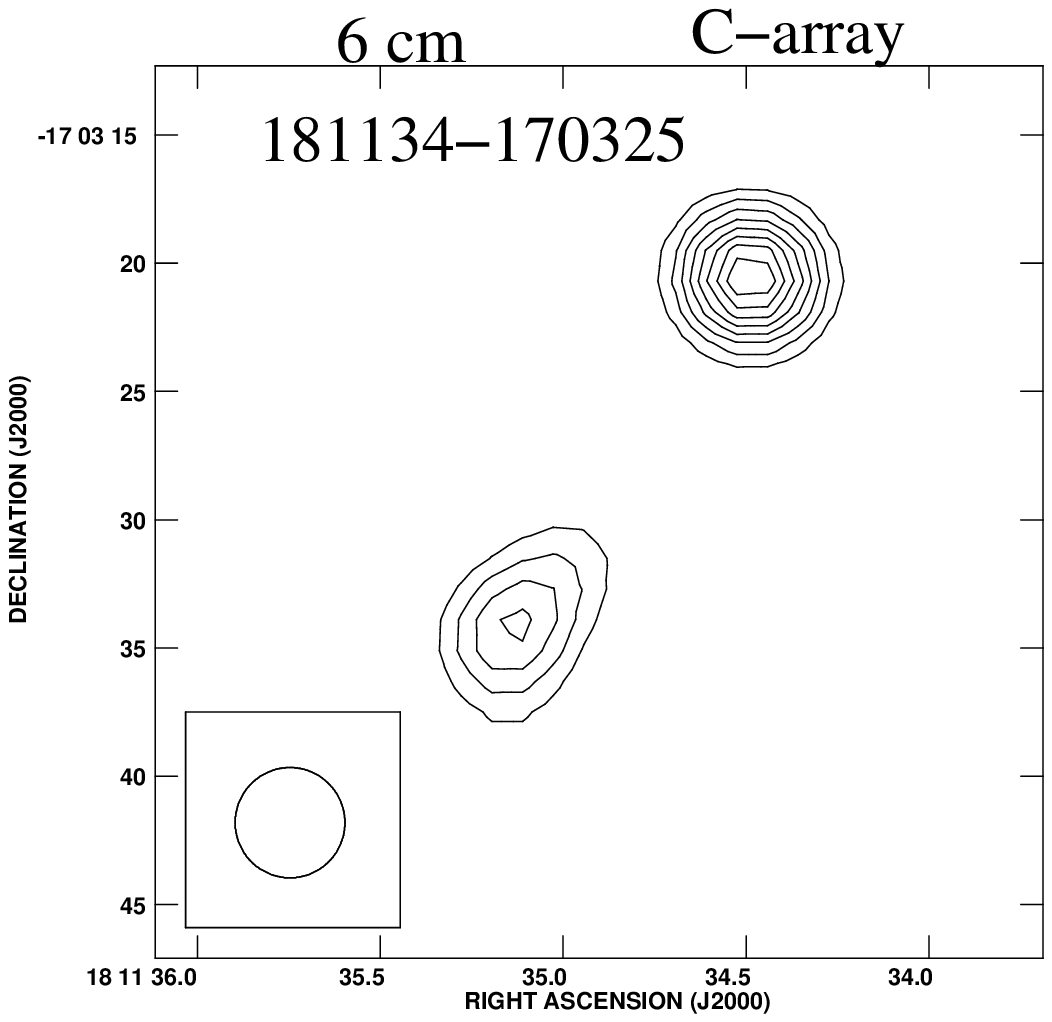}\\
\includegraphics[width=5cm]{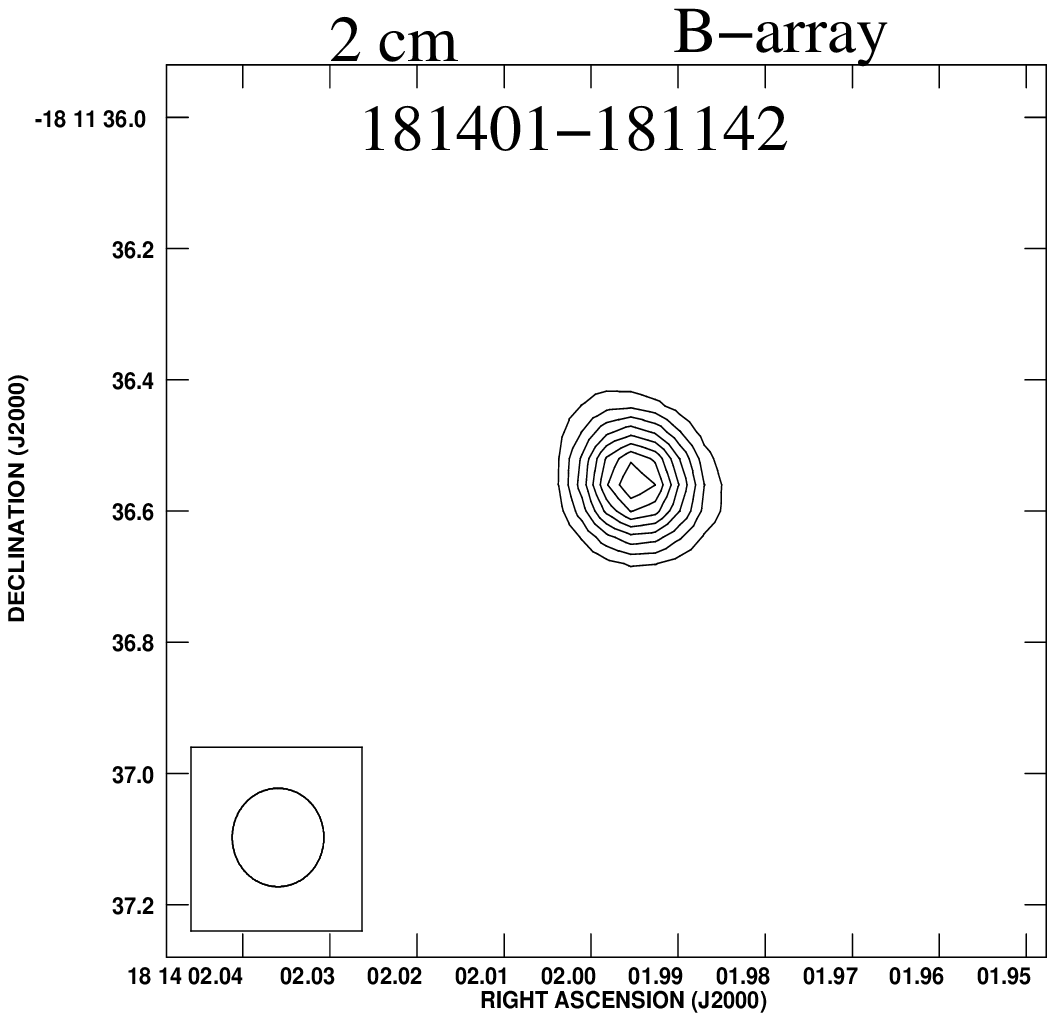} &
\includegraphics[width=5cm]{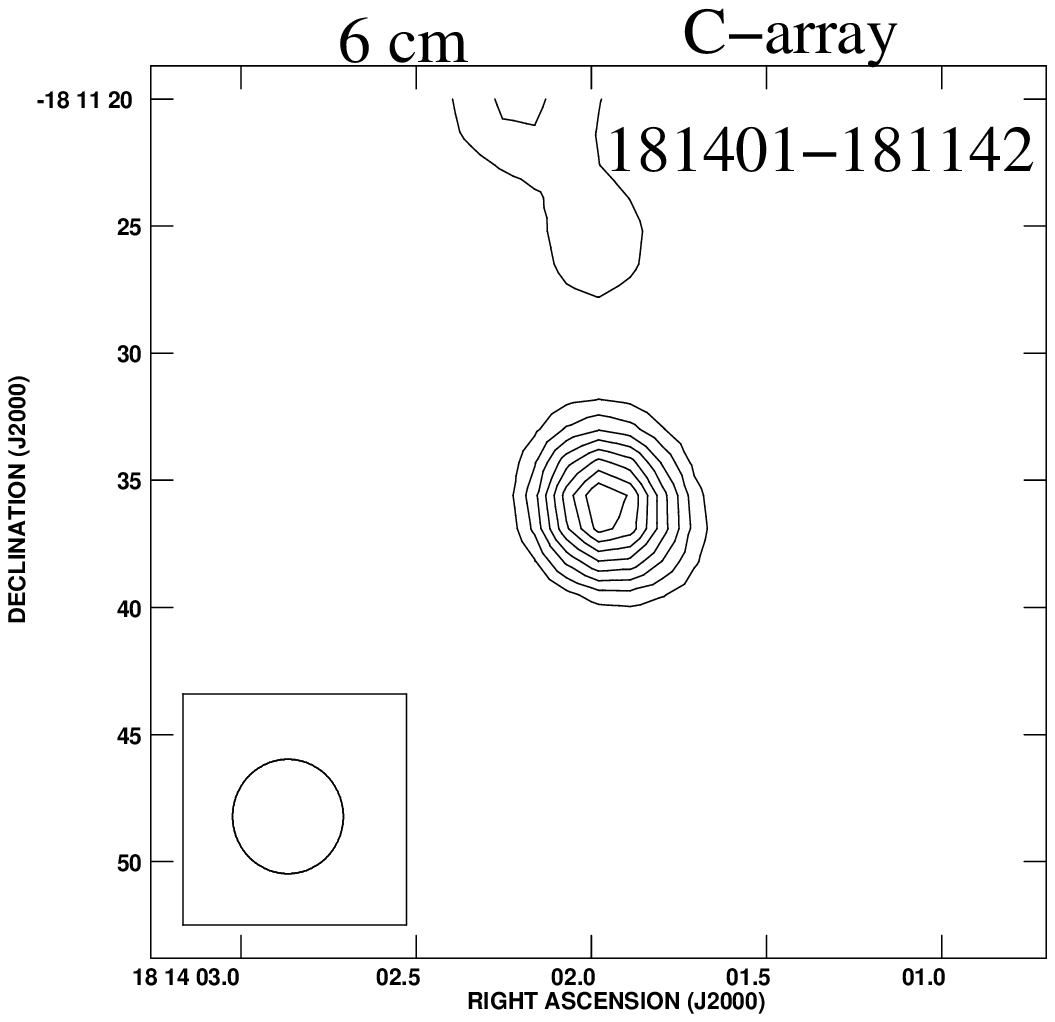}&
\includegraphics[width=5cm]{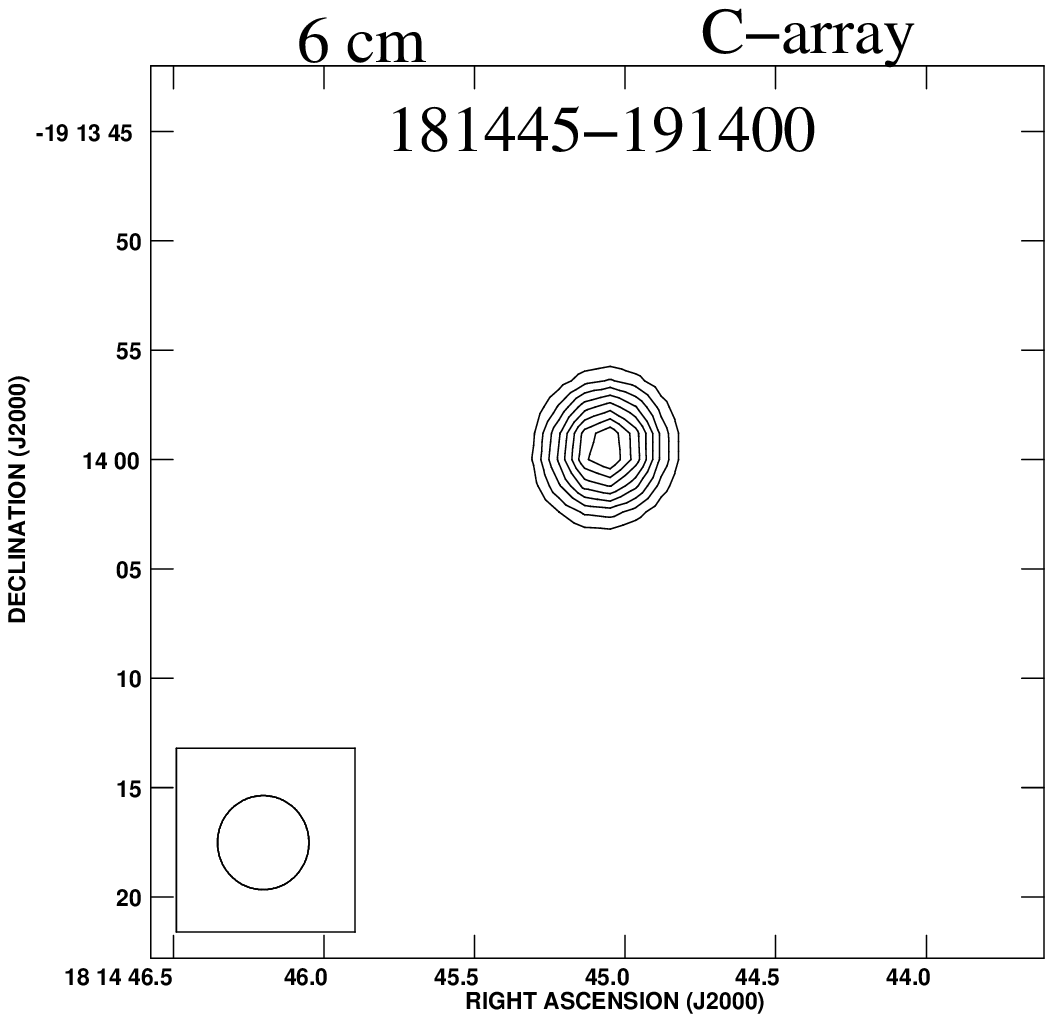}\\
\includegraphics[width=5cm]{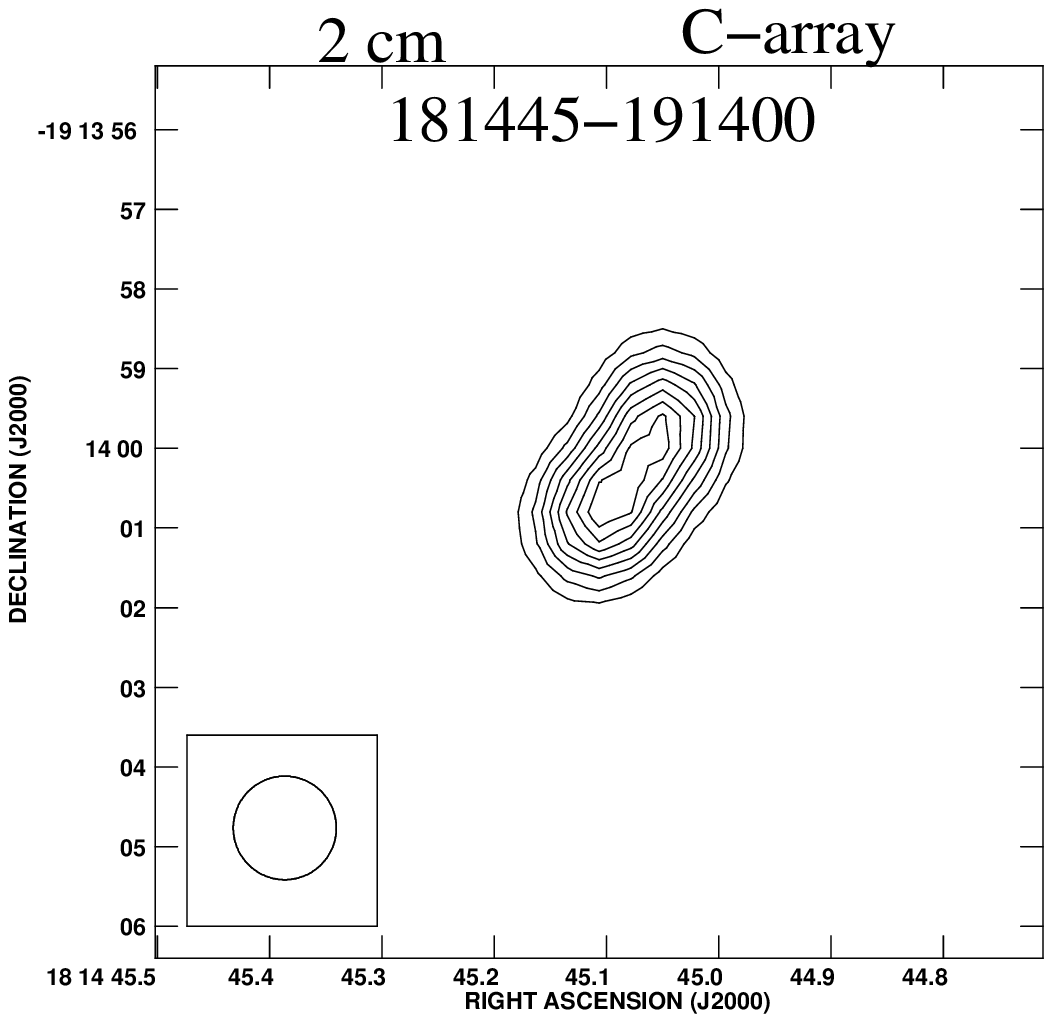}&
\includegraphics[width=5cm]{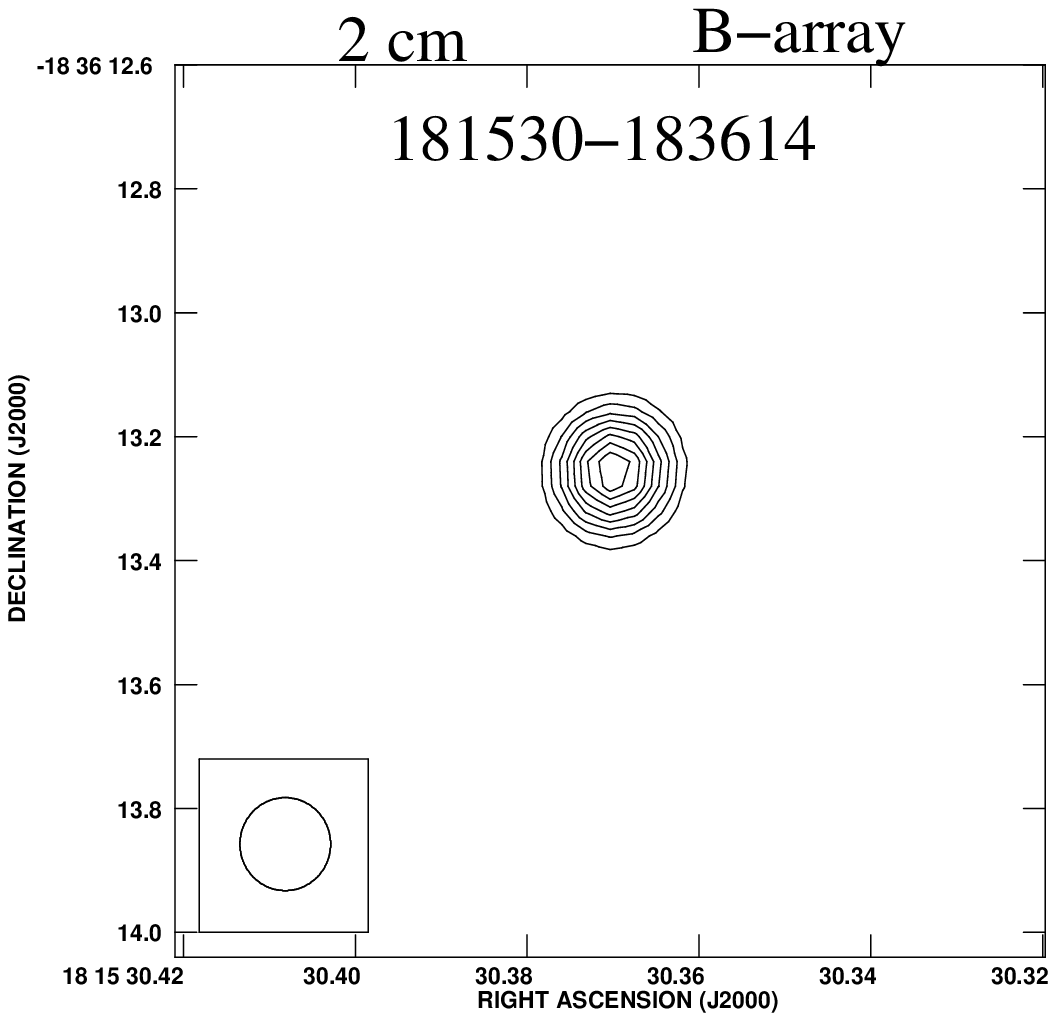}&
\includegraphics[width=5cm]{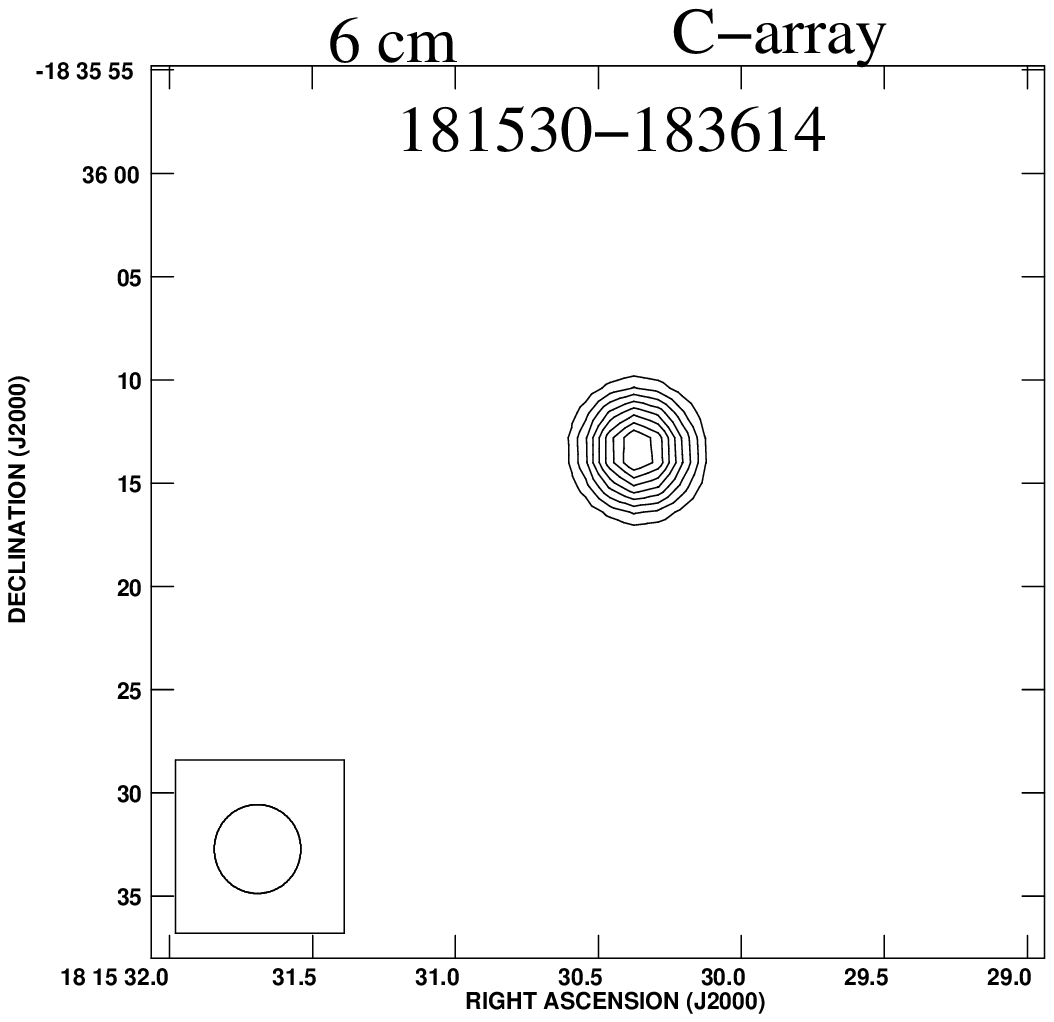}\\
\includegraphics[width=5cm]{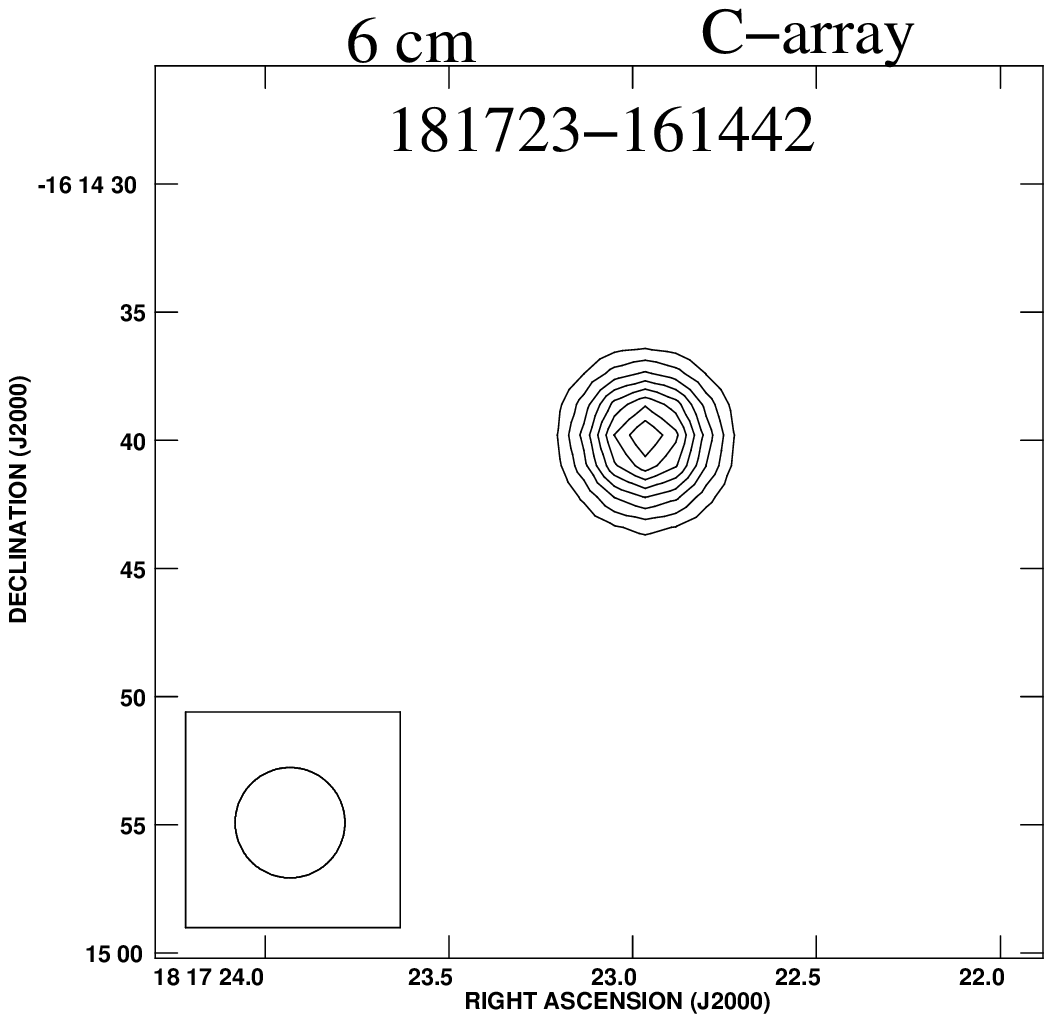} &
\includegraphics[width=5cm]{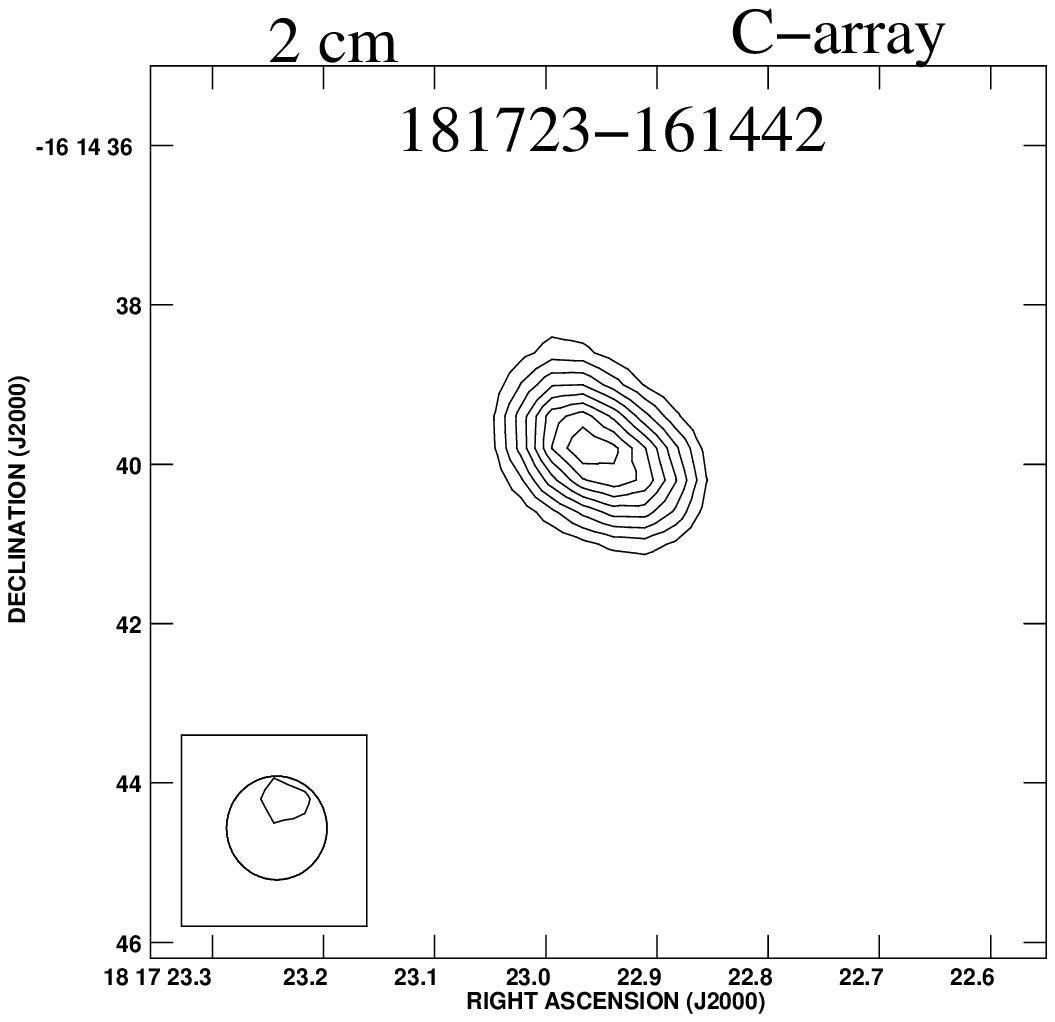}&
\includegraphics[width=5cm]{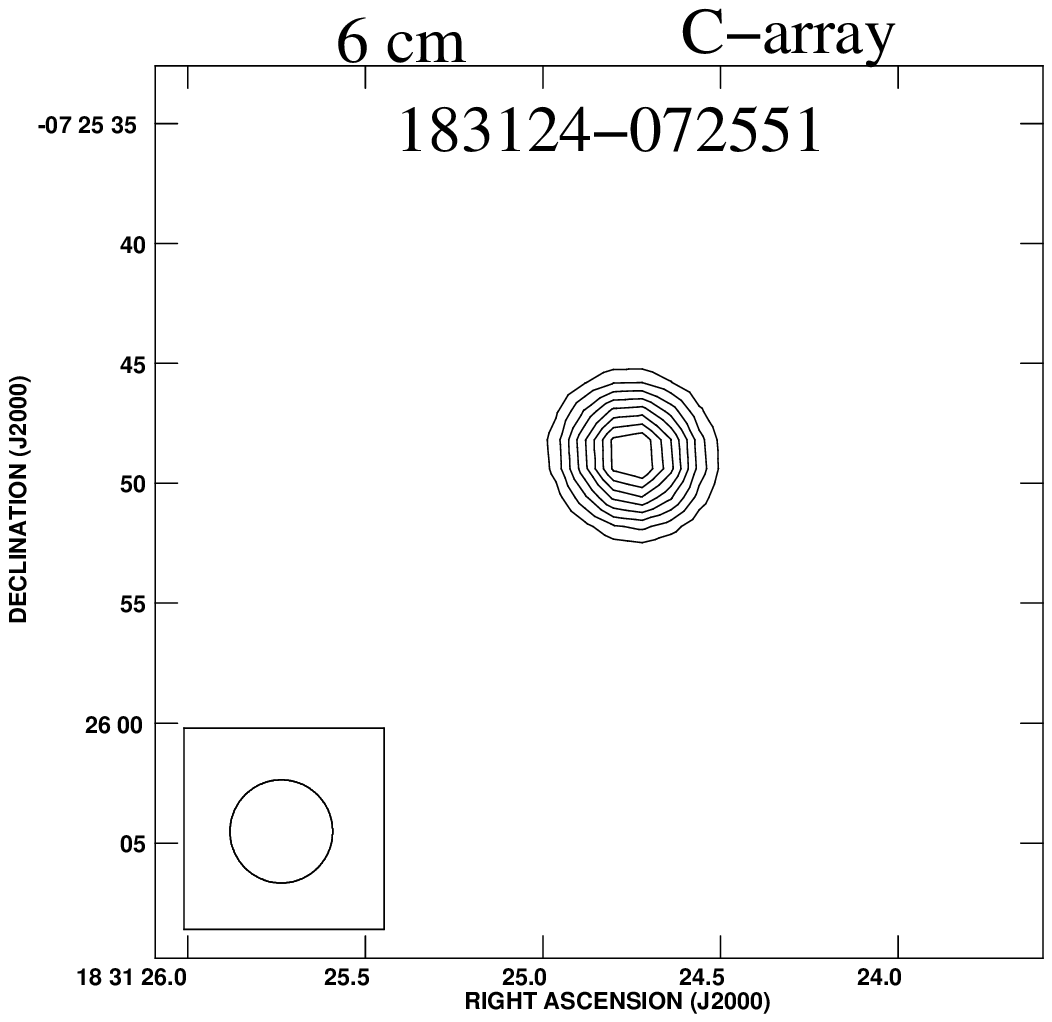}
\\

\end{tabular}
\end{figure*}

\begin{figure*}
\begin{tabular}{ccc}
\includegraphics[width=5cm]{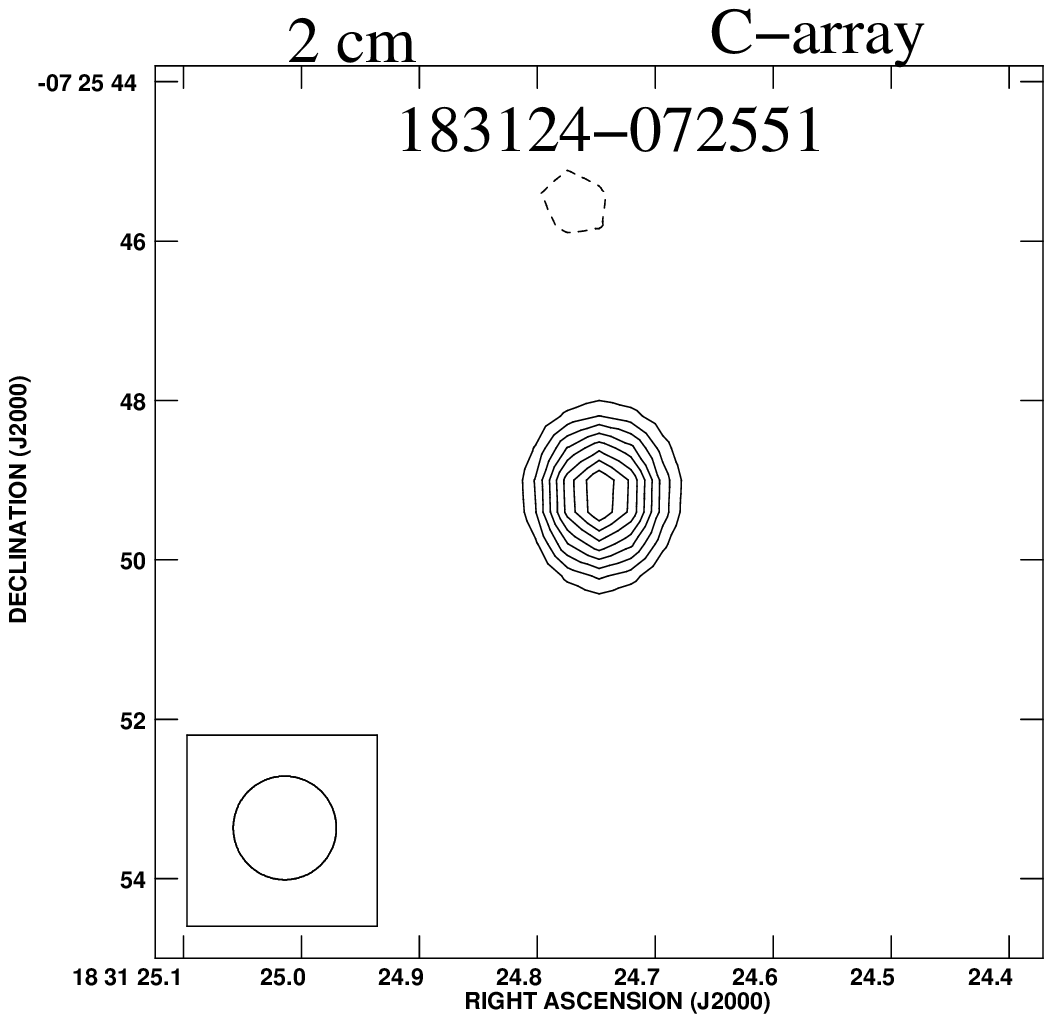} &
\includegraphics[width=5cm]{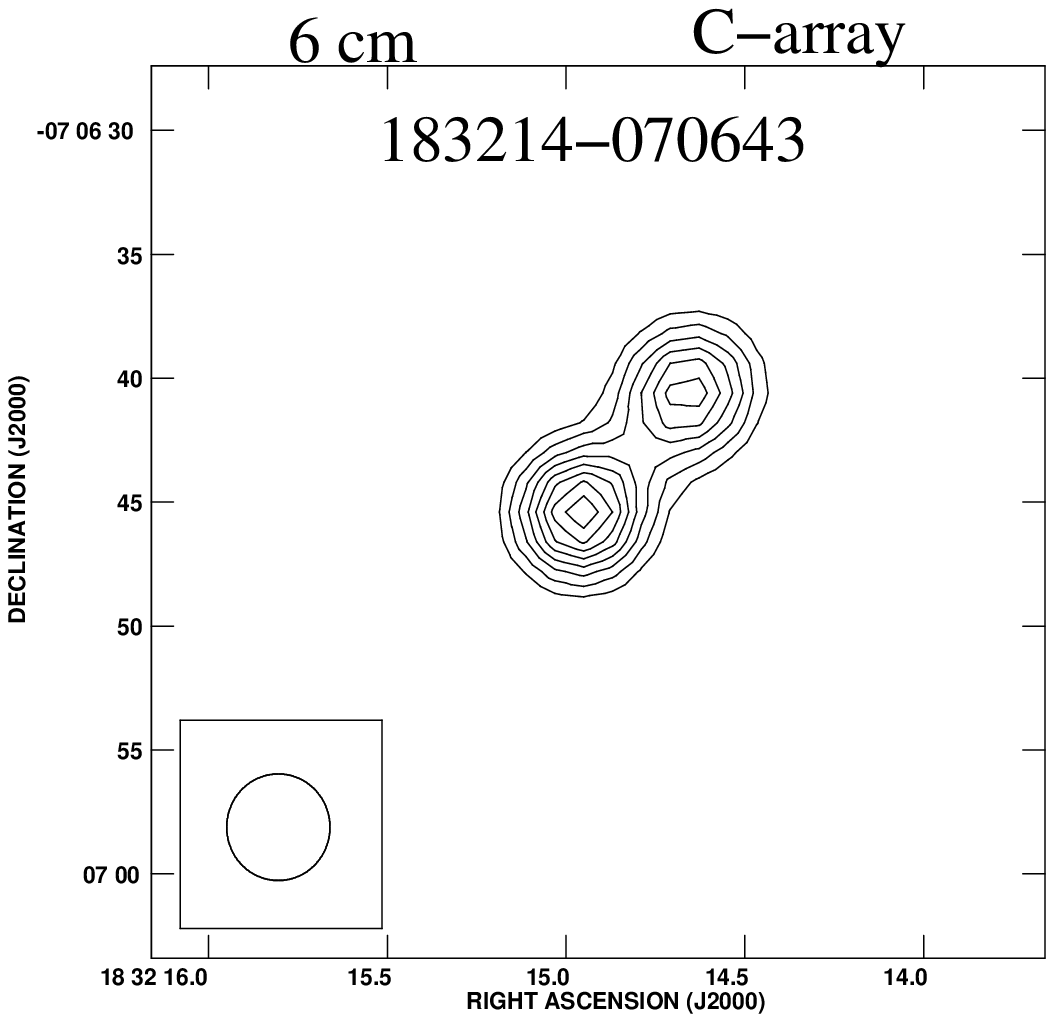}&
\includegraphics[width=5cm]{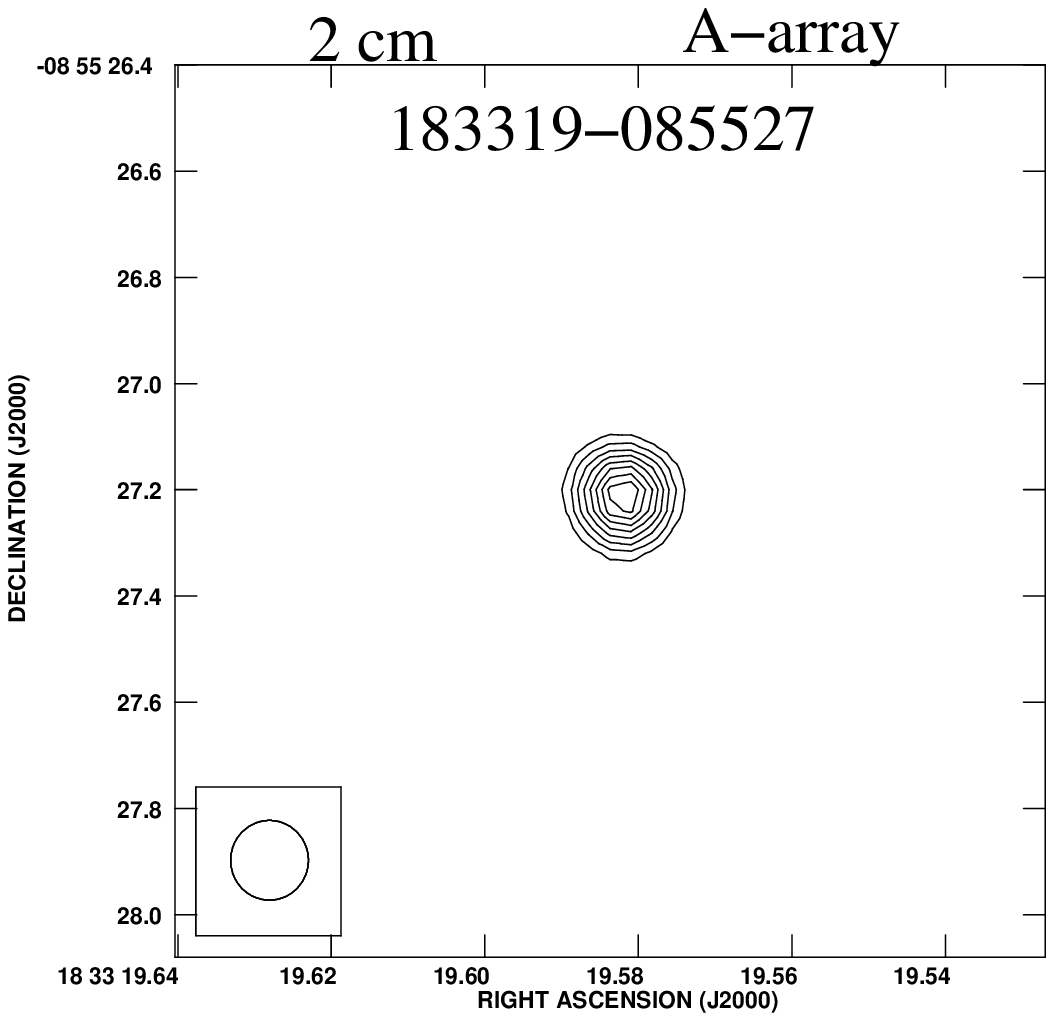}\\
\includegraphics[width=5cm]{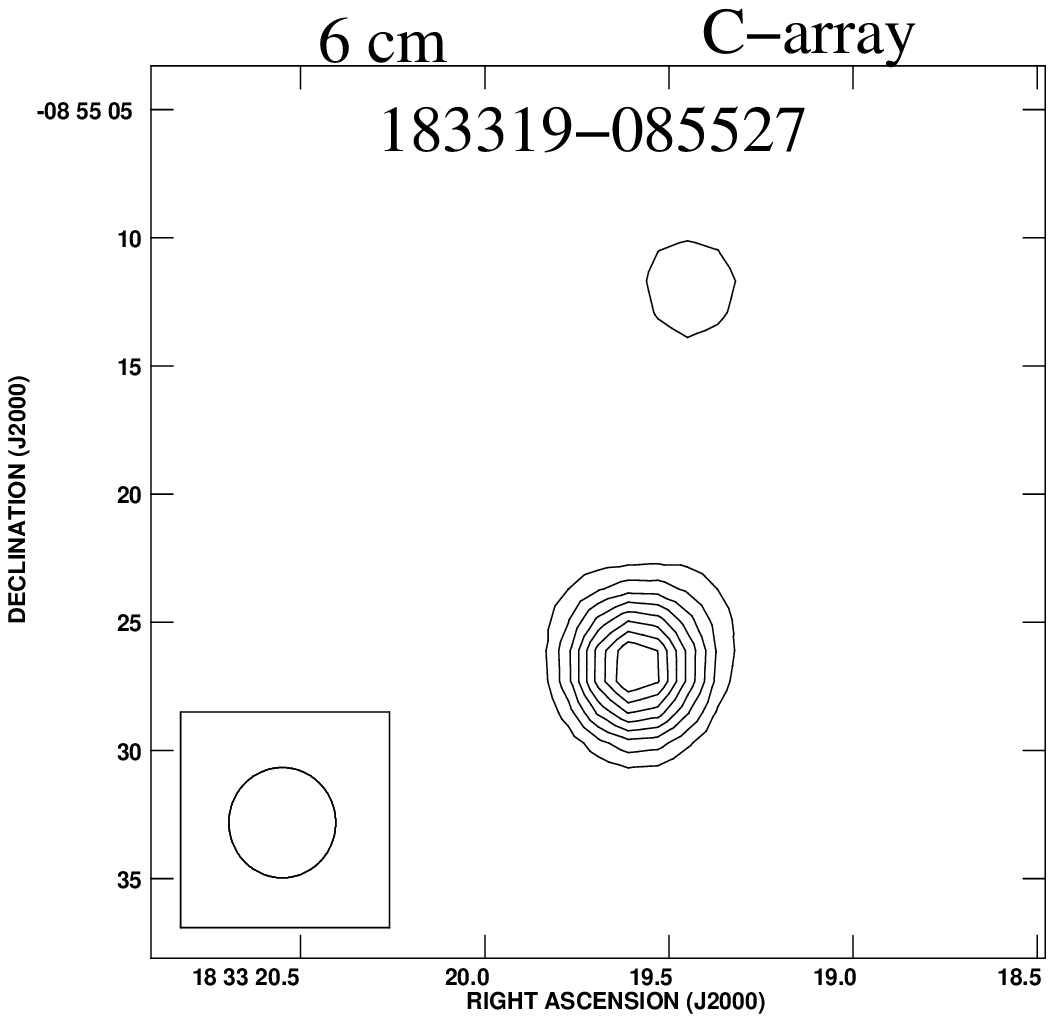} &
\includegraphics[width=5cm]{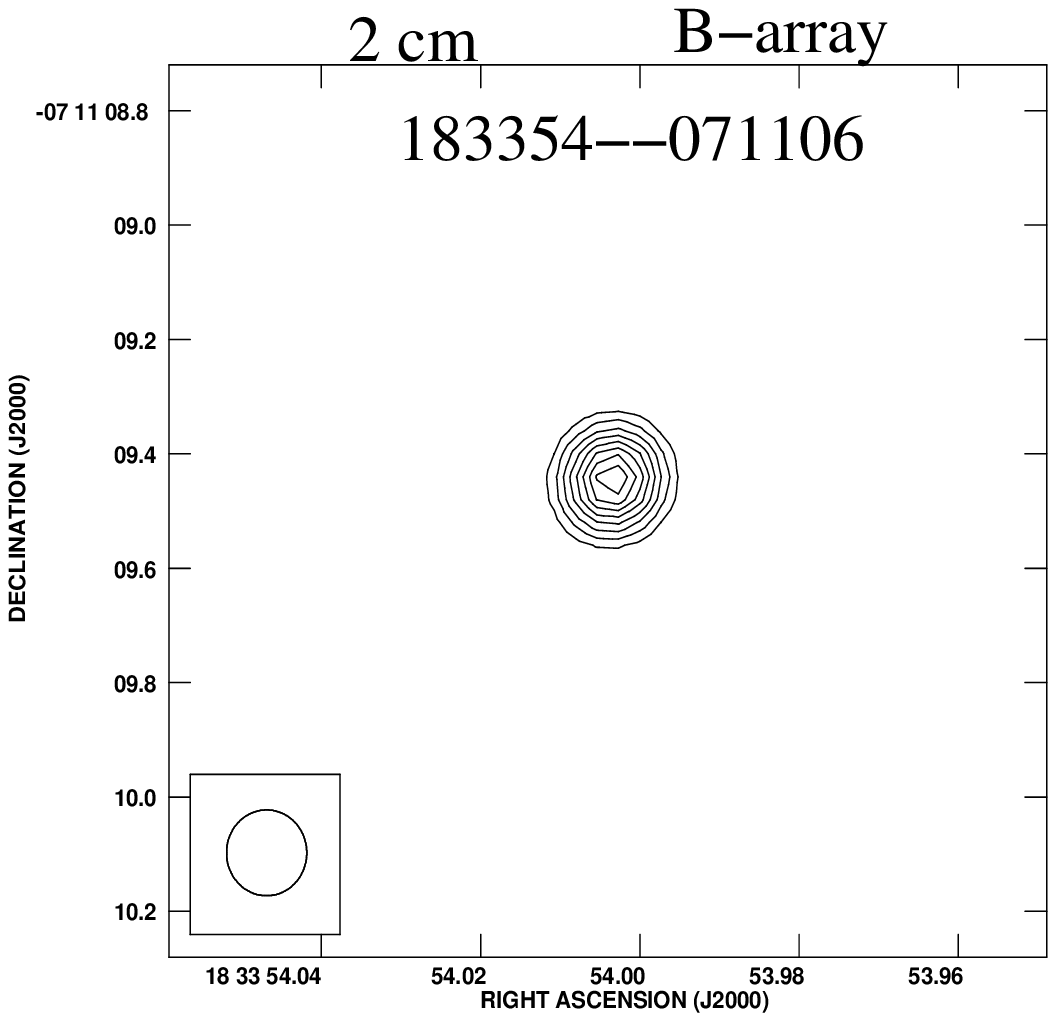}&
\includegraphics[width=5cm]{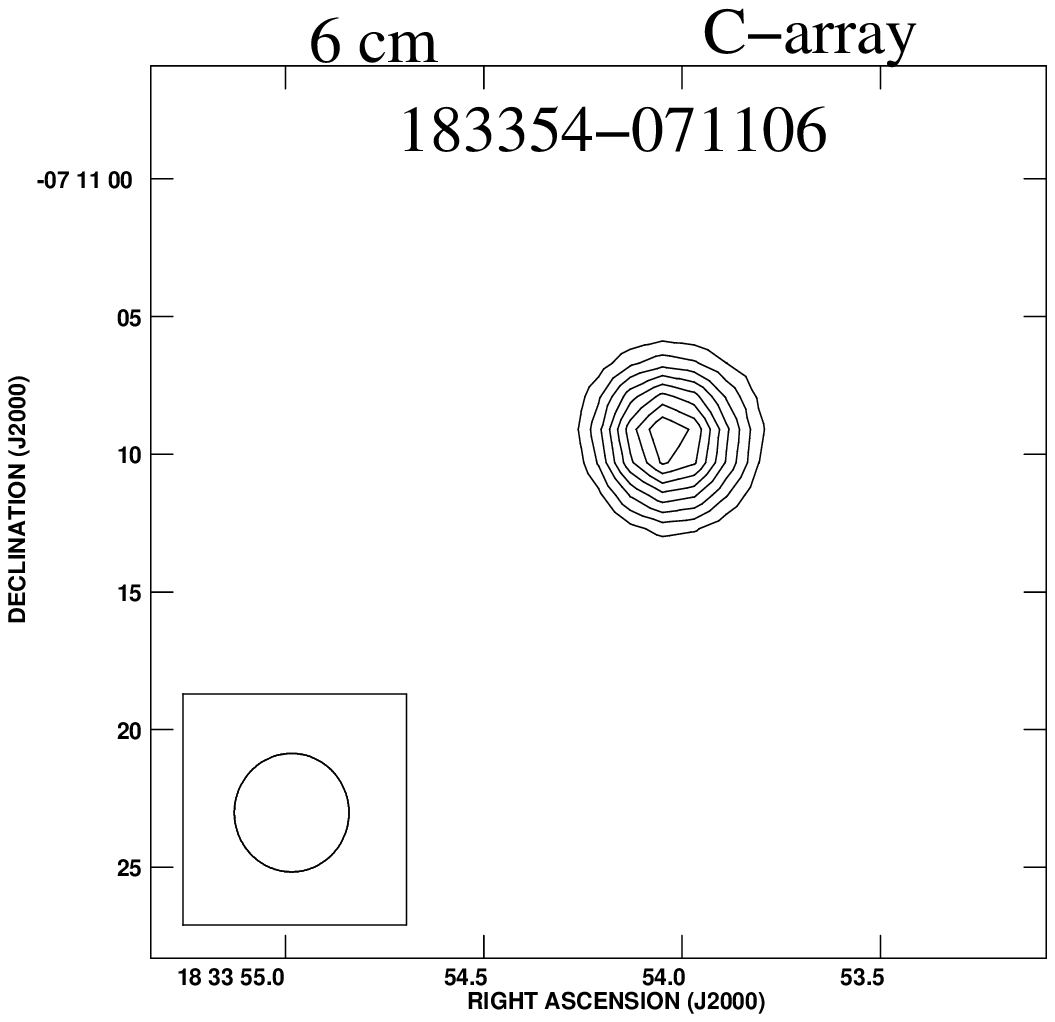}\\
\includegraphics[width=5cm]{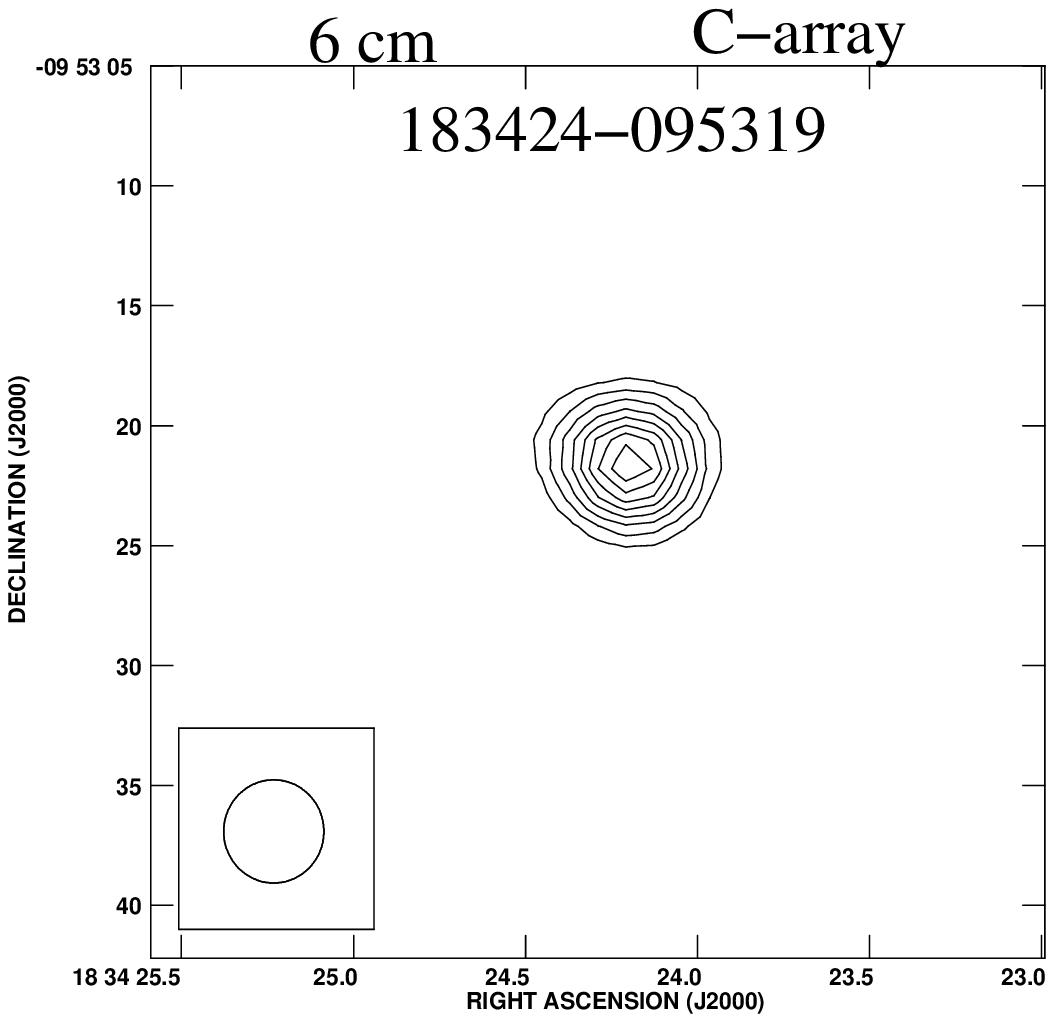} &
\includegraphics[width=5cm]{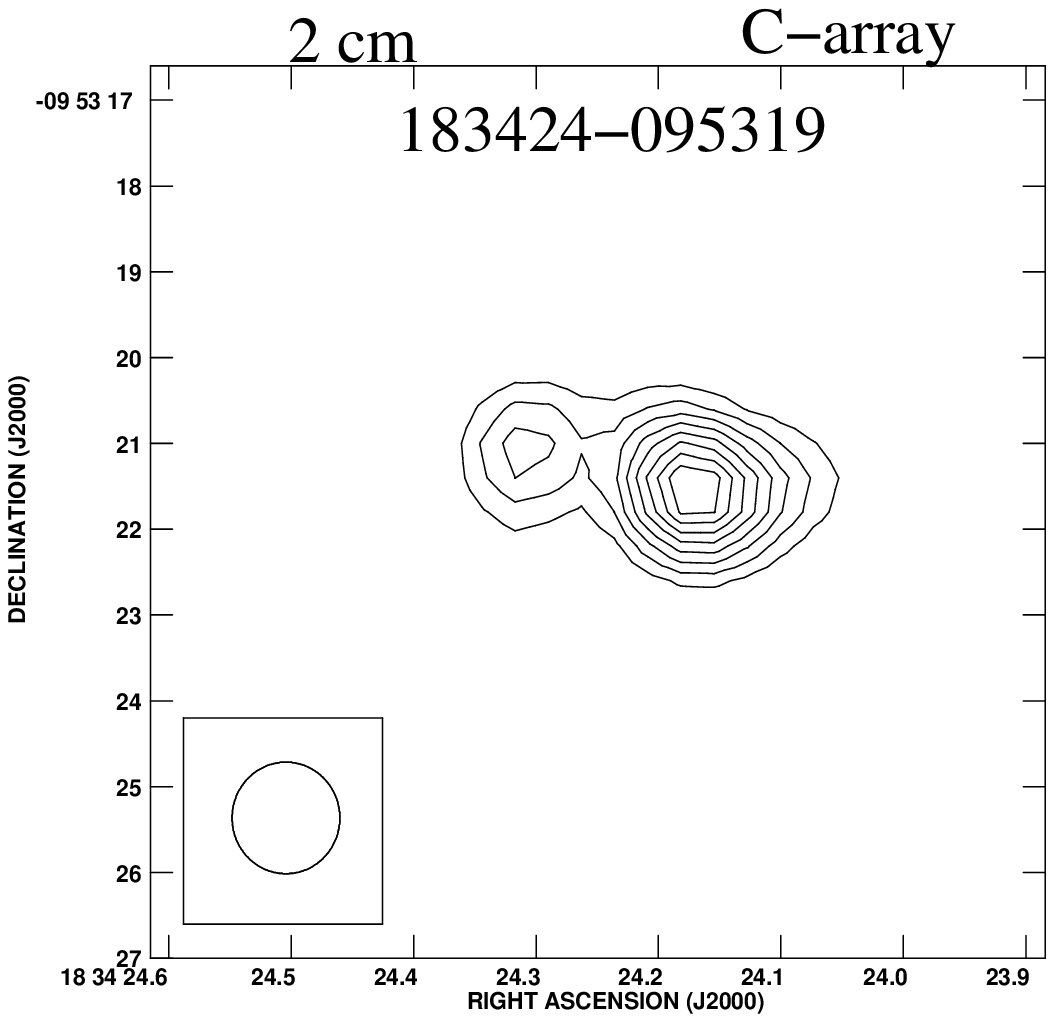}&
\includegraphics[width=5cm]{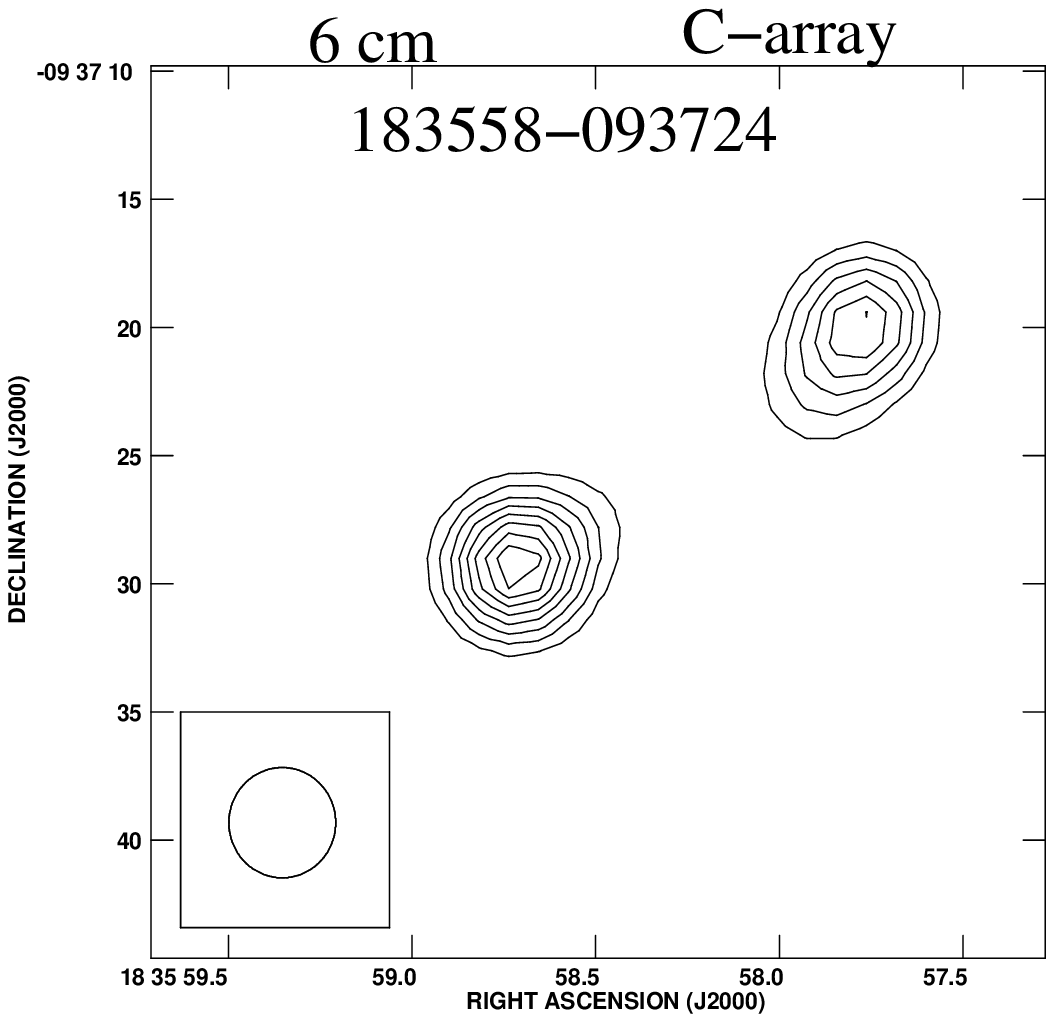}\\
\includegraphics[width=5cm]{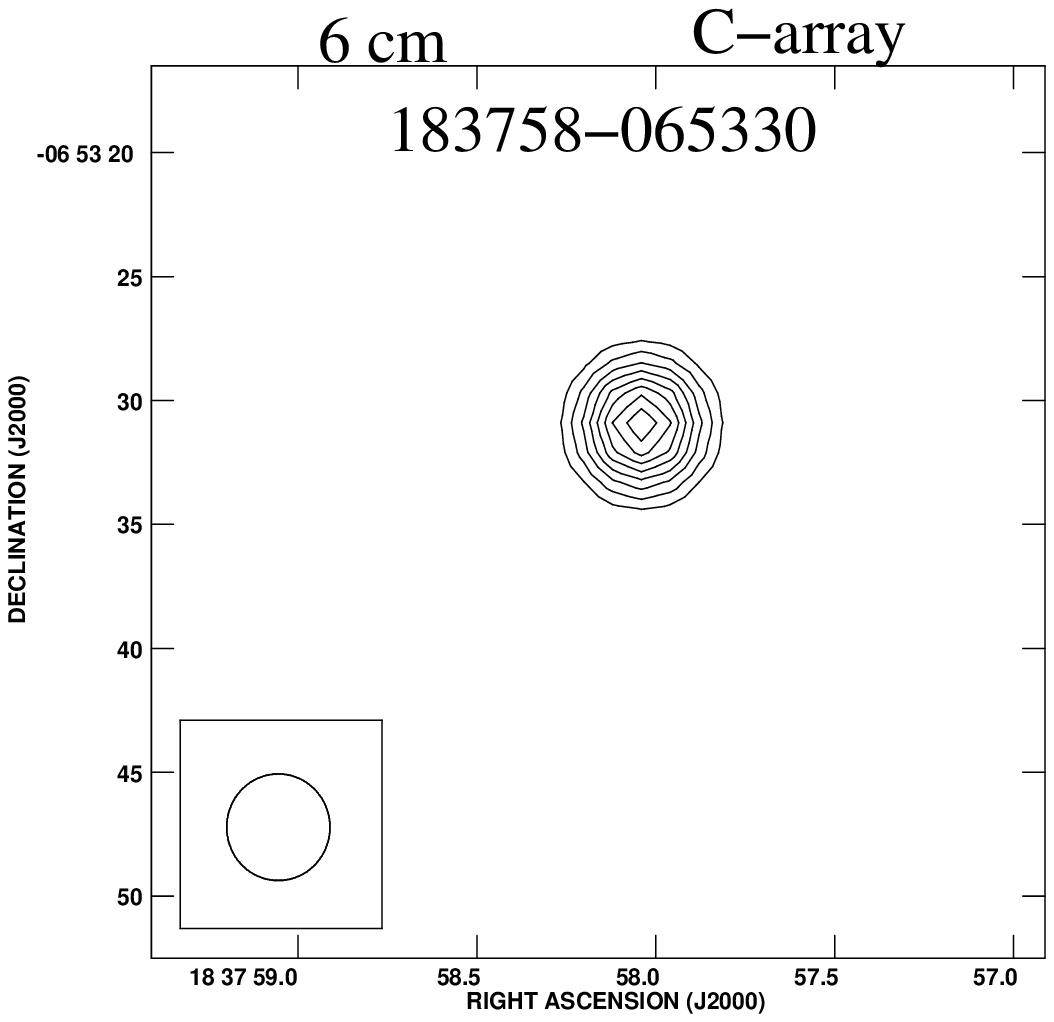} &
\includegraphics[width=5cm]{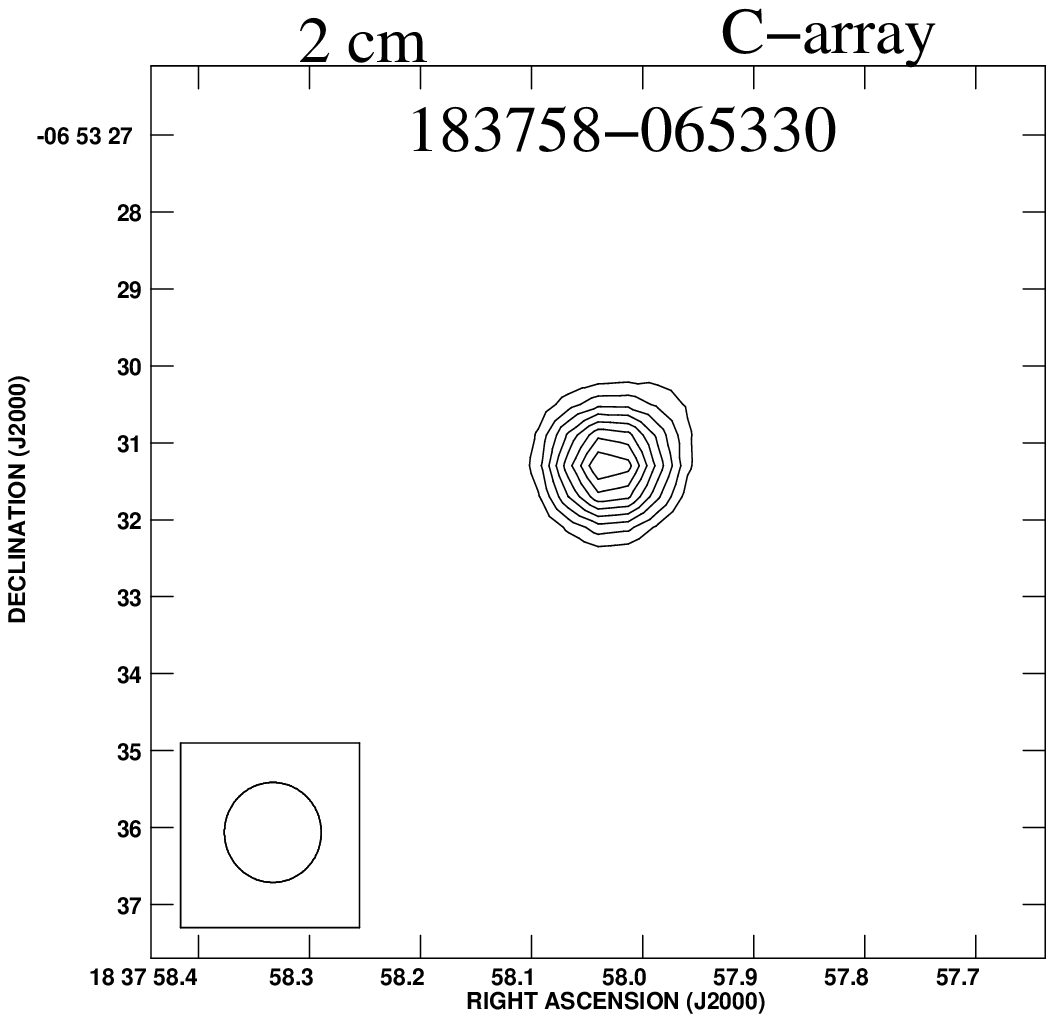}&
\includegraphics[width=5cm]{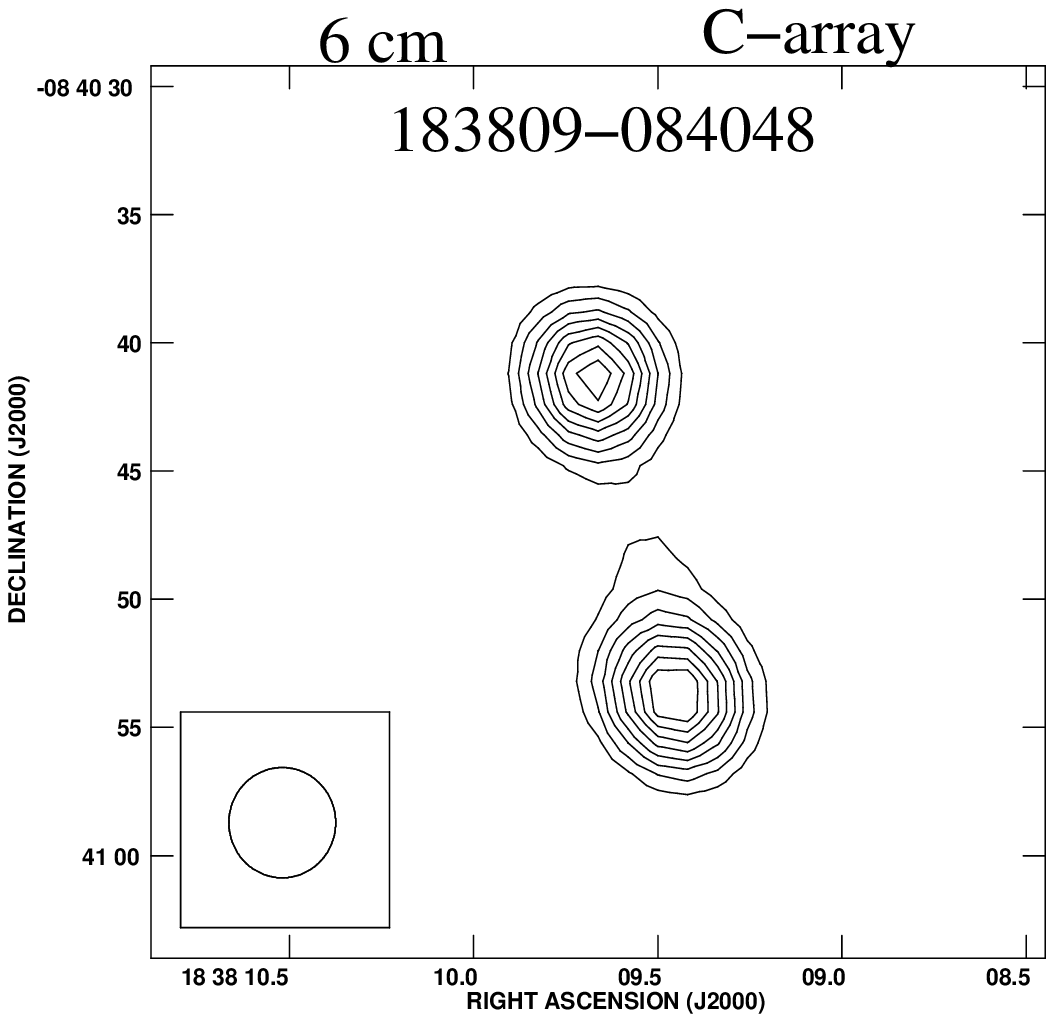}
\\

\end{tabular}
\end{figure*}
\begin{figure*}
\begin{tabular}{ccc}
\includegraphics[width=5cm]{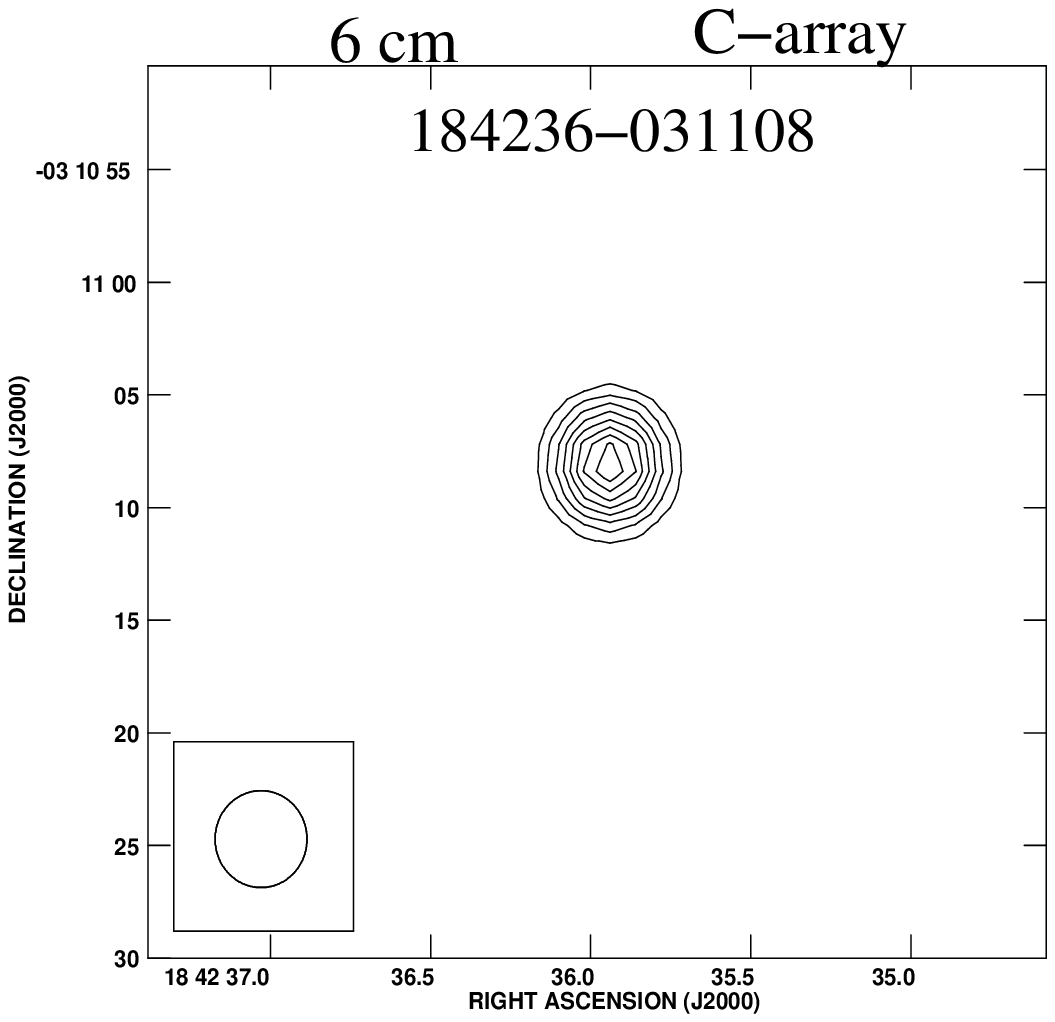} &
\includegraphics[width=5cm]{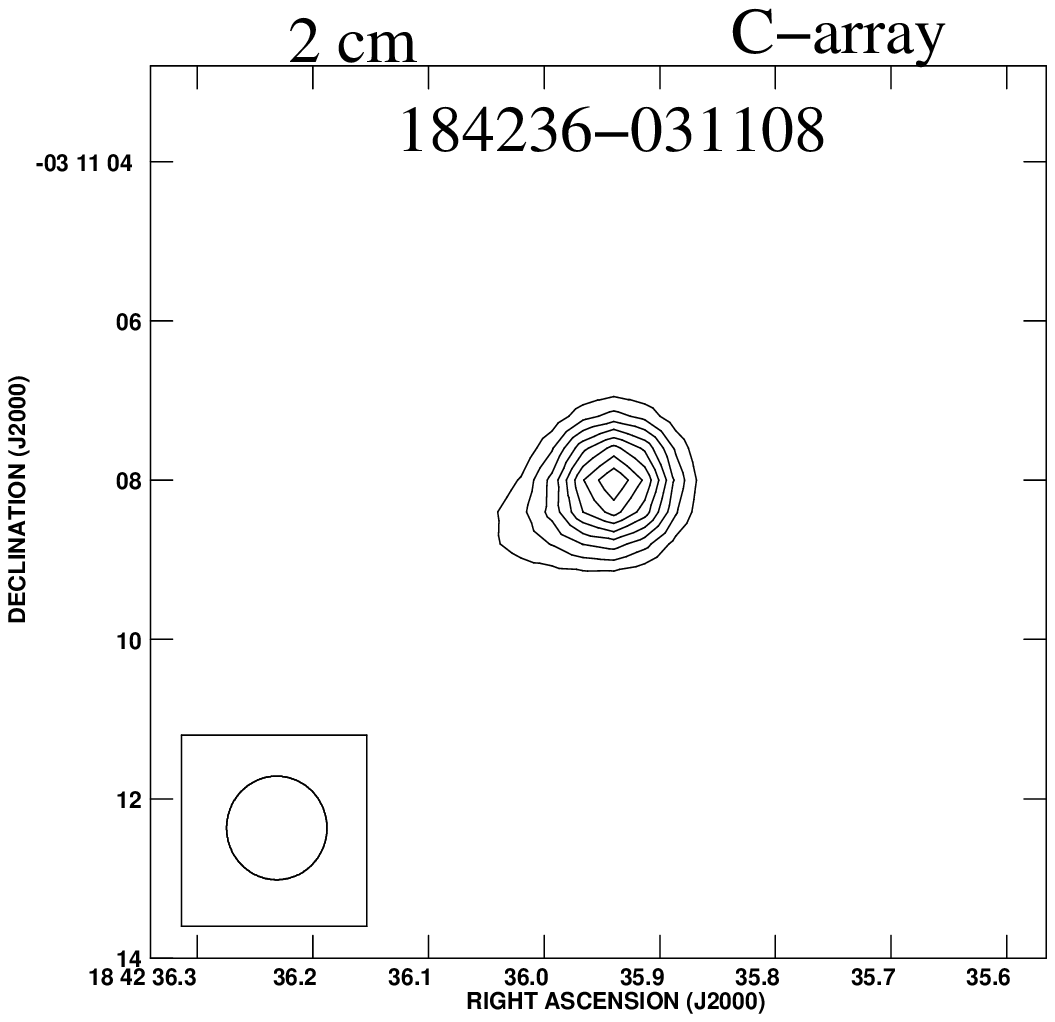}&
\includegraphics[width=5cm]{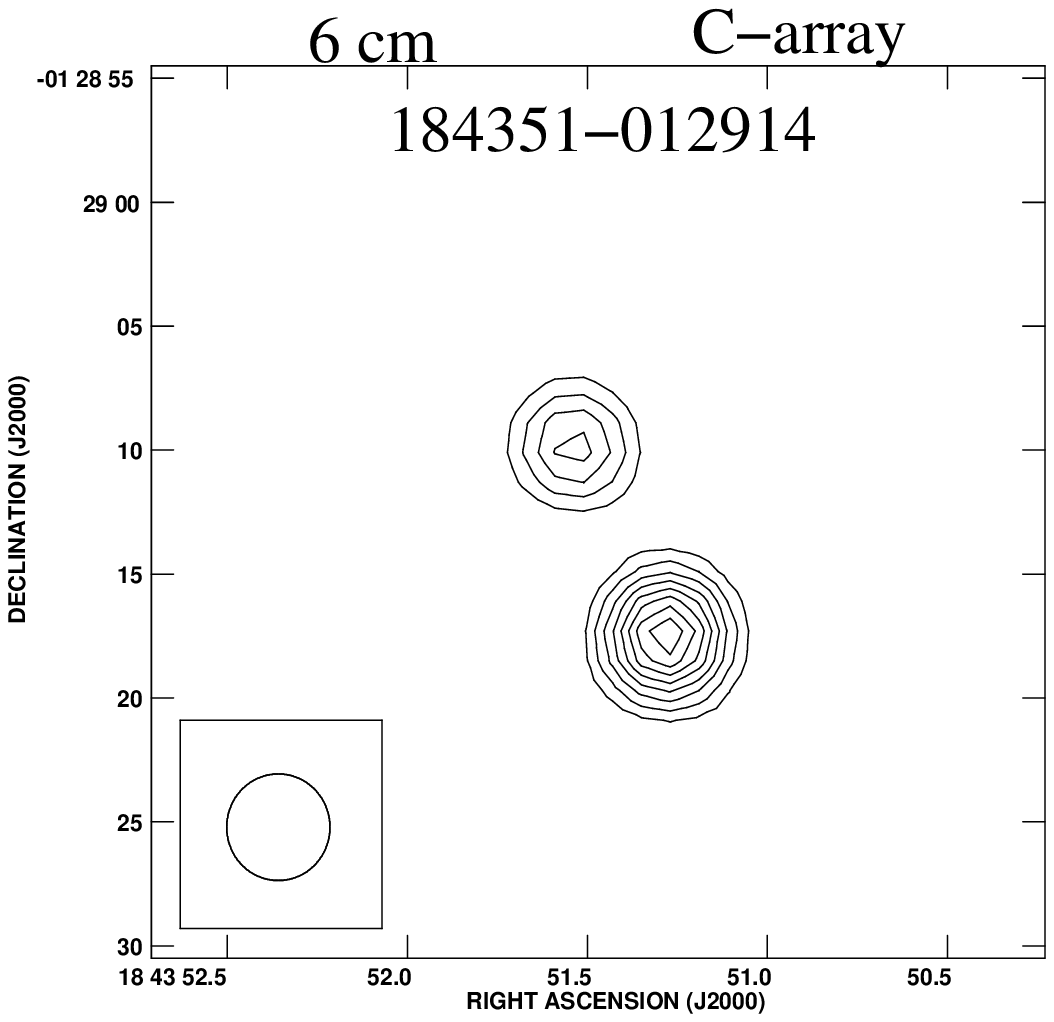}\\
\includegraphics[width=5cm]{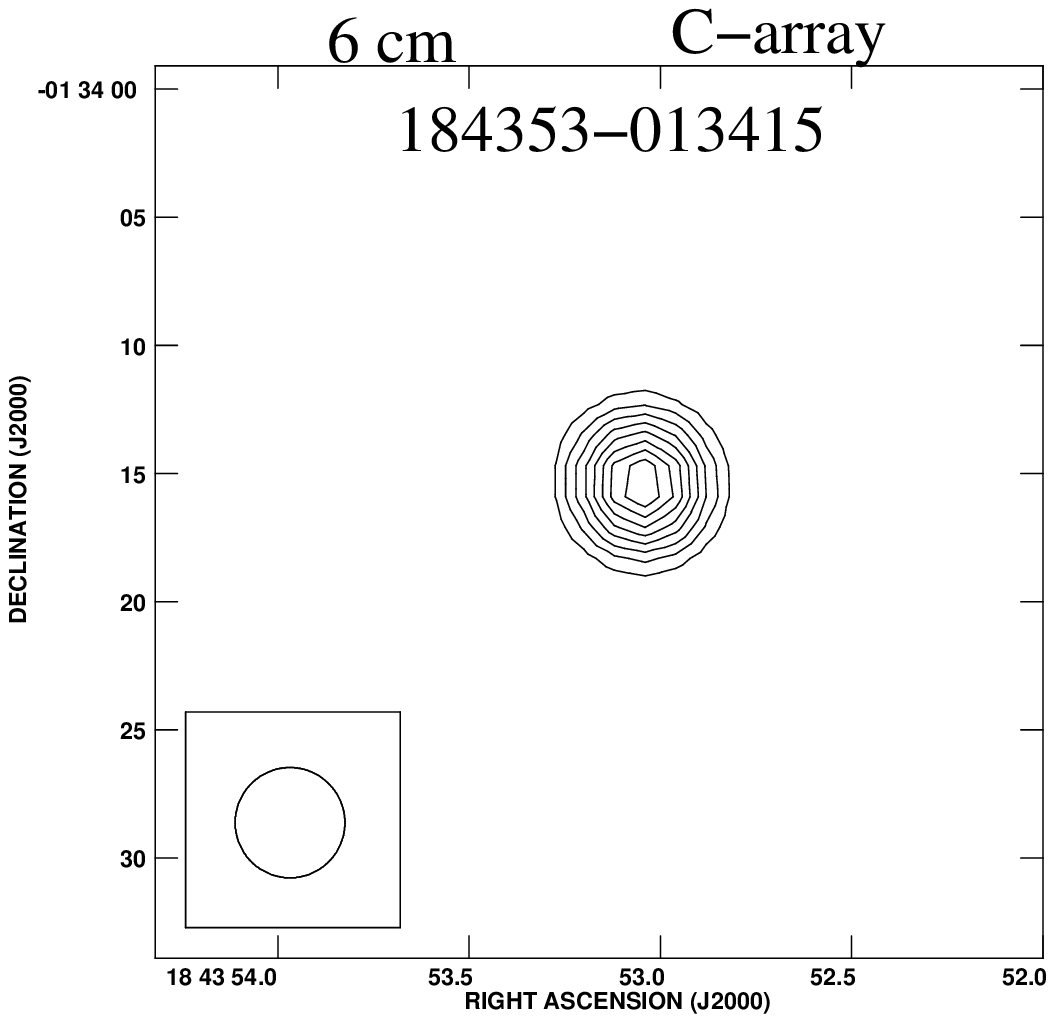} &
\includegraphics[width=5cm]{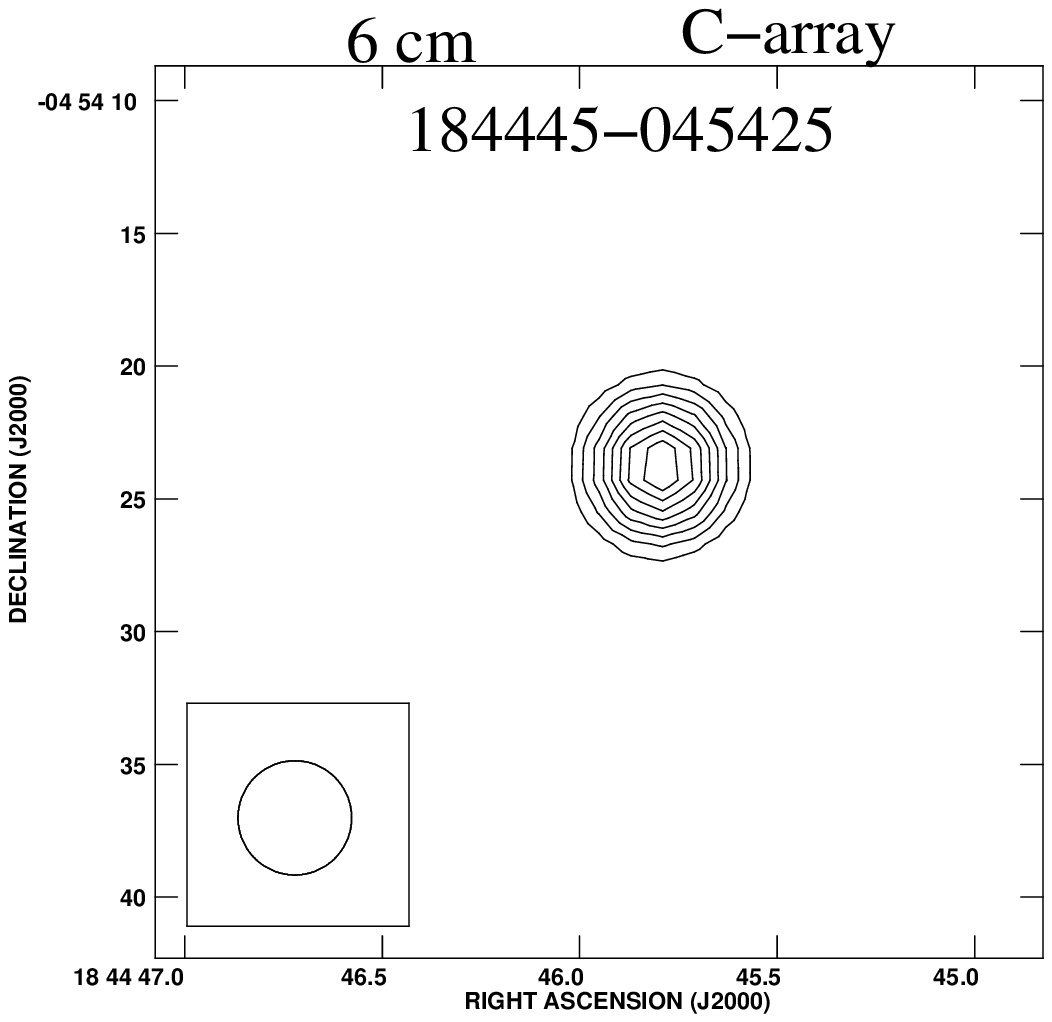}&
\includegraphics[width=5cm]{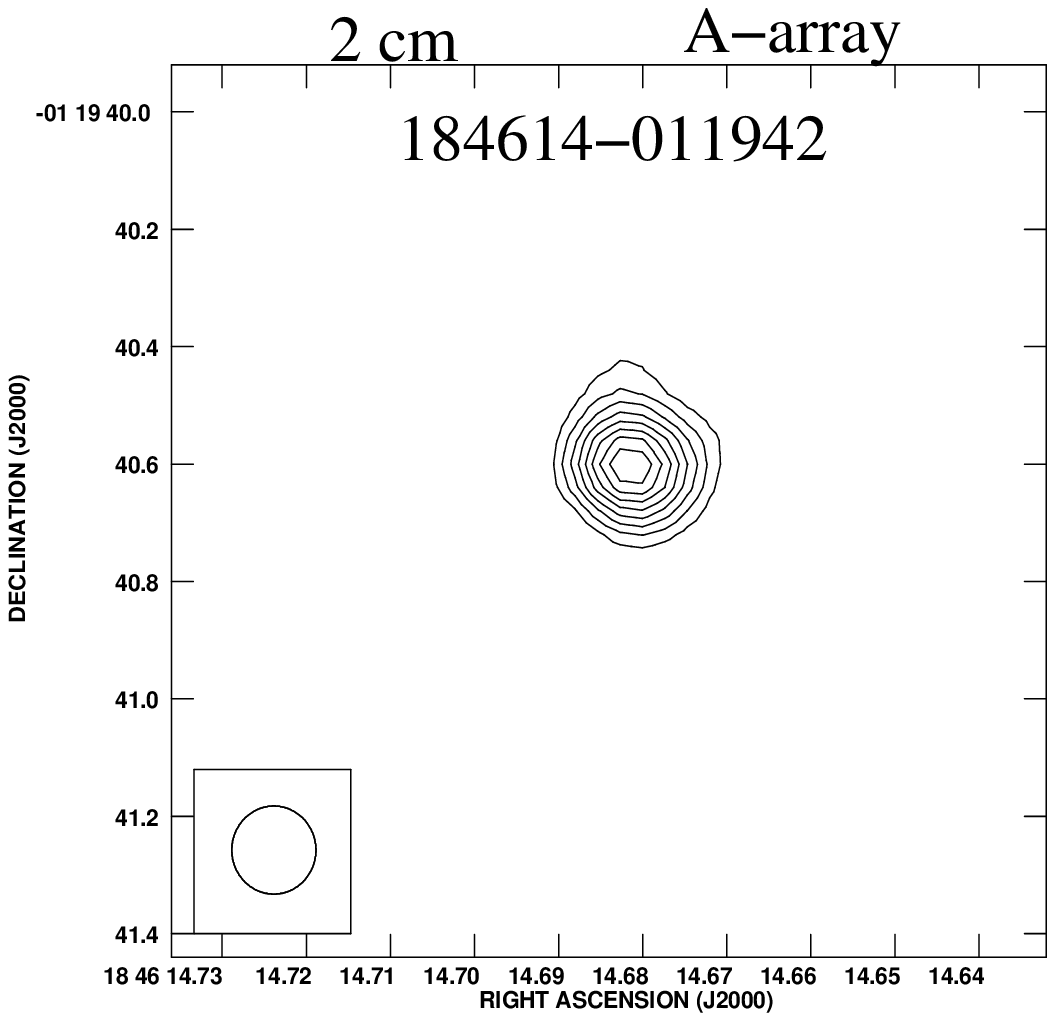}\\
\includegraphics[width=5cm]{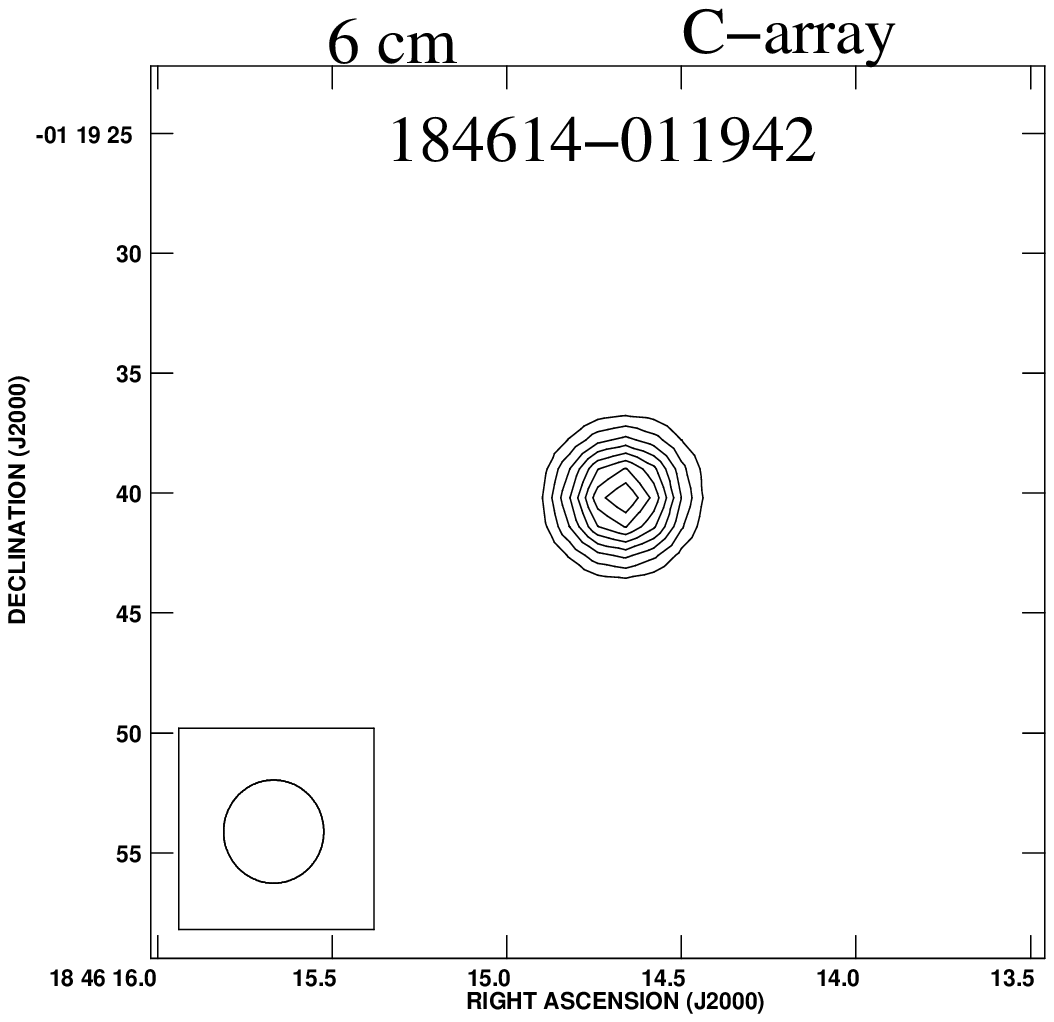} &
\includegraphics[width=5cm]{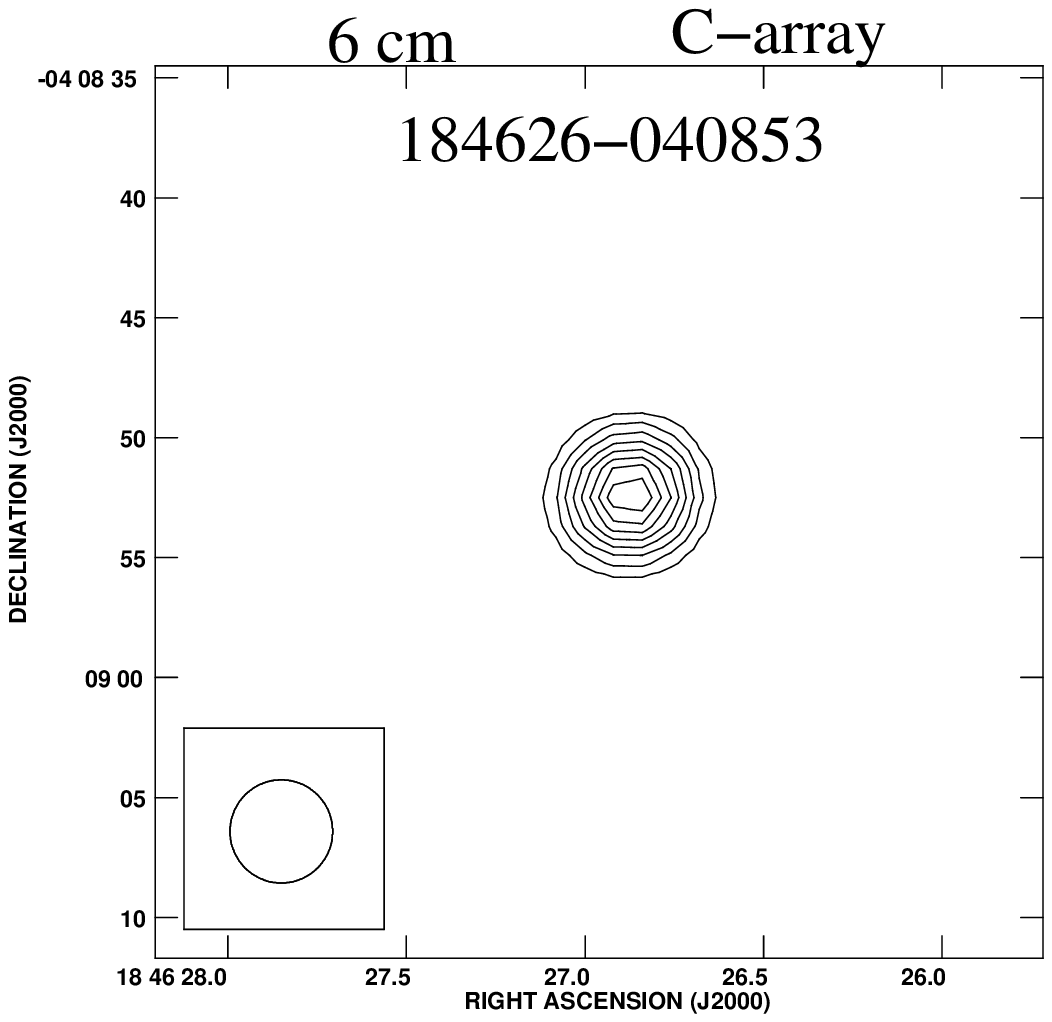}&
\includegraphics[width=5cm]{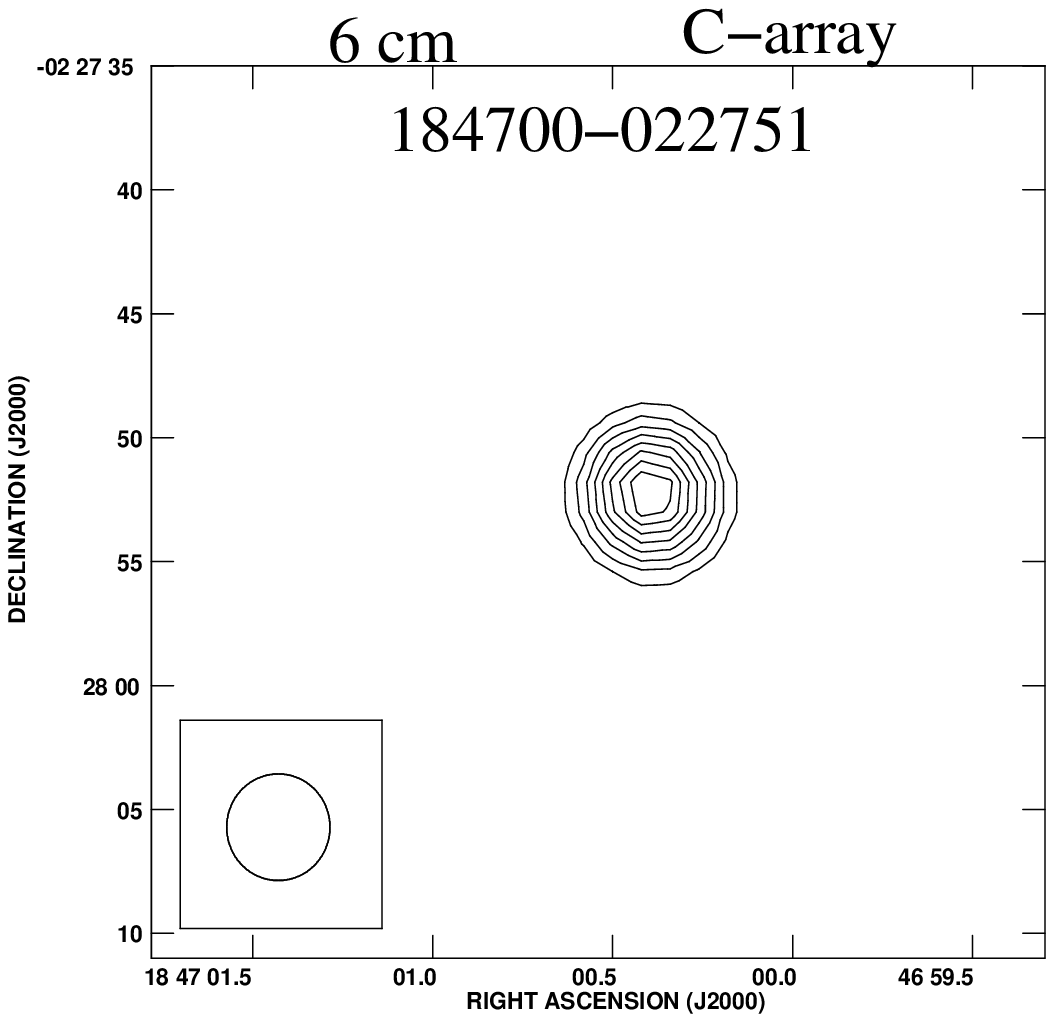}\\
\includegraphics[width=5cm]{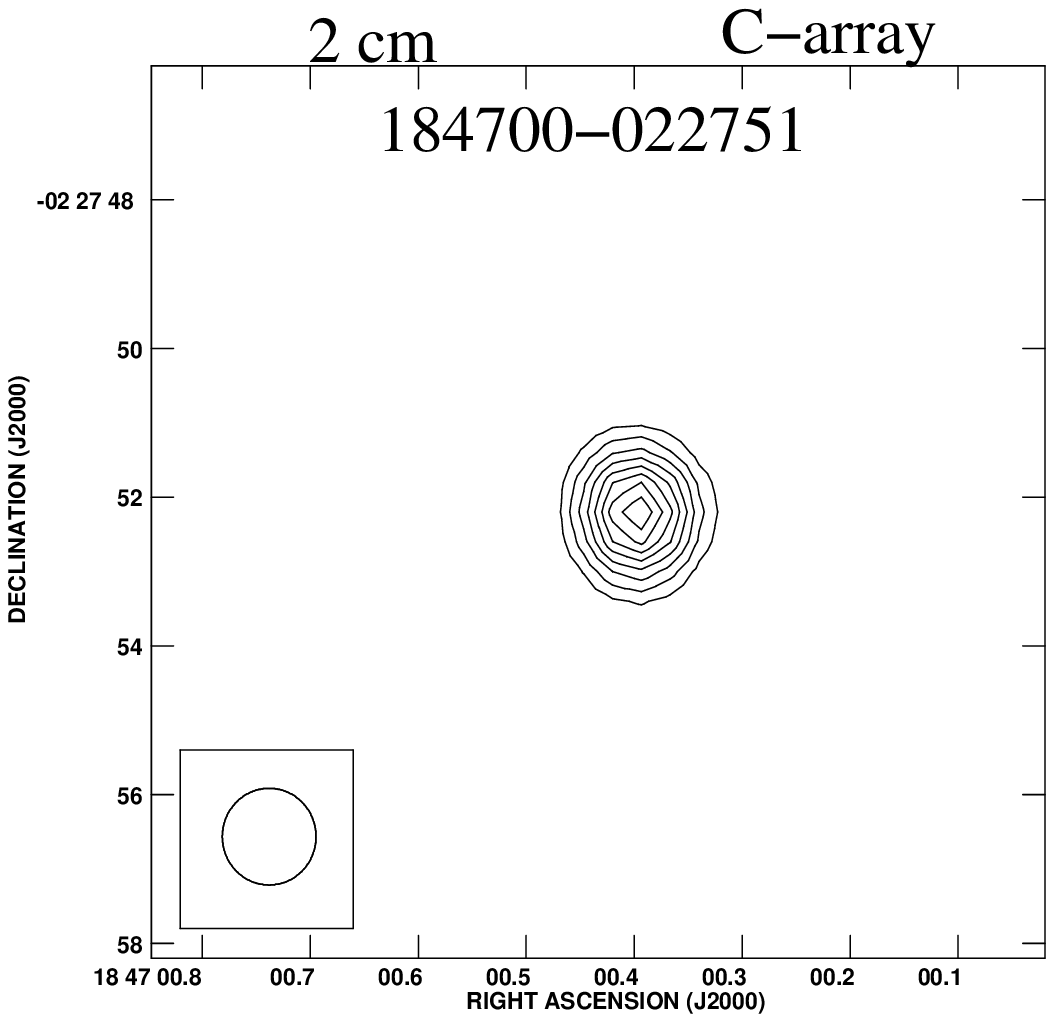} &
\includegraphics[width=5cm]{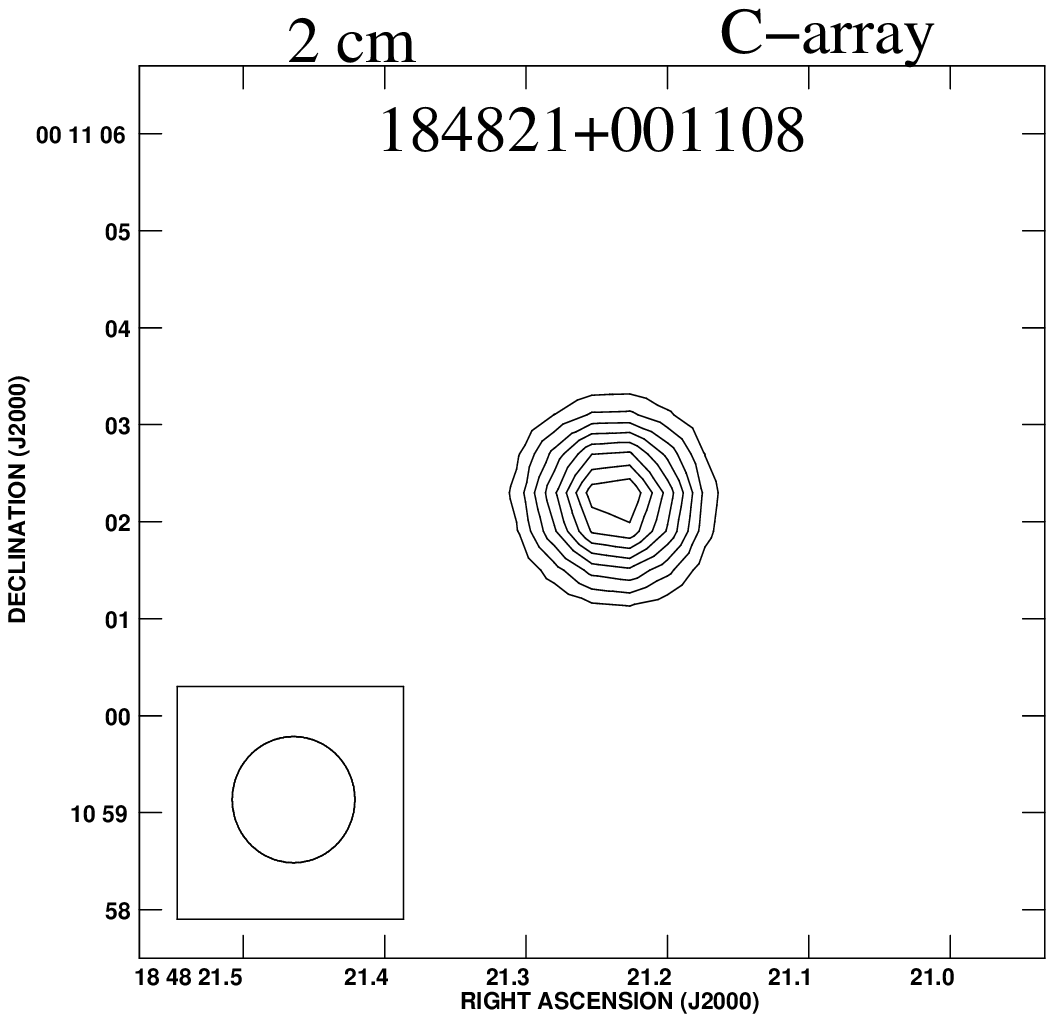}&
\includegraphics[width=5cm]{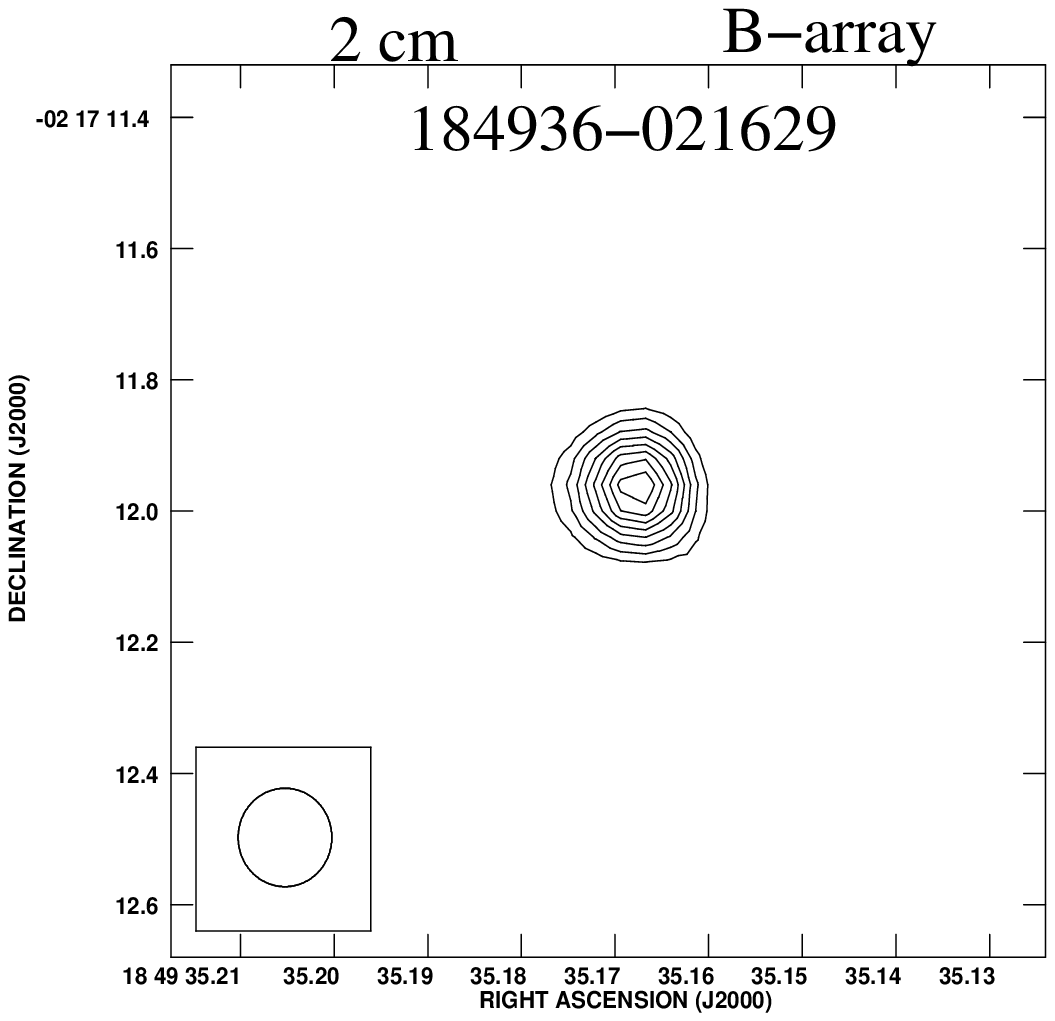}
\\

\end{tabular}
\end{figure*}
\begin{figure*}
\begin{tabular}{ccc}
\includegraphics[width=5cm]{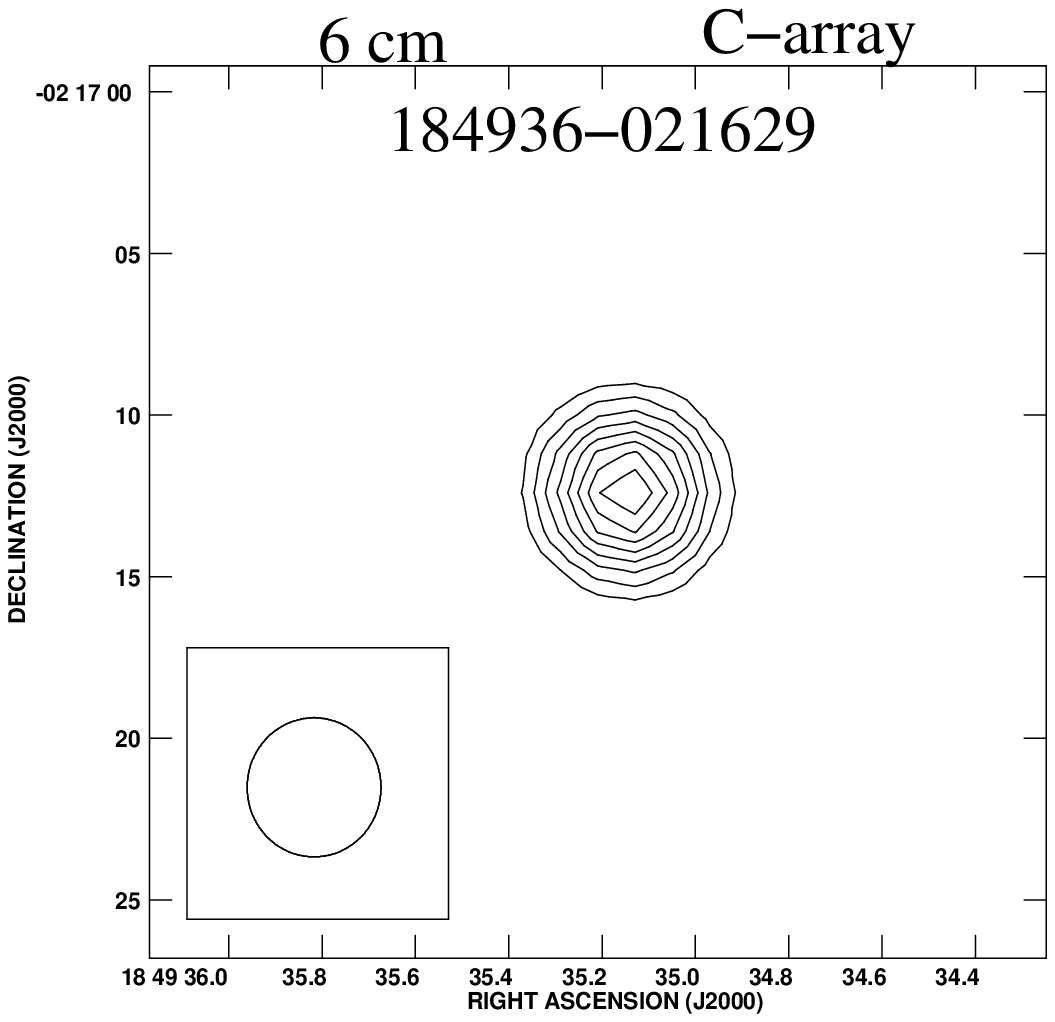} &
\includegraphics[width=5cm]{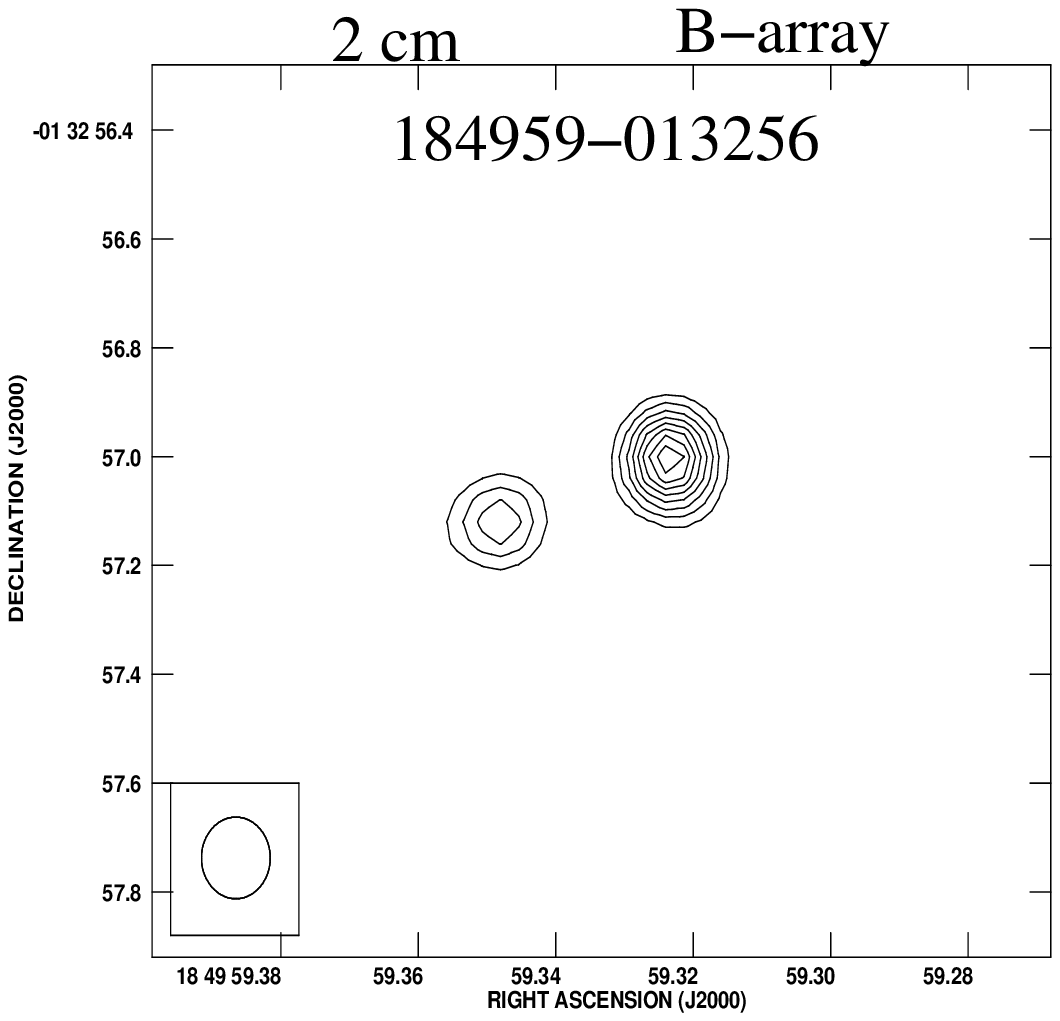}&
\includegraphics[width=5cm]{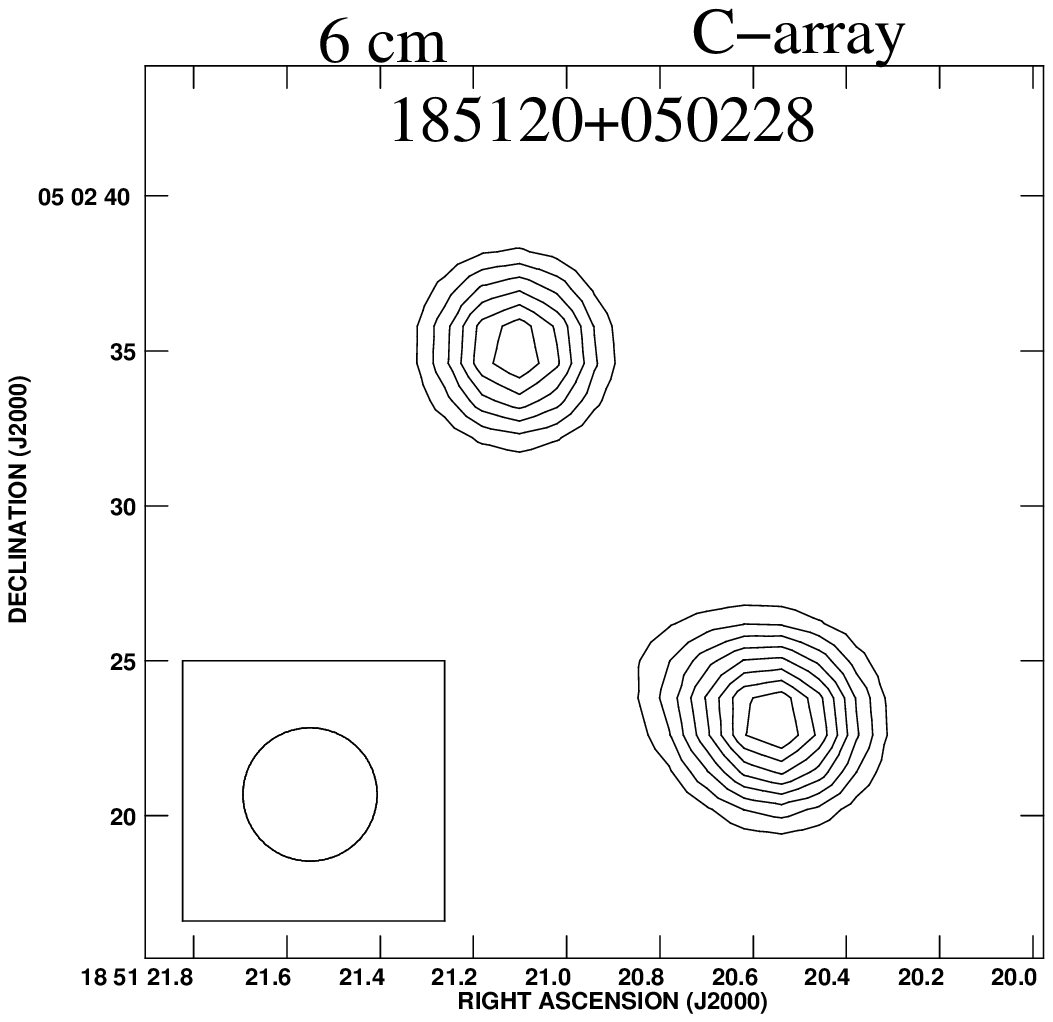}\\
\includegraphics[width=5cm]{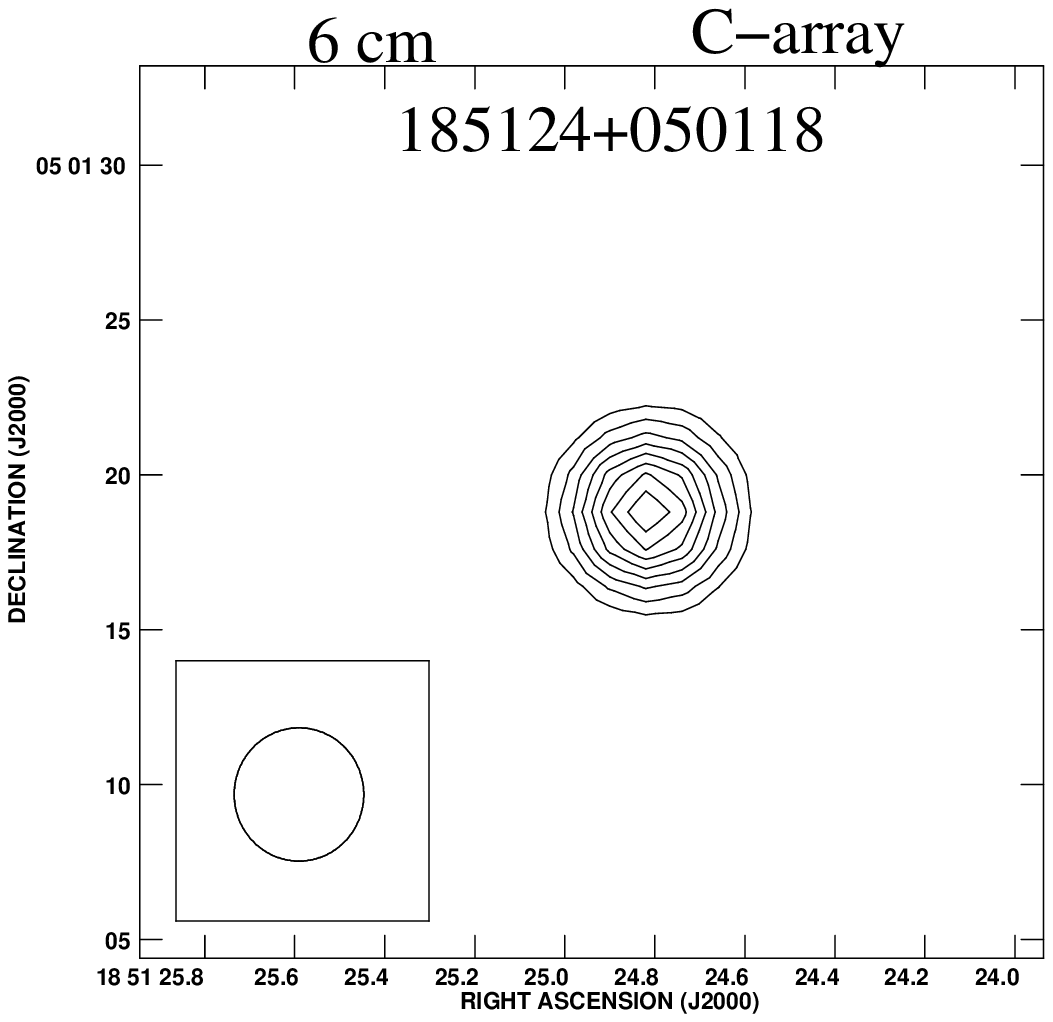} &
\includegraphics[width=5cm]{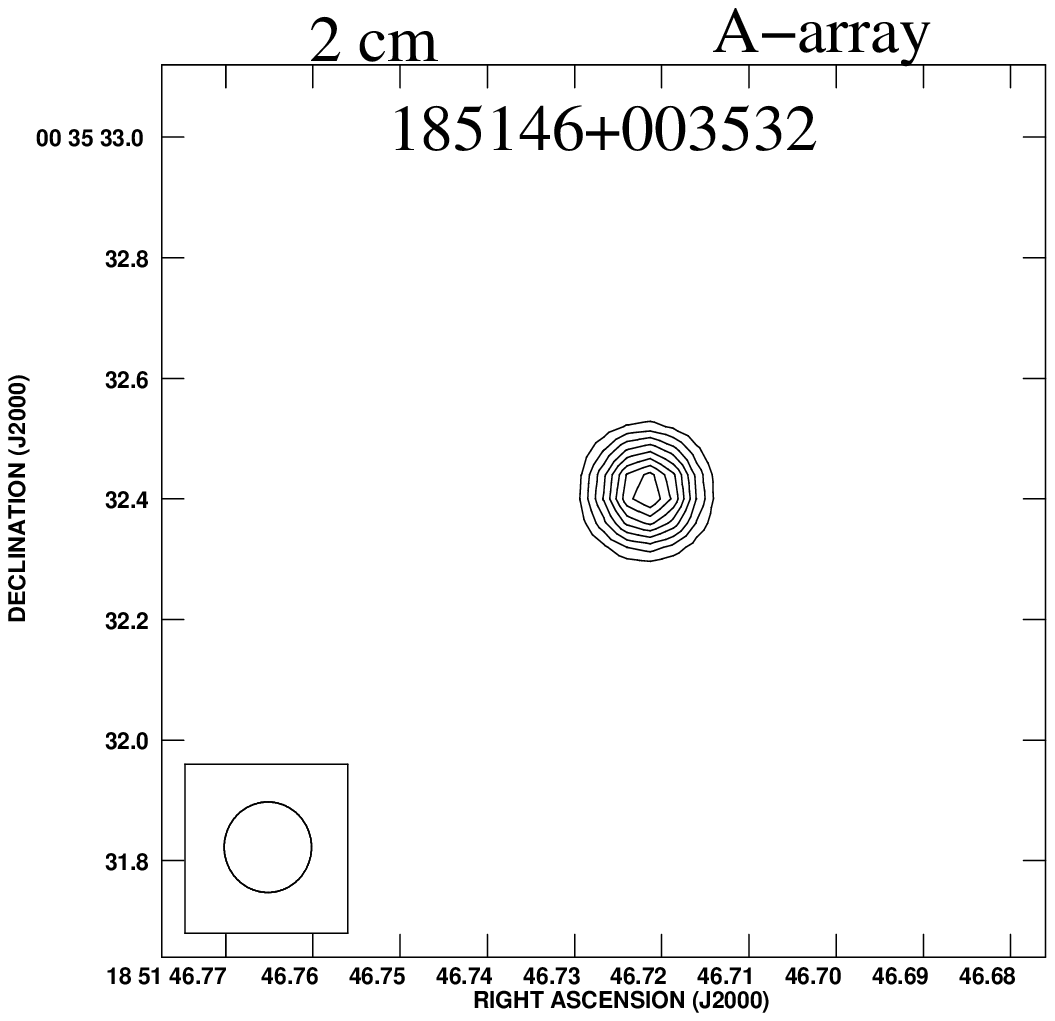}&
\includegraphics[width=5cm]{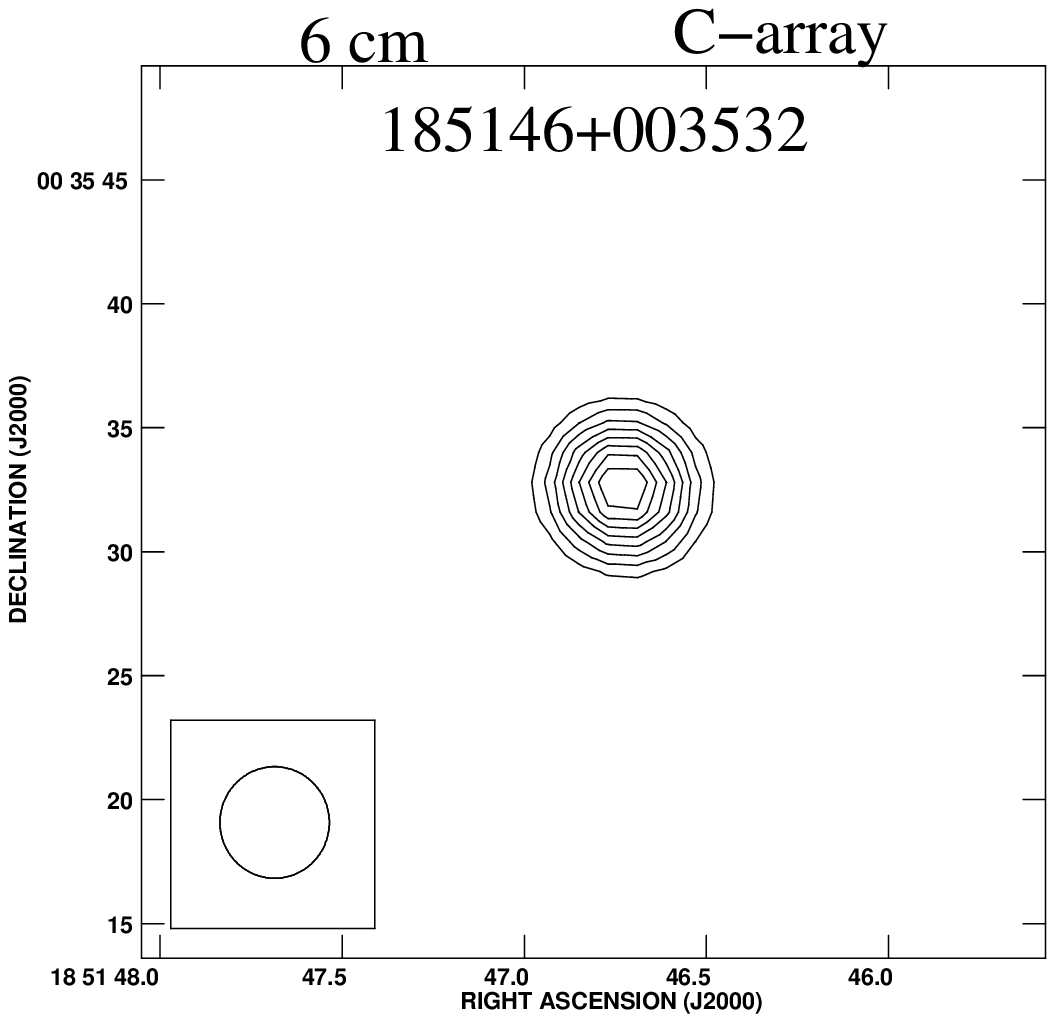}\\
\includegraphics[width=5cm]{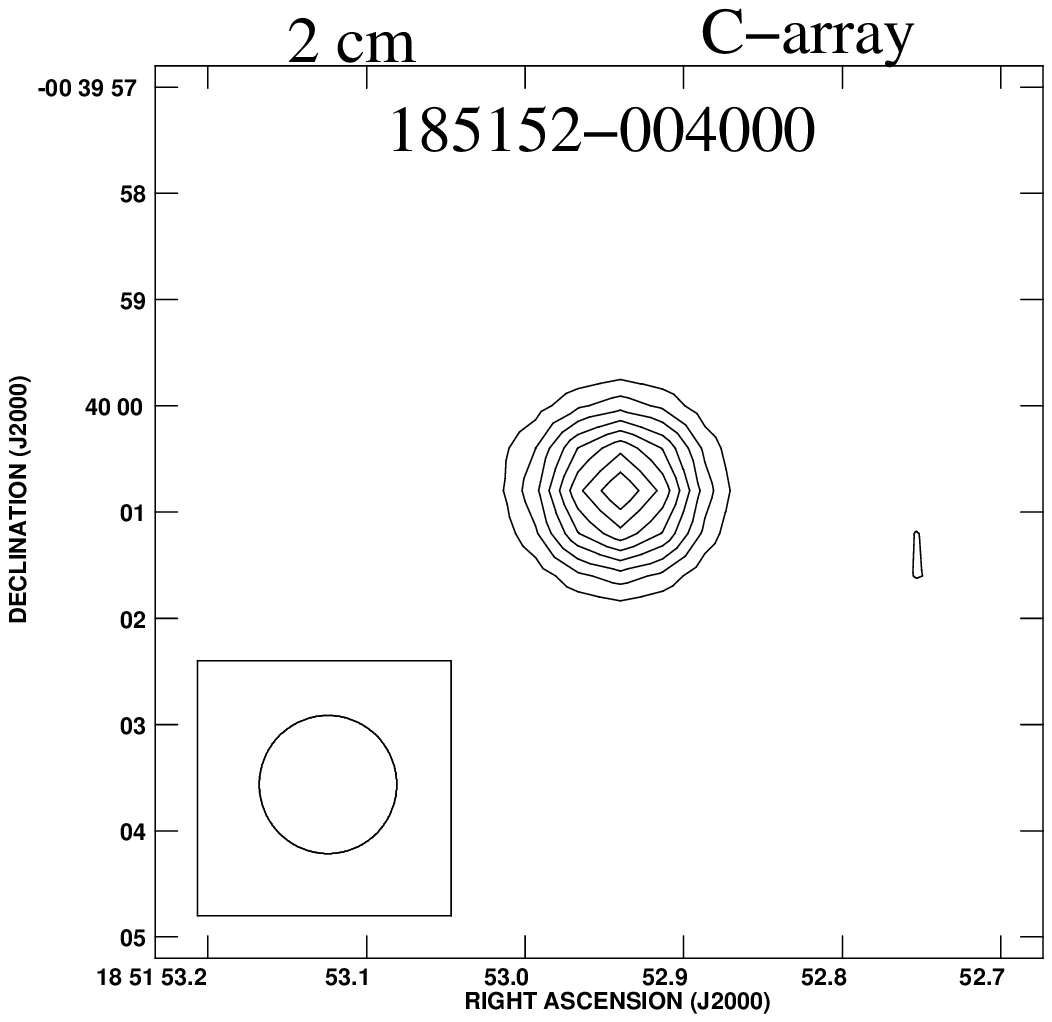} &
\includegraphics[width=5cm]{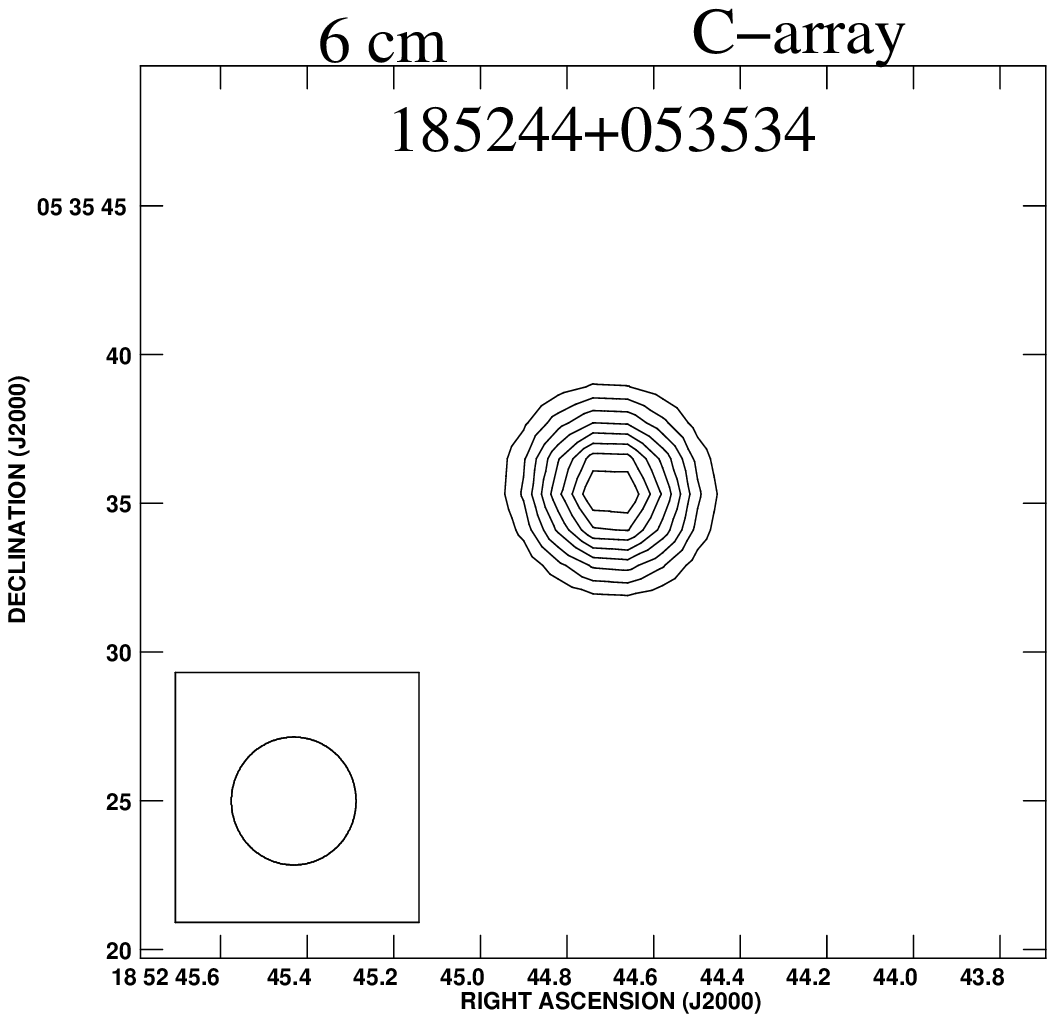}&
\includegraphics[width=5cm]{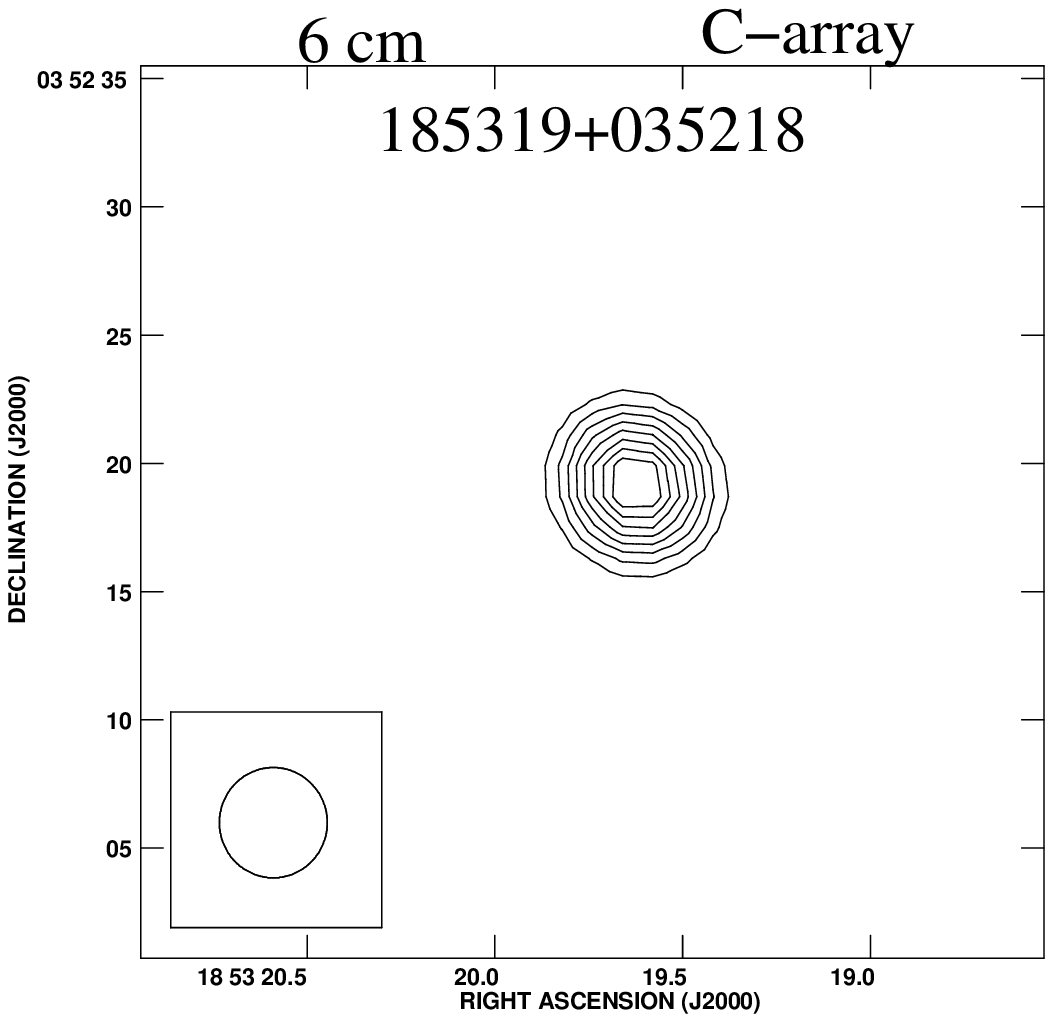}\\
\includegraphics[width=5cm]{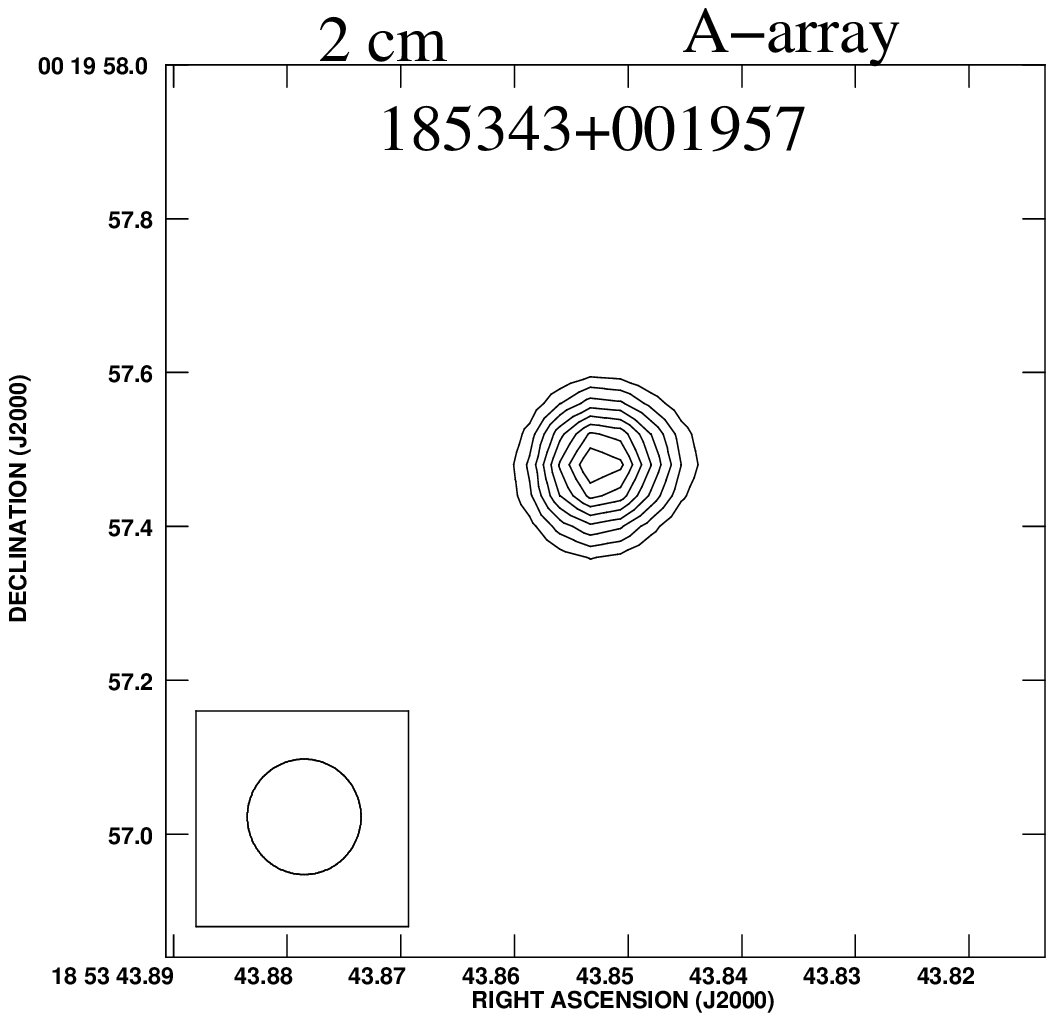} &
\includegraphics[width=5cm]{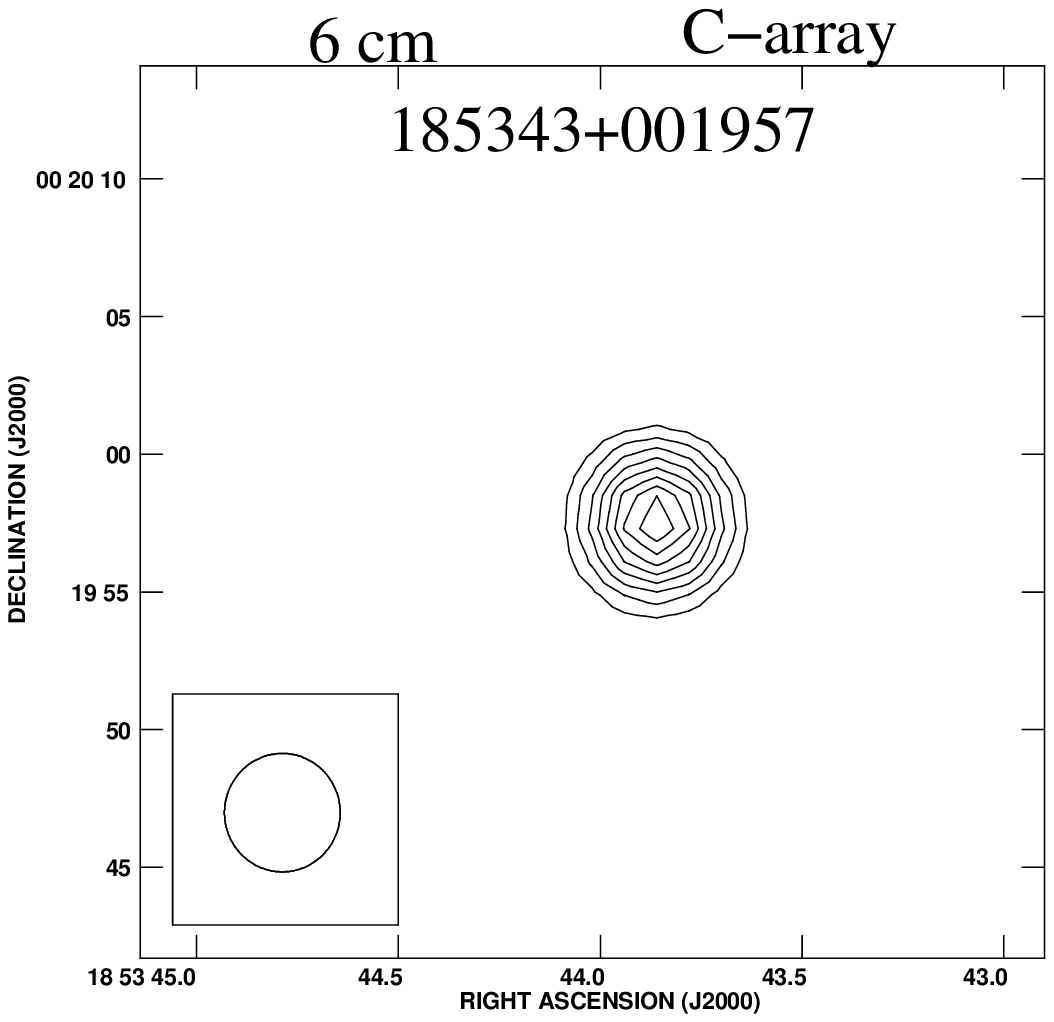}&
\includegraphics[width=5cm]{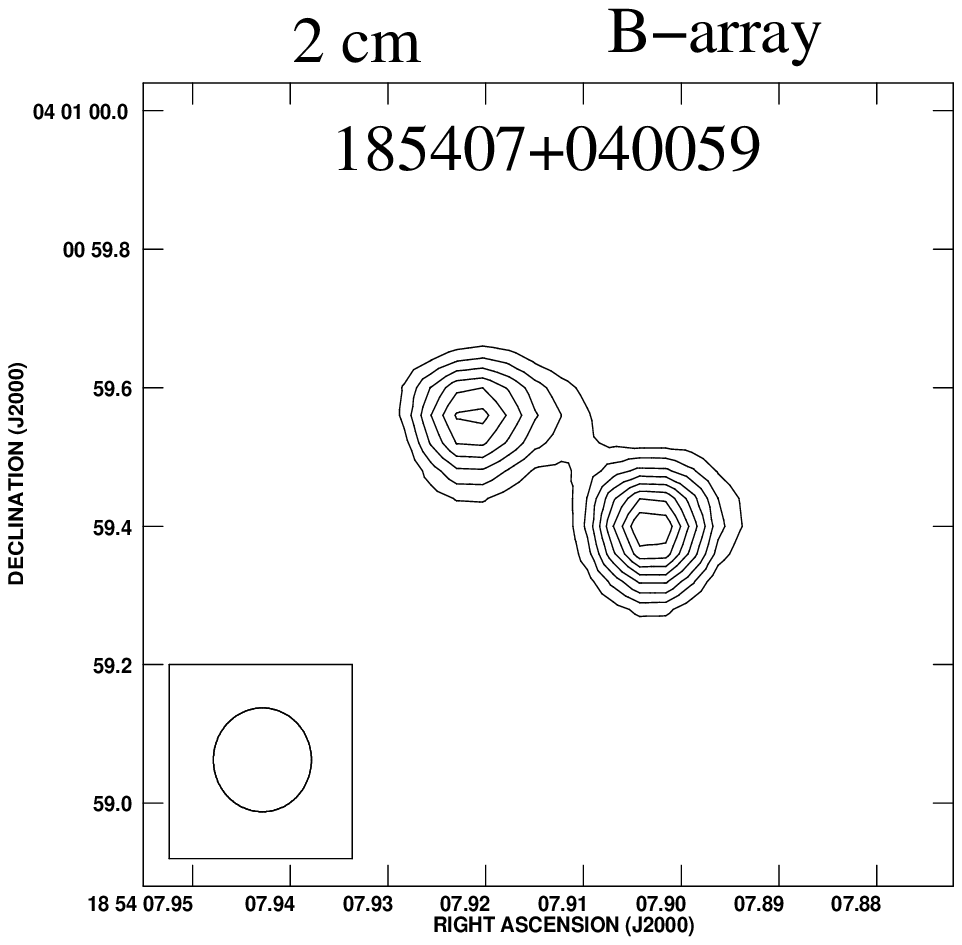}
\\

\end{tabular}
\end{figure*}
\begin{figure*}
\begin{tabular}{ccc}
\includegraphics[width=5cm]{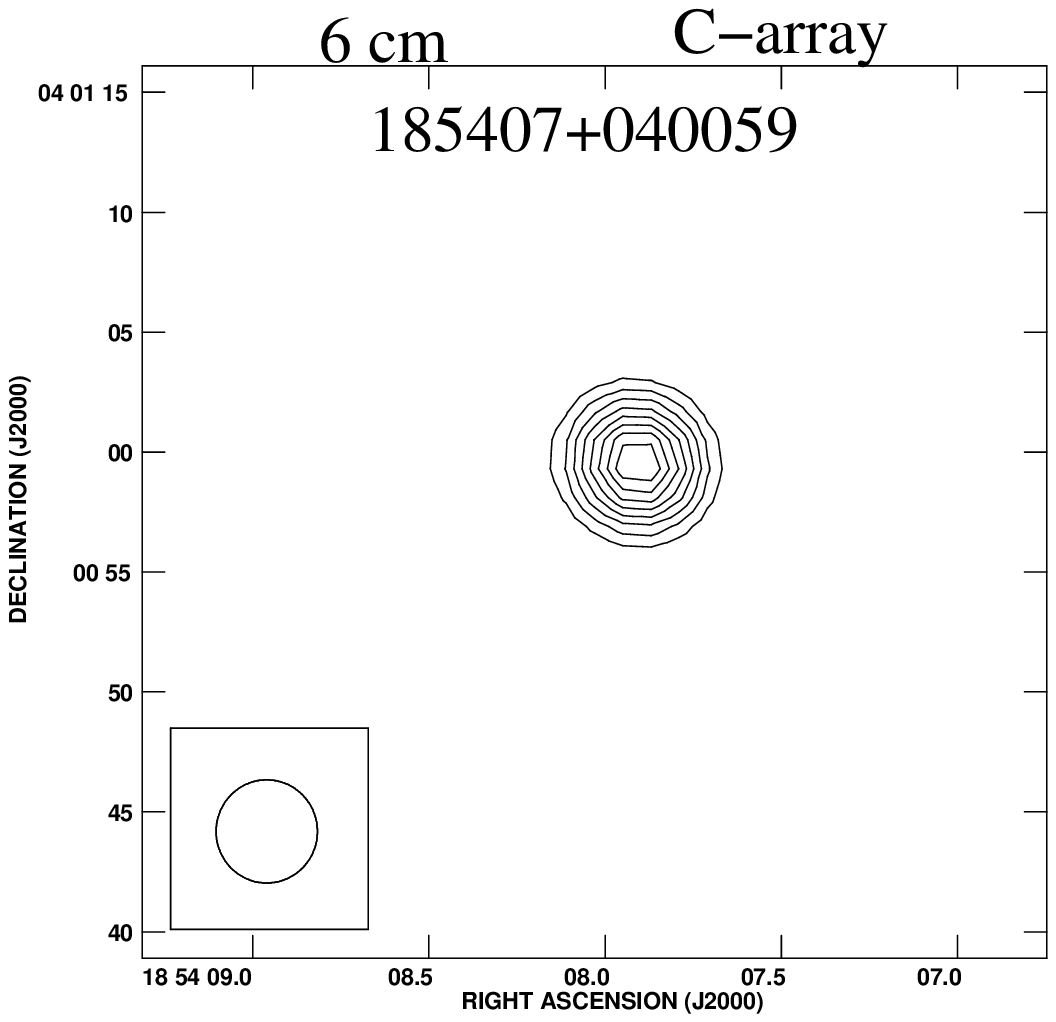} &
\includegraphics[width=5cm]{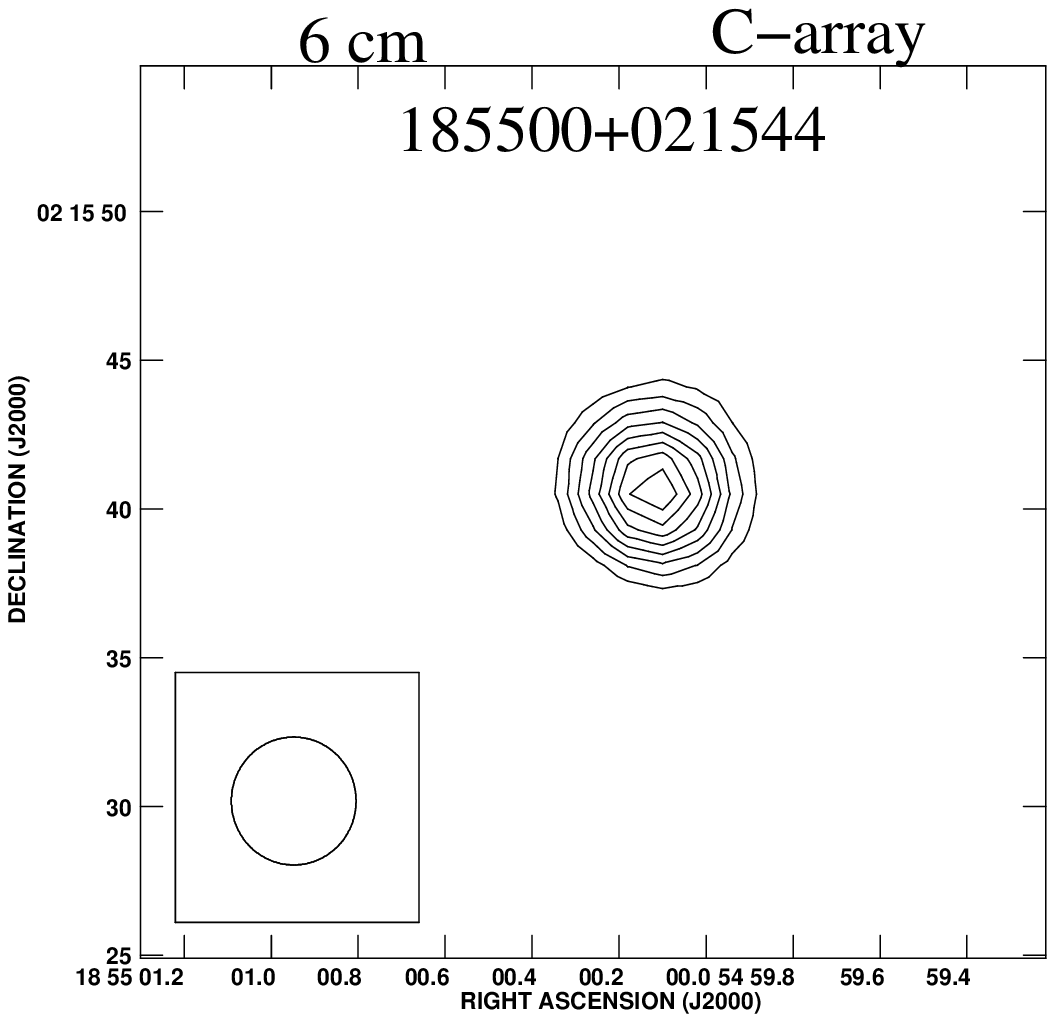}&
\includegraphics[width=5cm]{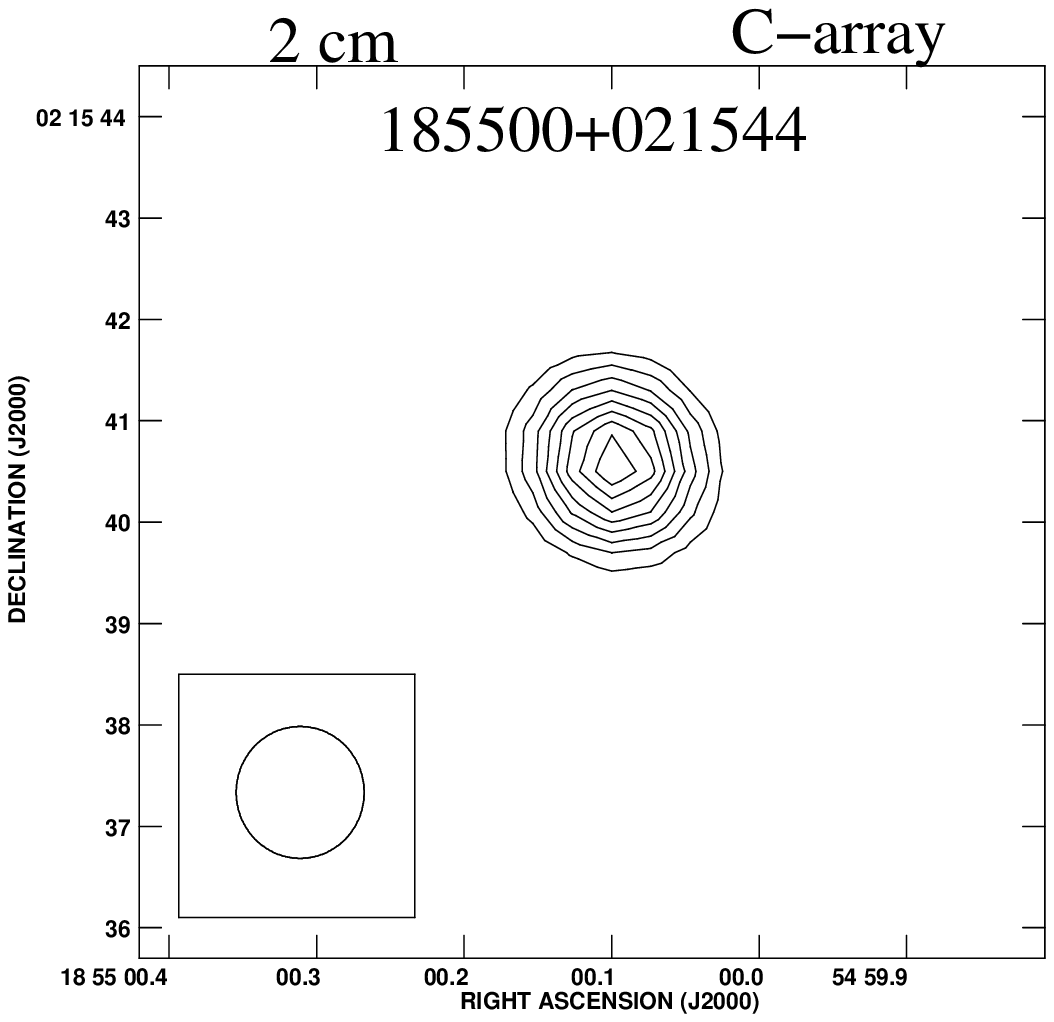}\\
\includegraphics[width=5cm]{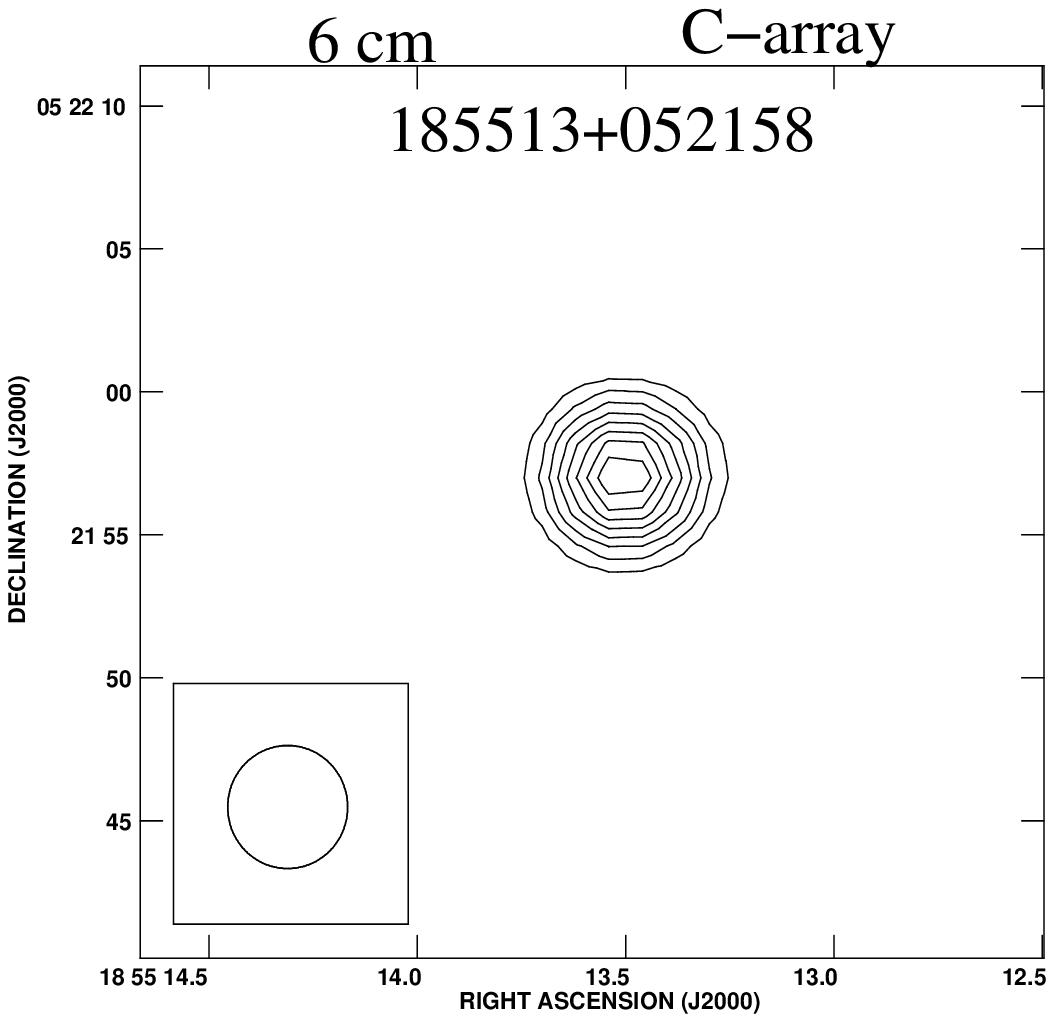} &
\includegraphics[width=5cm]{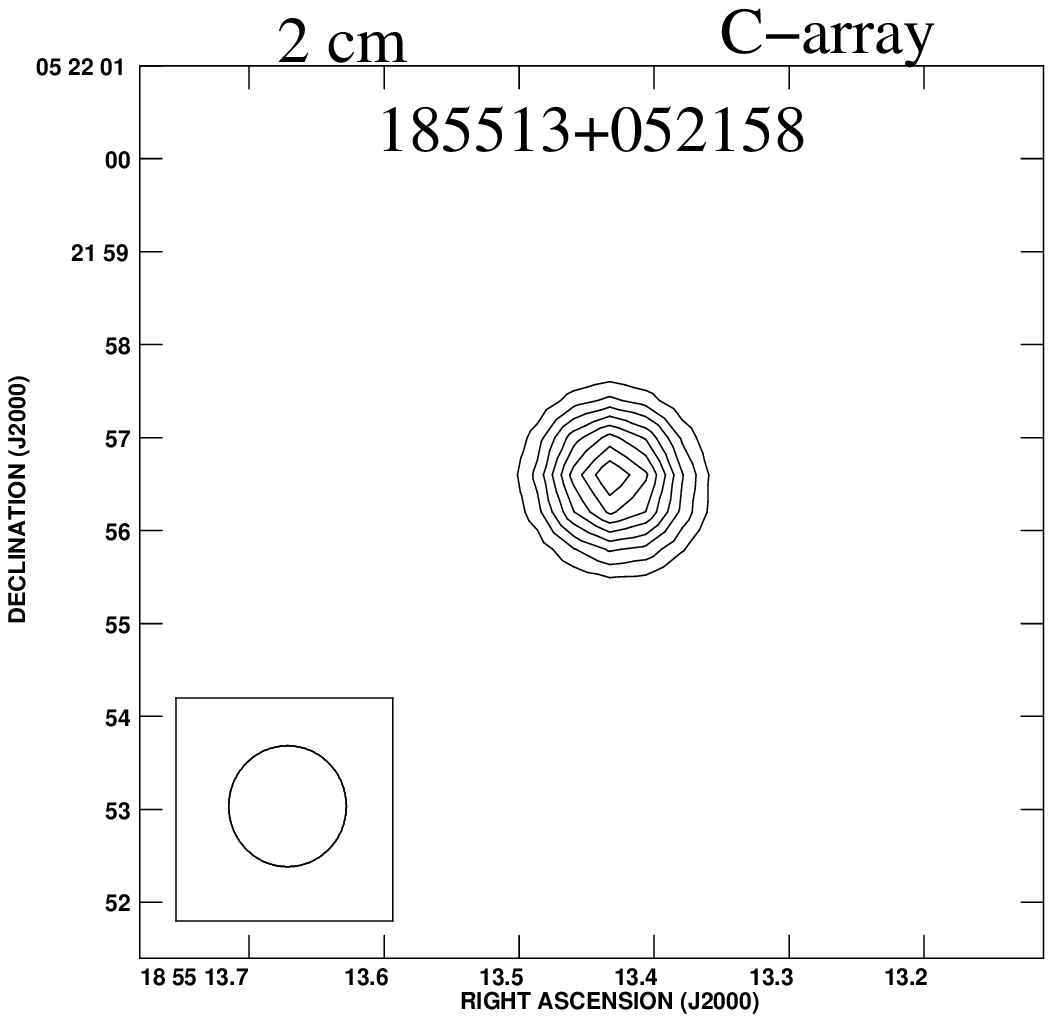}&
\includegraphics[width=5cm]{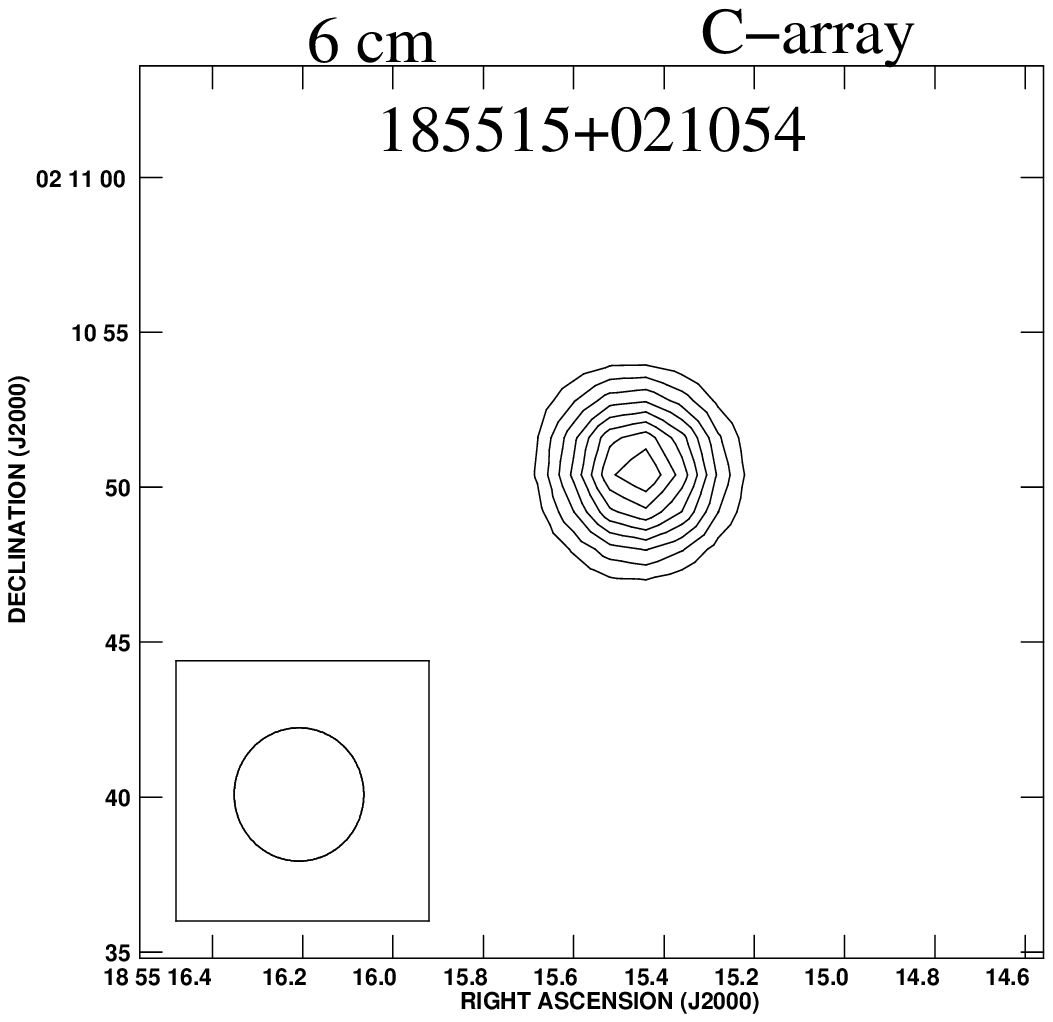}\\
\includegraphics[width=5cm]{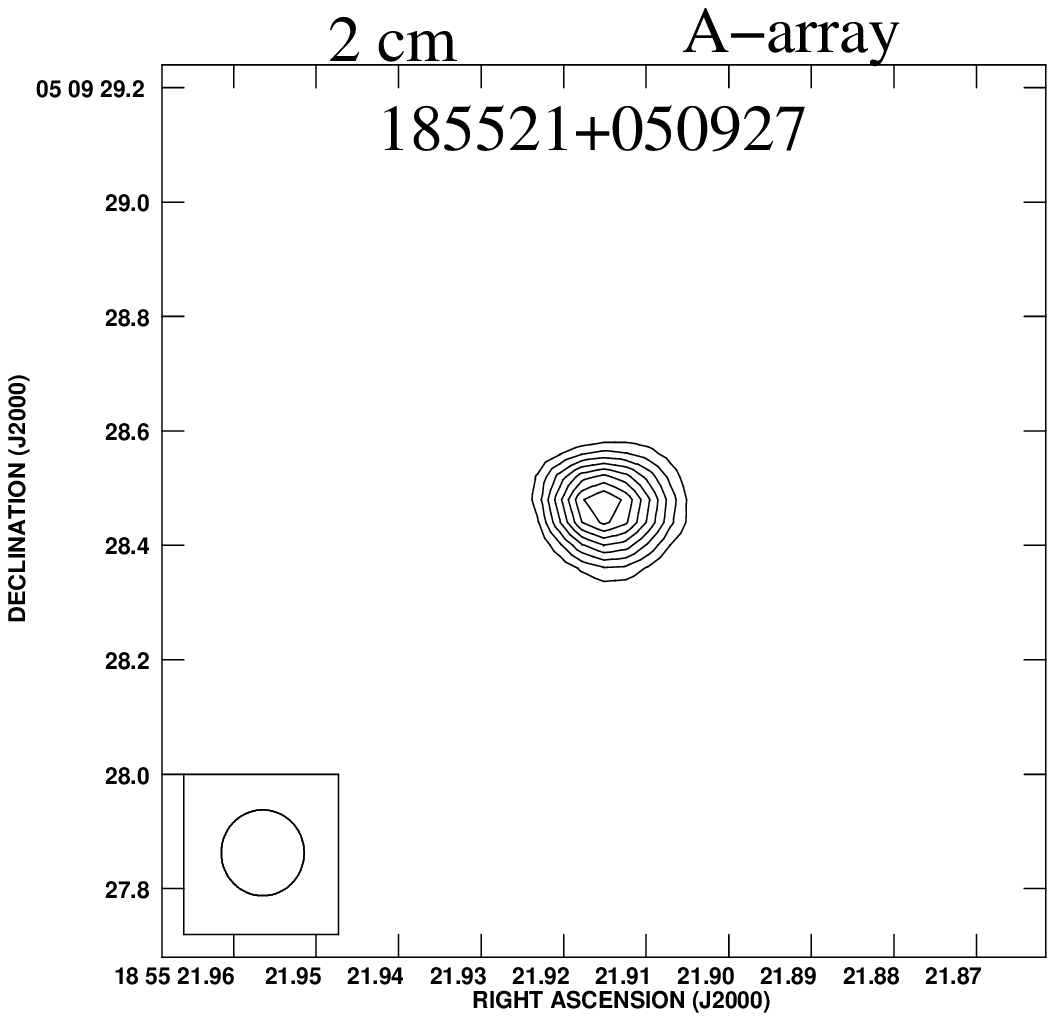} &
\includegraphics[width=5cm]{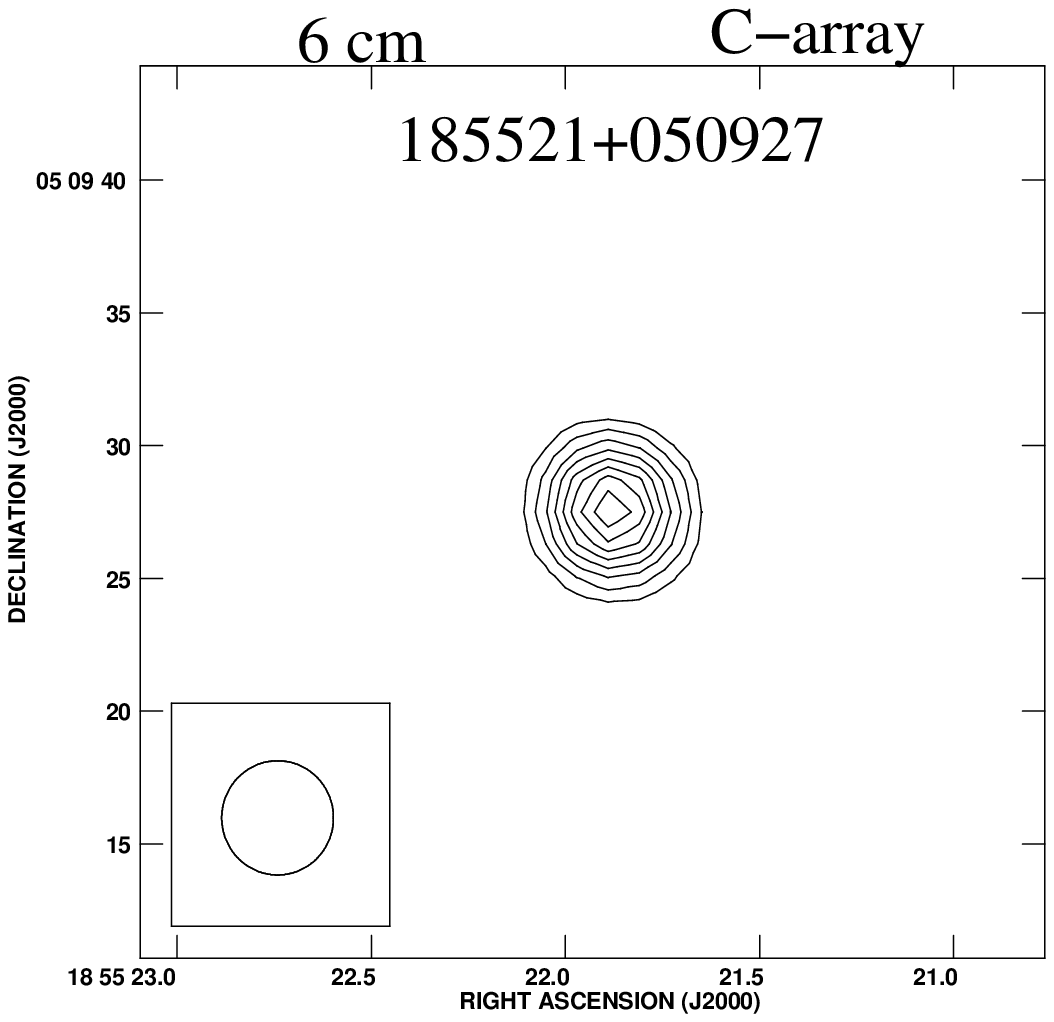}&
\includegraphics[width=5cm]{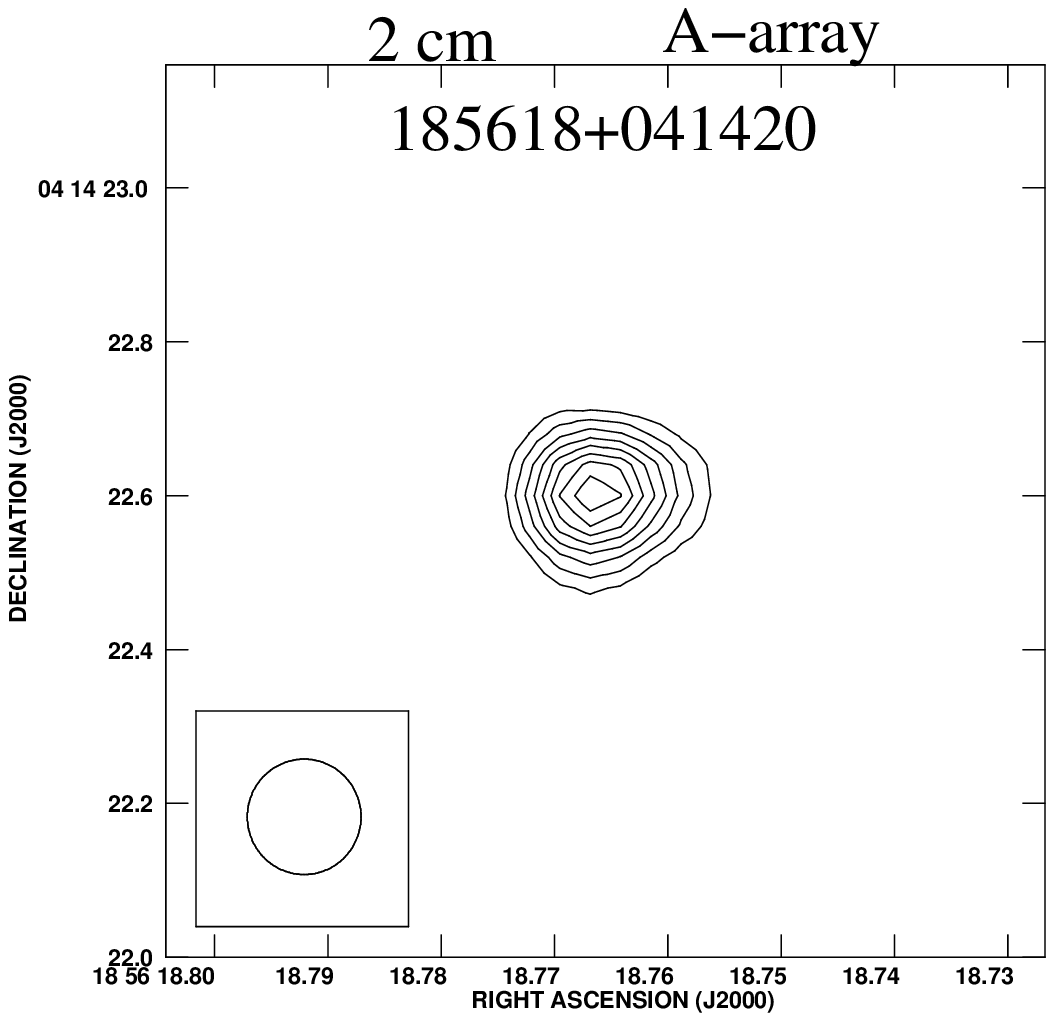}\\
\includegraphics[width=5cm]{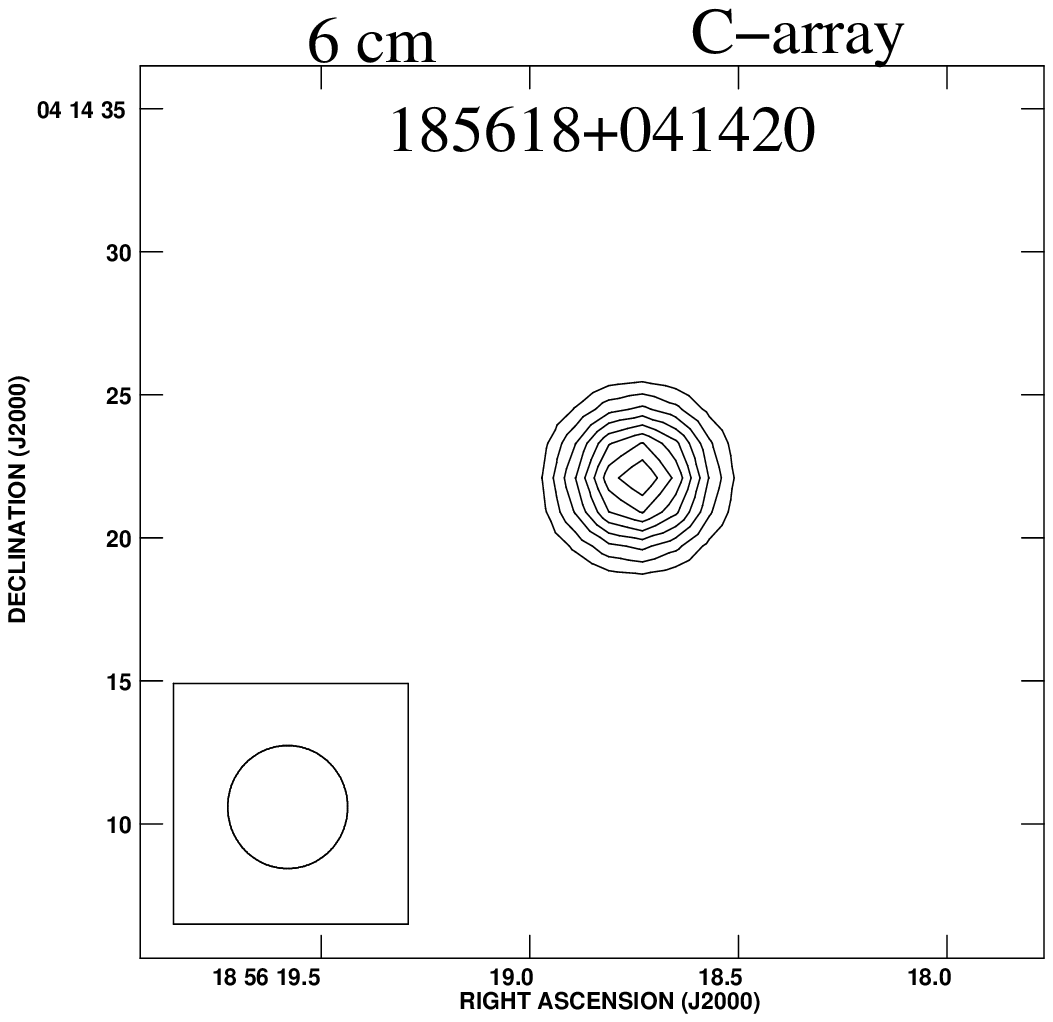} &
\includegraphics[width=5cm]{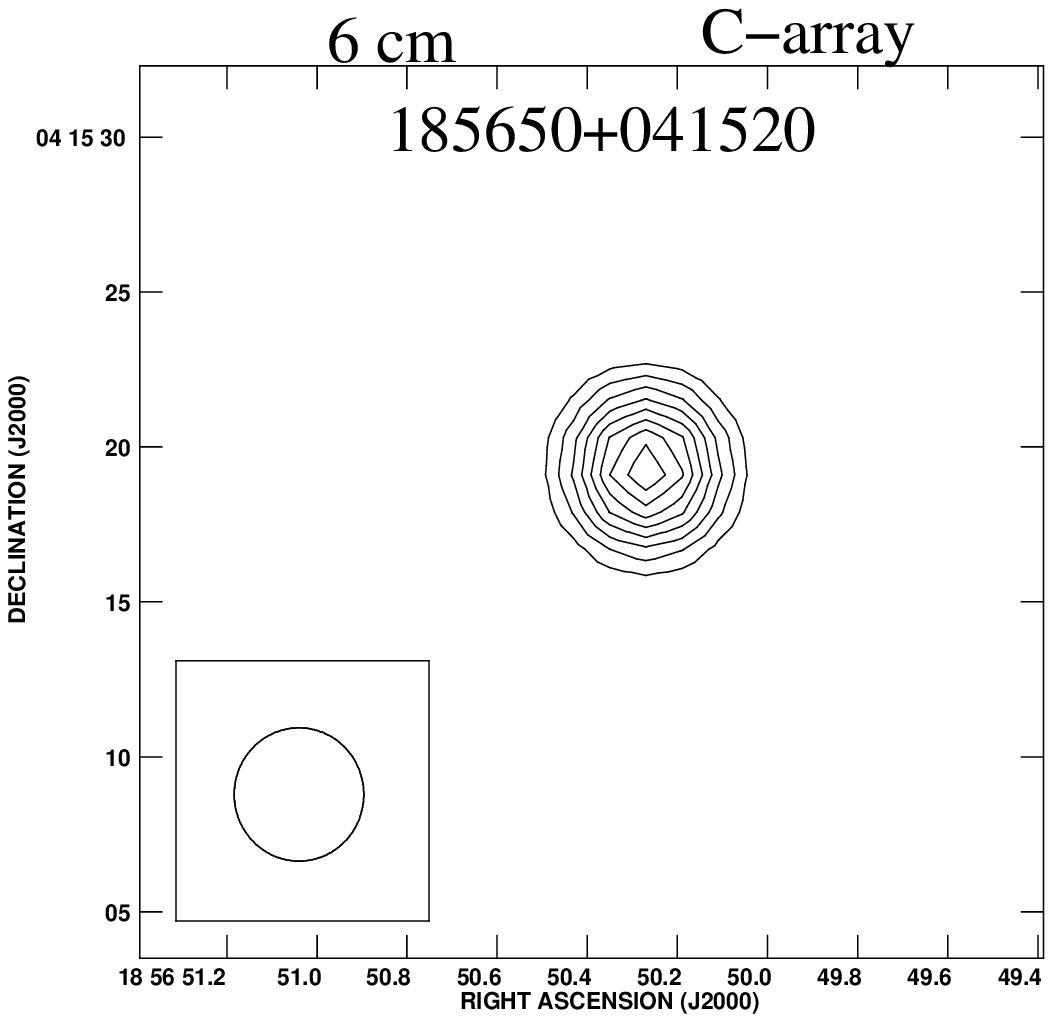}&
\includegraphics[width=5cm]{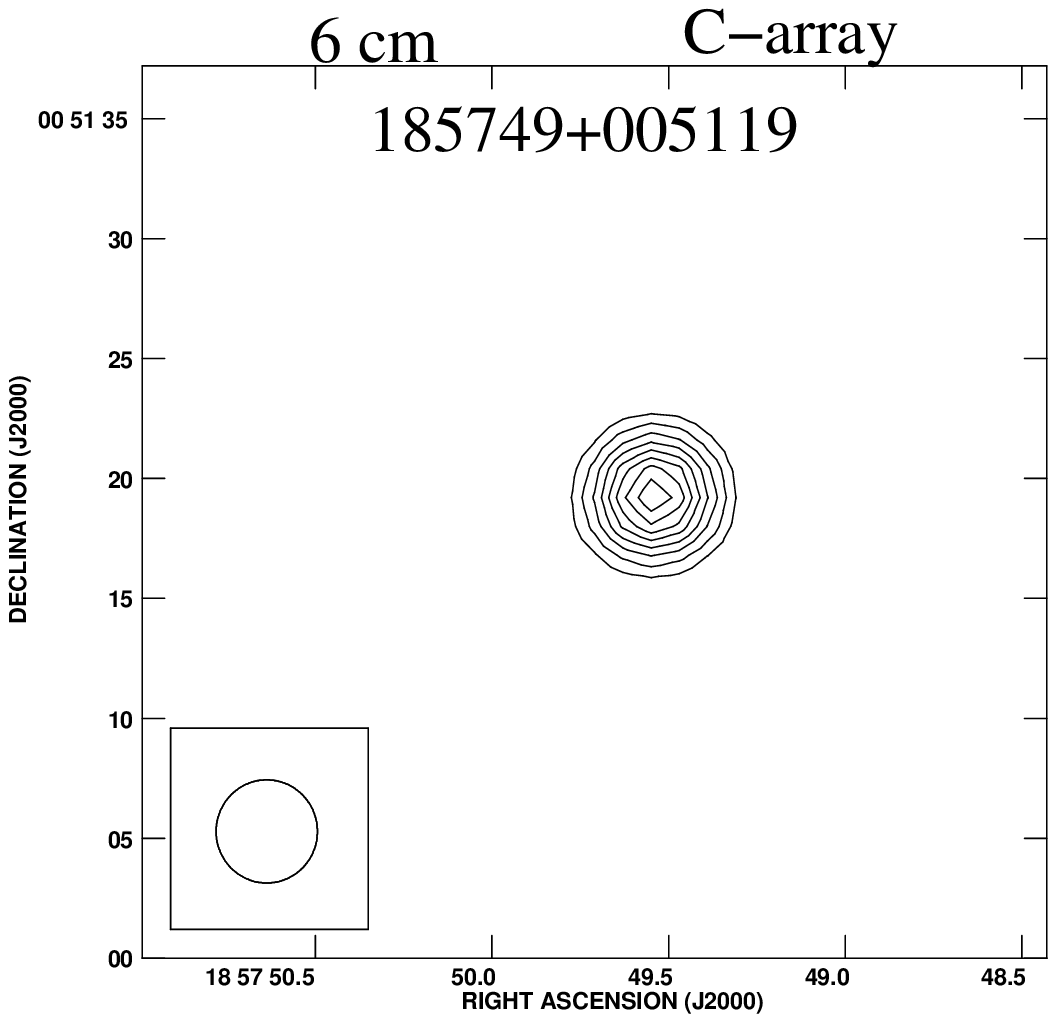}
\\

\end{tabular}
\end{figure*}
\begin{figure*}
\begin{tabular}{ccc}
\includegraphics[width=5cm]{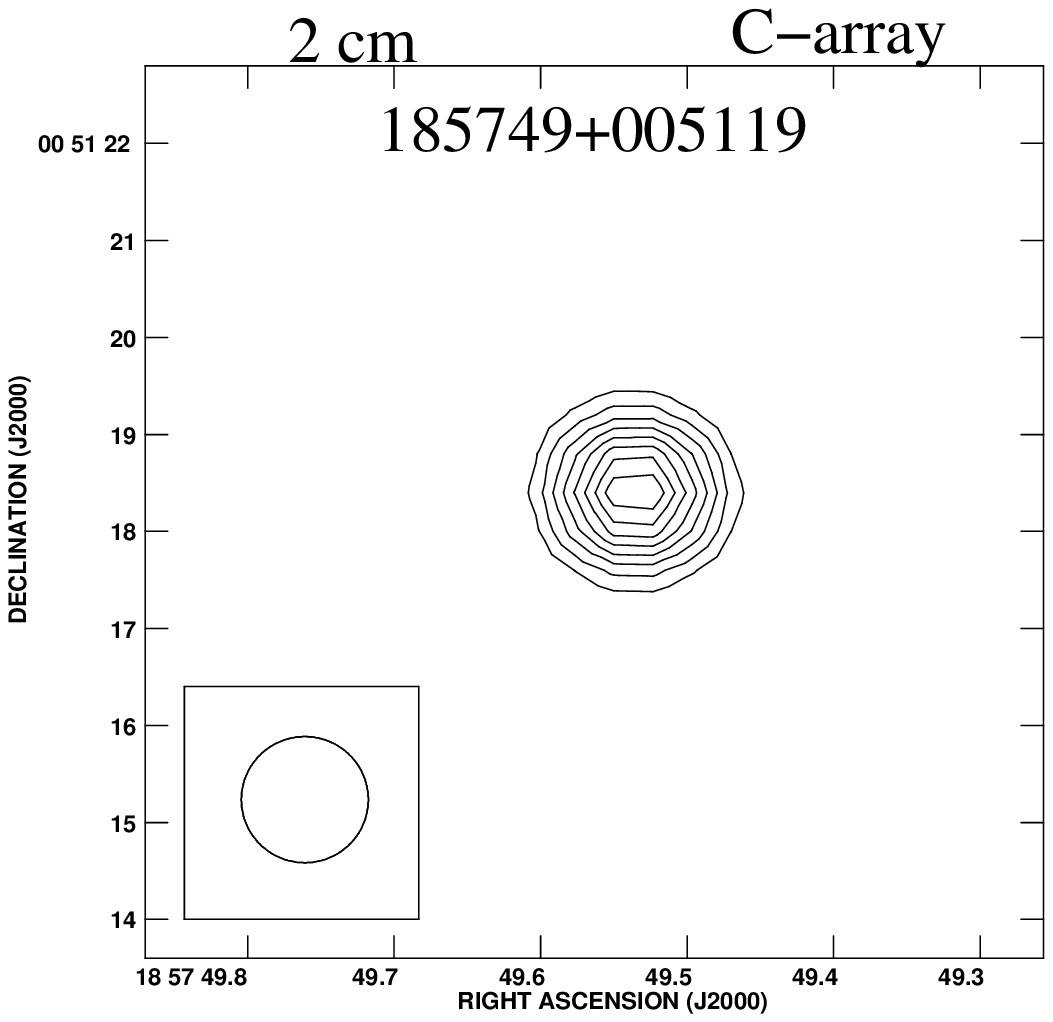} &
\includegraphics[width=5cm]{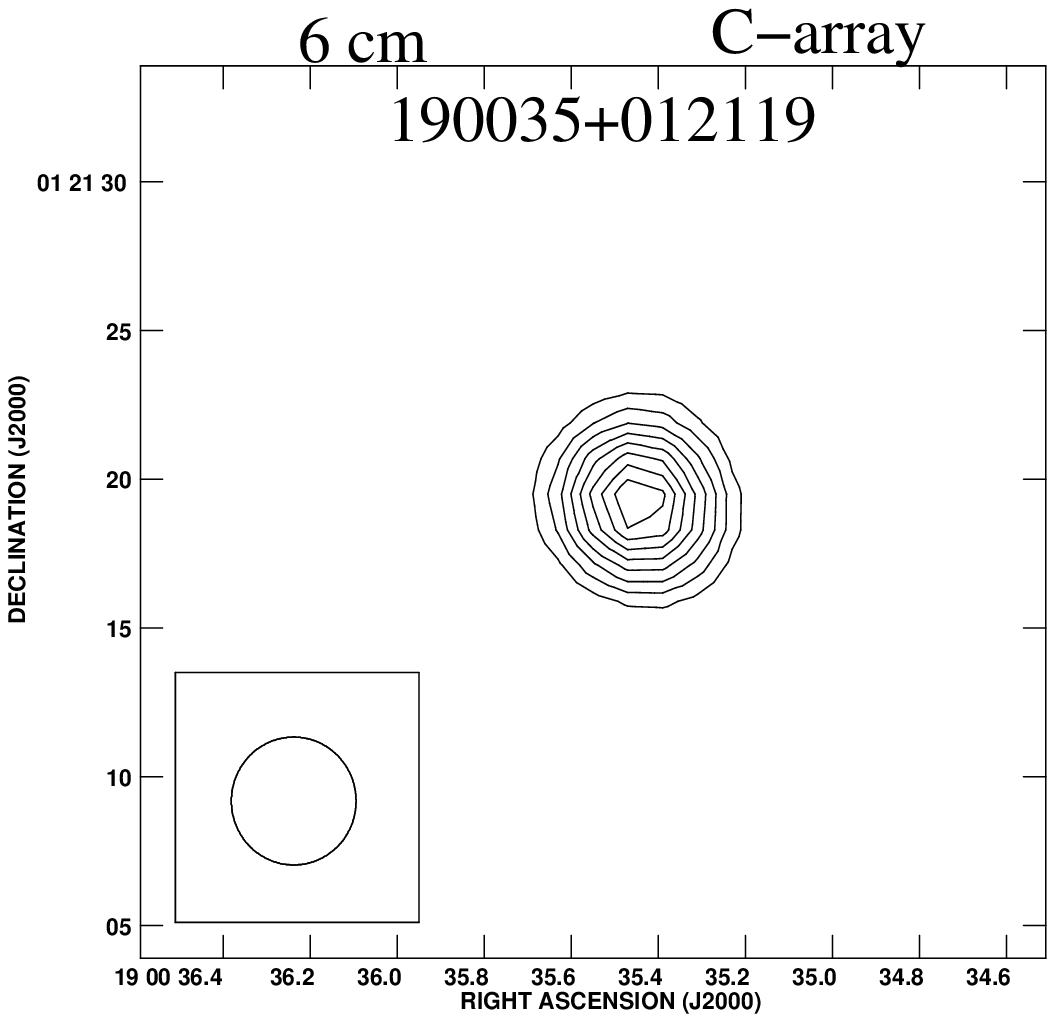}&
\includegraphics[width=5cm]{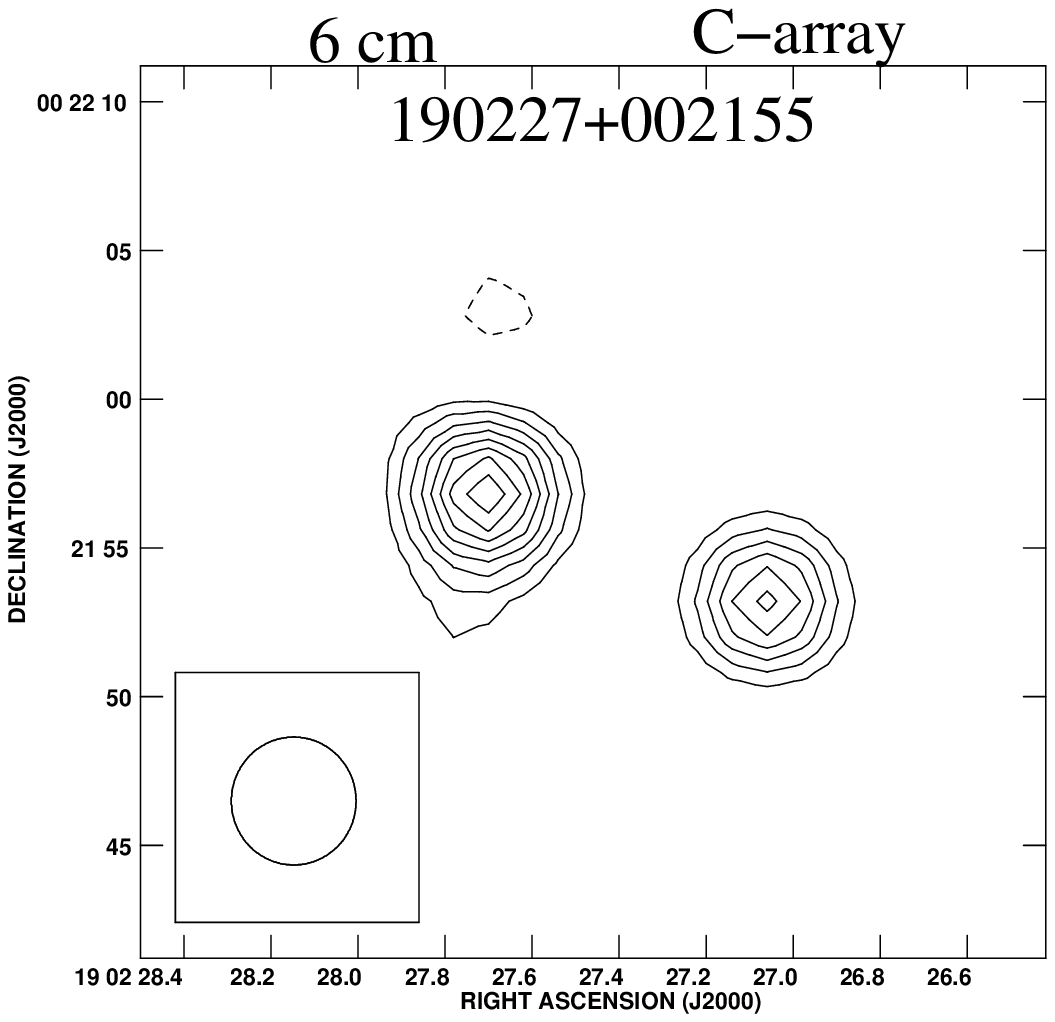}\\
\includegraphics[width=5cm]{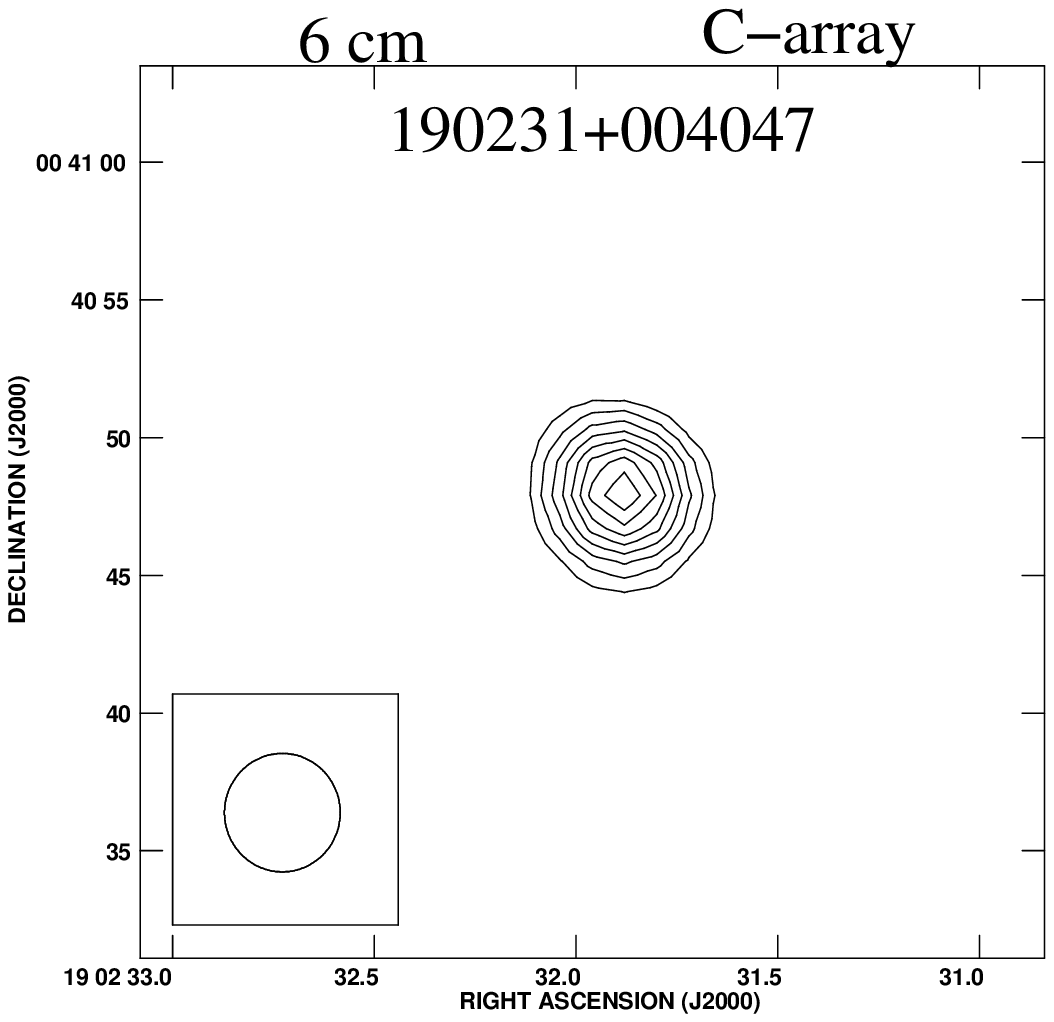} &
\includegraphics[width=5cm]{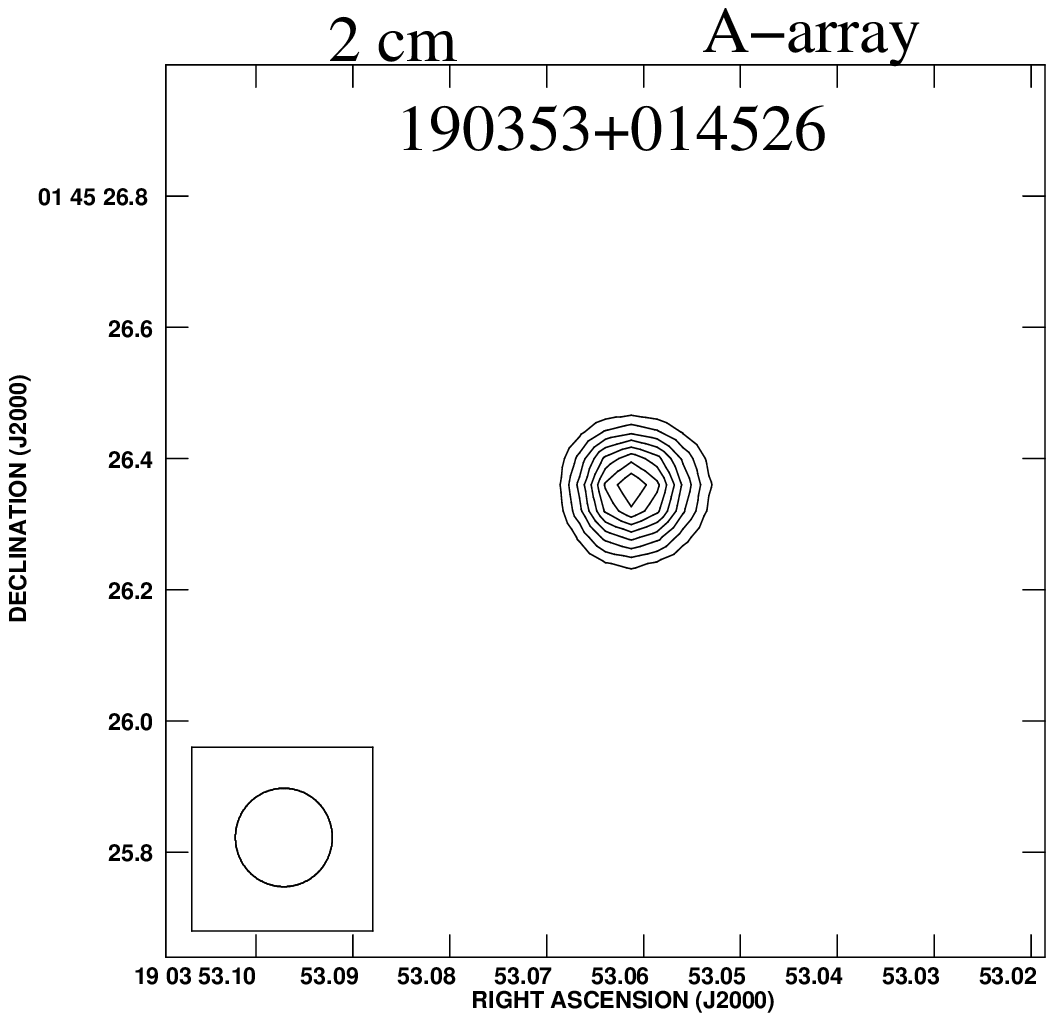}&
\includegraphics[width=5cm]{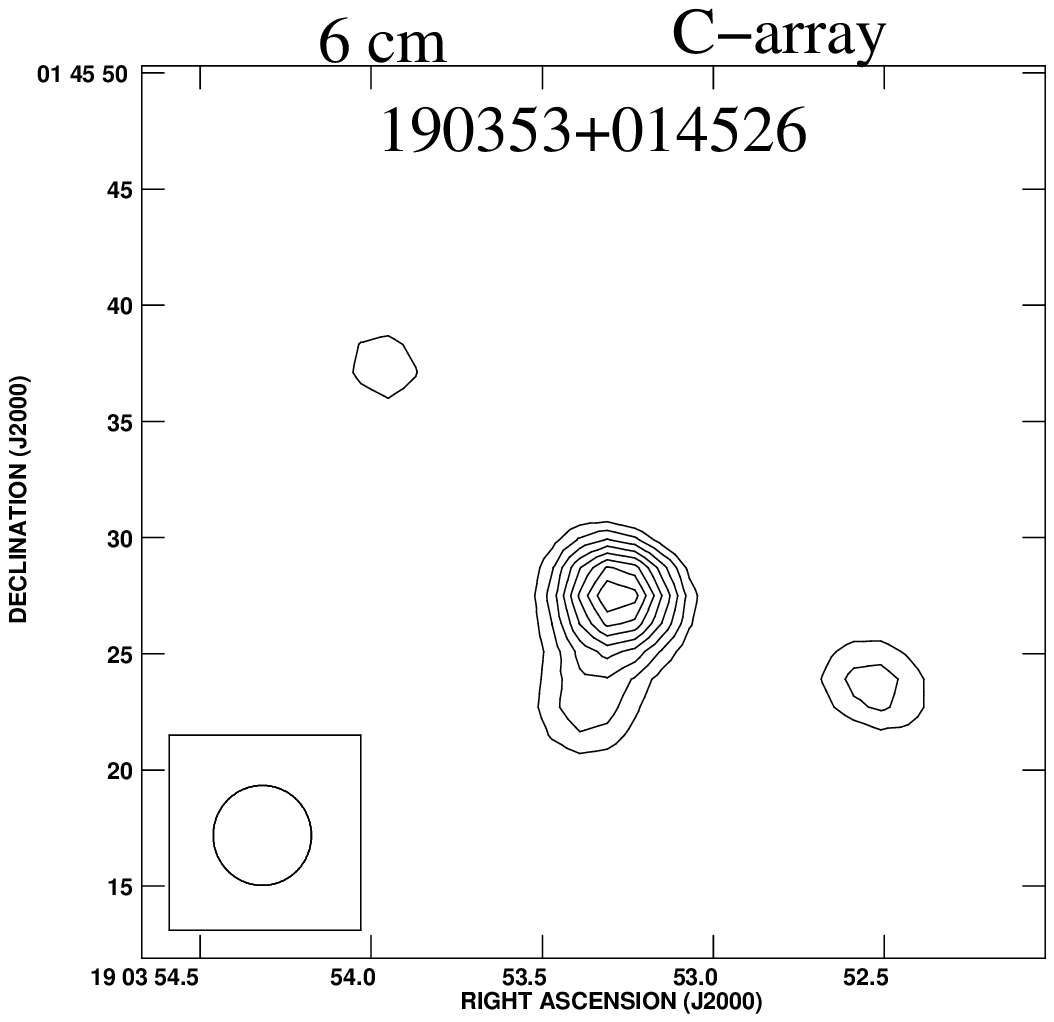}\\
\includegraphics[width=5cm]{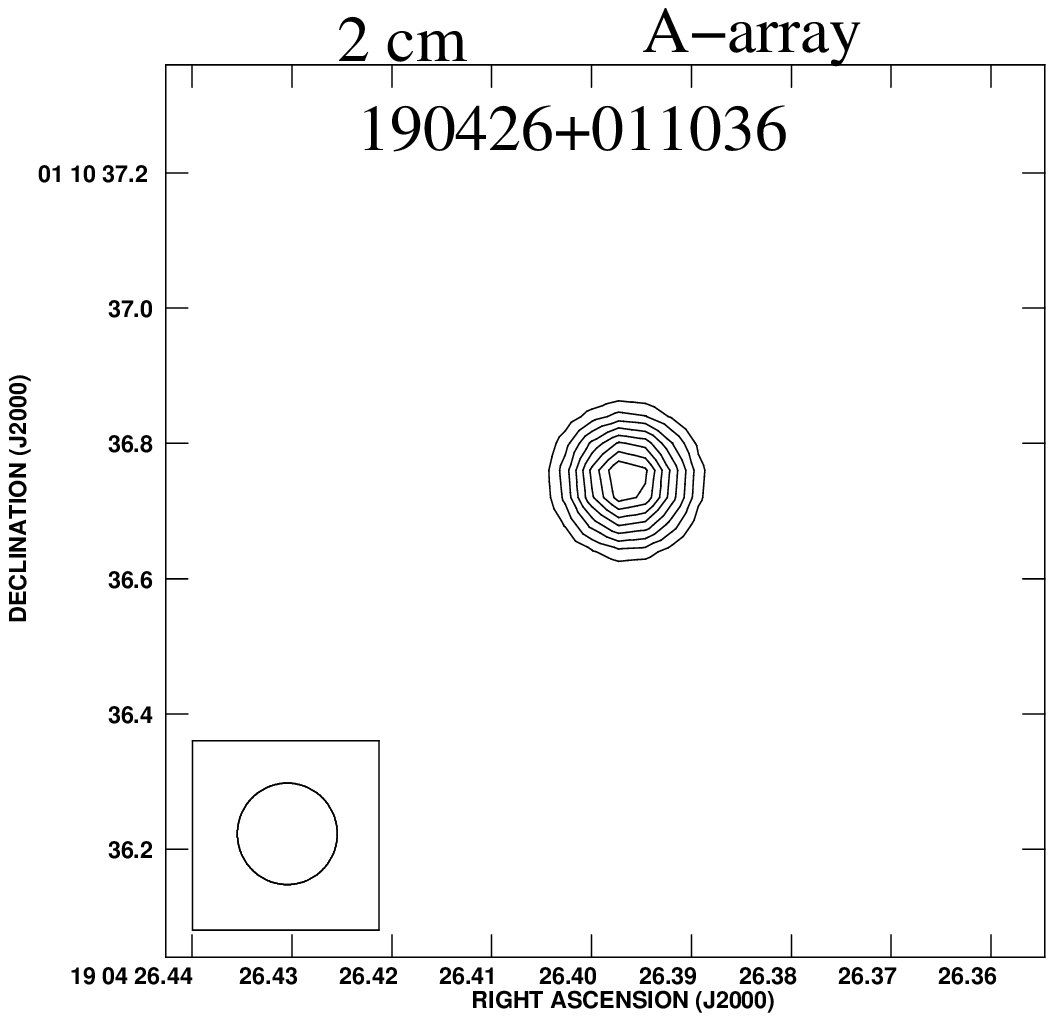} &
\includegraphics[width=5cm]{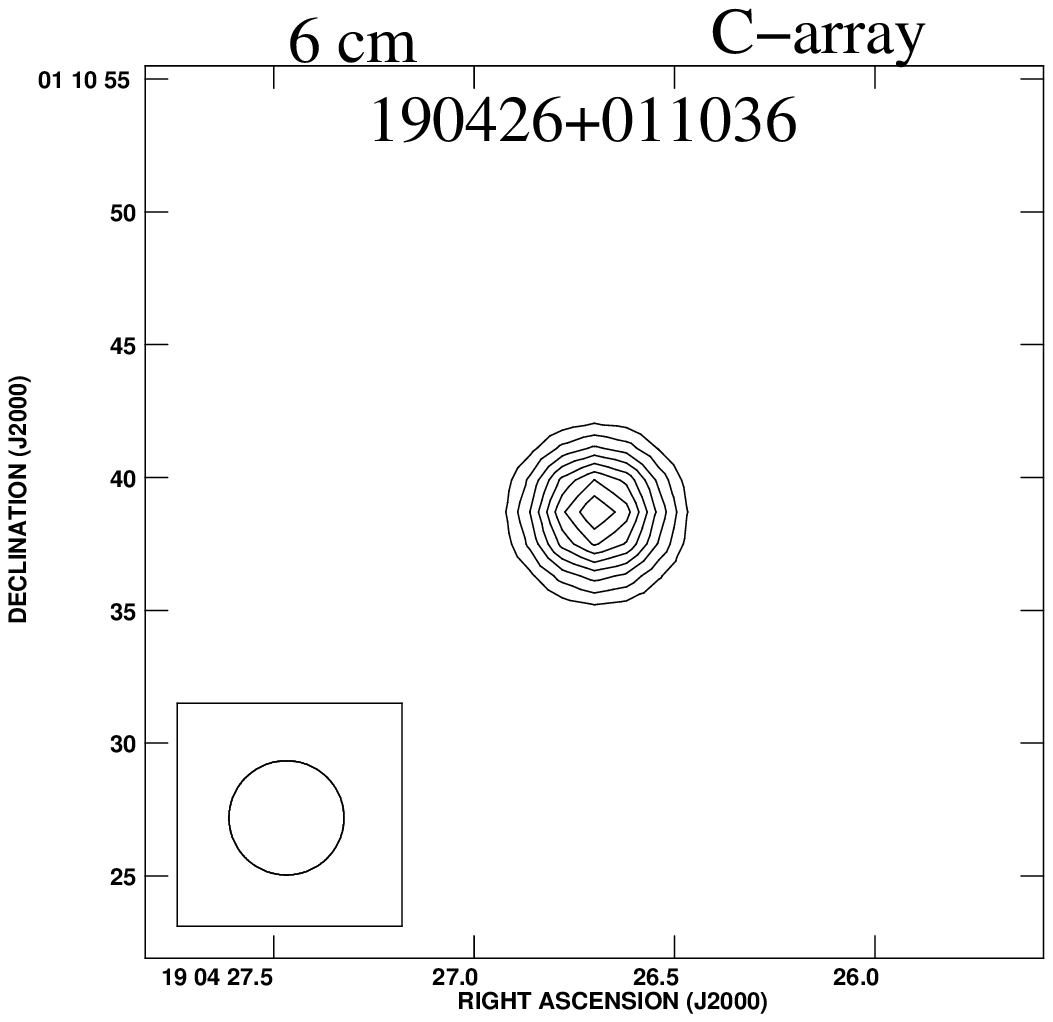}&
\includegraphics[width=5cm]{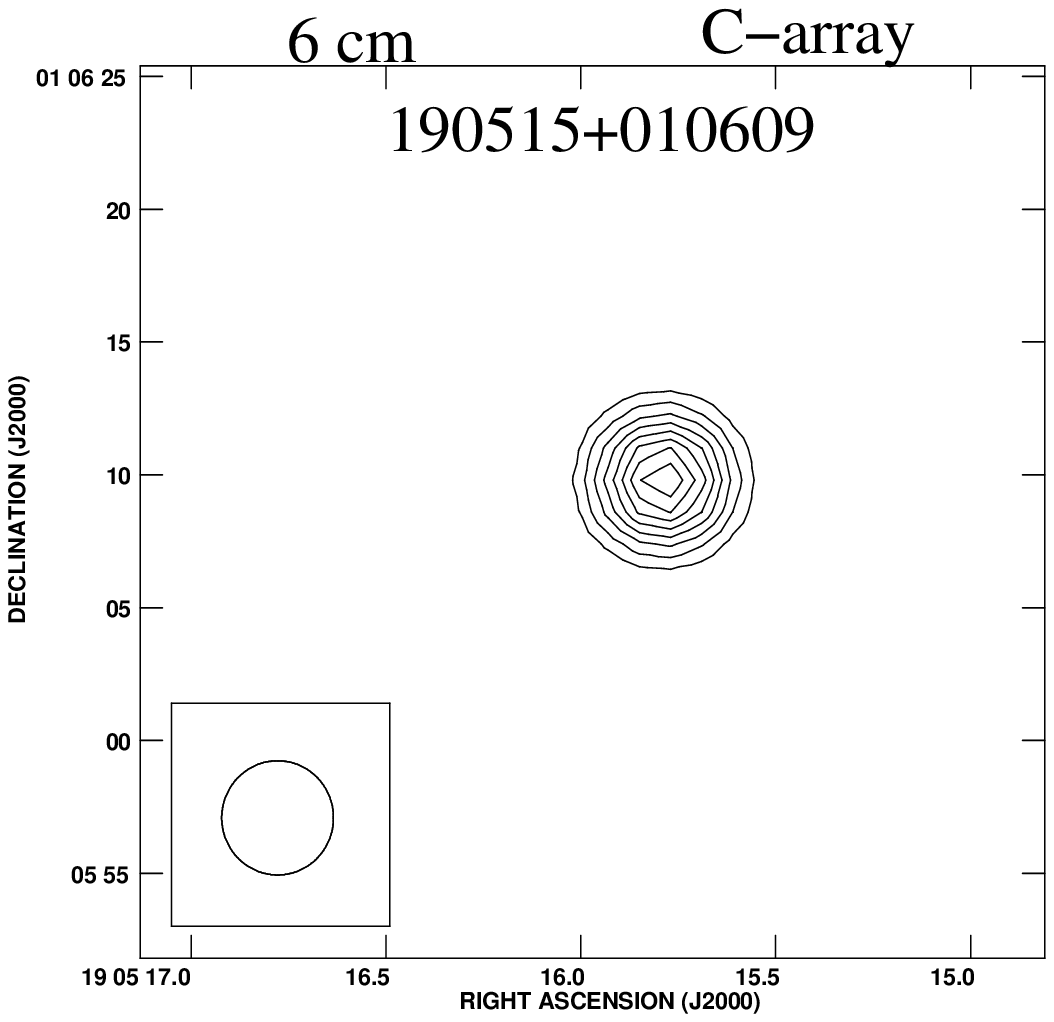}\\
\includegraphics[width=5cm]{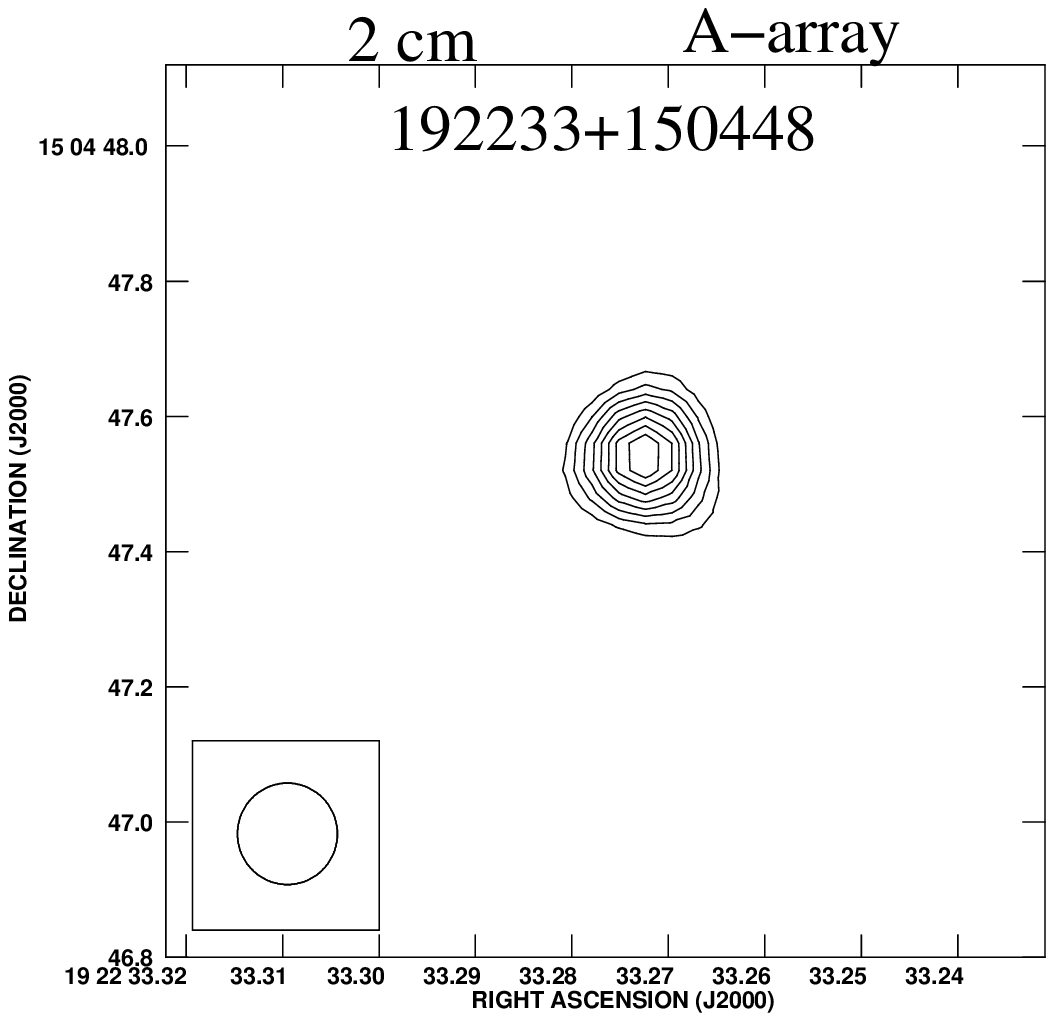} &
\includegraphics[width=5cm]{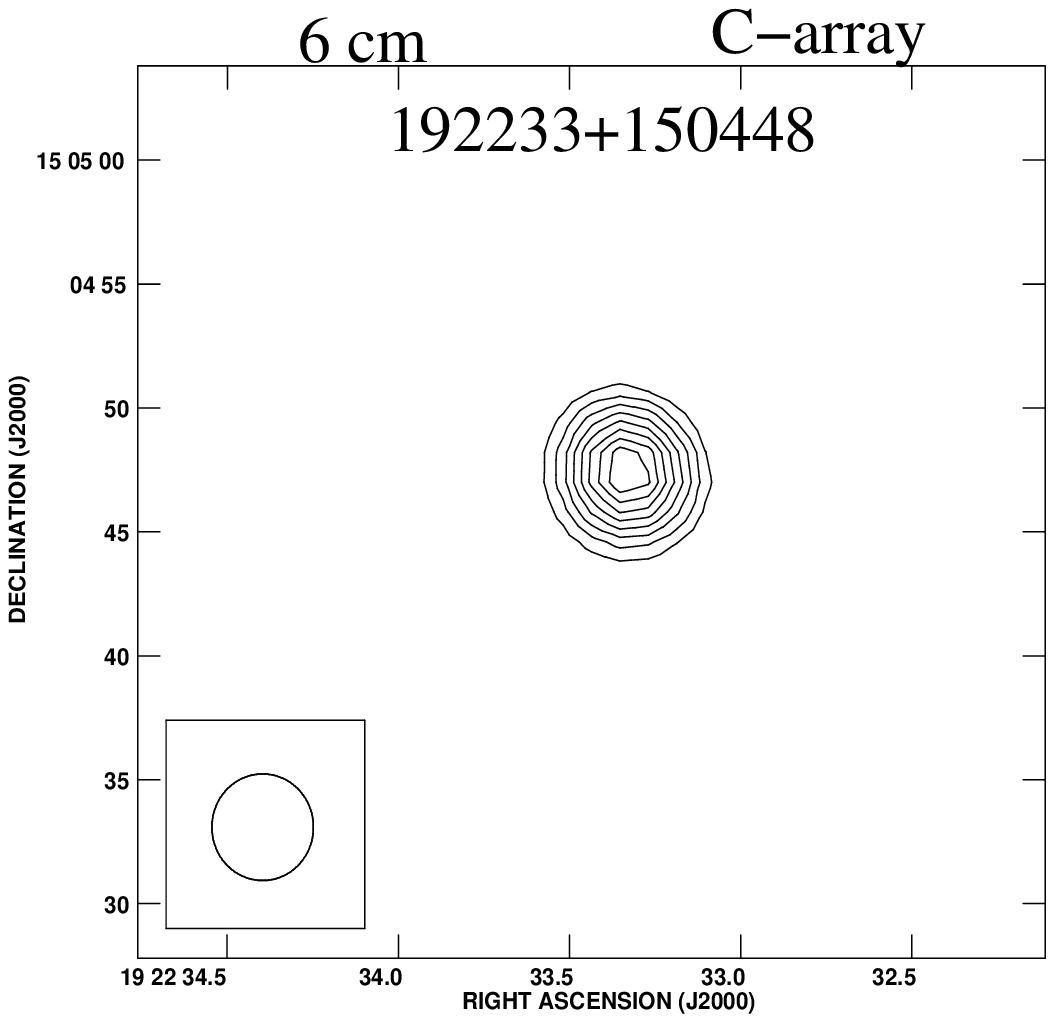}&
\includegraphics[width=5cm]{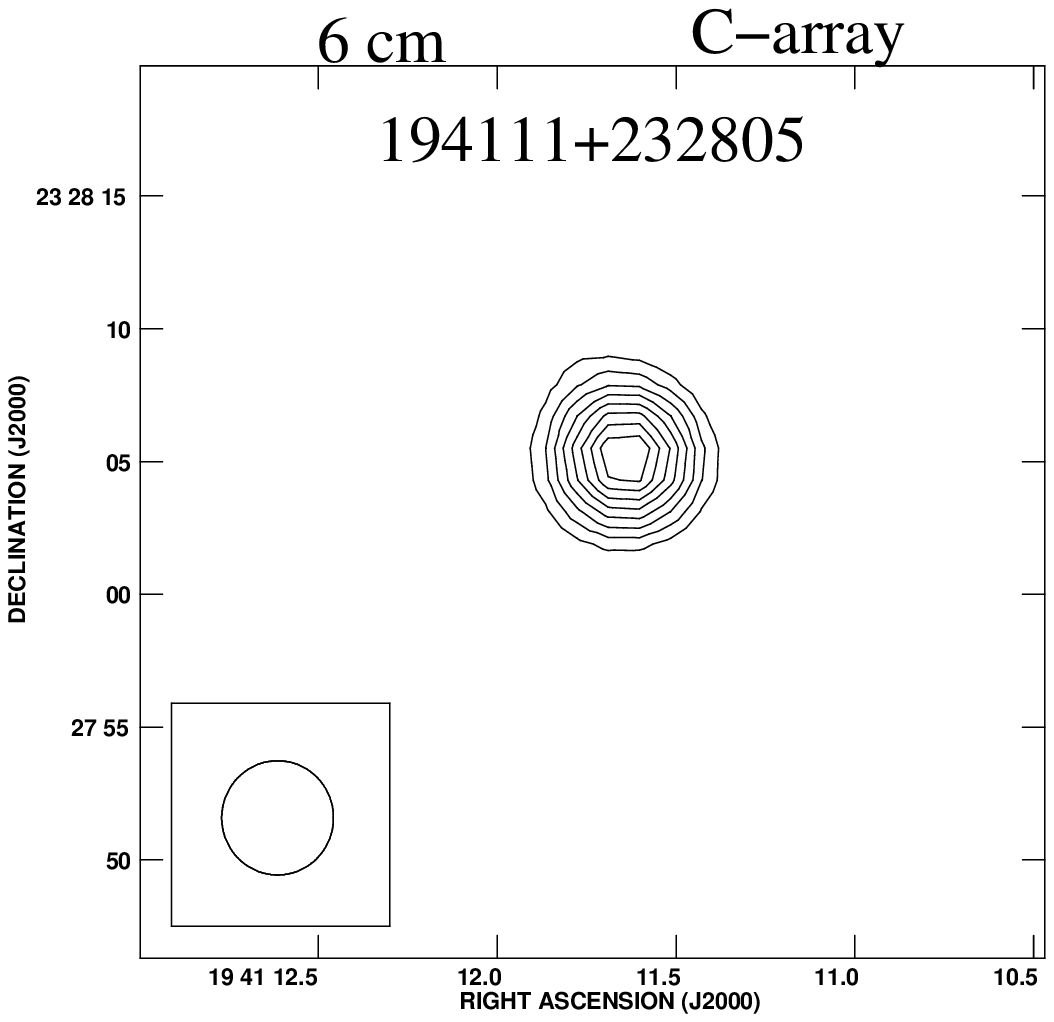}
\\

\end{tabular}
\end{figure*}
\begin{figure*}
\begin{tabular}{ccc}
\includegraphics[width=5cm]{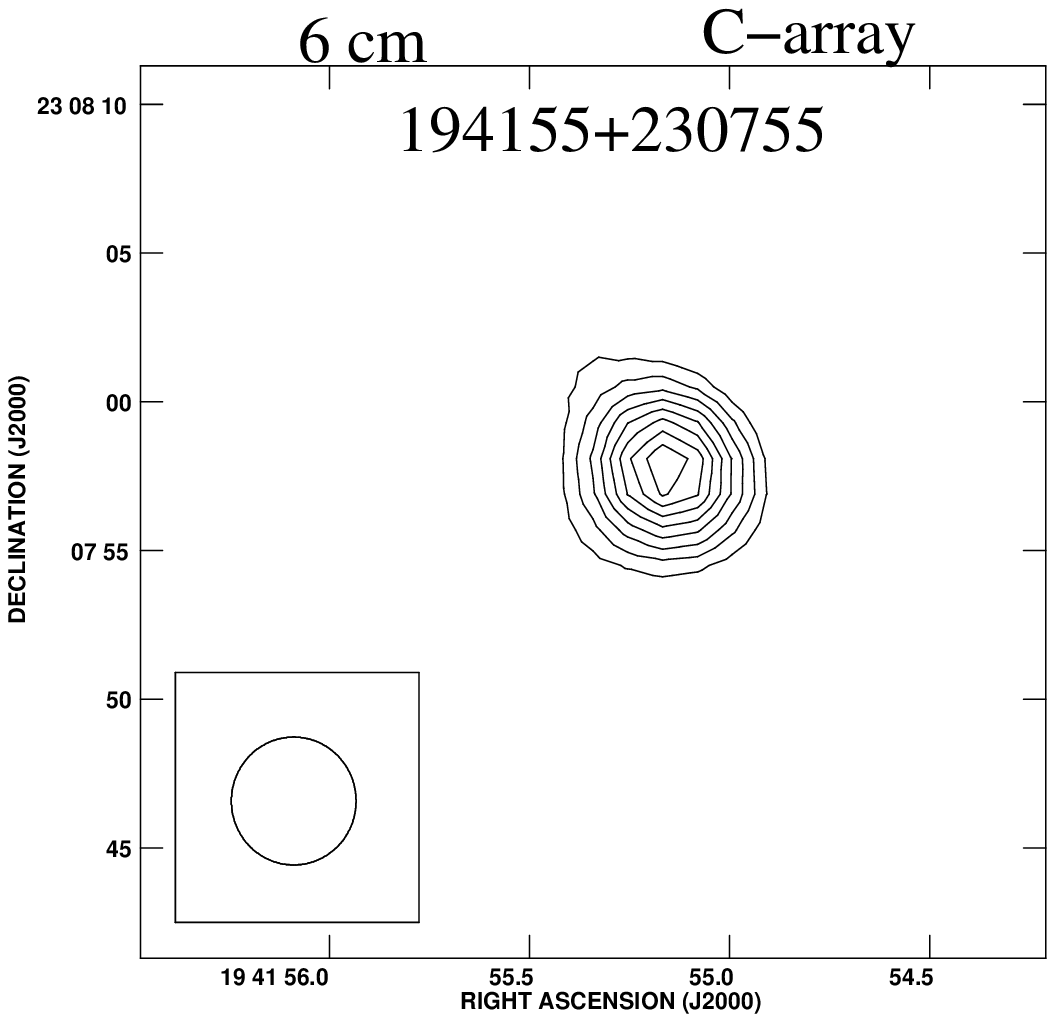} &
\includegraphics[width=5cm]{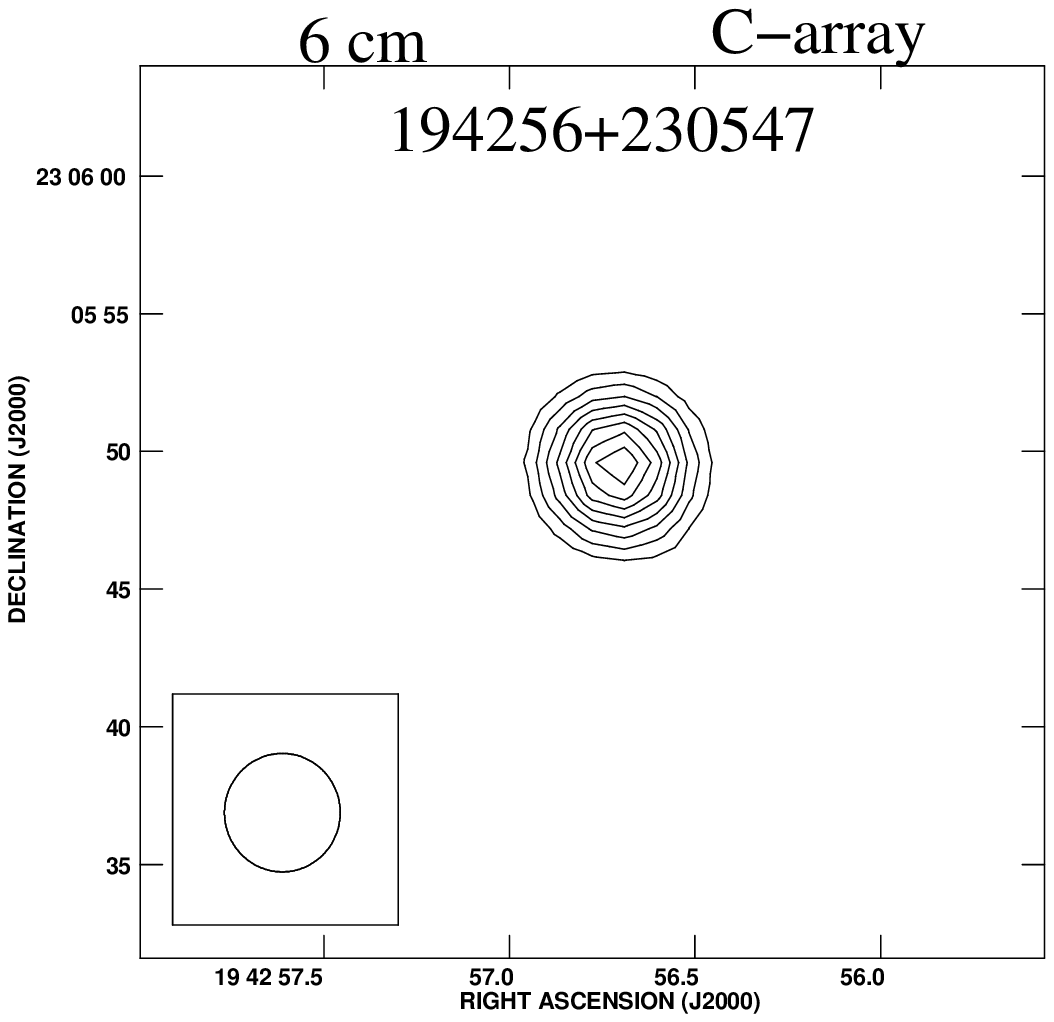}&
\includegraphics[width=5cm]{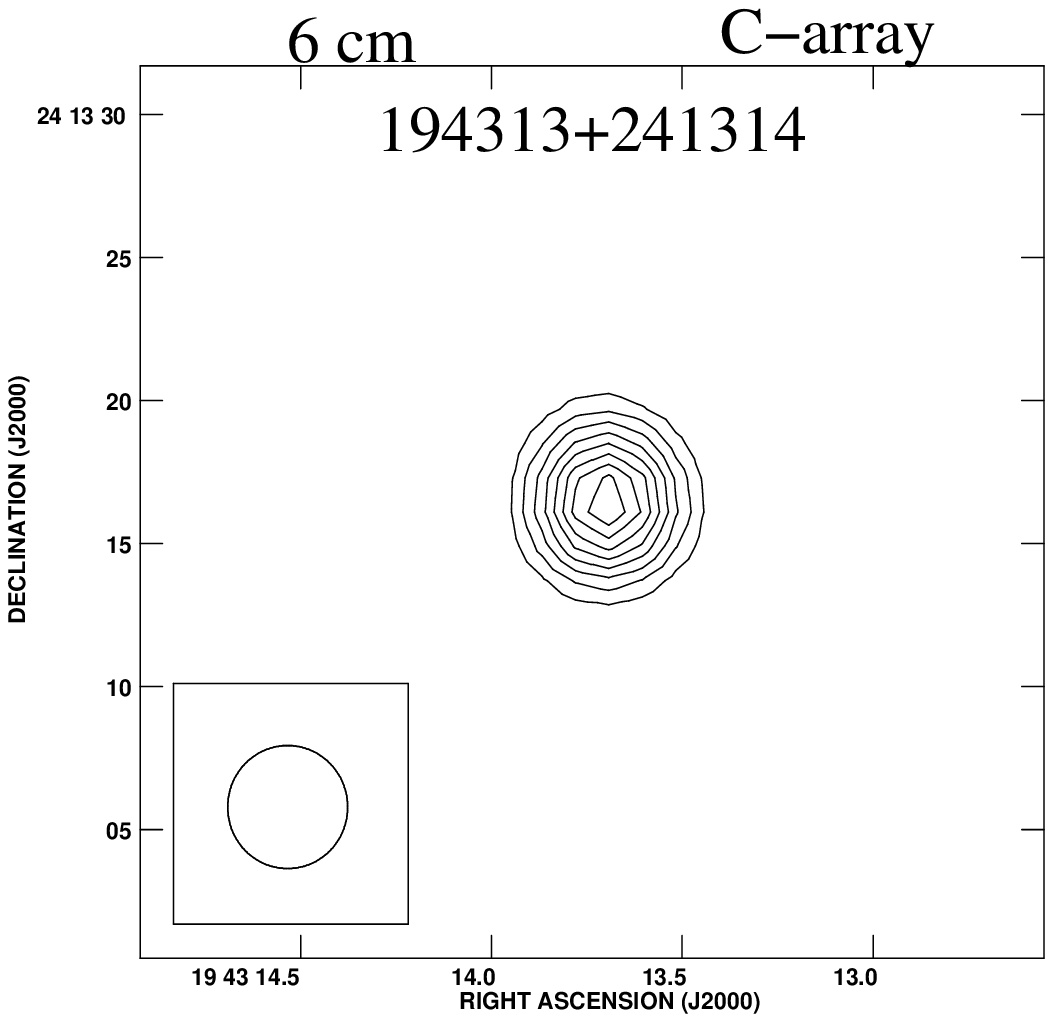}\\
\includegraphics[width=5cm]{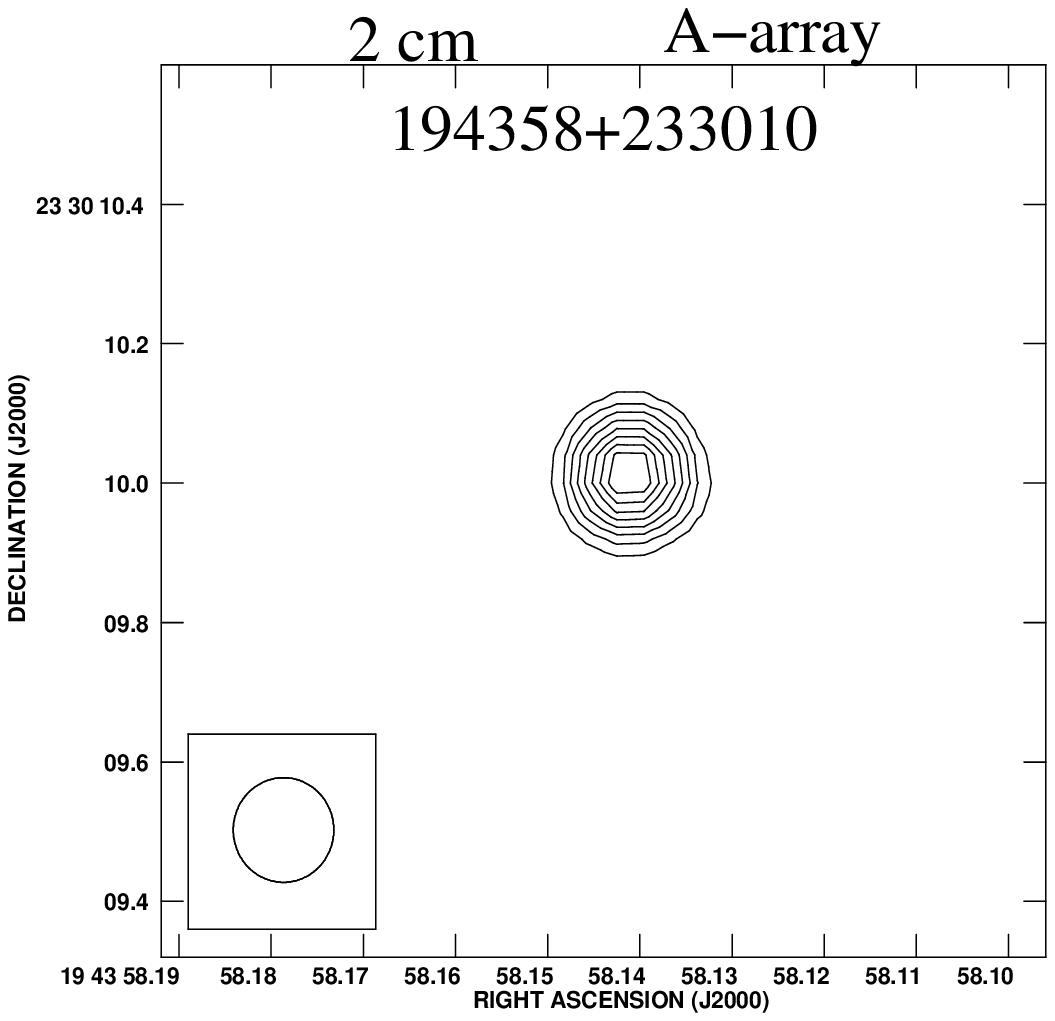} &
\includegraphics[width=5cm]{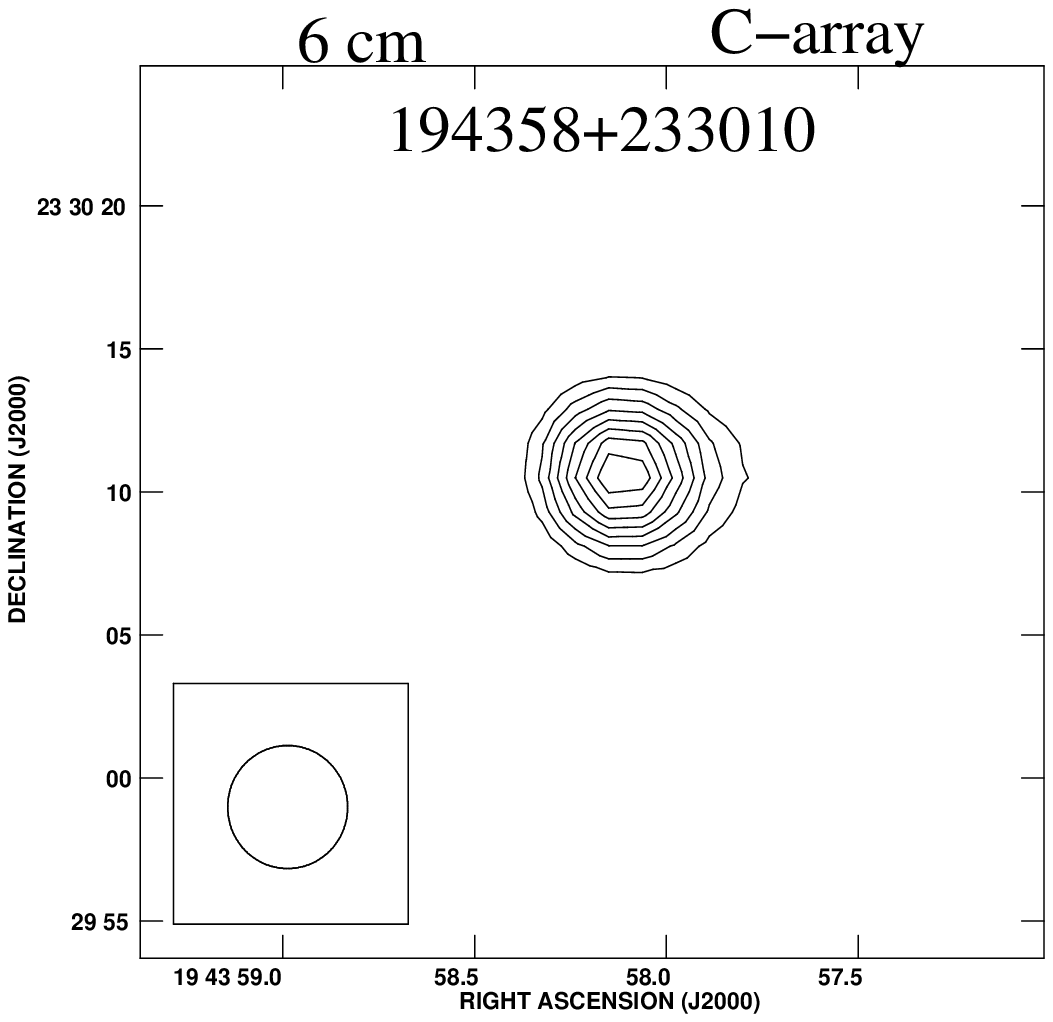}&
\includegraphics[width=5cm]{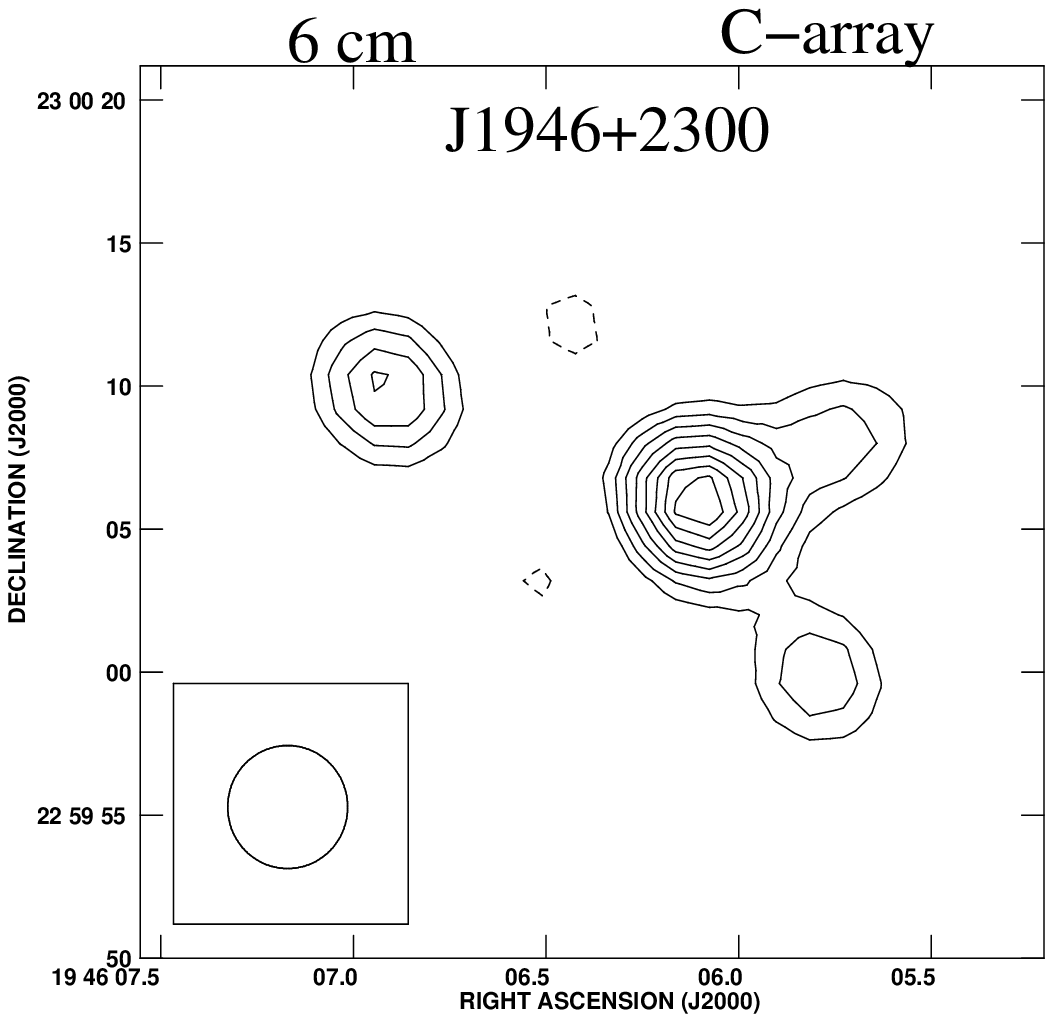}\\
\includegraphics[width=5cm]{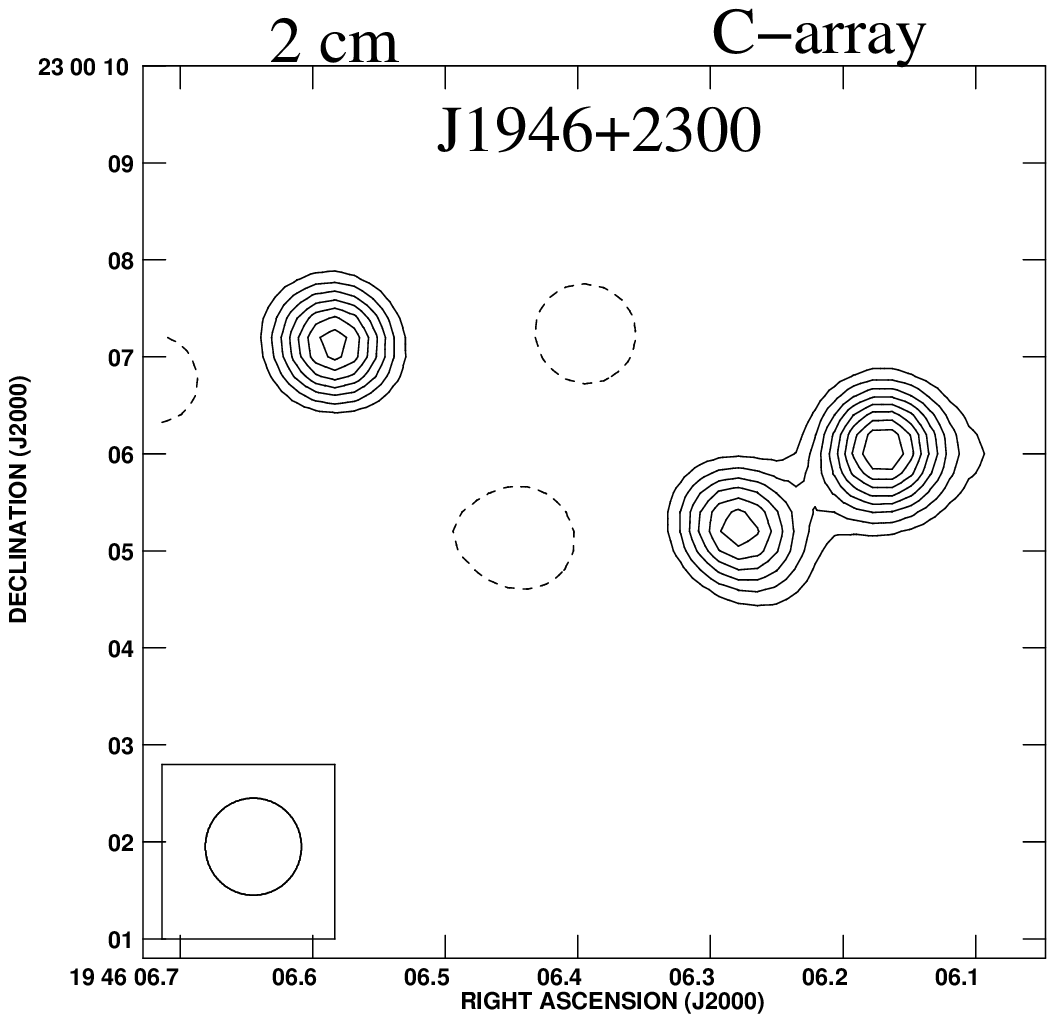} &
\includegraphics[width=5cm]{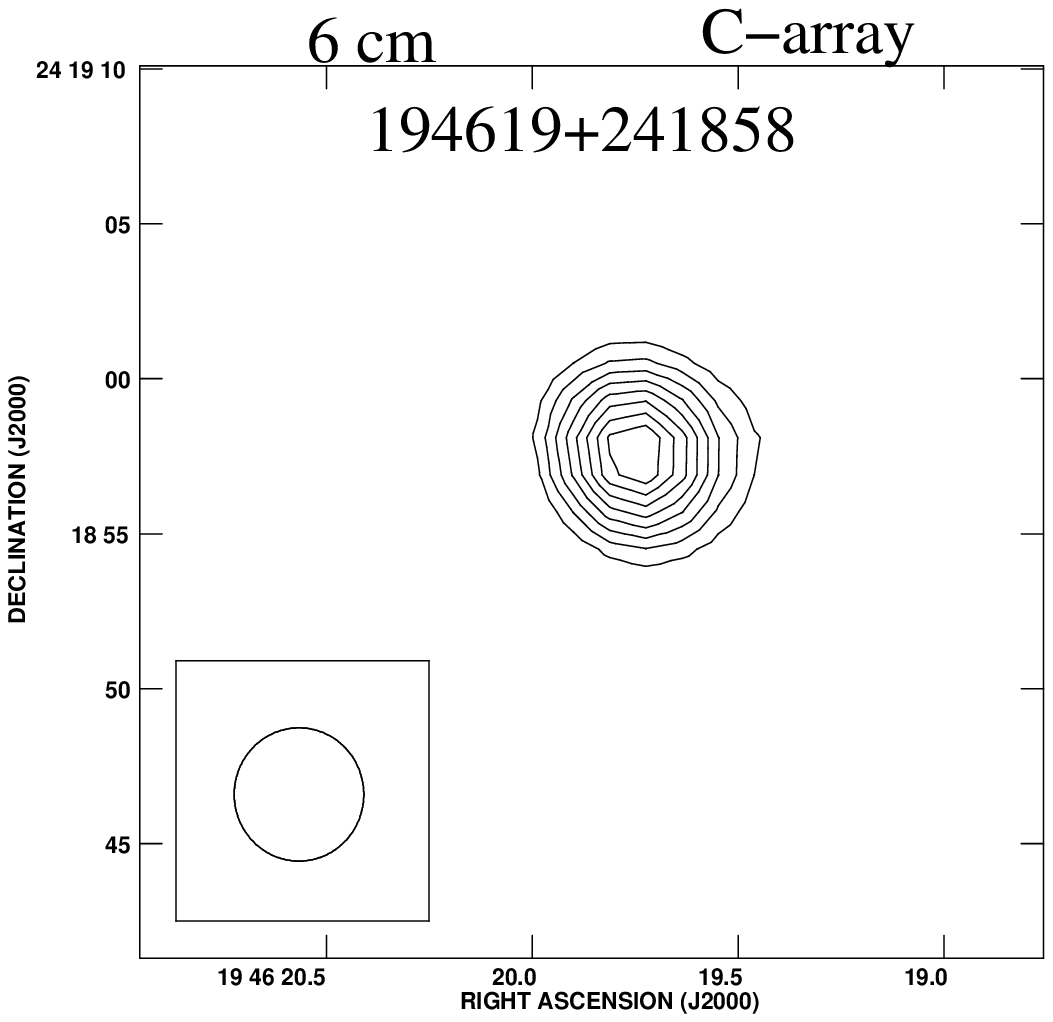}&
\includegraphics[width=5cm]{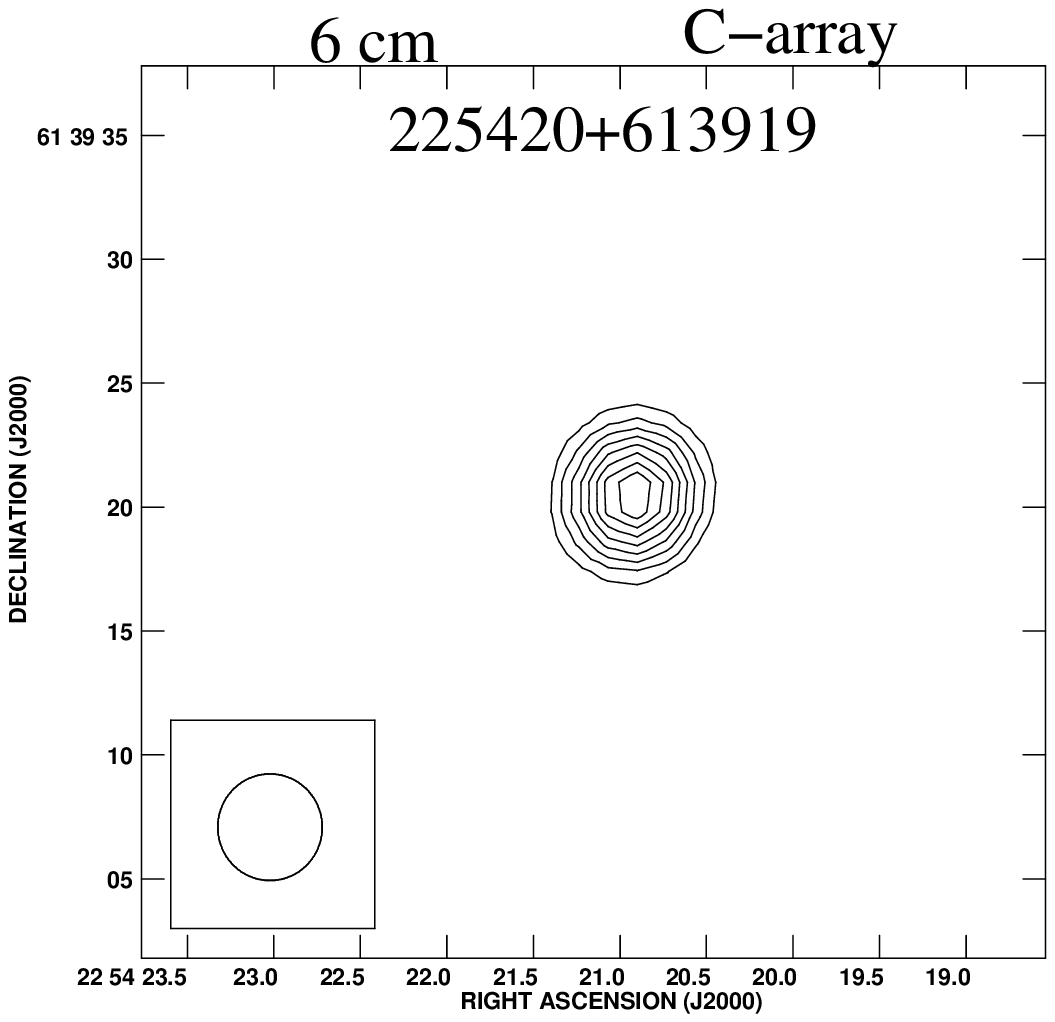}\\
\includegraphics[width=5cm]{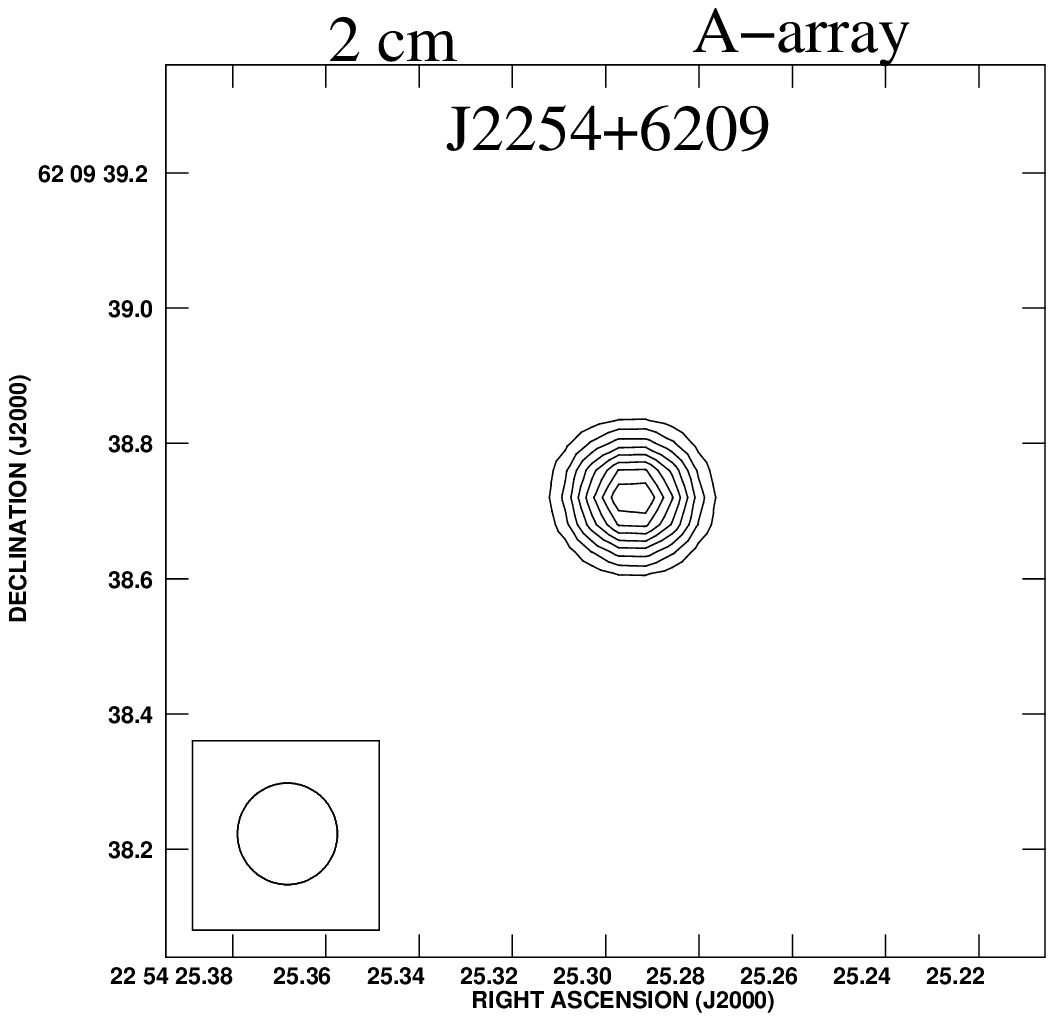} &
\includegraphics[width=5cm]{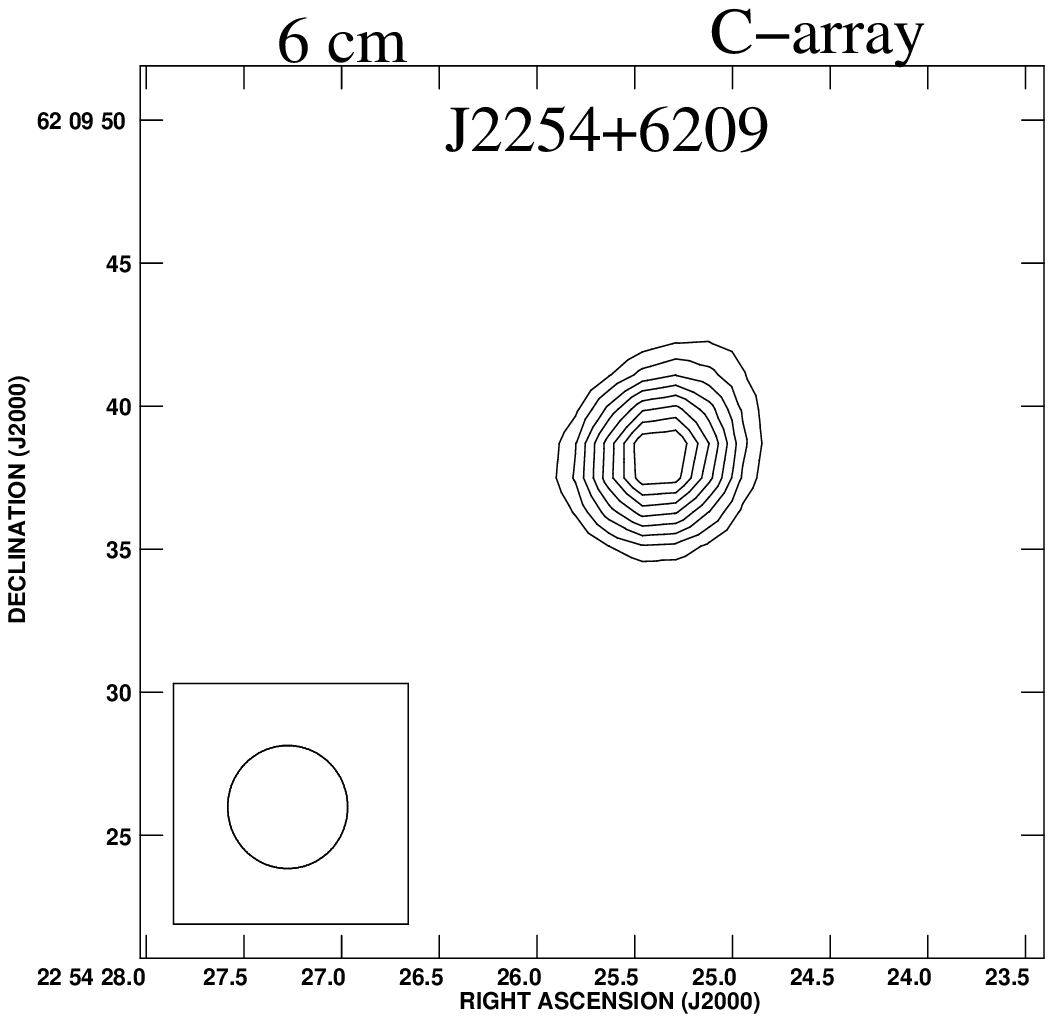}&
\includegraphics[width=5cm]{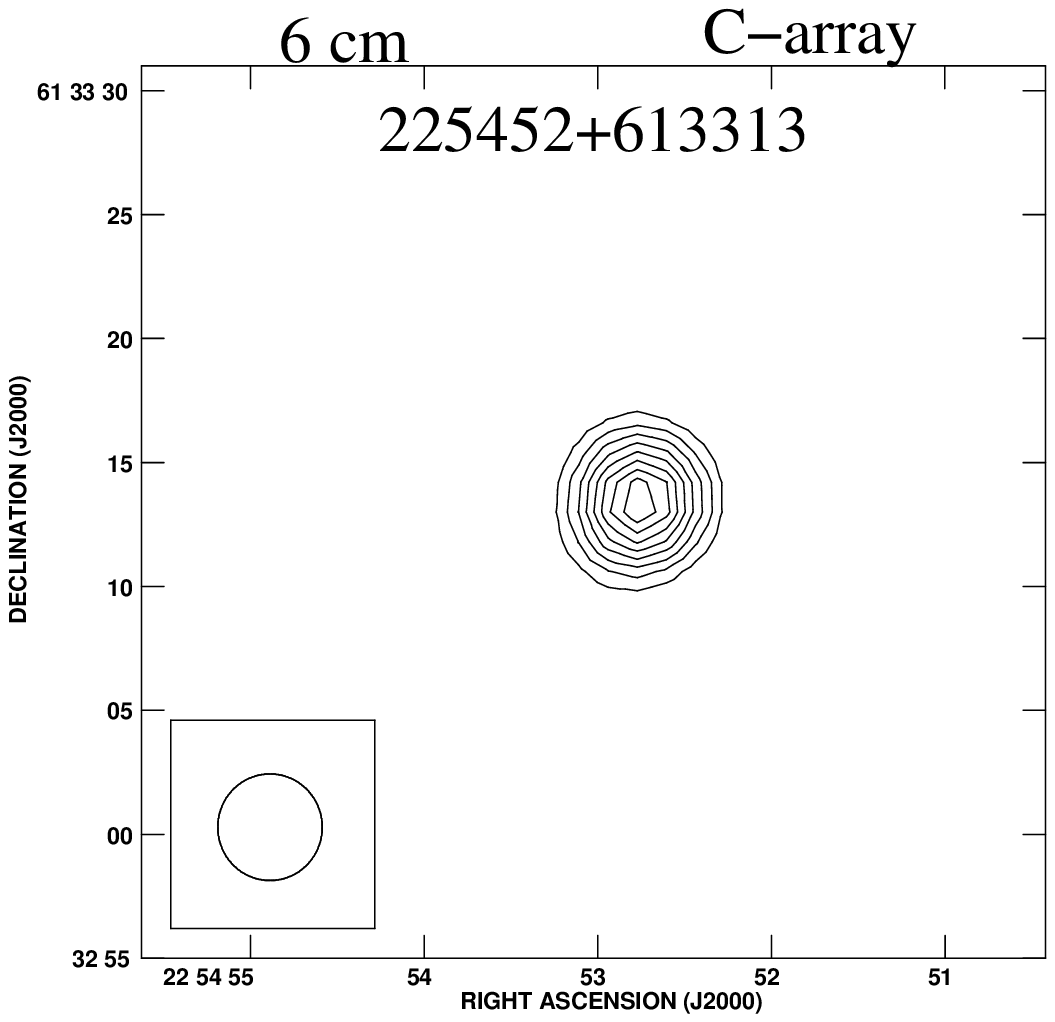}
\\

\end{tabular}
\end{figure*}
\begin{figure*}
\begin{tabular}{ccc}
\includegraphics[width=5cm]{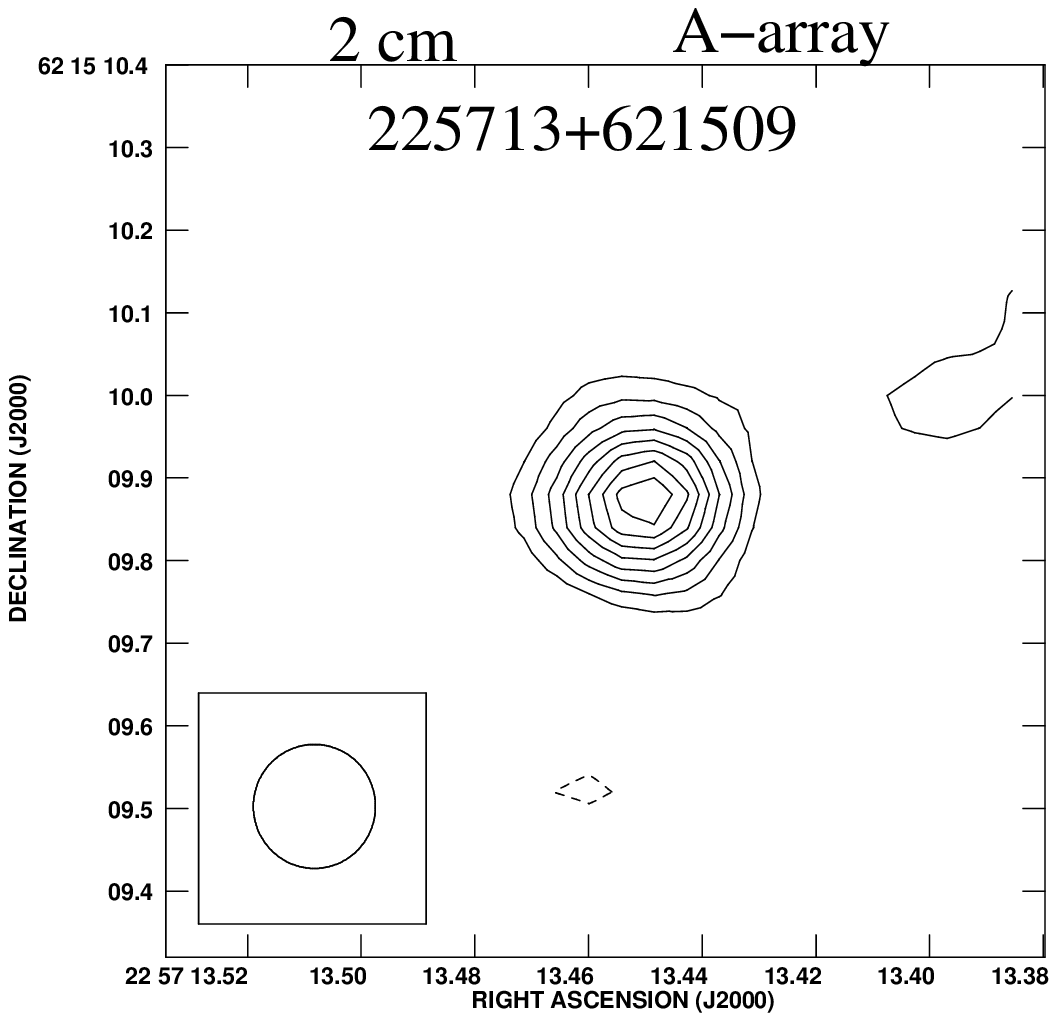} &
\includegraphics[width=5cm]{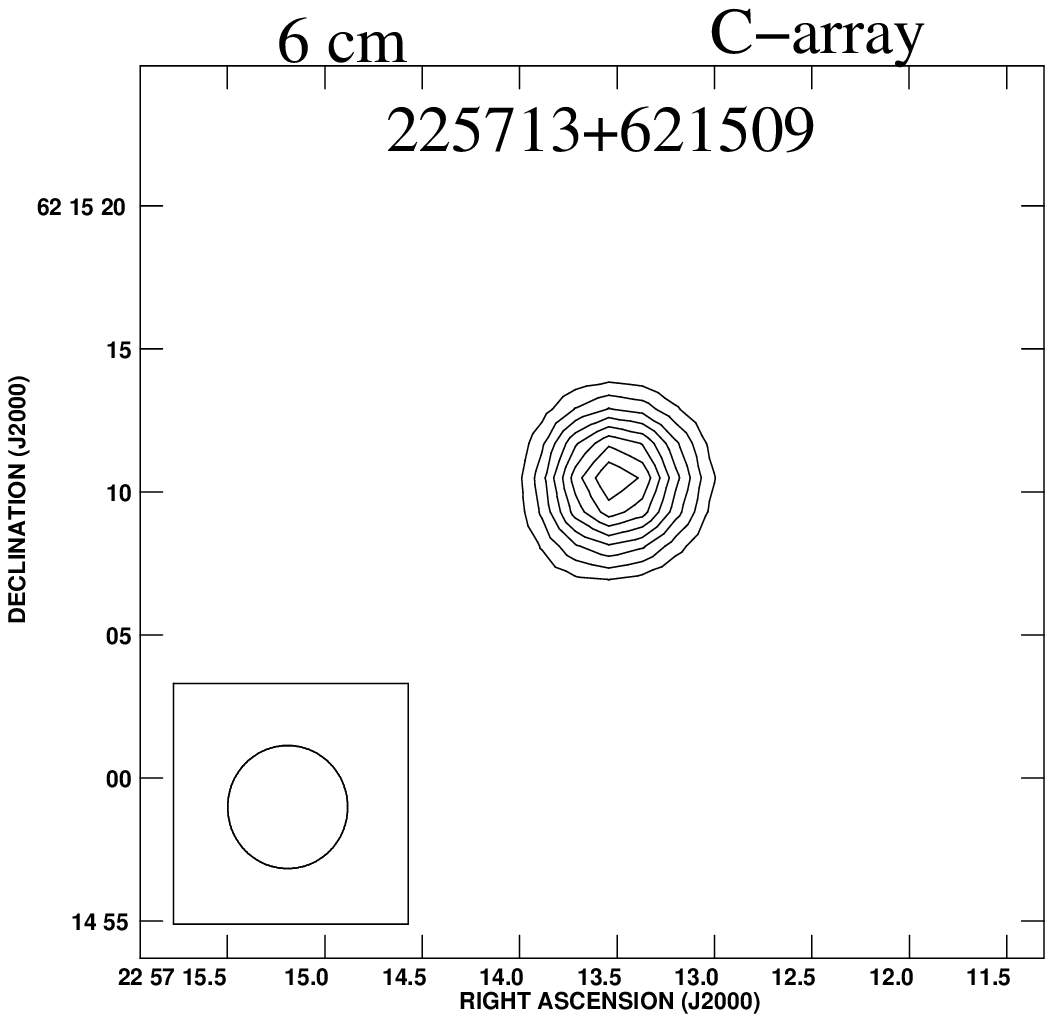} &
\includegraphics[width=5cm]{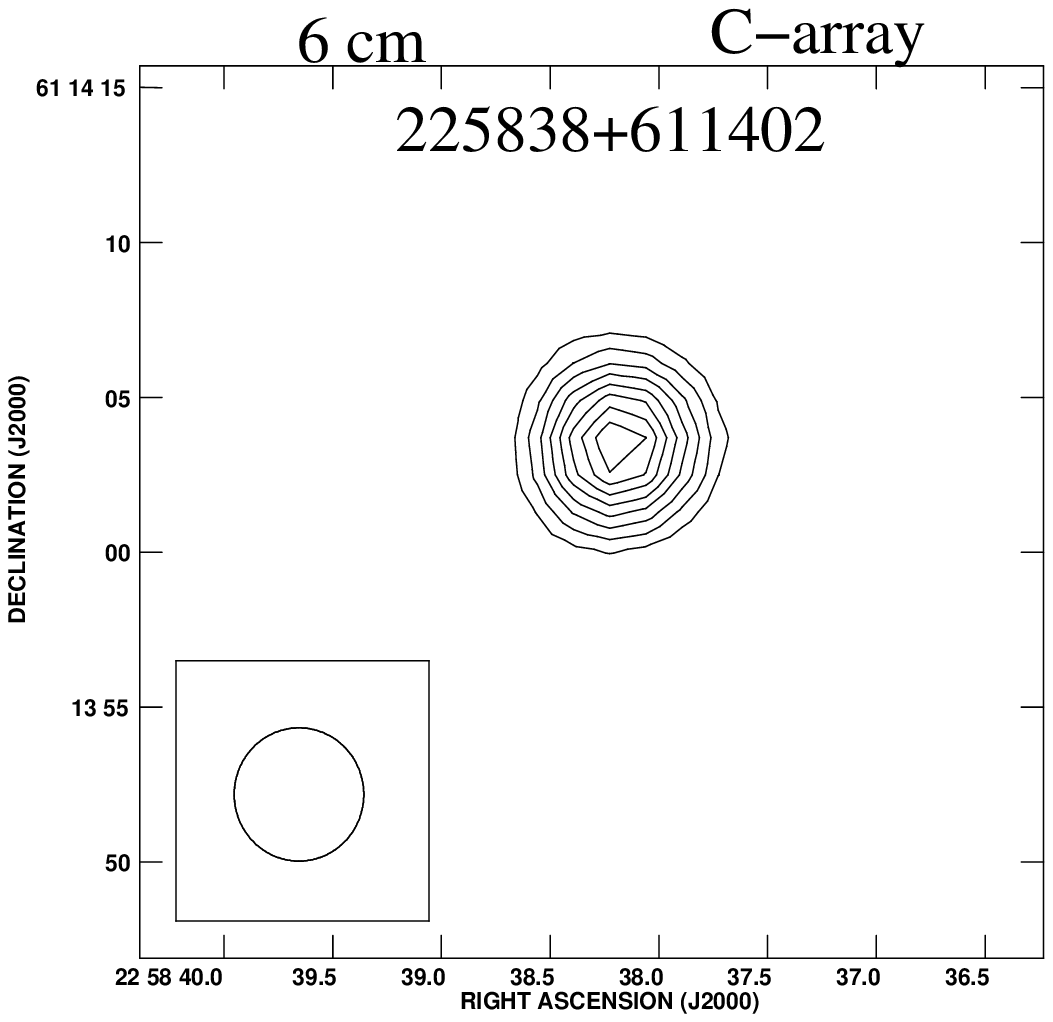}\\
\includegraphics[width=5cm]{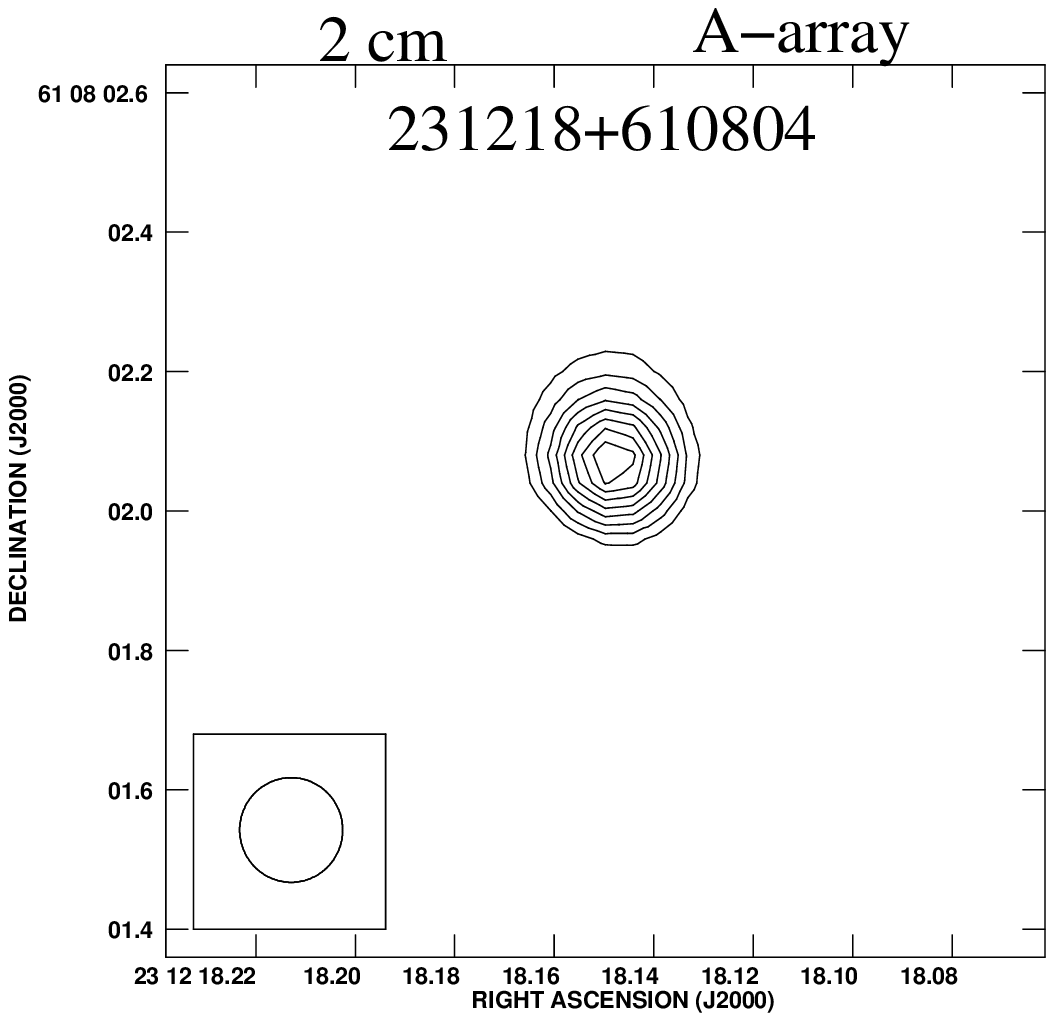} &
\includegraphics[width=5cm]{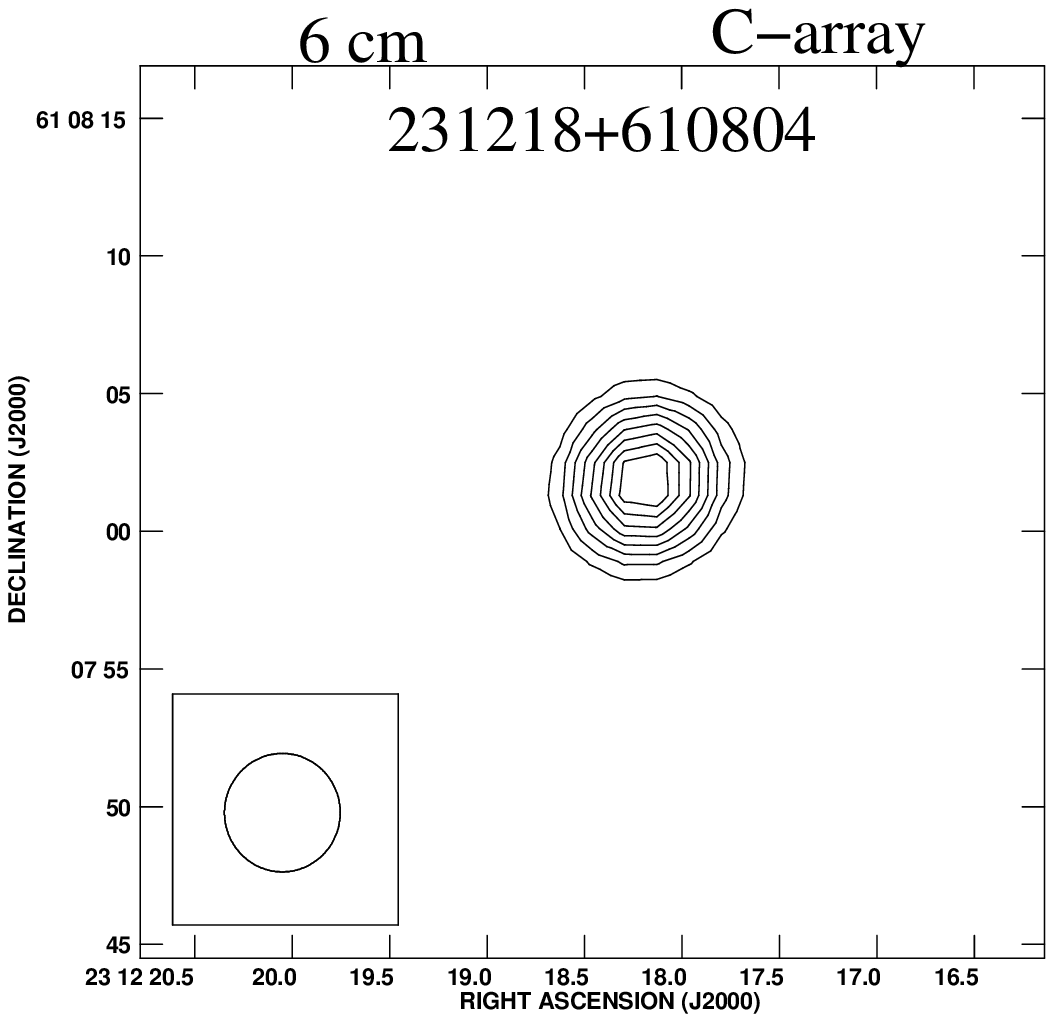} &
\includegraphics[width=5cm]{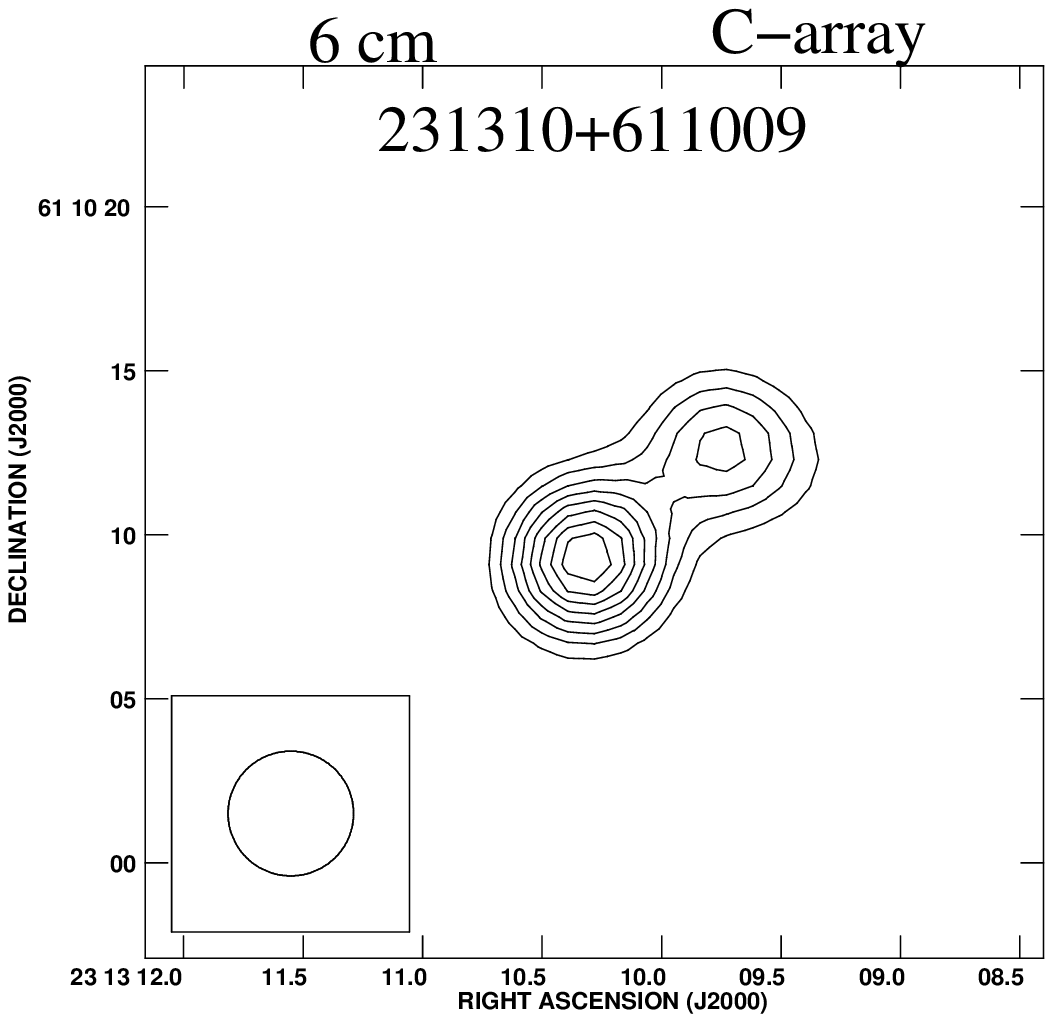}\\
\includegraphics[width=5cm]{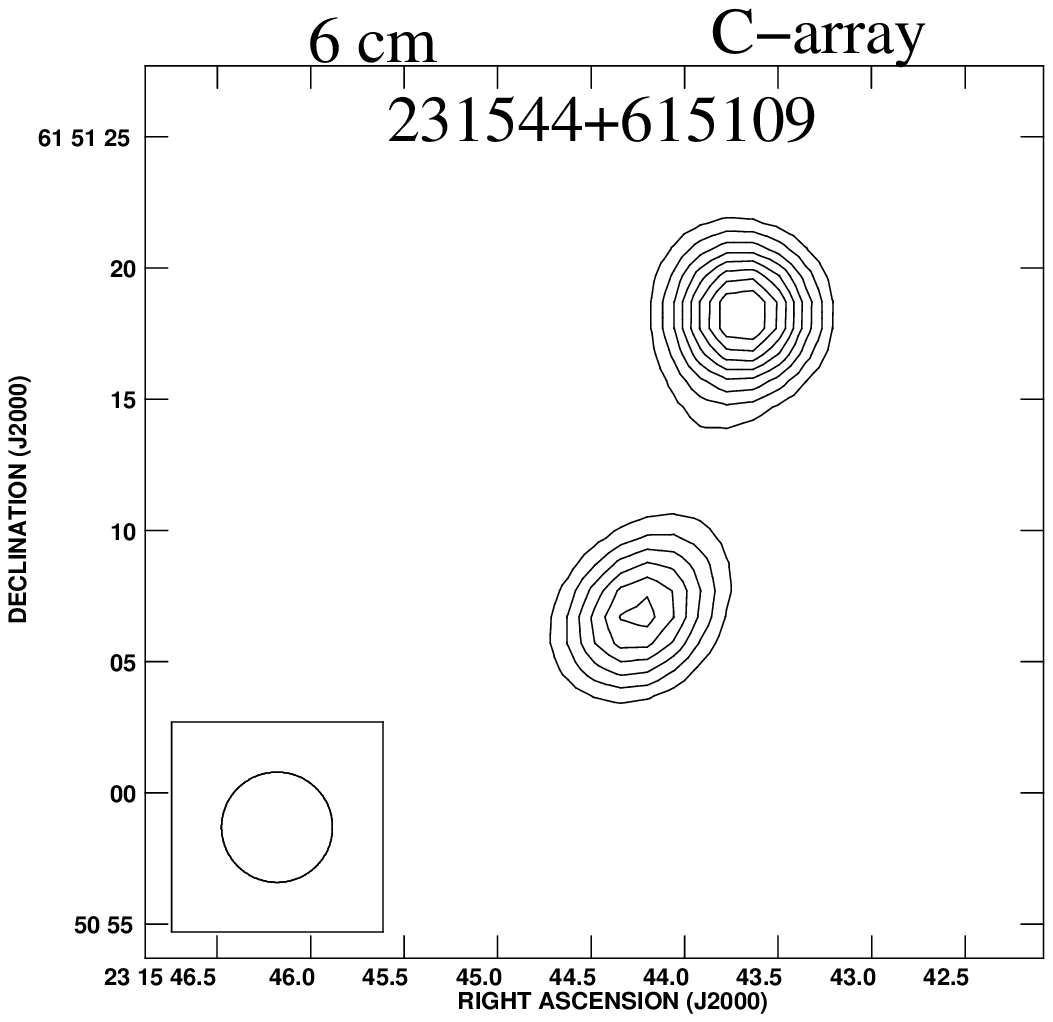} &
\includegraphics[width=5cm]{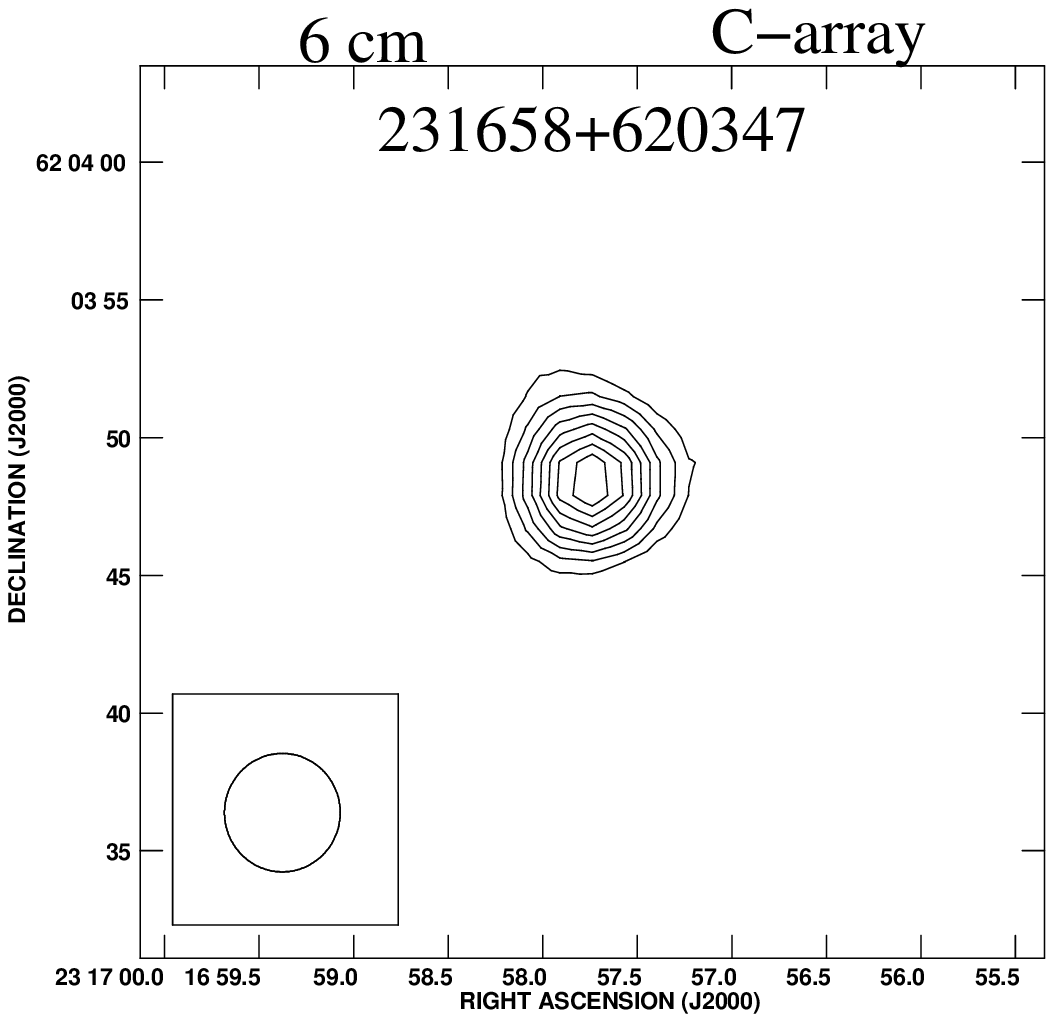} &

\\

\end{tabular}
\caption{Maps of the 2 and 6 cm continuum emission from
extragalactic radio sources with VLA-C, -B, and -A arrays. Contour
levels are -20, and 20 - 90\% in steps of 10\% of the peak
intensity.}
\end{figure*}

\end{document}